\documentclass[traditabstract]{aa} 
\usepackage{graphicx}
\usepackage{psfig}
\usepackage{txfonts}
\usepackage{natbib}
\usepackage[german,english]{babel}
\begin{document}
\title{The fundamental plane of EDisCS galaxies
\thanks{Based on observations collected at the European Southern
Observatory, Paranal and La Silla, Chile, as part of the
ESO LP 166.A-0162.}}
\subtitle{The effect of size evolution}

\author{
R.P. Saglia\inst{1,2},
P. S\'anchez-Bl\'azquez\inst{3,4},
R.~Bender\inst{2,1},
L.~Simard\inst{5},
V.~Desai\inst{6},
A.~Arag{\' o}n-Salamanca\inst{7},
B.~Milvang-Jensen\inst{8},
C.~Halliday\inst{9},
P.~Jablonka\inst{10,11}
S.~Noll\inst{12},
B.~Poggianti\inst{13},
D.~I.~Clowe\inst{14},
G.~De~Lucia\inst{15},
R.~Pell\'o\inst{16},
G.~Rudnick\inst{17},
T.~Valentinuzzi\inst{18},
S.~D.~M.~White\inst{19},
\and
D.~Zaritsky\inst{20}
}

   \offprints{R.P. Saglia}

   \institute{ \selectlanguage{german} Max-Planck Institut f\"ur
     extraterrestrische Physik, Giessenbachstra\ss e, D-85741
     Garching, Germany \\ 
     \email{saglia@mpe.mpg.de} \selectlanguage{english} \and
     Universit\"ats-Sternwarte M\"unchen, Scheinerstr. 1, D-81679
     M\"unchen, Germany 
     \and Departamento de Física Teórica, Universidad Autónoma de
     Madrid, 28049 Madrid, Spain
     \and Departamento de Astrof\'{\i}sica, Universidad de La Laguna,
     E-38205 La Laguna, Tenerife, Spain 
     \and Herzberg Institute of Astrophysics, National Research
     Council of Canada, Victoria, BC V9E 2E7, Canada 
     \and {\it Spitzer} Science Center, Caltech, Pasadena CA91125,
     USA 
     \and School of Physics and Astronomy, University of Nottingham,
     University Park, Nottingham NG7 2RD, United Kingdom 
     \and Dark Cosmology Centre, Niels Bohr Institute, University of
     Copenhagen, Juliane Maries Vej 30, DK-2100 Copenhagen, Denmark 
     \and Osservatorio Astrofisico di Arcetri, Largo Enrico Fermi 5,
     I-50125 Firenze, Italy 
     \and Observatoire de Gen\`eve, Laboratoire d'Astrophysique Ecole
     Polytechnique Federale de Lausanne (EPFL), CH-1290 Sauverny,
     Switzerland 
     \and GEPI, Observatoire de Paris, CNRS UMR 8111, Universit\'e
     Paris Diderot, F-92125, Meudon, Cedex, France 
     \and Institut f\"ur Astro- und Teilchenphysik, Universit\"at
     Innsbruck, Technikerstr.25/8, 6020 Innsbruck, Austria 
     \and Osservatorio Astronomico, vicolo dell'Osservatorio 5,
     I-35122 Padova Italy 
     \and Ohio University, Department of Physics and Astronomy,
     Clippinger Labs 251B, Athens, OH 45701, USA 
     \and INAF, Astronomical Observatory of Trieste, via Tiepolo 11,
     I-34143 Trieste, Italy 
     \and Laboratoire d'Astrophysique de Toulouse-Tarbes, CNRS,
     Universit\'e de Toulouse, 14 Avenue Edouard Belin,
     31400-Toulouse, France 
     \and The University of Kansas, Malott room 1082, 1251 Wescoe Hall
     Drive, Lawrence, KS 66045, USA 
     \and Astronomy Department, University of Padova, Vicolo
     dell'Osservatorio, 3 - 35122 Padova, Italy 
     \selectlanguage{german} \and Max-Planck-Institut f{\"u}r
     Astrophysik, Karl-Schwarzschild-Str. 1, Postfach 1317, D-85741
     Garching, Germany 
     \selectlanguage{english} \and Steward Observatory, University of
     Arizona, 933 North Cherry Avenue, Tucson, AZ 85721 
   }
  
 \authorrunning{R.P. Saglia et al.}

  \date{Received ; accepted }

  \abstract{We study the evolution of spectral early-type galaxies in
    clusters, groups and the field up to redshift 0.9 using the ESO Distant Cluster SUrvey (EDisCS)
    dataset.  We measure structural parameters (circularized
    half-luminosity radii $R_e$, surface brightness $I_e$, and
    velocity dispersions $\sigma$) for 154 cluster and 68 field
    galaxies. On average, we achieve precisions of 10\% in $R_e$, 0.1
    dex in $\log I_e$ and 10\% in $\sigma$. We sample $\approx 20$\%
    of cluster and $\approx 10$\% of field spectral early-type
    galaxies down to an I band magnitude in a 1
arcsec radius aperture of $I_1=22$. We study the
    evolution of the zero point of the fundamental plane (FP) and
    confirm results in the literature, but now also for the low
    cluster velocity dispersion regime. Taken at face value, the
    mass-to-light ratio varies as $\Delta \log
    M/L_B=(-0.54\pm0.01)z=(-1.61\pm0.01)\log(1+z)$ in clusters,
    independent of their velocity dispersion. The evolution is
    stronger ($\Delta \log
    M/L_B=(-0.76\pm0.01)z=(-2.27\pm0.03)\log(1+z)$) for field
    galaxies. A somewhat milder evolution is derived if a correction
    for incompleteness is applied.  A rotation in the FP with
      redshift is detected with low statistical significance.  The
    $\alpha$ and $\beta$ FP coefficients decrease with redshift, or,
    equivalently, the FP residuals correlate with galaxy mass and
    become progressively negative at low masses. The effect is visible
    at $z\ge0.7$ for cluster galaxies and at lower redshifts $z\ge
    0.5$ for field galaxies. We investigate the size evolution of our
    galaxy sample. In agreement with previous results, we find
      that the half-luminosity radius for a galaxy with a dynamical or
      stellar mass of $2\times 10^{11}M_\odot$ varies as 
        $(1+z)^{-1.0\pm0.3}$ for both cluster and field galaxies. At
    the same time, stellar velocity dispersions grow with redshift, as
    $(1+z)^{0.59\pm0.10}$ at constant dynamical mass, and as
    $(1+z)^{0.34\pm0.14}$ at constant stellar mass. The measured size
    evolution reduces to $R_e\propto (1+z)^{-0.5\pm0.2}$ and
      $\sigma\propto (1+z)^{0.41\pm0.08}$, at fixed dynamical masses,
      and $R_e\propto (1+z)^{-0.68\pm0.4}$ and $\sigma\propto
      (1+z)^{0.19\pm0.10}$, at fixed stellar masses, when the
    progenitor bias (PB, galaxies that locally are of spectroscopic
    early-type, but are not very old, disappear progressively from the
    EDisCS high-redshift sample; often these galaxies happen to be
    large in size) is taken into account. Taken together, the
    variations in size and velocity dispersion imply that the
    luminosity evolution with redshift derived from the zero point of
    the FP is somewhat milder than that derived without taking 
these variations 
      into account. When considering dynamical masses, the effects of
    size and velocity dispersion variations almost cancel out. 
For stellar masses, the maximum reduction of the inferred luminosity
evolution is by $-0.38$ units, from $L_B\propto (1+z)^{1.61}$ to
$L_B\propto (1+z)^{1.23}$ for cluster galaxies, and from
$L_B\propto (1+z)^{2.27}$ to $L_B\propto (1+z)^{1.89}$ for
field galaxies. Using simple stellar
    population models to translate the observed luminosity evolution
    into a formation age, we find that massive ($>10^{11}M_\odot$)
    cluster galaxies are old (with a formation redshift
      $z_f>1.5$) and lower mass galaxies are 3-4 Gyr younger, in
    agreement with previous EDisCS results from color and line
    index analyses. This confirms the picture of a progressive
    build-up of the red sequence in clusters with time. Field galaxies follow
    the same trend, but are $\approx 1 Gyr$ younger at a given
    redshift and mass. 
Taking into account the size and velocity dispersion evolution quoted 
above pushes all formation ages upwards by up to 2 Gyr.

}

   \keywords{Galaxies: elliptical and lenticular, cD --
                evolution -- formation -- fundamental parameters}

   \maketitle
\section{Introduction}
\label{sec_Intro}

Despite their apparent simplicity, the physical processes involved in
the formation of early-type galaxies (E/S0) remain unclear.  The
tightness of their scaling relations, such as the color-magnitude
relation, and their slow evolution with redshift, are indicative of a
very early and coordinated formation of their stars
\citep[e.g.,][]{vanDokkum00,Blakeslee03,Menanteau04}. However, in the
$\Lambda$CDM paradigm, these galaxies are expected to form through
mergers of smaller subsystems over a wide redshift range, managing to
obey these constraints \citep{Kauffmann96, Delucia06}.

A particularly interesting relation is that of the fundamental plane
(hereafter FP).  In the parameter space of central velocity dispersion
($\sigma$), galaxy effective radius (R$_{\rm e}$), and effective
surface brightness ($SB_{\rm e}=-2.5 \log I_e$), elliptical galaxies
occupy a plane, known as the FP \citep{Dressler87,
  Djorgovski87}, which exhibits very little scatter ($\sim$0.1
dex). The FP is usually expressed in the form:
\begin{equation}
\log R_{e} =\alpha \log \sigma + \beta SB_e +ZP,
\label{eq_FP}
\end{equation}
where the zero point, hereafter $ZP$, is computed from the mean values 
$\overline{\log R_{e}}$, $\overline{\log \sigma}$, and 
$\overline{SB_{e}}$ of the sample:
\begin{equation}
ZP=\overline{\log R_{e}} -\alpha \overline{\log \sigma} - \beta 
\overline{SB_{e}}.
\label{eq_FPZPint}
\end{equation}

Based on the assumption of homology, the existence of a FP
implies that the ratio of the total mass to luminosity (M/L) scales
with $\sigma$ and R$_{\rm e}$.  Since the galaxy M/L depends on both
the star formation history of the galaxies and the cosmology, the
study of the FP is a valuable tool for studying the evolution of the
stellar population in early-type galaxies.

Several studies of intermediate ($z\sim$0.3) and high-redshift
($z\sim$0.85) clusters of galaxies have used the ZP shift of the plane
to estimate the average formation redshifts of stars in early-type
galaxies \citep[e.g.,][]{Bender98, vanDokkum98a, Jorgensen99,
  Kelson00, vanDokkum03, Wuyts04, Jorgensen06}.  In general, they have
all found values compatible with a redshift formation greater than 3.
In the field, early studies found slow evolution, compatible with that
in clusters \citep[e.g.,][]{vanDokkum01a, Treu01,
  Kochanek00}. However, evidence of more rapid evolution in the field
has been found by other authors \citep{Treu02, Gebhardt03,
  Treu05a}. Taking into account the so-called progenitor bias (for
which lower redshift early-type samples contain galaxies that have
stopped their star formation only recently and that will not be
recognised as early-types at higher redshifts), forces a revision to
slightly lower formation redshifts \citep[$z\approx2$]{vanDokkum01b}.

The current view is that both the evolution of early-type galaxies
with redshift and the dependence of this evolution on environment is
different for galaxies of different mass.  These differences manifest
themselves as an evolution in the FP coefficient $\alpha$ at
increasing redshift, from 1.2 (in the B band) at redshift 0.0 to 0.8
at z$\sim$0.8-1.3 \citep{vanderWel04, Treu05a, Treu05b, vanderWel05,
  Alighieri05, Holden05, Jorgensen06}.  However, this change in the
slope has not been observed at 0.2$<$z$<$0.8
\citep[e.g.,][]{vanDokkum96, Kelson00, Wuyts04, vanderMarel07b,
  MacArthur08}.  If interpreted as a M-M/L ratio relation, this
rotation of the FP indicates that there is a greater
evolution in the luminosity of low-mass galaxies with redshift.  
This interpretation
was however questioned by \citet{vanderMarel07b}.  Dynamical
models provide little evidence of a difference in M/L evolution
between low- and high-mass galaxies, and the steepening of the FP may
be affected by issues other than M/L evolution, such as an increasing
importance of internal galaxy rotation at lower luminosities, not
captured by the simple aperture-corrected velocity dispersion used in
Eq. \ref{eq_FP} \citep{Zaritsky08}, superimposed on the well known
change with redshift in the fraction of S0 galaxies contributing to
the early-type population \citep{Dressler97, Desai07, Just10}.  This
so-called rotation of the FP, or change in the tilt of the FP, was
originally found in field samples, but \citet{Jorgensen06} claimed
that is also exists for cluster galaxies at z=0.89.

Most studies of evolution with redshift in cluster early-type galaxies
have concentrated on single clusters.  It remains unclear whether
early-type galaxies in clusters at the same redshift share the same
FP, or whether the FP coefficients vary systematically as a function
of the global properties of the host cluster (e.g., richness, optical and
X-ray luminosity, velocity dispersions, concentration, and subclustering).  
\citet{Donofrio08} demonstrated that the universality of the
FP has yet to be proven and that to avoid causing any biases by comparing the
FP relation of clusters at different redshifts a larger
number of clusters should be studied.

Furthermore, the ZP evolution of the FP with redshift has been
interpreted as an evolution in the M/L ratio.  However, this may not
be entirely true if there is a structural evolution in the size of the
galaxies.  At face value, observations seem to show that the most
massive ($M_*>10^{11}M_{\sun}$) spheroid-like galaxies at z$>$1.5,
irrespective of their star-formation activity \citep{Perez08} were
much smaller (a factor of $\sim$4) than their local counterparts
\citep{Daddi05, Trujillo06, Trujillo07, Longhetti07, Zirm07, Toft07,
  Cimatti08, vanDokkum08, Buitrago08, Saracco09, Damjanov09,
  Ferreras09}. \citet{vanDokkum10} argue that the growth in size with
decreasing redshift is due to the progressive build-up of the outer
($R>5$ kpc) stellar component of galaxies, while the inner core is
already in place at redshift $\approx 2$. We note also that these
conclusions have been questioned by \citet{Mancini10}, who find
evidence for galaxies as large as local ones at redshifts higher than
1.4. Complementing our discussion above about the evolution of the
zero point of the FP, if galaxy size were to vary with redshift, we should
expect an accompanying partial revision of the importance of the effect 
when taking into account the
progenitor bias \citep{Valentinuzzi10a}. Finally, if a variation in
galaxy size with redshift were to occur, we should expect an
accompanying increase in the central velocity dispersion with redshift
\citep{Cenarro09, vanDokkum09}. Both the evolution
in size and velocity dispersion are predicted by theoretical models
that take into account internal feedback 'puffing' mechanisms
\citep{Biermann79, Fan08} or the effect of merging \citep{Khochfar06,
  Hopkins09}.  As one can read from Eq. \ref{eq_FPZPint}, a change in
$\overline{\log R_{e}}$, $\overline{\log \sigma}$, and
$\overline{SB_{e}}$ with redshift due to structural evolution will
change the amount of stellar population evolution needed to explain
the ZP variation and therefore needs to be taken into account when
deriving constraints on the formation epoch of early-type galaxies.

In this paper, we present the evolution of the FP in a
sample of 154 spectral early-type galaxies in 28 clusters or groups and 62 in
the field using spectra and images from the ESO Distant Cluster Survey
of galaxies \citep[EDisCS]{White05}.  The clusters have redshifts
between $\sim$0.4 and 0.9 and velocity dispersions between 166 and
1080 km/s \citep{Halliday04, Clowe06, Milvang08}.  Our clusters have
generally lower velocity dispersions than those typically studied at
similar redshifts and represent an intermediate-redshift sample for
which a majority of the clusters may be progenitors of typical
low-redshift clusters \citep[see][]{Poggianti06, Milvang08}.

The paper is organized as follows. Section \ref{sec_data} presents the
data set.  In particular, Sect. \ref{sec_sigma} describes the
measurements of the galaxy velocity dispersions. Section
\ref{sec_photometry} describes the measurement of the structural
parameters, their errors, and the photometric calibration. Section
\ref{sec_selection} characterizes the statistical properties of the
sample. Section \ref{sec_FP} presents the FP of EDisCS galaxies. We
start in Sect. \ref{sec_FPclus} with the FP for 25 clusters and
discuss the evolution of the FP zero point as a function of redshift
and cluster velocity dispersion. Section \ref{sec_mass} considers the
differences between the FP of galaxies in clusters and the field and
the dependence on galaxy mass. Section \ref{sec_rotation} discusses
the related problem of the rotation of the FP.  In
Sect. \ref{sec_sizeev}, we consider the size evolution of galaxies and
how this affects the stellar population time-dependence implied by the
evolution of the FP.  In Sect.  \ref{sec_conclusions}, we draw our
conclusions. Appendix \ref{app_ReCir} explains in detail
  how we compute circularized half-luminosity radii. Throughout the
paper, we assume that $\Omega_M=0.3$, $\Omega_\Lambda=0.7$, and
$H_0=70$km/s/Mpc.

\section{Data analysis}
\label{sec_data}

The sample of galaxies analyzed in this paper consists of
spectroscopic early-type objects.  We considered the flux-calibrated
spectra reduced in \citet{Halliday04} and \citet{Milvang08} of
galaxies with early spectral type (1 or 2). This indicates the total
absence (type 1) or the presence of only weak (with equivalent width
smaller than 5 \AA) [OII] lines \citep{Sanchez09}. We derive galaxy
velocity dispersions from these spectra (Sect. \ref{sec_sigma}). We
match this dataset with HST and VLT photometry
(Sect. \ref{sec_photometry}).  The HST images \citep{Desai07} provide
visual classification and structural parameters for 70\% of our
galaxies. For the remaining 30\%, we use VLT photometry, where no
visual classification is available \citep{Simard09}. Approximately
70\% of the galaxies with HST photometry have early-type morphology
(Sect. \ref{sec_selection}).

\subsection{Velocity dispersions}
\label{sec_sigma}

Velocity dispersions were measured in all galaxy spectra using the IDL
routine pPXF \citep{Cappellari04}. This routine is based on a maximum
penalized likelihood technique that employs an optimal template, and
also performs well when applied to spectra of low signal-to-noise
ratio \citep{Cappellari09}. The algorithm works in pixel space,
estimating the best fit to a galaxy spectrum by combining stellar
templates that are convolved with the appropriate mean galaxy velocity
and velocity dispersion.  The results depend critically on how well
the spectra are matched by the template.  To compile an optimal
template, we use 35 synthetic spectra from the library of single
stellar population models of \citet{Vazdekis10}, which uses the new
stellar library MILES \citep{Sanchez06}. These spectra have been
degraded to the wavelength-dependent resolution of the EDisCS spectra,
determined from the widths of the lines in the arc lamp spectra, slit
by slit,  and matching well the widths of the sky lines on the science 
spectra.  

The library contains spectra spanning an age range from 0.13
to 17 Gyr and metallicities from $[$Z/H$]=-0.68$ to $[$Z/H$]=+0.2$.
Operating in pixel space, the code allows the masking of regions of
the galaxy spectra during the measurements. We use this to mask
regions affected by skyline residuals.  Although the code allows the
measurement of the higher Gauss-Hermite order moments
\citep{Bender94}, we only fit the velocity and $\sigma$, which
stabilises the fits in our spectra of low signal-to-noise ratio.
Errors were calculated by means of Monte Carlo simulations in which
each point was perturbed with the typical observed error, following a
Gaussian distribution. Because the template mismatch affects the
measurement of the velocity and $\sigma$ determined with pPXF, a new
optical template was used in each simulation. The errors were assumed
to be the standard deviation in measurements inferred from 20
simulations. Owing to limitations caused by the instrumental
resolution, only velocity dispersions larger than 100 km/s are
reliable and unbiased. Therefore, galaxies with smaller $\sigma$,
  as well as velocity dispersions with uncertainties larger than 20\%,
  the approximate intrinsic scatter of the local FP (see
  Introduction), will not be considered further.

We note that the velocity dispersions measured here are $\approx 10$\%
lower than those given in \citet{Sanchez09}. The difference is caused
by the fact that the instrumental resolution in that paper was assumed
to be constant with wavelength at the value of 6 \AA. In reality,
  this is just the best resolution possible with our setup, that
  extends up to 8 \AA.  The change is important here, but does not
affect any of the results presented in \citet{Sanchez09}.

\begin{figure}
\begin{tabular}{cc}
\vbox{\psfig{file=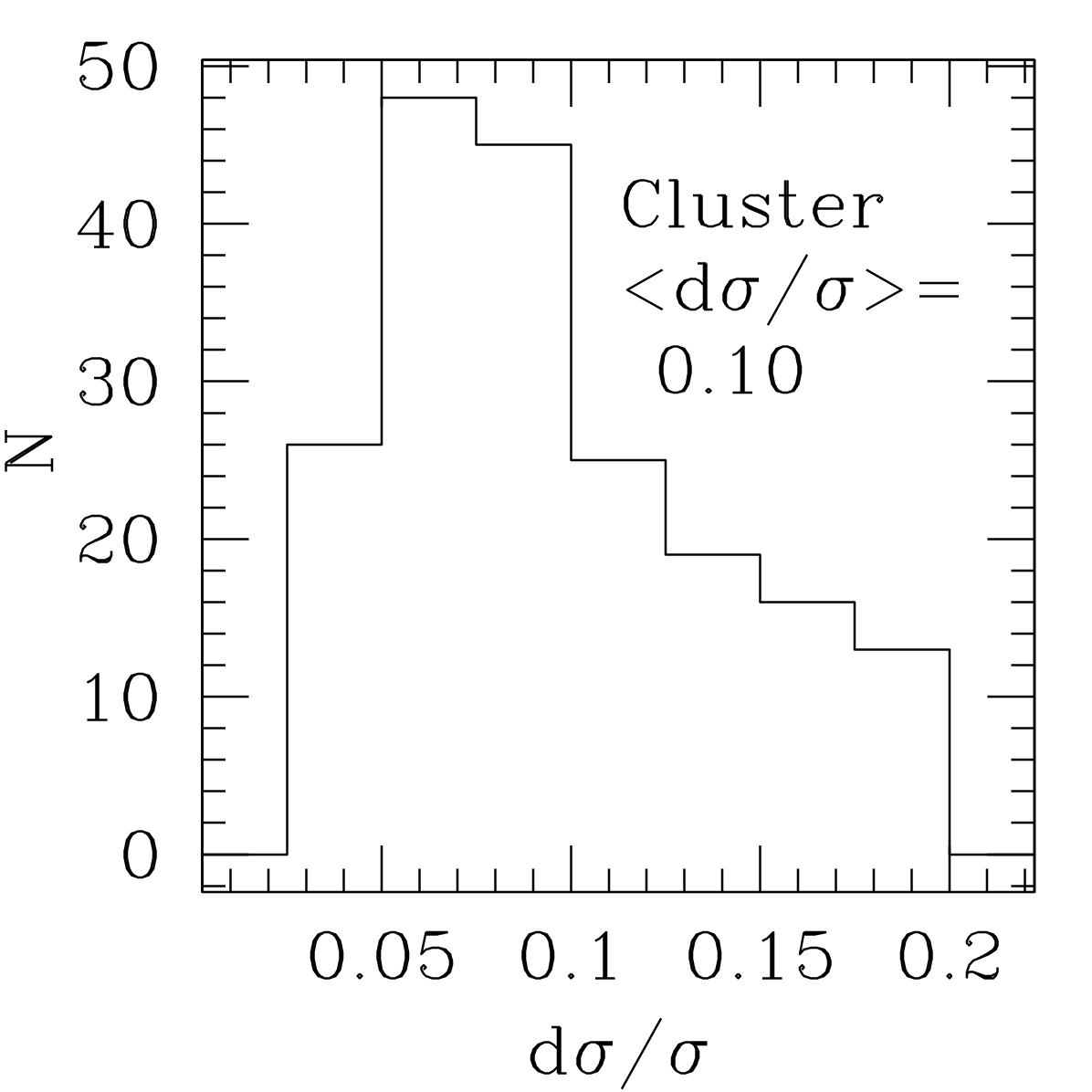,angle=0,width=4cm}}&
\vbox{\psfig{file=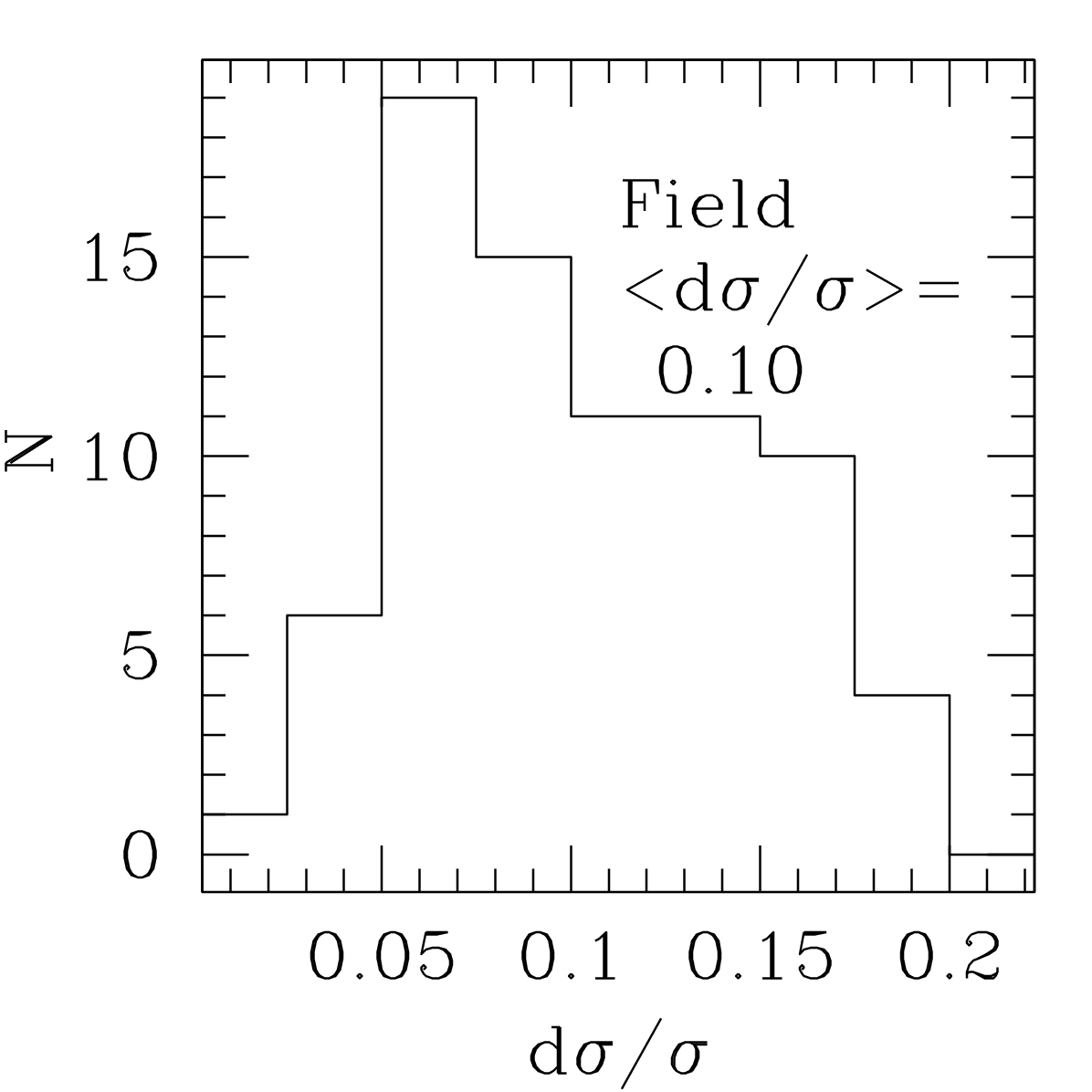,angle=0,width=4cm}}\\
\vbox{\psfig{file=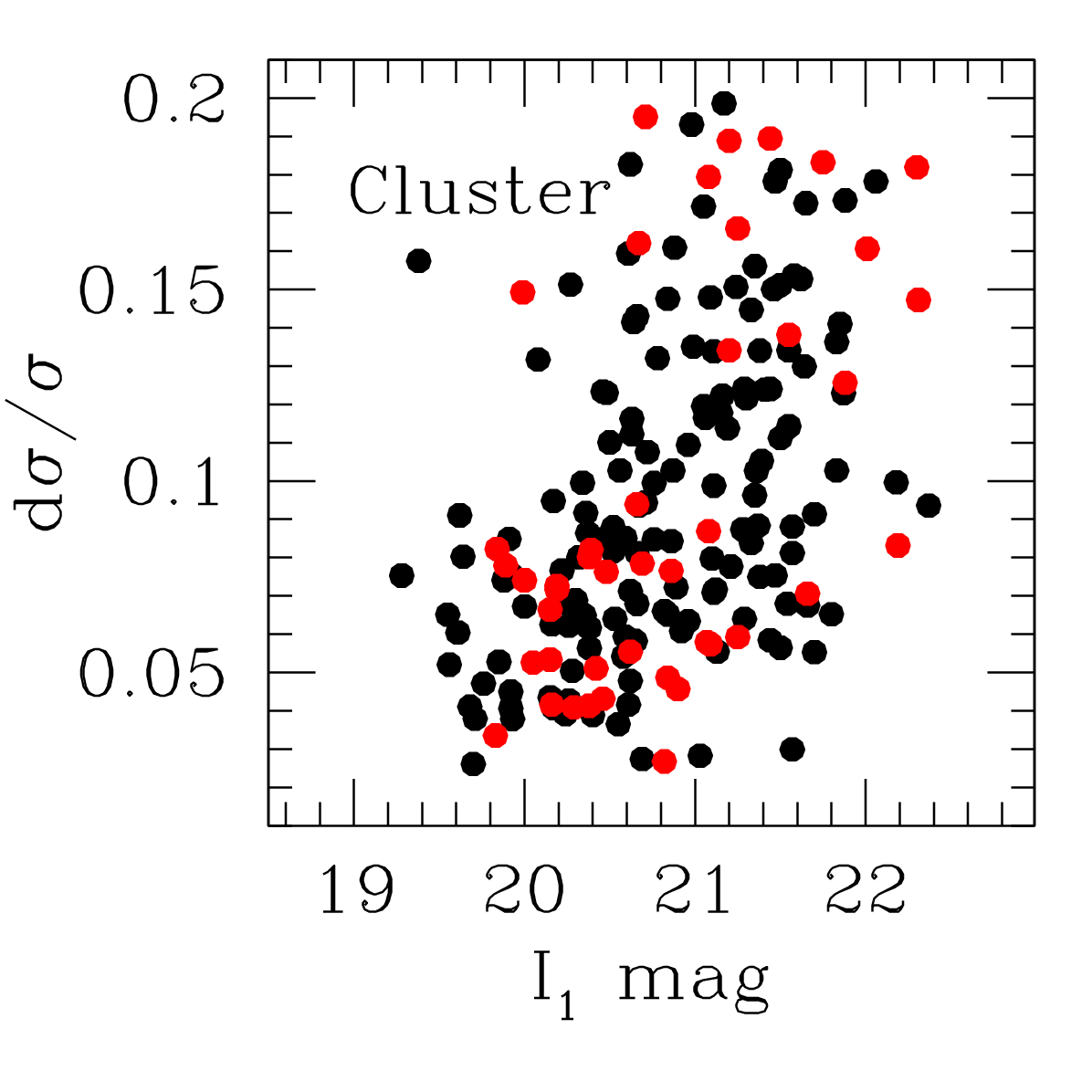,angle=0,width=4cm}}&
\vbox{\psfig{file=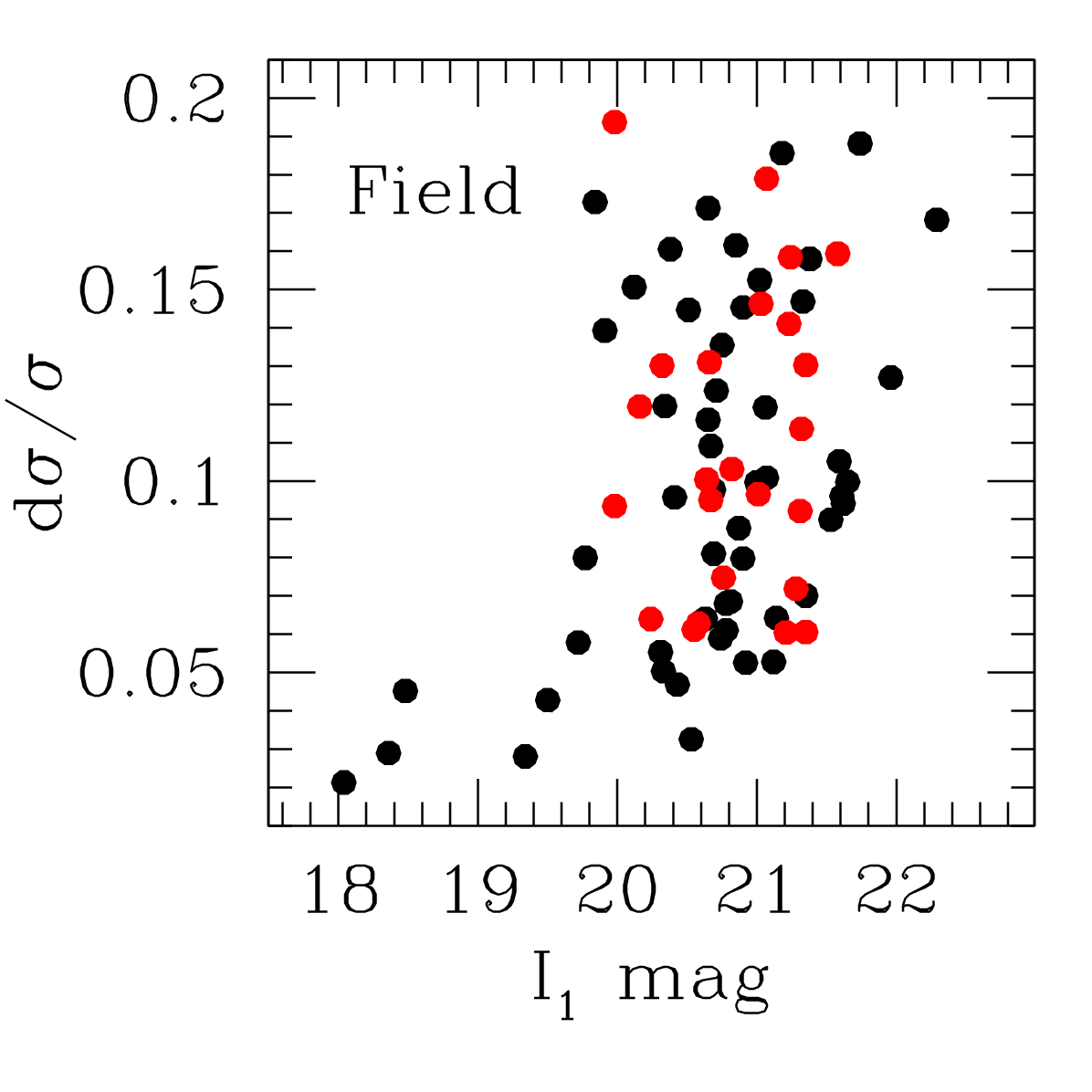,angle=0,width=4cm}}\\
\end{tabular}
\caption{The velocity dispersion errors. 
First row: the histograms of statistical
errors on velocity dispersions. Second row: the statistical errors as a 
function of apparent I band magnitude in a 1 arcsec radius aperture. 
Colors code the spectroscopic 
type (black: 1; red: 2). 
  \label{fig_dsighis}}
\end{figure}

We measured velocity dispersions for 192 cluster and 78 field galaxies. 
Figure \ref{fig_dsighis} shows the histograms of the statistical and
systematic errors.  The statistical errors are on average 10\% and are
a function of magnitude.

The systematic errors are more difficult to estimate, as they
  depend on the template mismatch, continuum variations, and filtering
  schemes. They have been extensively studied in the past
  \citep{Cappellari04} and can be as large as 5-10\%.  To check
  the size of systematic errors, we derived the galaxy velocity
  dispersions using the FCQ method of \citet{Bender94}, which is less
  prone to template mismatching systematics and operates in Fourier
  space. We focused on the G band region at $z\approx 0.5$, the Mgb
  region at lower redshifts, or the largest available continuous range
  redder than the 4000 \AA\ break, similar to the approach of \citet{Ziegler05}.
The two methods agree well, with
68\% of the values differing by less than the combined 1-$\sigma$
error, and 96\% by less than 3-$\sigma$, but smaller errors are
derived using the pixel fitting approach, partially because most of
each spectrum can be used. This allow us to conclude that our residual
systematic errors are always smaller than the statistical ones.

\begin{figure}
\psfig{file=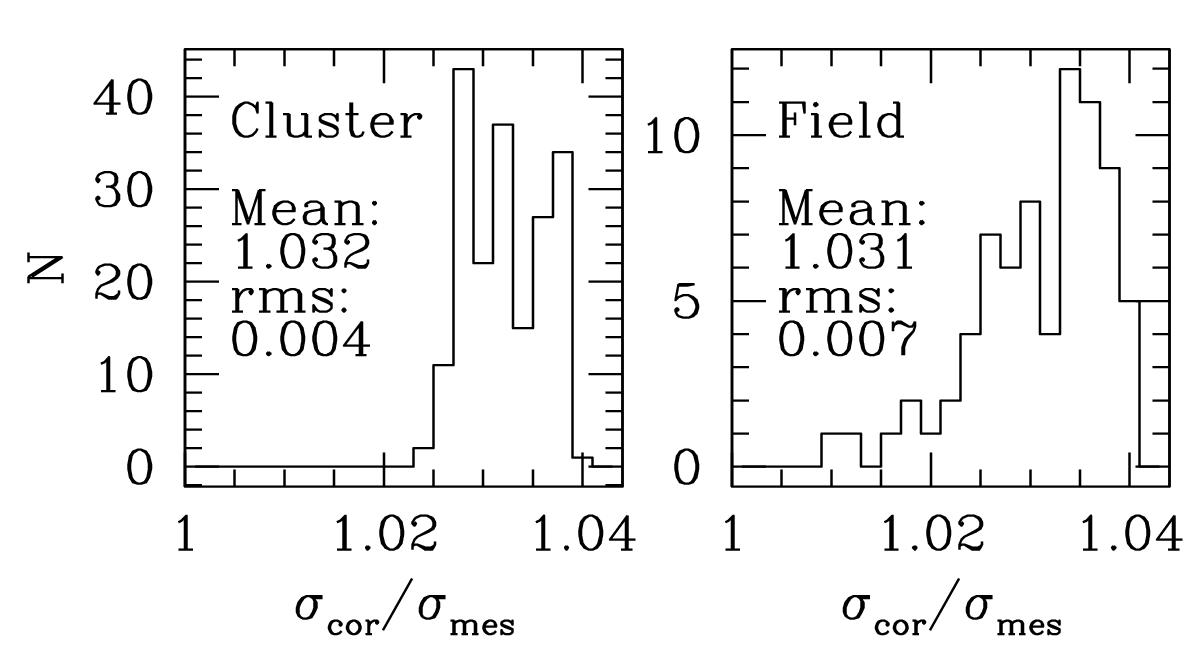,angle=0,width=8cm}
\caption{The histogram of the fractional aperture corrections 
$\sigma_{cor}/\sigma_{mes}$
for cluster (left) and field (right) galaxies.}
\label{fig_sigcorhis}
\end{figure}

Finally, an aperture correction following \citet{Jorgensen95}
\begin{equation}
\log \sigma_{\rm cor} = \log \sigma_{\rm mes}+0.04*\log({\rm Ap}/3.4 {\rm kpc}),
\label{eq_aperture}
\end{equation}
where Ap represents the average aperture of our observations, 1.15 arcsec,
scaled with the distances of the objects, 
was applied to the measured velocity dispersions $\sigma_{\rm mes}$ to place
them on the Coma cluster standard aperture system of 3.4 kpc. Figure
\ref{fig_sigcorhis} shows that, on average, this correction amounts to
3\% with $\approx 0.5$\% spread.  From this point on, we drop the $cor$
and indicate with $\sigma$ the aperture-corrected value of the
velocity dispersion.

\begin{figure}
\begin{tabular}{cc}
\vbox{\psfig{file=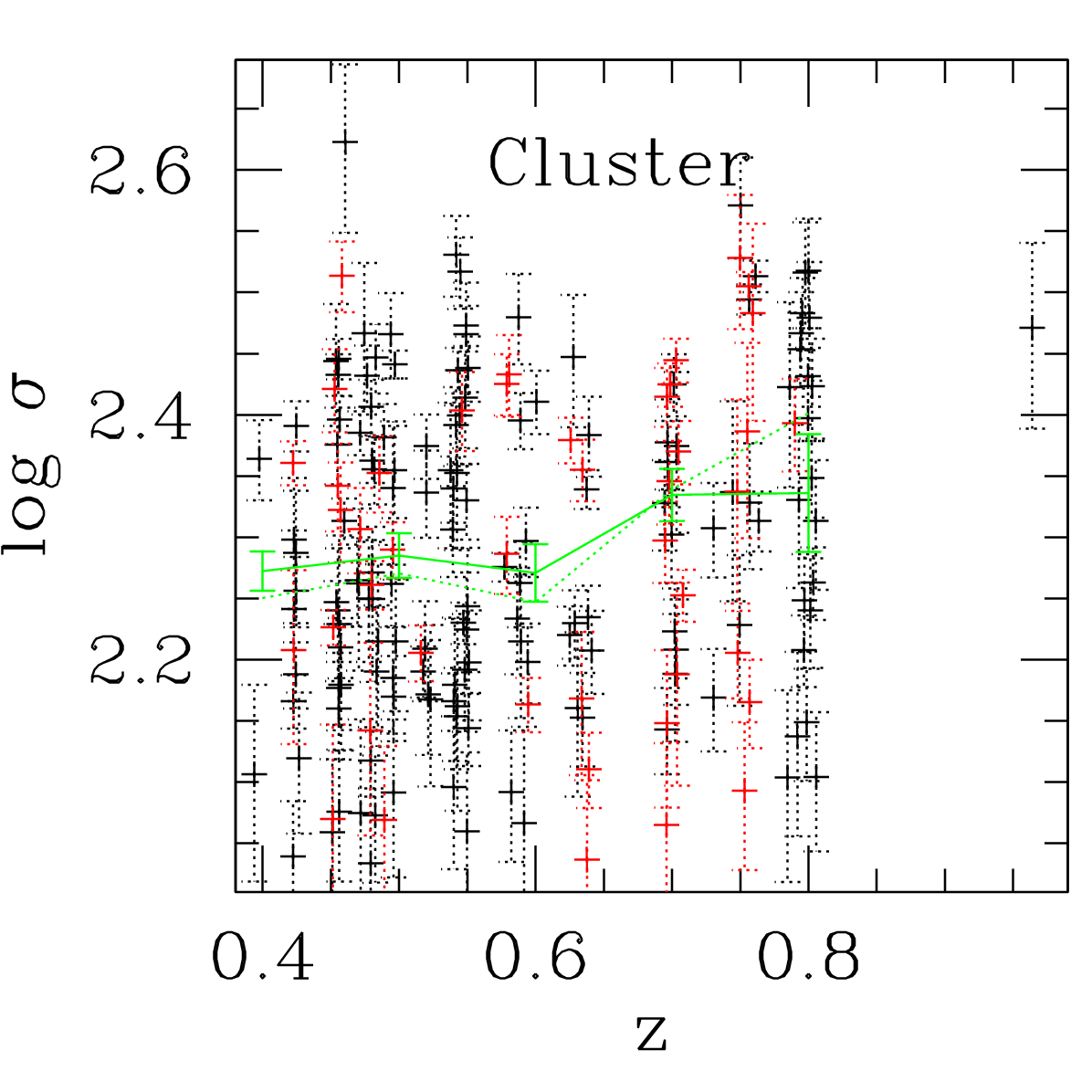,angle=0,width=4cm}}&
\vbox{\psfig{file=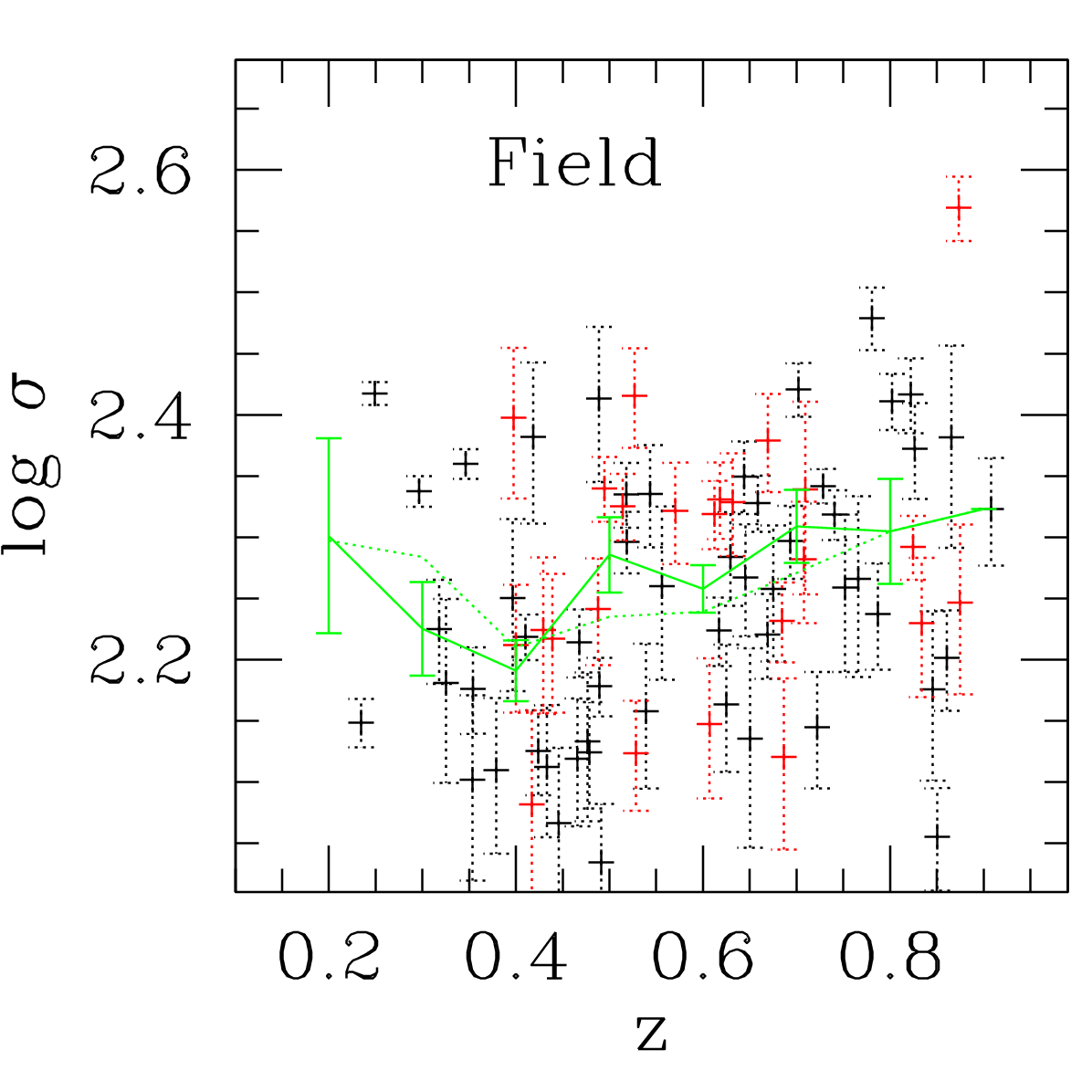,angle=0,width=4cm}}\\
\vbox{\psfig{file=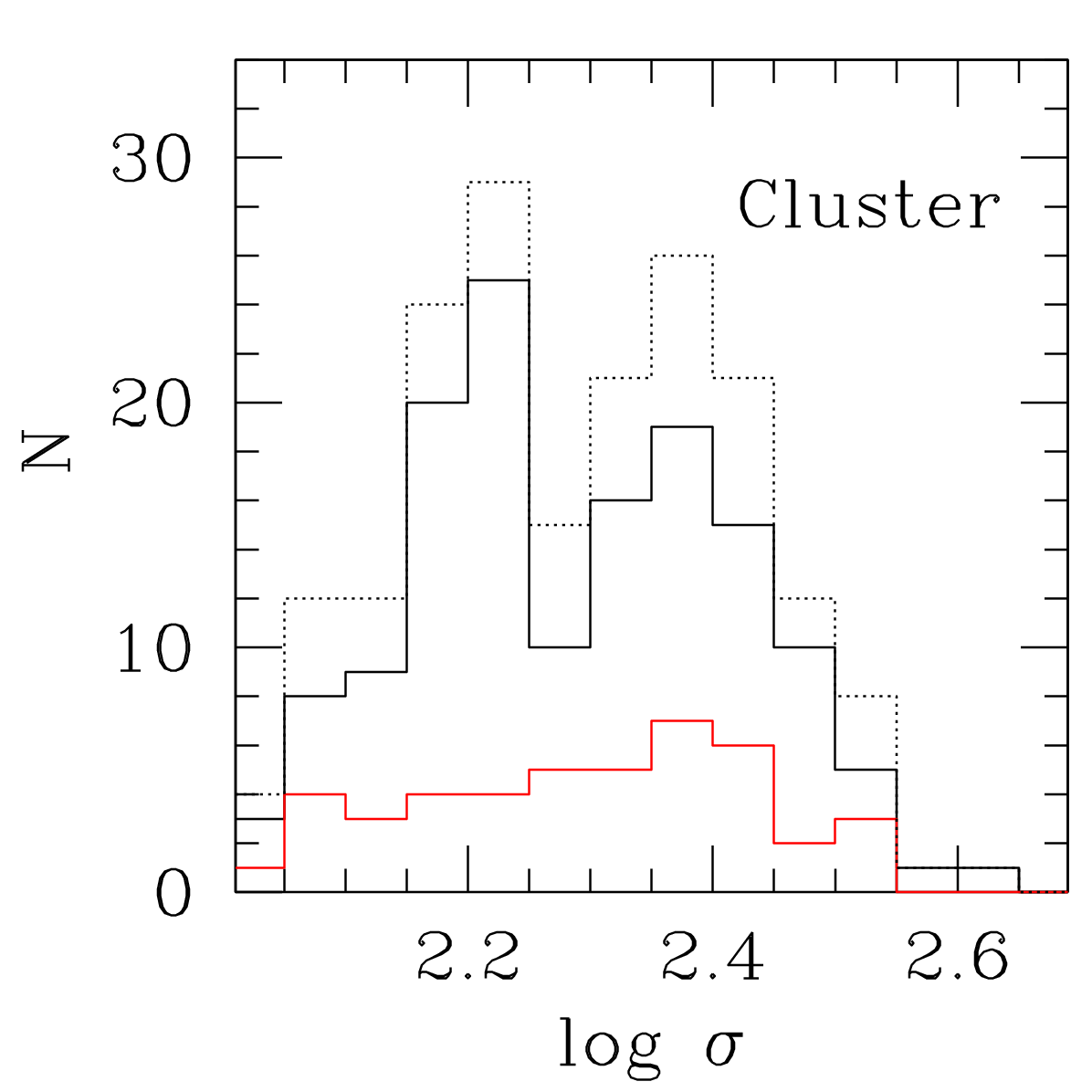,angle=0,width=4cm}}&
\vbox{\psfig{file=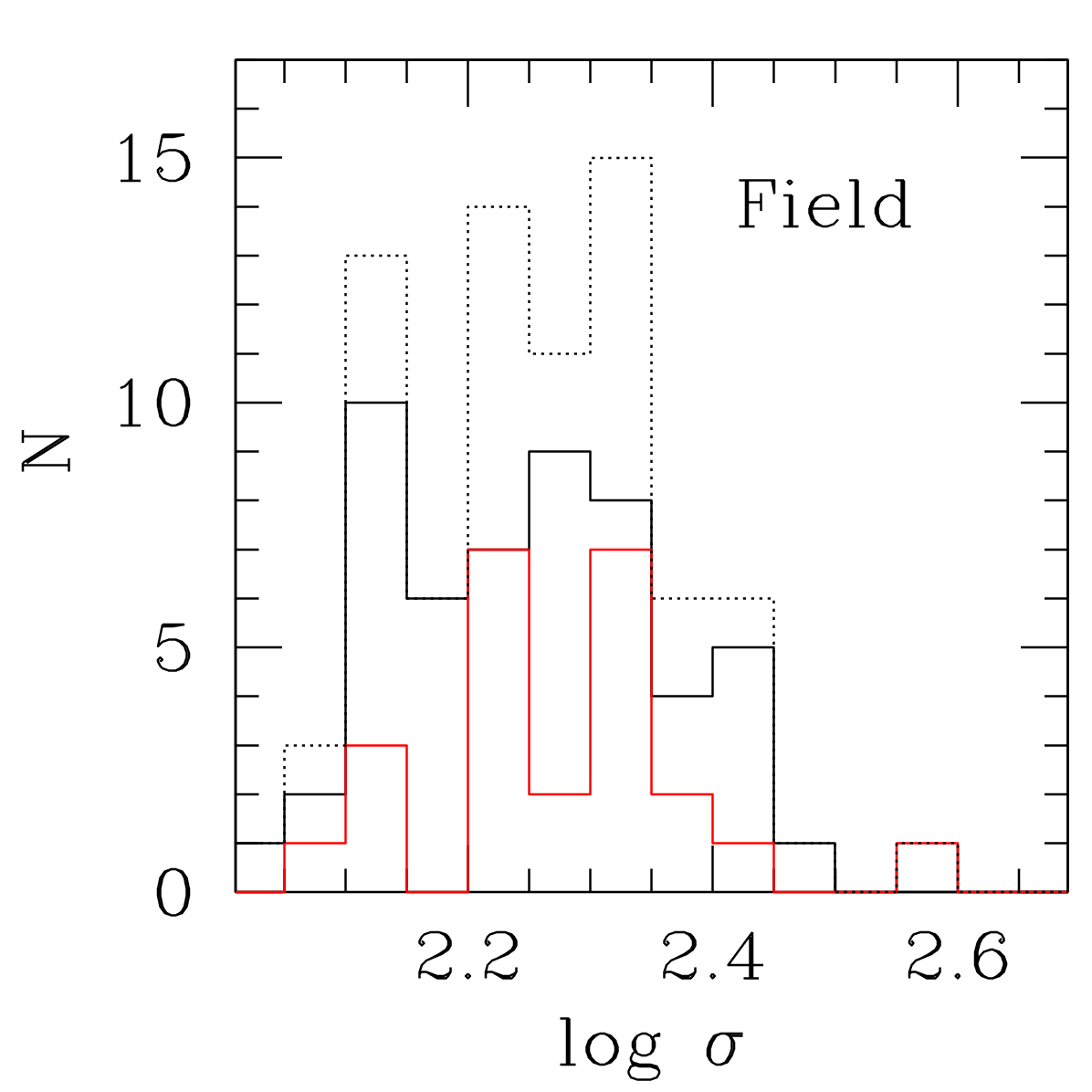,angle=0,width=4cm}}\\
\end{tabular}
\caption{The velocity dispersions of the galaxy sample.
Top: the measured galaxy velocity dispersions as a function of 
redshift in clusters (left) and the field (right). The green lines show 
the mean values in 0.1 redshift bins and the relative errors. The dotted lines
show the mean values weighting each galaxy with the inverse of its 
completeness value. Bottom: the histogram 
of galaxy velocity dispersions in clusters (left) and  the field (right).
Colors code the spectral type (black: 1; red:2). The dotted lines show the 
histogram for the entire sample irrespective of spectral type.
\label{fig_sigsample}}
\end{figure}

Figure \ref{fig_sigsample} presents the velocity dispersions as a
function of redshift and their distribution. On average, the galaxy
velocity dispersion is $\approx 200$ km/s, with a mildly increasing
trend with redshift. Weighting each galaxy with the inverse of its
completeness value (see Sect. \ref{sec_selection}) in general changes
the mean by no more than its error.

\subsection{Photometry}
\label{sec_photometry}

The photometric part of the FP, i.e., the half-luminosity radius $R_e$
and average effective surface brightness $\langle SB_e\rangle=-2.5\log
\frac{L}{2\pi R_e^2}$, where $L$ is the total luminosity, was derived
by fitting either HST ACS images \citep{Desai07} or I-band VLT images
\citep{White05} using the GIM2D software
\citep{Simard02}. \citet{Simard09} provide an extensive description of
the methods and tests performed to assess the accuracy of the derived
structural parameters, using exhaustive Monte Carlo simulations. To
summarize, a two-component two-dimensional fit was performed, adopting
an $R^{1/4}$ bulge plus an exponential disk convolved to the PSF of
the images. From the parameters of the fit, we measured the
(circularized) $R_e$ and effective surface brightness from curves
  of growth constructed from the best fit models using the procedure
described in Appendix \ref{app_ReCir}.

Historically, effective radii were derived from fits to curves of
  growths, constructed from photoelectric photometry using circular
  apertures of increazing sizes \citep{Burstein87}.  Our procedure
  reproduces this approach and is identical to that followed by
  \cite{Gebhardt03} to study the evolution of the FP of field galaxies
  with redshift. We prefer it to less sophisticated approaches (such as
  the straight $R^{1/4}$ fit often used in the literature) as it provides
  far superior fits to the images. As \cite{Gebhardt03} do, we note 
  that in the past a variety of methods have been adopted to measure
  the structural parameters that enter in the FP: curve of growth,
  isophotal photometry or 2-dimensional fitting, pure $R^{1/4}$,
  Sersic or bulge+disk (B+D) functions. The derived effective radii
  and surface brightness, however, when combined in 
  Eq. \ref{eq_FP} of the FP, deliver the same ZP to a high
  degree of accuracy \citep{SBD93}. This has been proven for a large
  set of local clusters, including the Coma cluster
  \citep{Saglia97b,Dejong2004}, and remains valid for the present
  data set (see below). This justifies the comparisons with FP samples
  from the literature presented below.

\begin{figure}
\psfig{file=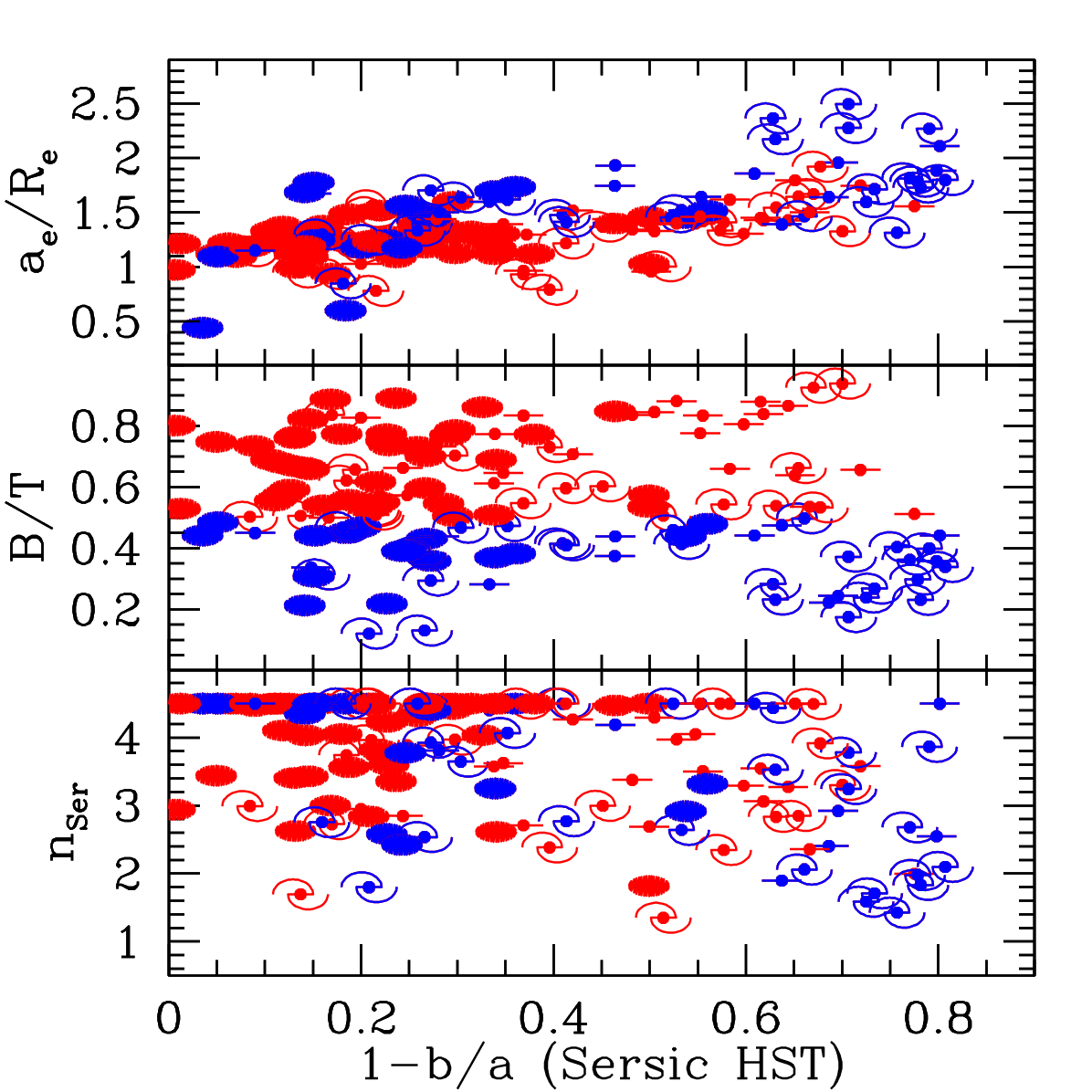,angle=0,width=8cm}
\caption{The properties of the bulge+disk fits to galaxies with HST 
photometry and a measured velocity 
dispersion. We plot the ratio $a_e/R_e$ between the semi-major effective scale 
length $a_e$ of the best-fitting Sersic profile to the circularized effective 
radius $R_e$ of the best fits B+D model (top), the bulge-to-total ratio 
$B/T$ (middle), and the Sersic index $n_{Ser}$ (bottom) as a function of the 
ellipticity $1-b_e/a_e$ (where $b_e$ is the semi-minor effective scale length) 
of the Sersic fit. Objects with $B/T>0.5$ are plotted in red, the remainder
in blue. Symbols code the morphology: filled ellipses
  show $T\le-4$, filled circles crossed by a line $-3\le T\le 0$,
  spirals $T>0$. 
\label{fig_checkFits} 
}
\end{figure}

  We later use effective radii to probe the size evolution of
  galaxies.  Without doubt, the scale length along the major axis of a
  pure disk galaxy is the correct measurement of its size, and our
  circularized $R_e$ progressively underestimates the effective
  semi-major axis length as the inclination increases (see
  Fig. \ref{fig_checkFits}). However, for a pure bulge the inverse is
  true, and our $R_e$ then averages out projection effects, producing
  the equivalent circularized size of each spheroid.

On the other hand, the resolution and signal-to-noise
  ratio of the images considered here is too low to allow us to perform an
  accurate and unbiased determination of the sizes of the bulge and
  the disk components separately for our galaxies. Since the percentage
  of disk-dominated, highly inclined objects in the galaxy sample
  considered here is low, as it is in the low redshift comparison, we
  conclude that our choice is reasonable. In particular, the mean
  axial ratios of our sample and the low redshift comparison are
  identical, as discussed in \citet{Valentinuzzi10b}.

  We now consider the quantitative question of the extent to which our
  procedure for computing structural parameters is equivalent to other
  approaches discussed in the literature.

  In analogy with procedures followed for local galaxies
  \citep{Saglia97a}, where systematic errors are gauged by 
  comparing different photometric fits, we assess the robustness
  of the structural parameters to the chosen $R^{1/4}$ bulge plus
  exponential disk surface brightness model by considering a second  
two-dimensional fitting approach to the HST images. We fit a 
single-component Sersic profile (with $0.5\le n_{Ser}\le 4.5$) to the HST
  ACS imaging in the F814W band, available for 10 of the EDisCS
  clusters.  Again, the circularized half-luminosity radius $R_e(Ser)$
  is computed from curves of growth constructed from the best fit model 
as described in Appendix \ref{app_ReCir}.
   
Figure \ref{fig_checkFits} summarizes the results of our B+D and
  Sersic fits.  The galaxies of our HST sample have on average a
  flattening $1-b_e/a_e$ of 0.37 (0.33 without spirals), with some
  disk-dominated, nearly edge-on spiral galaxies reaching
  $1-b_e/a_e\approx 0.8$. As a consequence, our circularized effective
  radii are on average 39\% (33 \% without spirals) smaller than the
  effective semi-major lengths $a_e$. On average, our objects are
  bulge-dominated ($\langle B/T\rangle=0.59$, 0.64 without spirals) and
  reasonably well described by a de Vaucouleurs law ($\langle
  n_{Ser}\rangle=3.7$, 3.9 without spirals).

\begin{figure}
\psfig{file=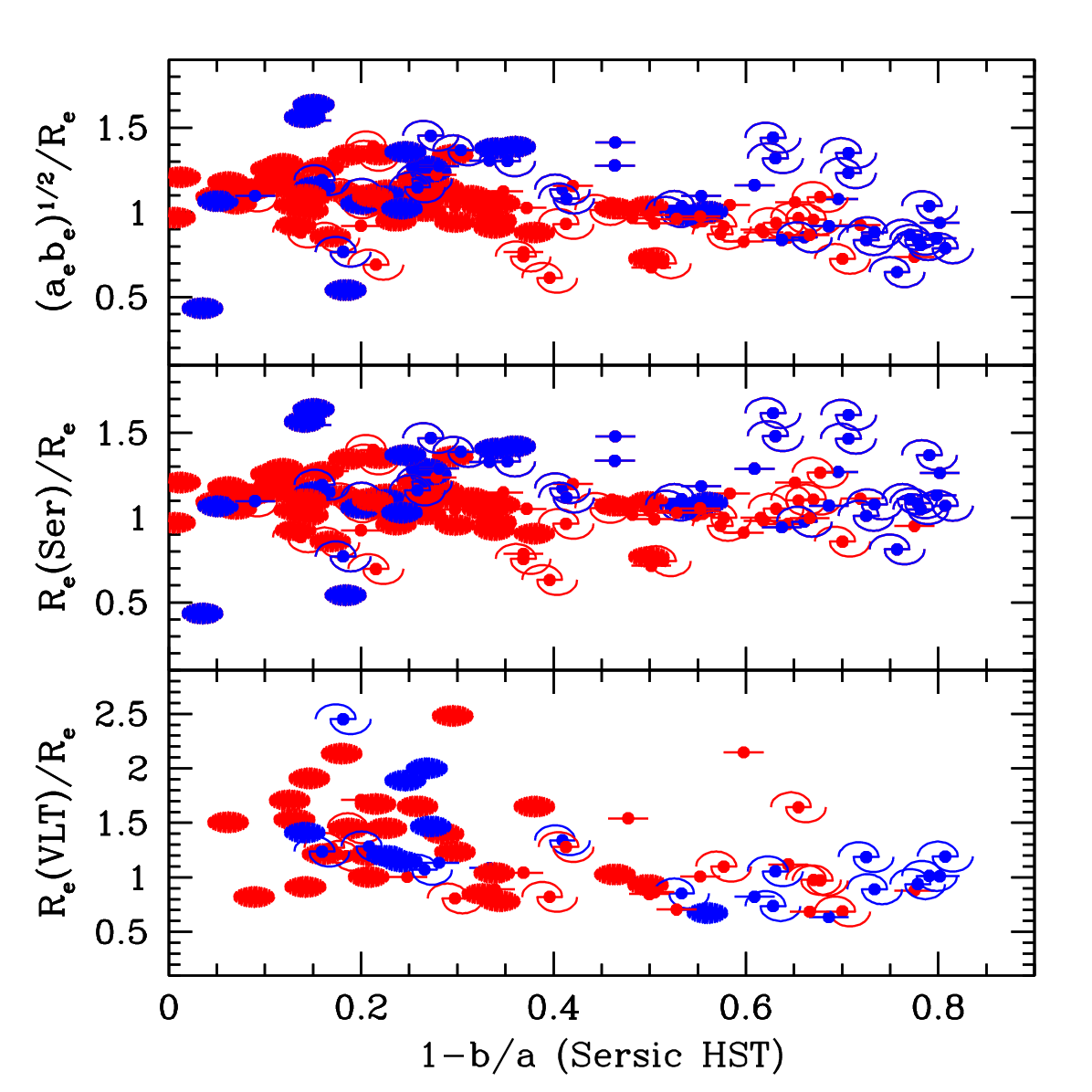,angle=0,width=8cm}
\caption{ The comparison between different estimations of the 
half-luminosity radii of all galaxies with HST photometry and a 
measured velocity 
dispersion. We plot the ratio of the harmonic radius $(a_eb_e)^{1/2}$ to 
the circularized effective radius $R_e$ of the best fits HST B+D model (top), 
the ratio of the circularized effective radius $R_e(Ser)$ of the Sersic 
fit to $R_e$ (middle), and the ratio of the circularized effective 
radius $R_e(VLT)$ of the best fits VLT B+D model to $R_e$ (bottom)
 as a function of $1-b_e/a_e$. Symbols 
and color coding are as in Fig. \ref{fig_checkFits}.
\label{fig_checkRe} 
}
\end{figure}

\begin{figure}
\psfig{file=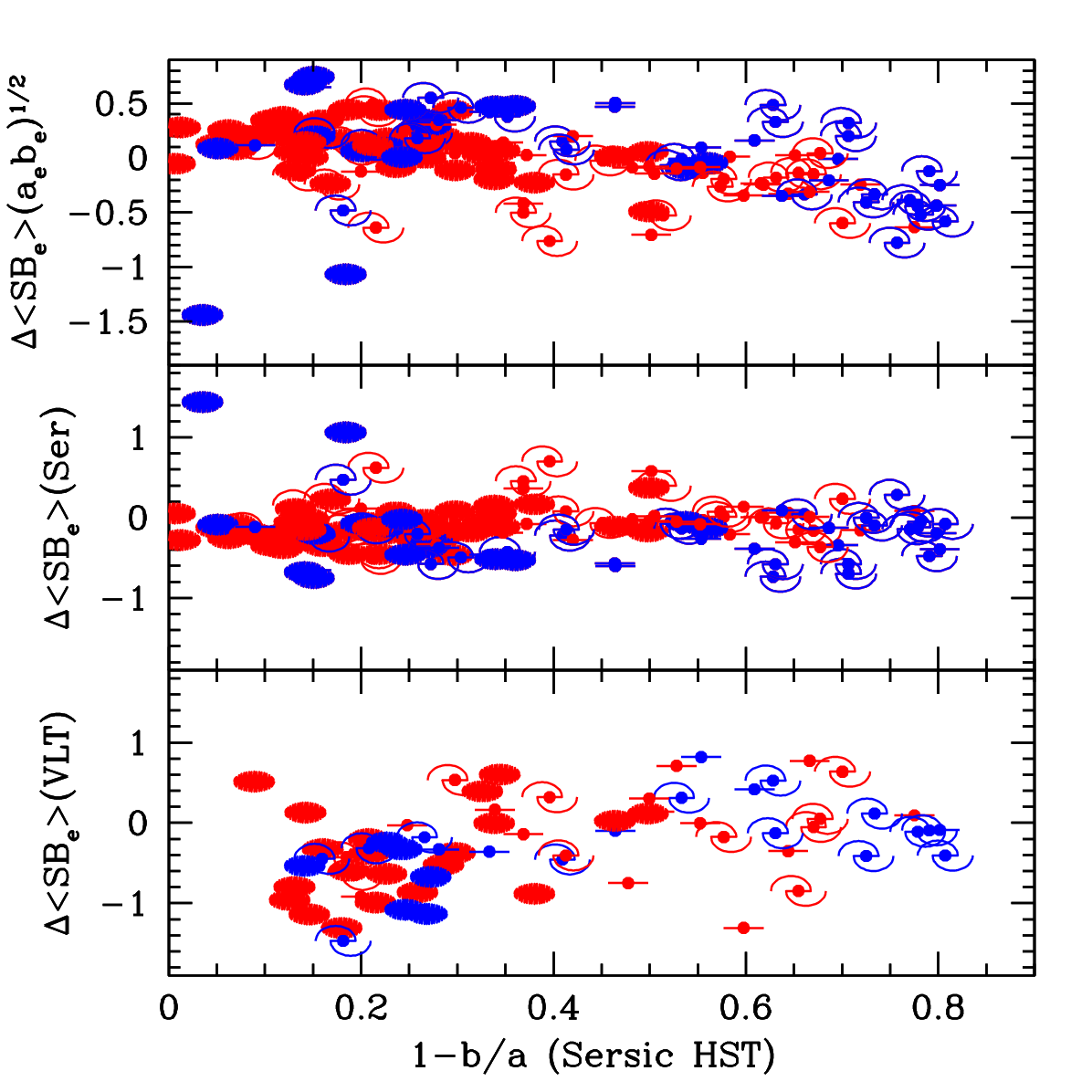,angle=0,width=8cm}
\caption{ The comparison between different estimates of the effective 
surface brightness of all galaxies with HST photometry and a measured velocity 
dispersion. We plot the difference  $\Delta \langle SB_e(a_eb_e)\rangle$ 
between the average surface brightness within  $(a_eb_e)^{1/2}$ and $R_e$ 
(top), the difference   $\Delta \langle SB_e(Ser)\rangle$ between the 
average surface brightness within $R_e(Ser)$ and $R_e$ (middle), and the 
difference between the average surface brightness within $R_e(VLT)$  
and $R_e$ (bottom)  as a function of $1-b_e/a_e$. Symbols and color coding 
are as in Fig. \ref{fig_checkFits}.
\label{fig_checkSBe} 
}
\end{figure}

\begin{figure}
\psfig{file=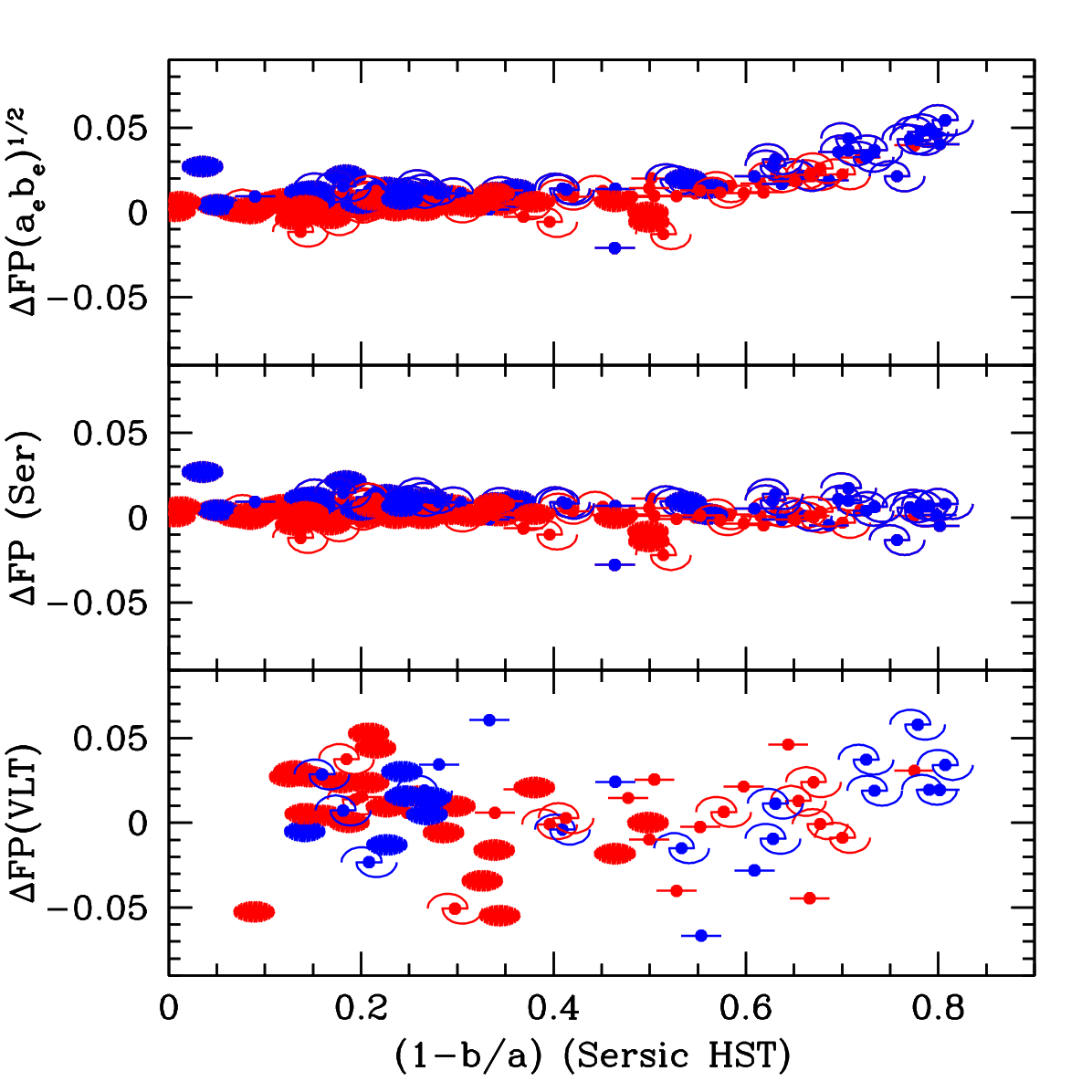,angle=0,width=8cm}
\caption{ The comparison between different estimations of the quantity
$FP=\log R_e -0.27 \langle SB_e\rangle$ for  all galaxies with HST photometry 
and a measured velocity dispersion. We plot 
$\Delta FP(a_eb_e)^{1/2}=\log((a_eb_e)^{1/2}/R_e)-0.27
(\langle SB_e\rangle(a_eb_e)^{1/2})-\langle SB_e\rangle)$  (top), 
$\Delta FP=\log(R_e(Ser)/R_e)-0.27(\langle SB_e\rangle(Ser)-\langle SB_e\rangle)$ (middle), 
and $\Delta FP=\log(R_e(VLT)/R_e)-0.27(\langle SB_e\rangle(VLT)-\langle SB_e\rangle)$ (bottom)
 as a function of $1-b_e/a_e$. 
Symbols and color coding are as in Fig. \ref{fig_checkFits}.
\label{fig_checkFP} 
}
\end{figure}

  Figures \ref{fig_checkRe}, \ref{fig_checkSBe} and \ref{fig_checkFP}
  (top and middle panels) assess the robustness of the derived
  structural parameters derived for the galaxies with measured
  velocity dispersions.  For this purpose, we also consider the
  harmonic radius $R_{har}=(a_eb_e)^{1/2}$, often used in the
  literature as a proxy for $R_e$ (sometimes fixing the Sersic index
  to 4, the $R^{1/4}$ law) and the related average surface brightness
  $\langle SB_e^{har}\rangle$, where $a_e$ and $b_e$ are the effective
  semi-major and minor axis of the Sersic fits.  The evaluated harmonic and
  circularized Sersic radii are on average very similar to our adopted
  $R_e$, as well as the resulting effective surface brightness. When
  combined into the quantity orthogonal to the FP $\log
  R_e-0.27\langle SB_e\rangle$, they show minimal systematic
  differences and scatter. As discussed in Appendix \ref{app_ReCir},
  only at high flattening (i.e. for almost edge-on disk-dominated
  galaxies) do the harmonic quantities show the expected stronger
  deviations. 

\begin{figure}
\begin{tabular}{cc}
\vbox{\psfig{file=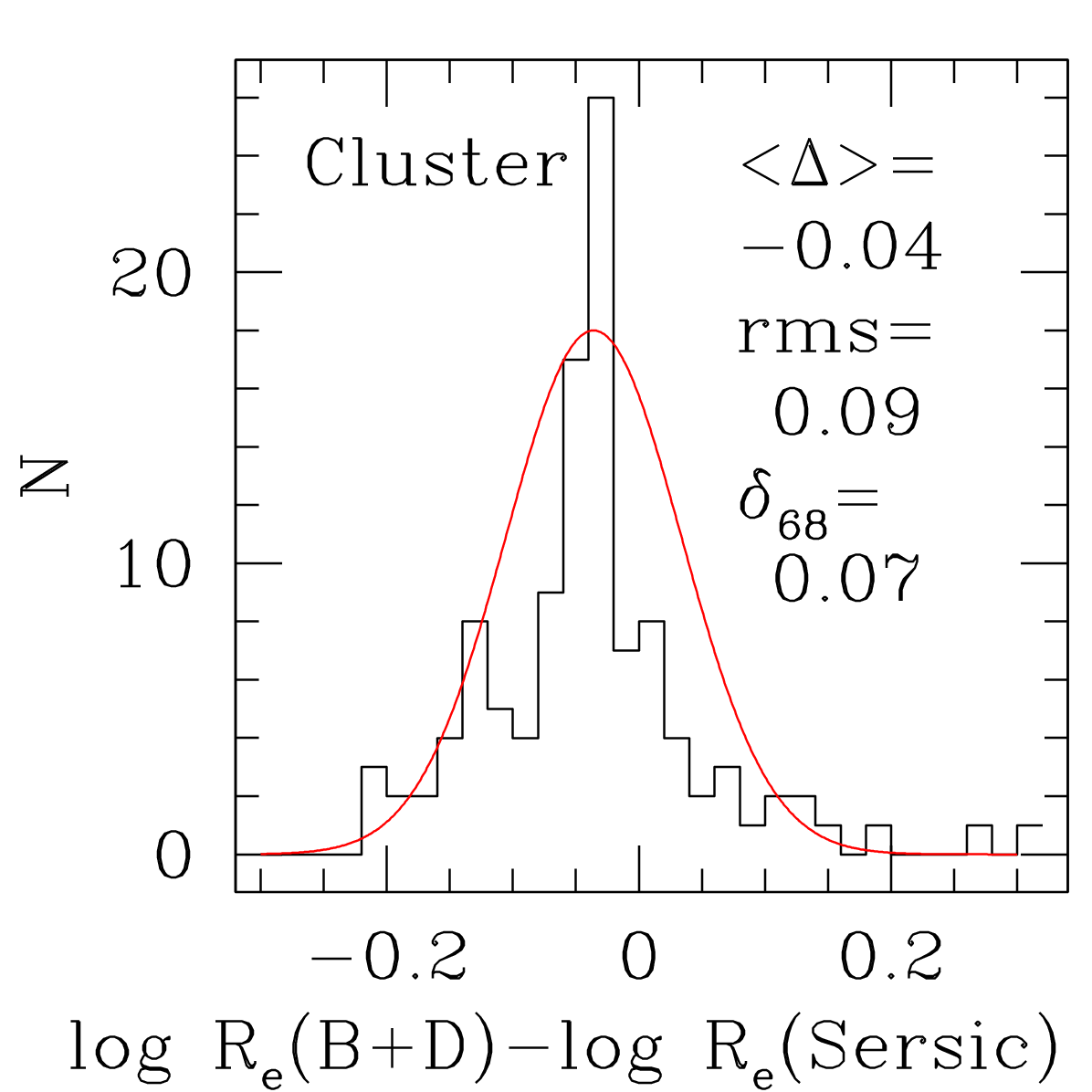,angle=0,width=4cm}}&
\vbox{\psfig{file=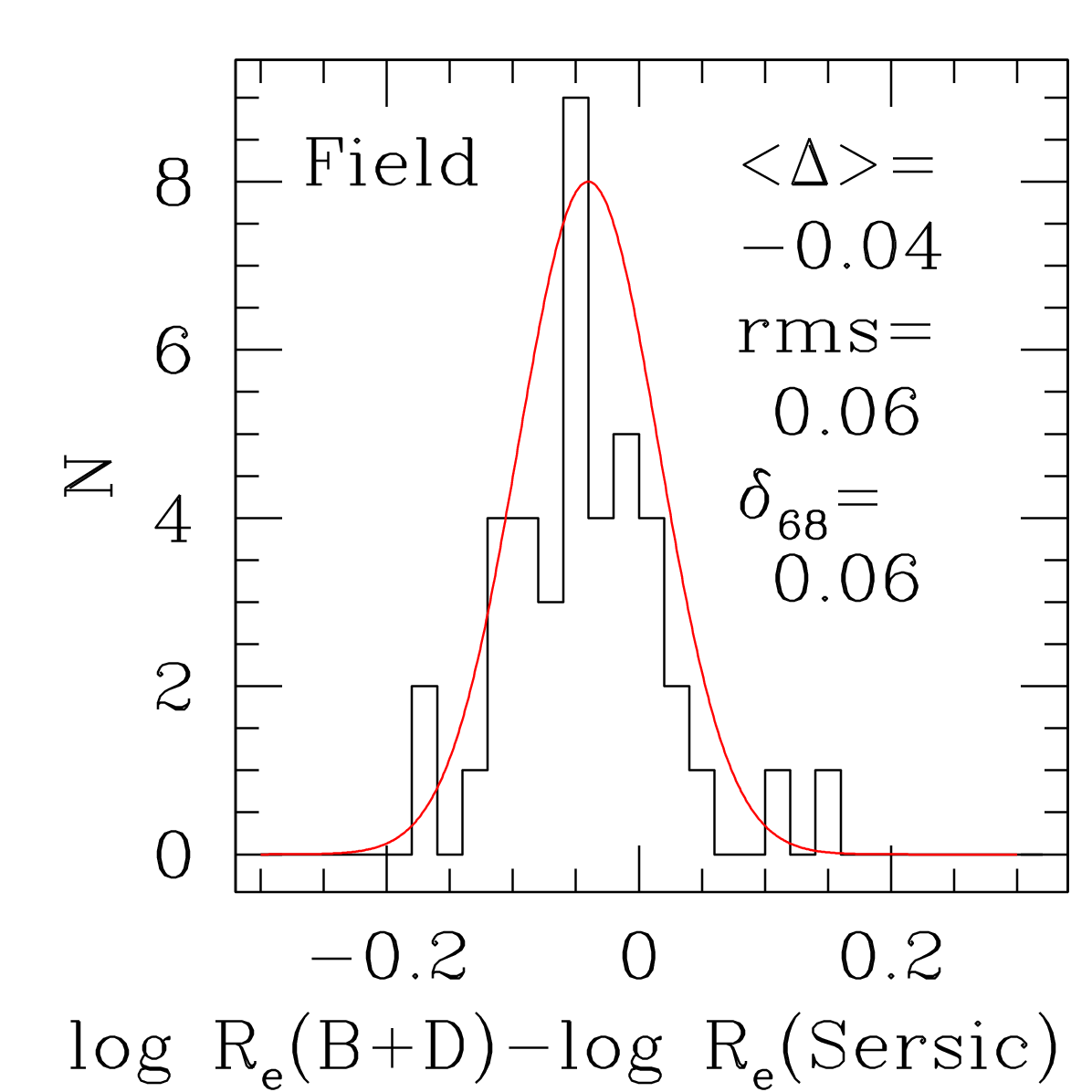,angle=0,width=4cm}}\\
\vbox{\psfig{file=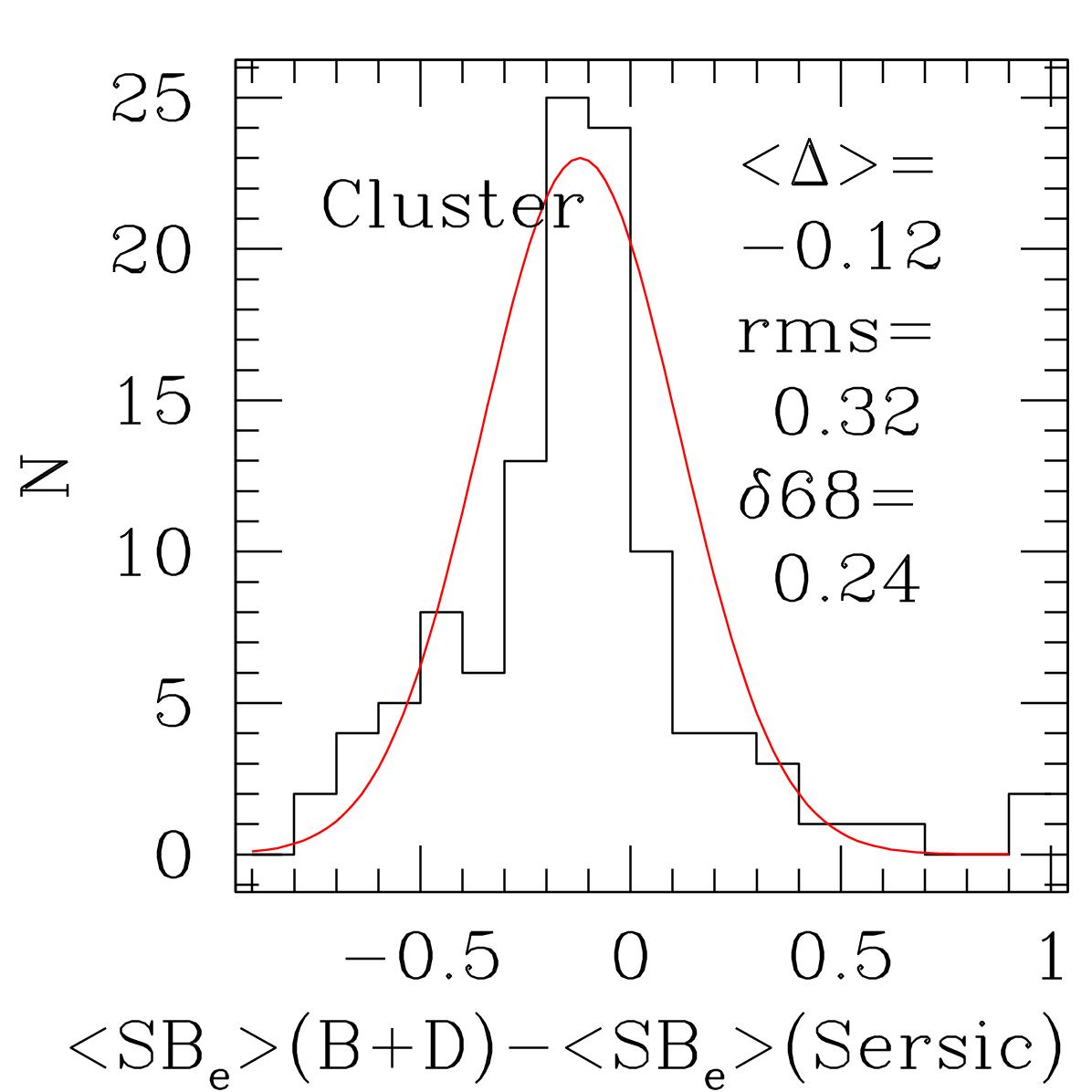,angle=0,width=4cm}}&
\vbox{\psfig{file=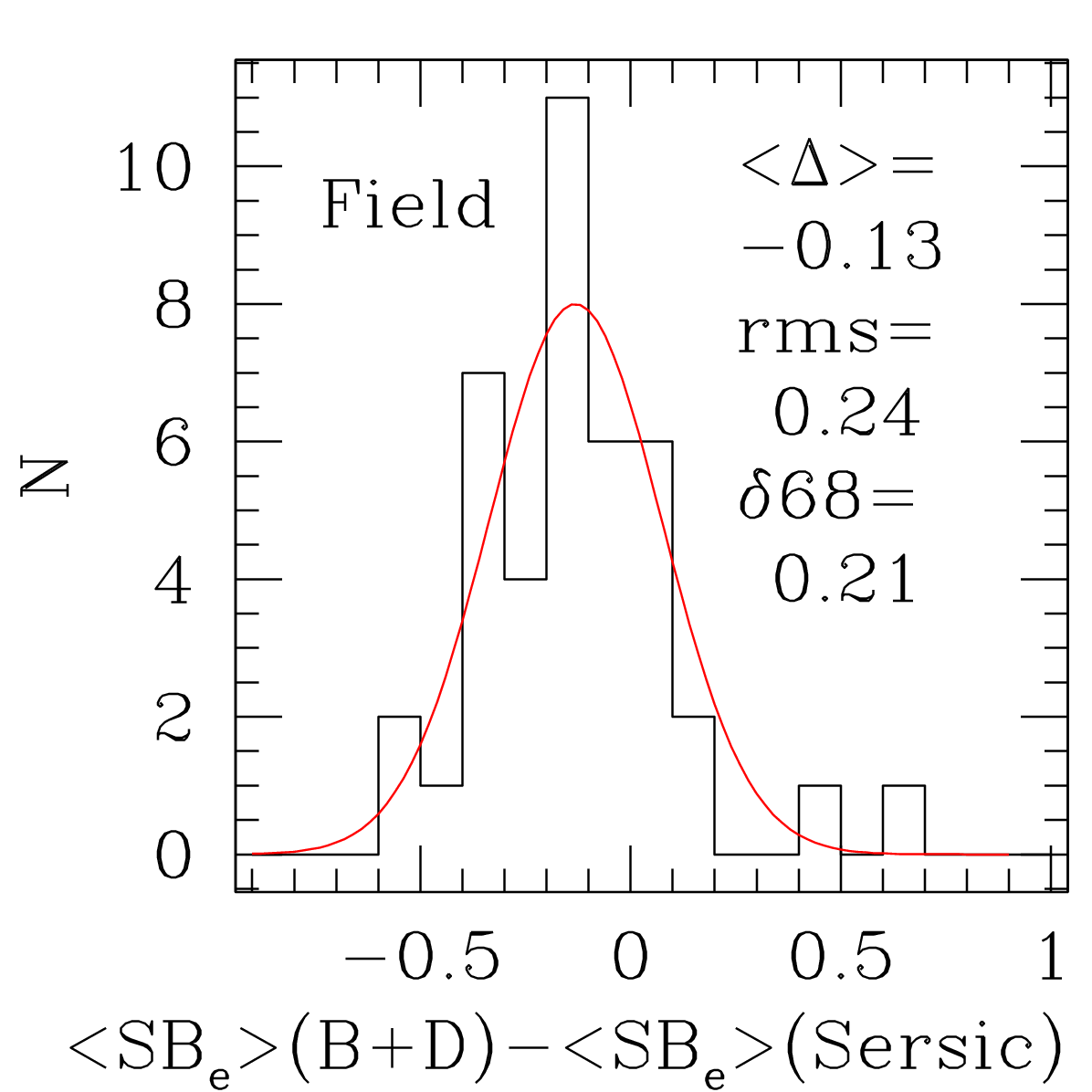,angle=0,width=4cm}}\\
\vbox{\psfig{file=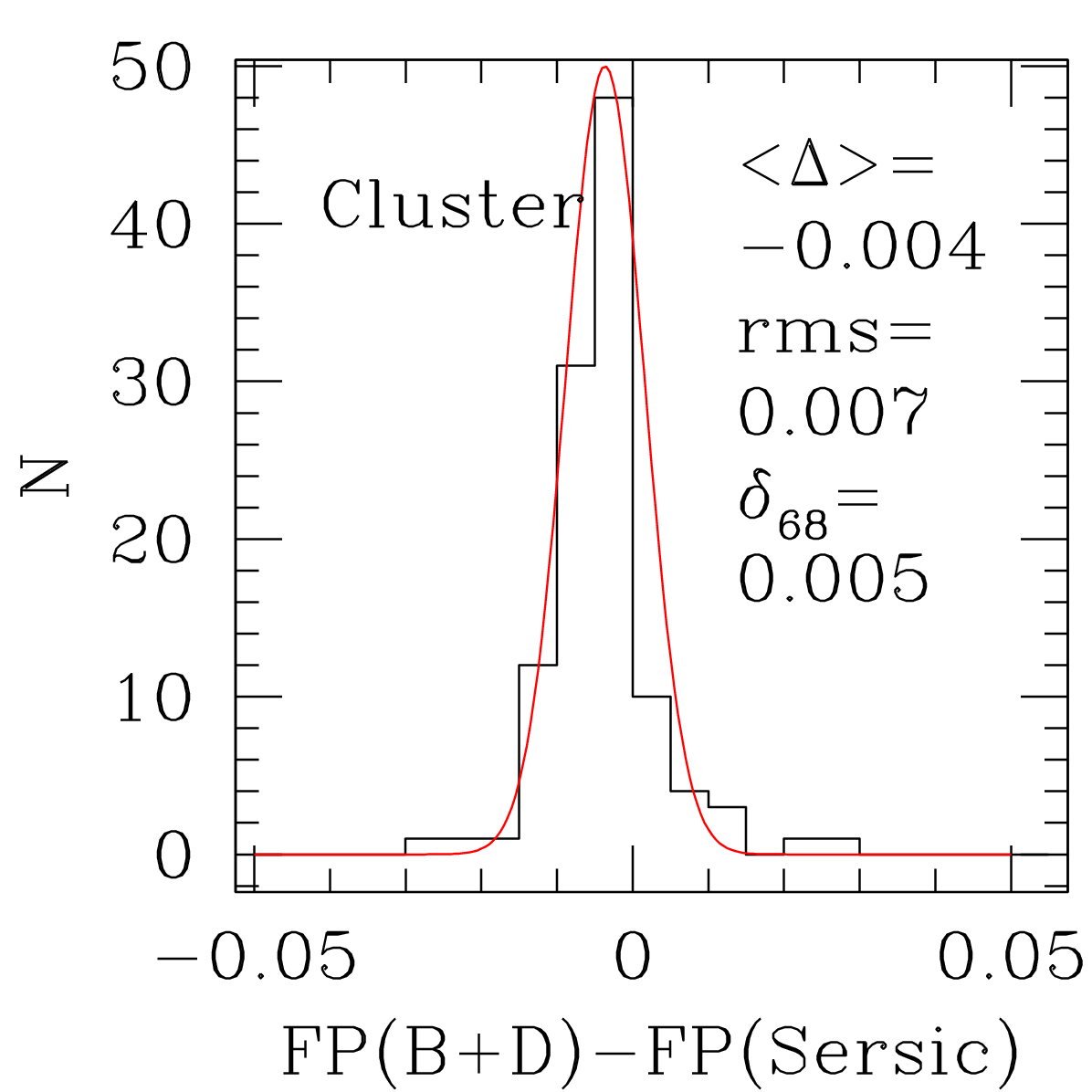,angle=0,width=4cm}}&
\vbox{\psfig{file=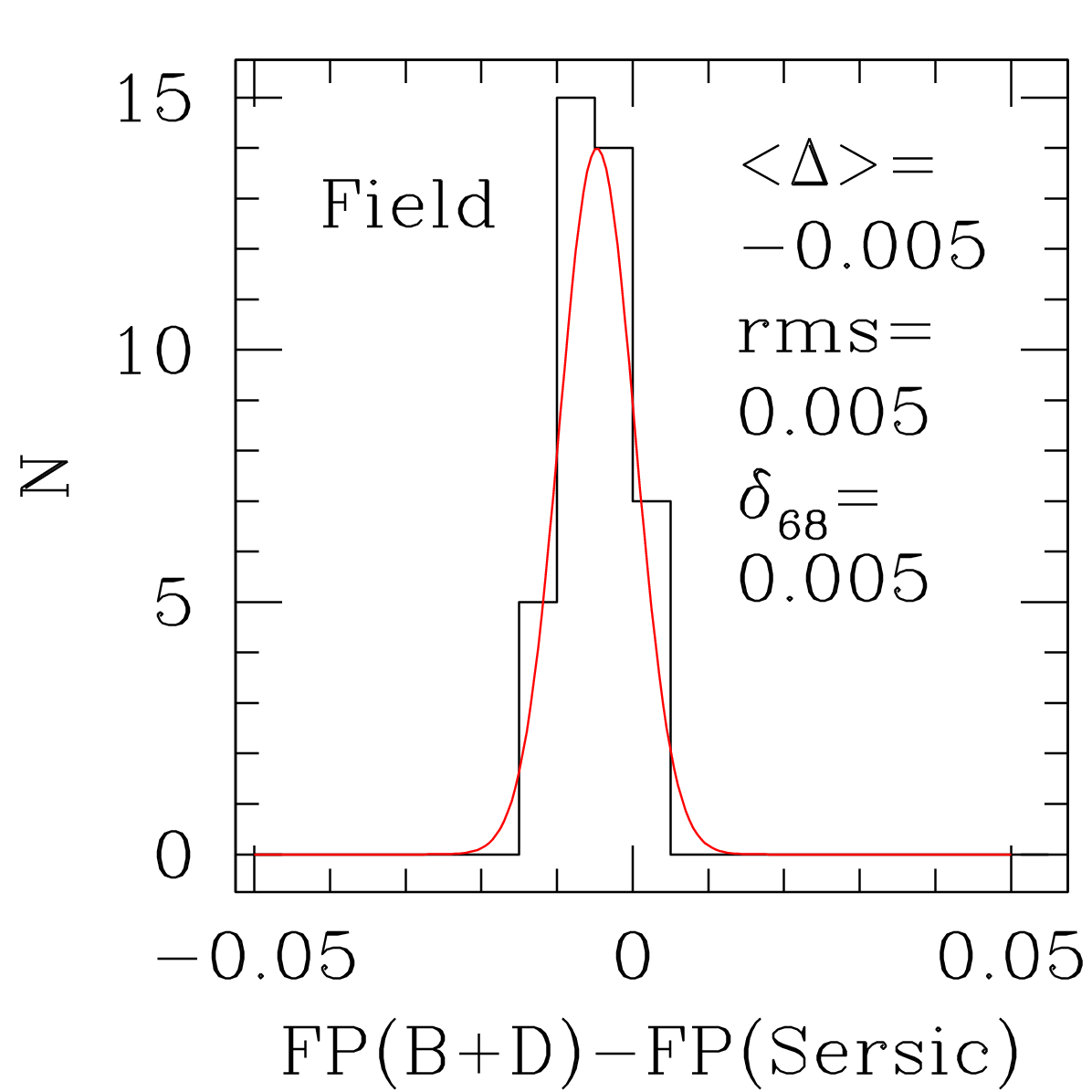,angle=0,width=4cm}}\\
\end{tabular}
\caption{The quality of the photometry parameters derived from HST images
for cluster (left) and field (right) galaxies.  
We show histograms of the differences between structural parameters
  derived from bulge plus disk (B+D) and Sersic GIM2D fits to the HST
  ACS images of the galaxies with measured velocity dispersions. The 
mean, rms, and the widths at the 68\% of the distributions are given.
\label{fig_HSTPhot} 
}
\end{figure}

Figure \ref{fig_HSTPhot} quantifies the differences $\delta \log
R_e=\log R_e(B+D)- \log R_e(Sersic)$, $\delta \langle SB_e
\rangle=\langle SB_e\rangle(B+D) -\langle SB_e\rangle(Sersic)$, and the
direction orthogonal to the FP, $\delta FP=\delta \log R_e-0.27\delta
\langle SB_e\rangle$ by showing their histograms, separately for
cluster and field galaxies.

In summary, the median
  differences are small (the Sersic $R_e$ are 9\% larger, the Sersic
  effective surface brightnesses are $\approx 0.13$ mag brighter). The
  widths at the 68\% of the distributions are $\delta_{68} \log
  R_e\sim 0.07$, $\delta_{68} \langle SB_e \rangle\sim 0.24$, and
  $\delta_{68} FP\sim0.005$ for cluster and (slightly smaller for)
  field galaxies with measured velocity dispersions. Given the quality
  of our HST ACS images, we conclude that we measure the structural
  parameters of galaxies with a precision similar to that of local galaxies
\citep{Saglia97b,Dejong2004}.


 For the remaining clusters with only ground-based images, we
  derive the structural parameters as described above
  \citep{Simard09}, i.e., by fitting an $R^{1/4}$ bulge plus an
  exponential disk 2D model to the I-band VLT deep images that were
  obtained in excellent seeing conditions. Circularized
  half-luminosity radii are derived from curves of growth constructed
  from the best fits as described in Appendix \ref{app_ReCir}. In
general, simulations show that the structural parameters derived from
the fits to VLT images are of reasonably good precision when nearly-isolated
galaxies (i.e., those for which the segmentation area has little
contamination by nearby objects) are considered. Statistical errors
smaller than 0.27 mag in total magnitudes and smaller than 0.36 dex in
$\log R_e$ are derived, in addition to systematic errors smaller than
0.15 mag and 0.2 dex, respectively, if bright objects (Imag$<22.5$)
are examined \citep{Simard09}. The galaxies in our sample are
typically at least one magnitude brighter than this limit.

\begin{figure}
\begin{tabular}{cc}
\vbox{\psfig{file=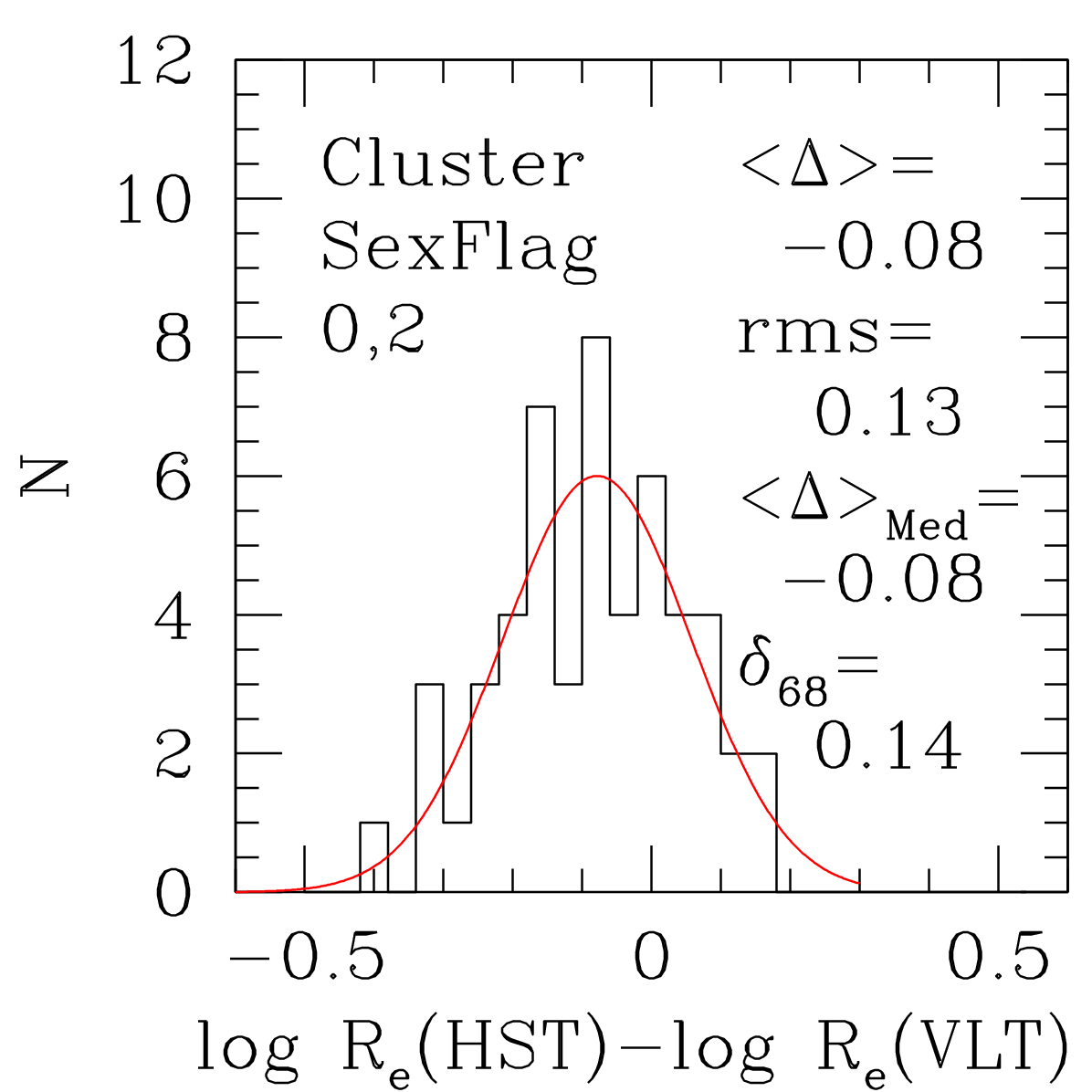,angle=0,width=4cm}}&
\vbox{\psfig{file=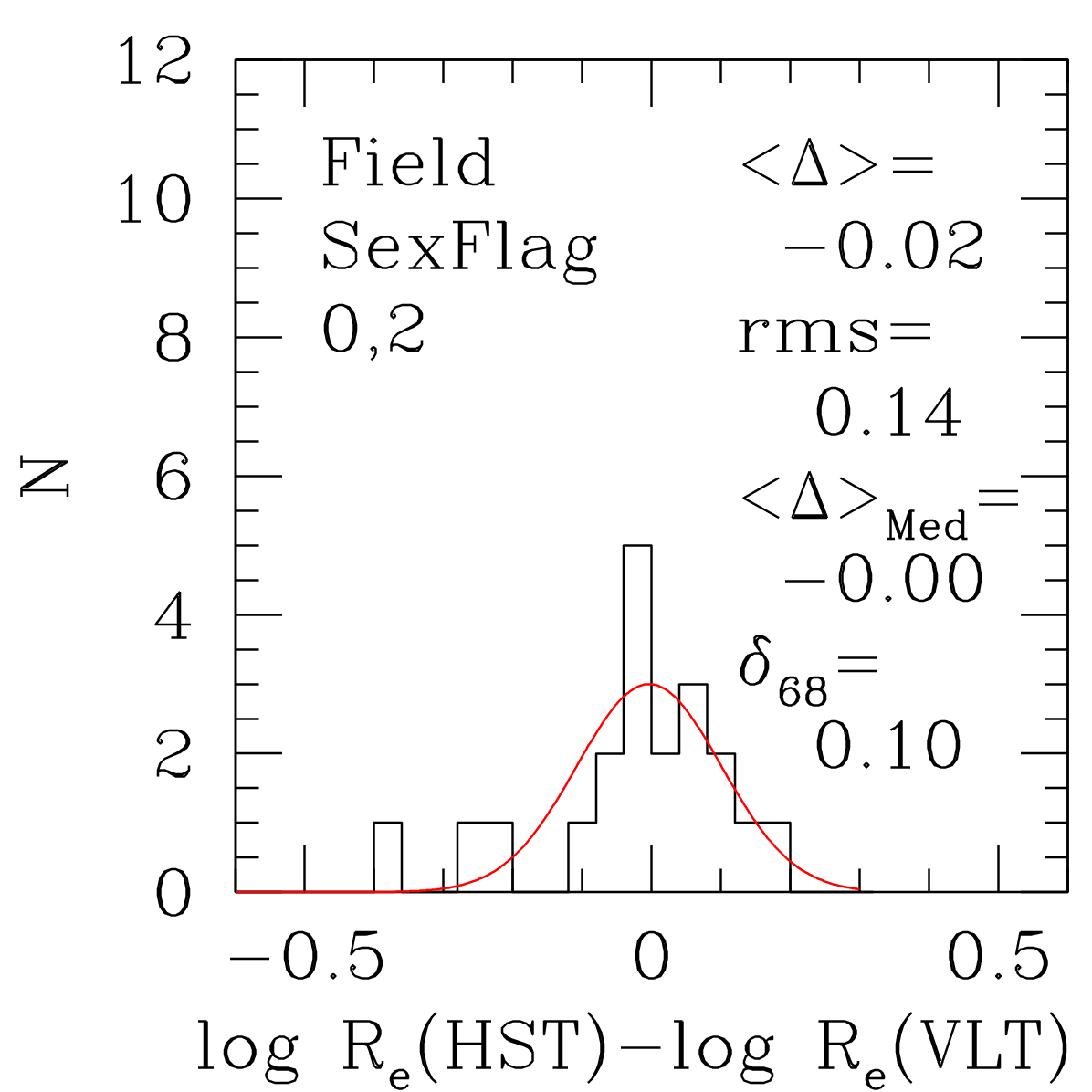,angle=0,width=4cm}}\\
\vbox{\psfig{file=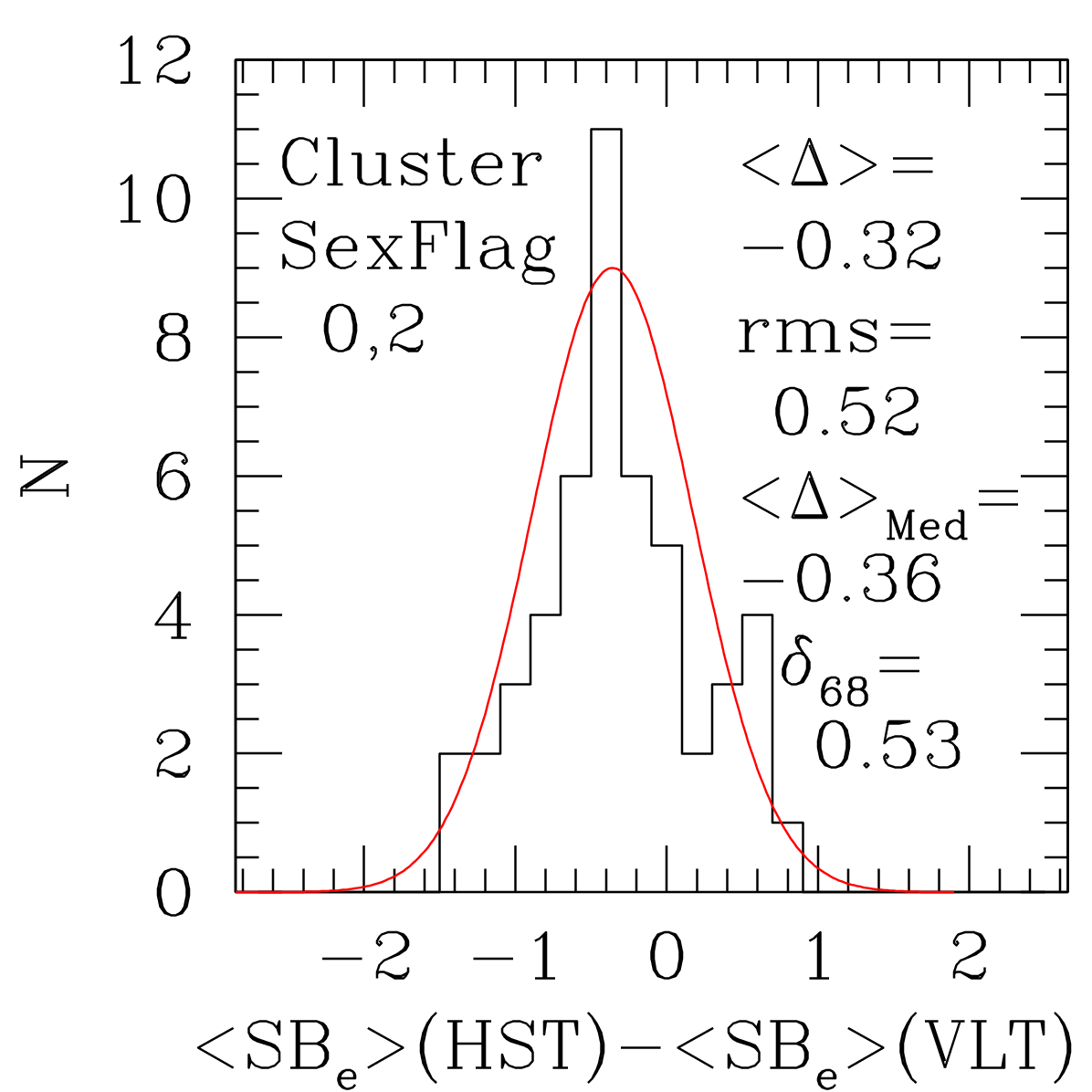,angle=0,width=4cm}}&
\vbox{\psfig{file=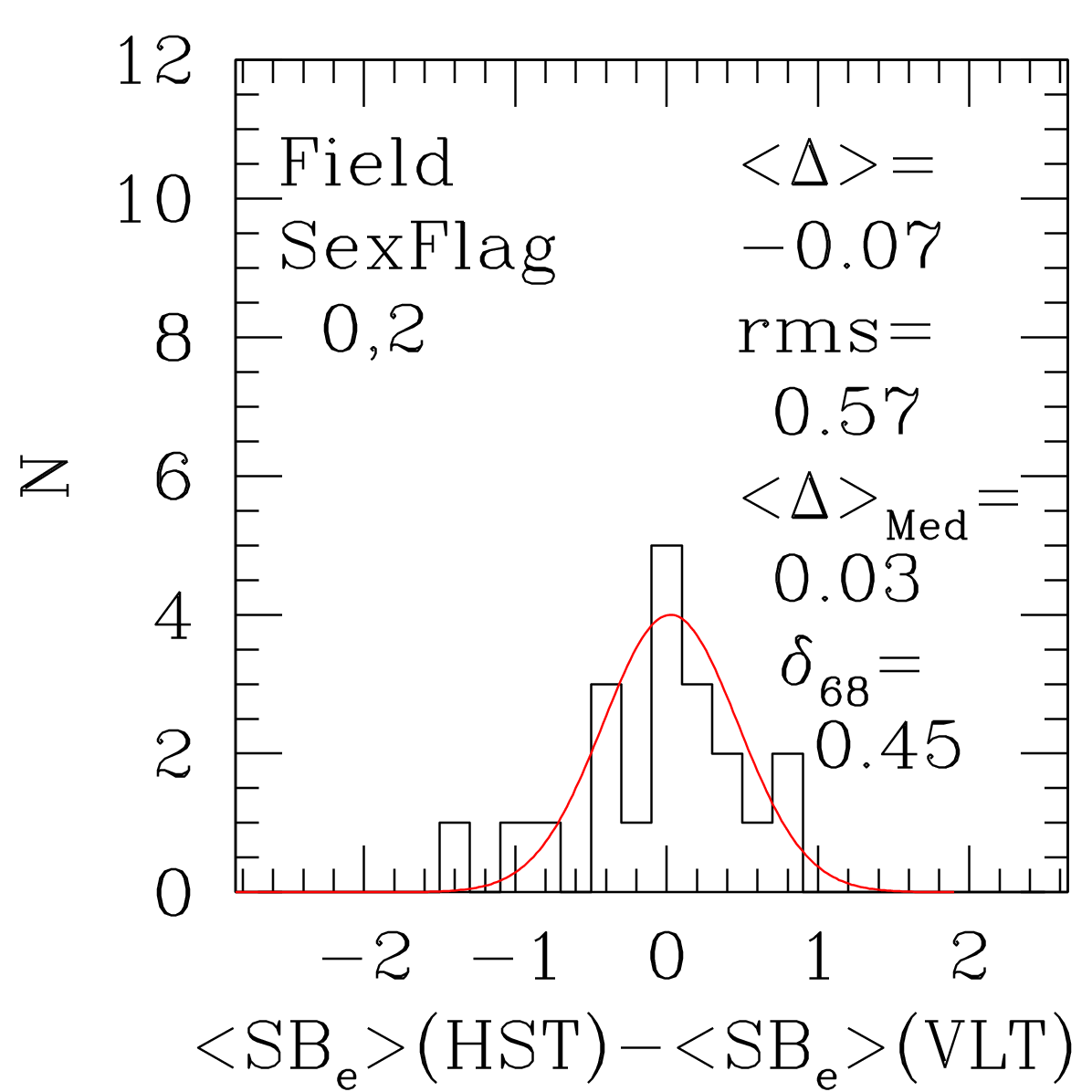,angle=0,width=4cm}}\\
\vbox{\psfig{file=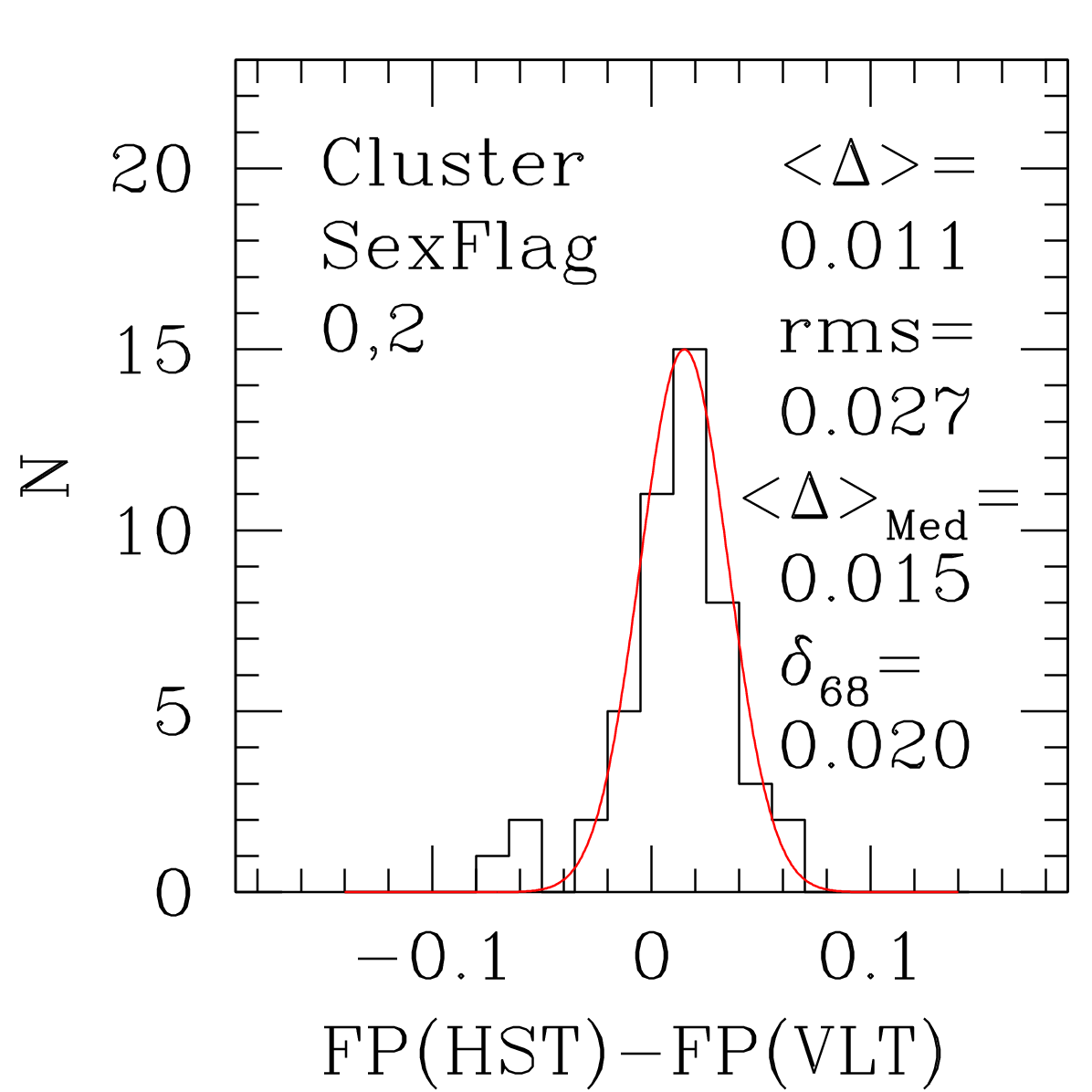,angle=0,width=4cm}}&
\vbox{\psfig{file=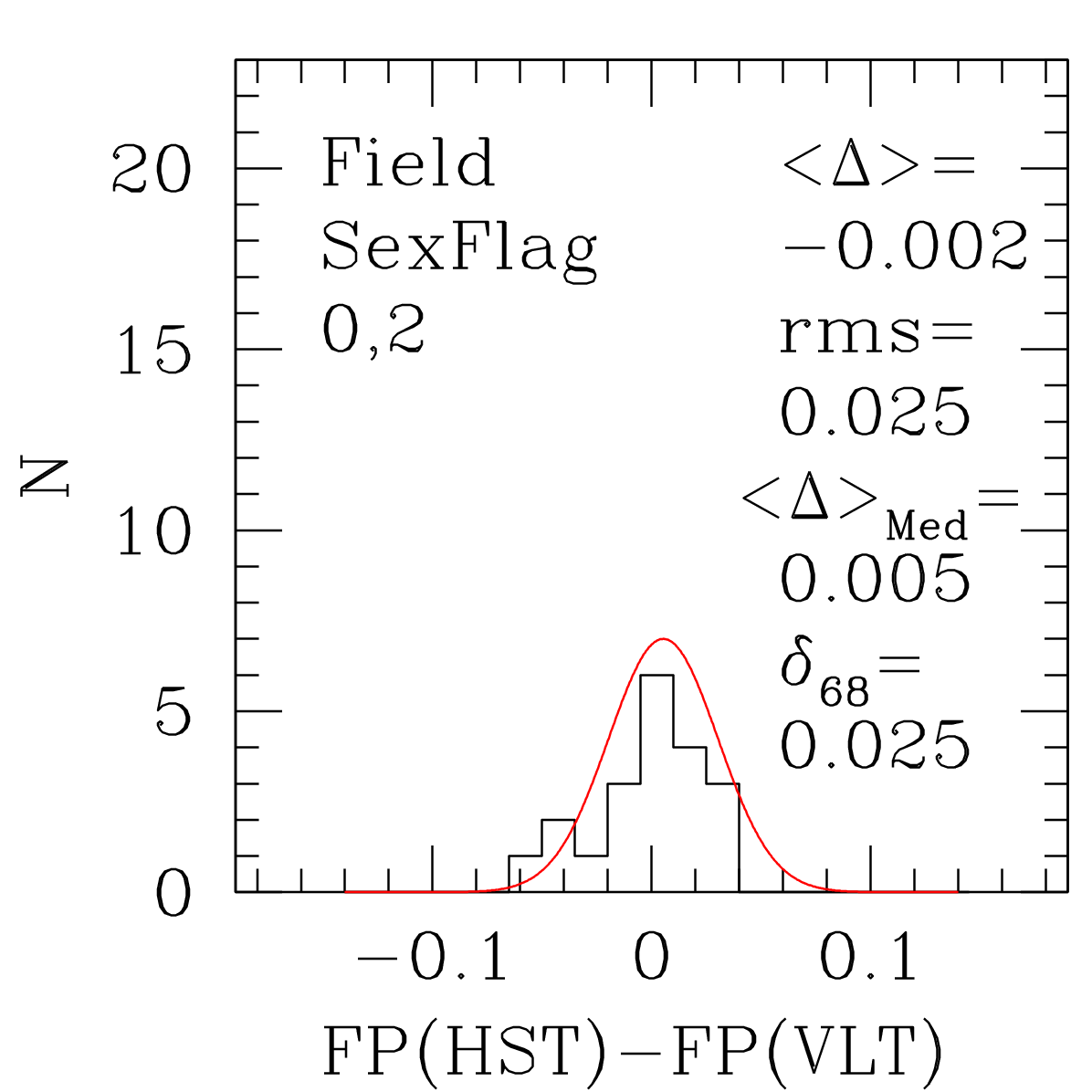,angle=0,width=4cm}}\\
\end{tabular}
\caption{{The quality of the photometry parameters derived from VLT images
for cluster (left) and field (right) galaxies. }
Histograms of the differences between structural parameters
  derived from bulge plus disk GIM2D fits to the HST ACS and VLT I
  band images of the isolated, undisturbed galaxies with measured
  velocity dispersions. The mean, rms and the widths at the 68\% of
  the distributions are given. \label{fig_VLTPhot} }
\end{figure}

{The bottom panels of Figs. \ref{fig_checkRe}, \ref{fig_checkSBe},
  and \ref{fig_checkFP} show the comparison of the VLT-derived
  structural parameters with the HST derived structural parameters as
  a function of galaxy flattening, while} Fig. \ref{fig_VLTPhot} shows
the histograms of the differences $\delta \log R_e=\log R_e(HST)- \log
R_e(VLT)$, $\delta \langle SB_e \rangle=\langle SB_e\rangle(HST)
-\langle SB_e\rangle(VLT)$, and $\delta FP=\delta \log R_e-0.27\delta
\langle SB_e\rangle$ for objects with measured velocity dispersions
where HST images are also available.  For cluster objects that are
isolated or have only relatively small companions \citep[SExtractor
flags 0 or 2,][]{BA96}, the comparison is reasonable, with median
$\langle\delta \log R_e\rangle_{med}\sim-0.08$, $\delta_{68} \log
R_e\sim 0.14$, (i.e., VLT half-luminosity radii are on average
  $20$\% larger than HST $R_e$ with $\le 25$\% scatter), and median
difference $\langle\delta \langle SB_e \rangle\rangle_{med}
\sim-0.32$, $\delta_{68} \langle SB_e \rangle\sim 0.53$ (i.e., VLT
effective surface brightnesses are on average $ 0.32$ mag
brighter than those from HST $\langle SB_e\rangle$ with $\le 0.53$ mag
scatter). The errors $\delta \log R_e$ and $\delta \langle SB_e
\rangle$ are correlated, with minimal scatter in the direction almost
orthogonal to the FP, i.e., $\delta FP=\delta \log R_e-0.27\delta
\langle SB_e \rangle$ and $\delta_{68} FP\sim 0.025$ and there is a small median
shift. No trend with redshift is seen.  These values agree with or are
of higher precision than those derived from simulations (see above).
Very similar results are obtained for field objects. Therefore, the
VLT dataset can be merged with the HST-based one to study the
evolution of the FP (Sect. \ref{sec_FP}).

The systematic and random errors increase dramatically if objects with
sizable companions (VLT SExtractor flag 3) are considered. In these
cases, the VLT segmentation areas fitted by GIM2D are heavily
contaminated by the companions. As a consequence, $R_e(VLT)$ and VLT
total magnitudes are systematically larger and brighter, respectively,
than those derived from HST fits. There are 38 cluster and 10 field
galaxies with early spectral type and measured velocity dispersion
that have only VLT imaging and a SExtractor flag equal to 3. Given the 
already sizeable systematics in $R_e$ detected for the 'isolated' objects, 
we refrain from attempting an iterative fit  and  just
exclude the affected galaxies from the FP analysis.

In Sect.  \ref{sec_sizeev}, we use the half-luminosity radii discussed
above to constrain the size evolution of our galaxies.   The
  high-precision ($\approx 10$\% systematic) HST half-luminosity radii
  are certainly good enough and our results are based on this dataset
  only.  A number of caveats have to be kept in mind when considering
  the VLT radii.  According to the Monte Carlo simulations discussed
by \citet[Fig.1]{Simard09}, the VLT radii of the largest galaxies of
the sample (larger than 1.8 arcsec) might underestimate the true
radii by up to 40\%. But only 2.5\% of our sample has
$R_e>1.8"$. Sizes below 0.1 arcsec are probably unreliable because of a
lack of resolution, but only 3\% of cluster galaxies and 5\% of field
galaxies fall into this category. Finally, if galaxies have strong color
gradients, our half-luminosity radii, derived from I band images
(i.e., approximately rest-frame V band at redshift 0.5 and rest-frame
B band at redshift 0.8) might be affected differentially with
redshift. However, we do not detect any significant trend with
redshift in the sizes derived from our VLT B and V band images
relative to the ones used here from the I band images.  Despite
  all these systematic differences between HST and VLT $R_e$ radii (on
  average 20\%), Sect.  \ref{sec_sizeev} shows that the size evolution
  derived from VLT $R_e$ radii is very similar.

As a last step, effective surface brightnesses were calibrated as
follows. Corrections to rest-frame Johnson B band were applied based
on the spectroscopic redshift $z$ and an interpolation of the
best-fit spectral energy distribution, according to our
photometric redshift procedure \citep{Rudnick09, Pello09}. Moreover,
the Tolman correction $(1+z)^4$ was taken into account. Finally, to be
able to compare our results with those of \citet{Wuyts04} and related
papers, we transformed effective surface brightness to surface
brightness at $R_e$ using the conversion factor valid for a pure
$R^{1/4}$ law, i.e. $I_e=\langle I_e\rangle/3.61$ and $\log \langle
I_e\rangle(L_\odot/pc^2) =-0.4(\langle SB_e\rangle-27)$.

\begin{figure}
\begin{tabular}{cc}
\vbox{\psfig{file=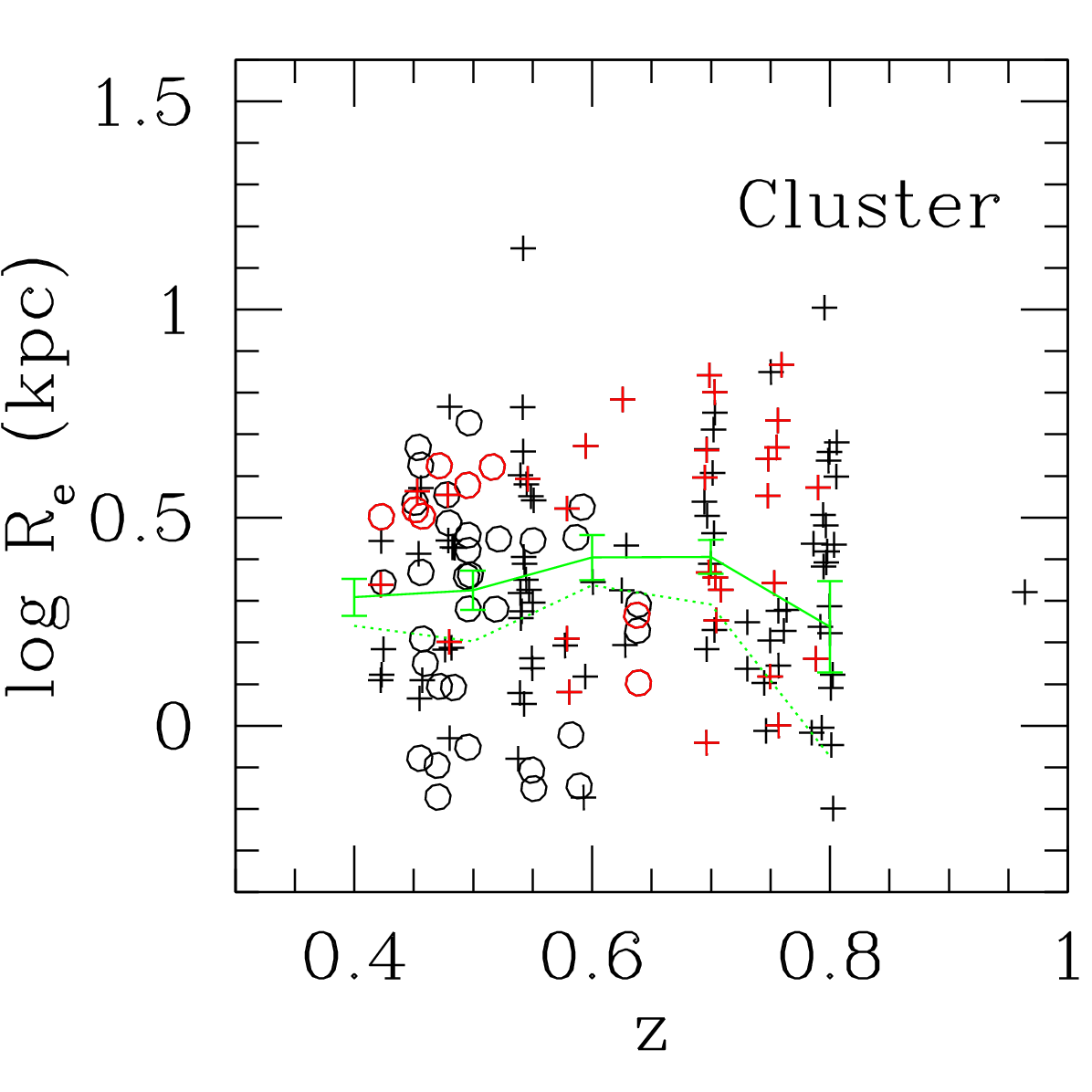,angle=0,width=4cm}}&
\vbox{\psfig{file=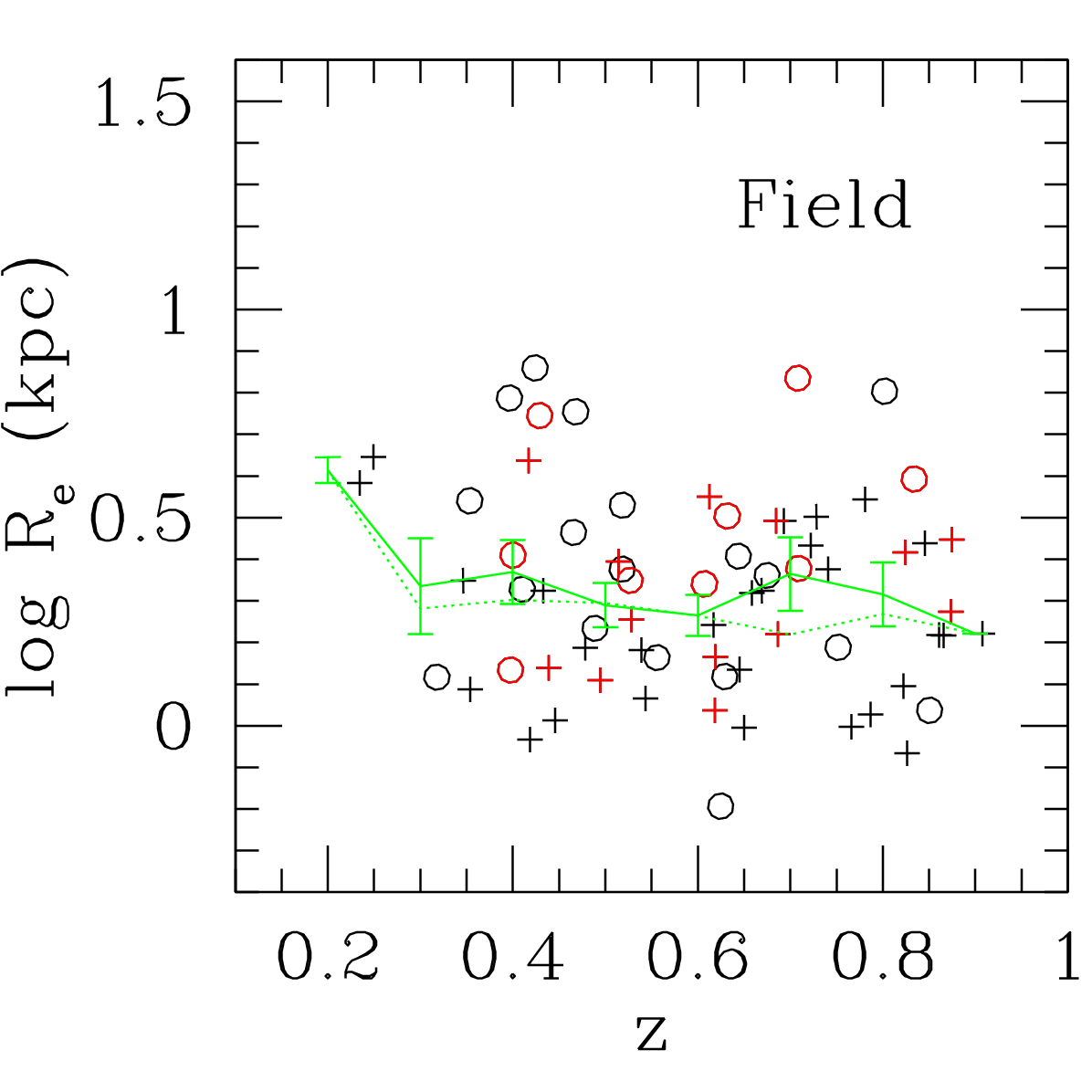,angle=0,width=4cm}}\\
\vbox{\psfig{file=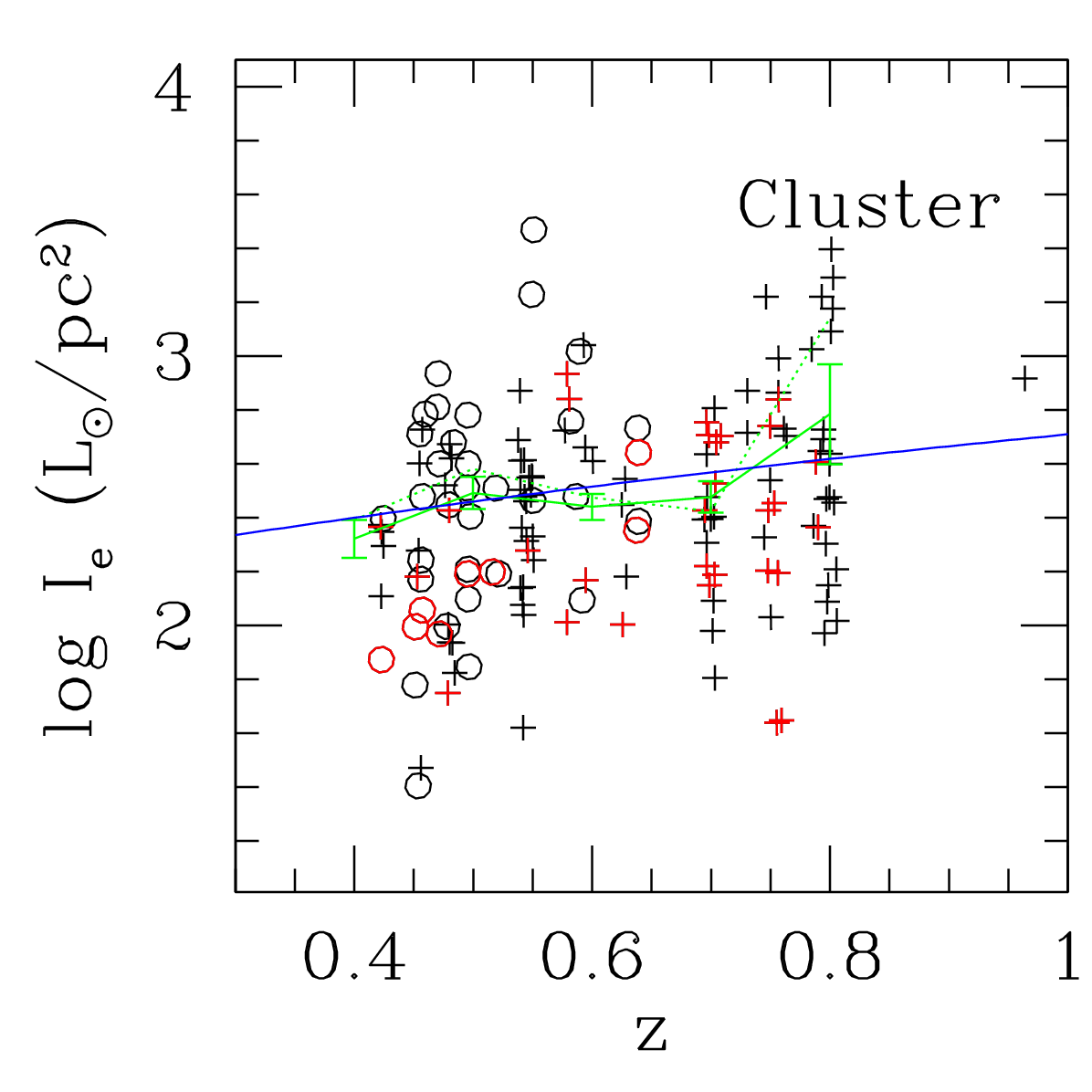,angle=0,width=4cm}}&
\vbox{\psfig{file=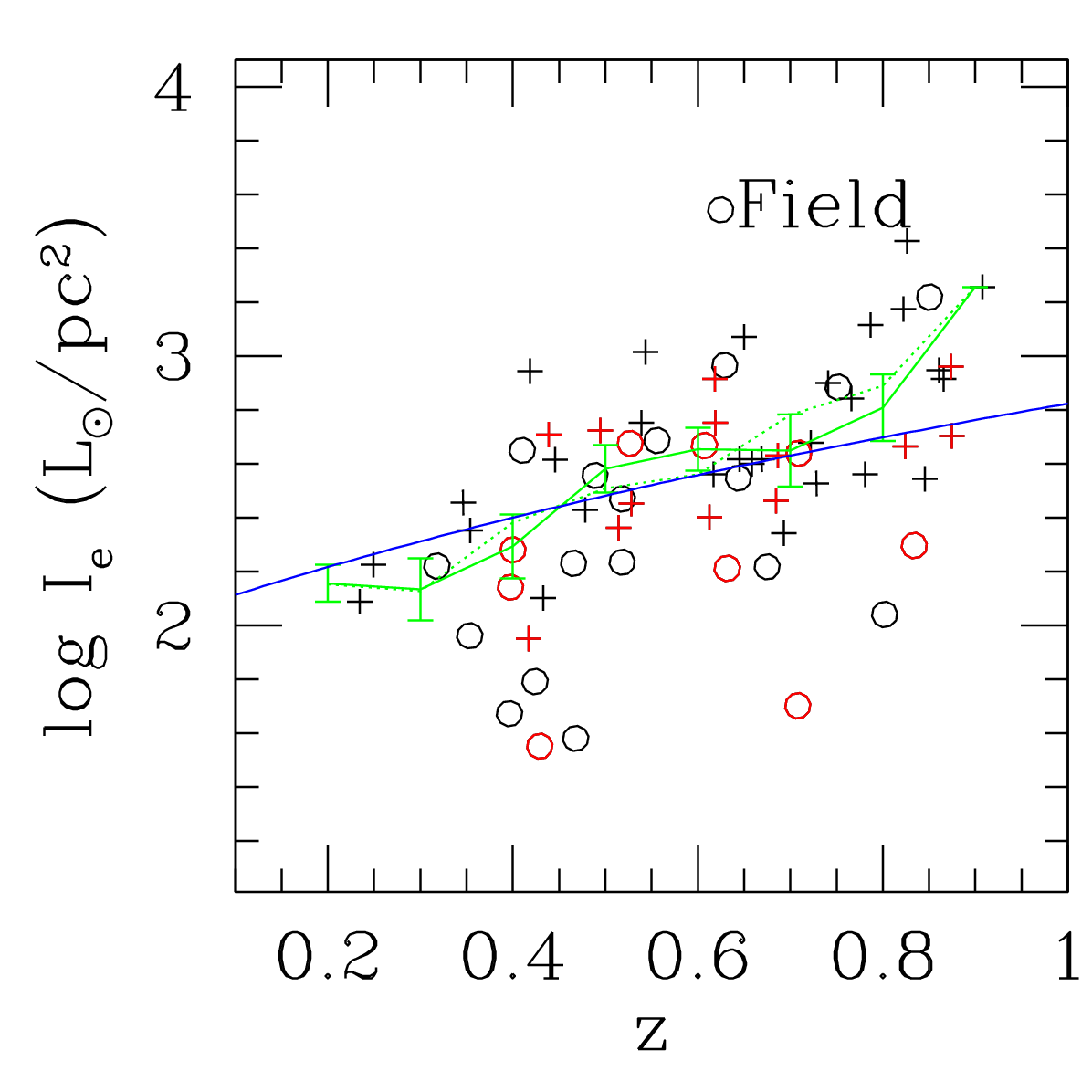,angle=0,width=4cm}}\\
\vbox{\psfig{file=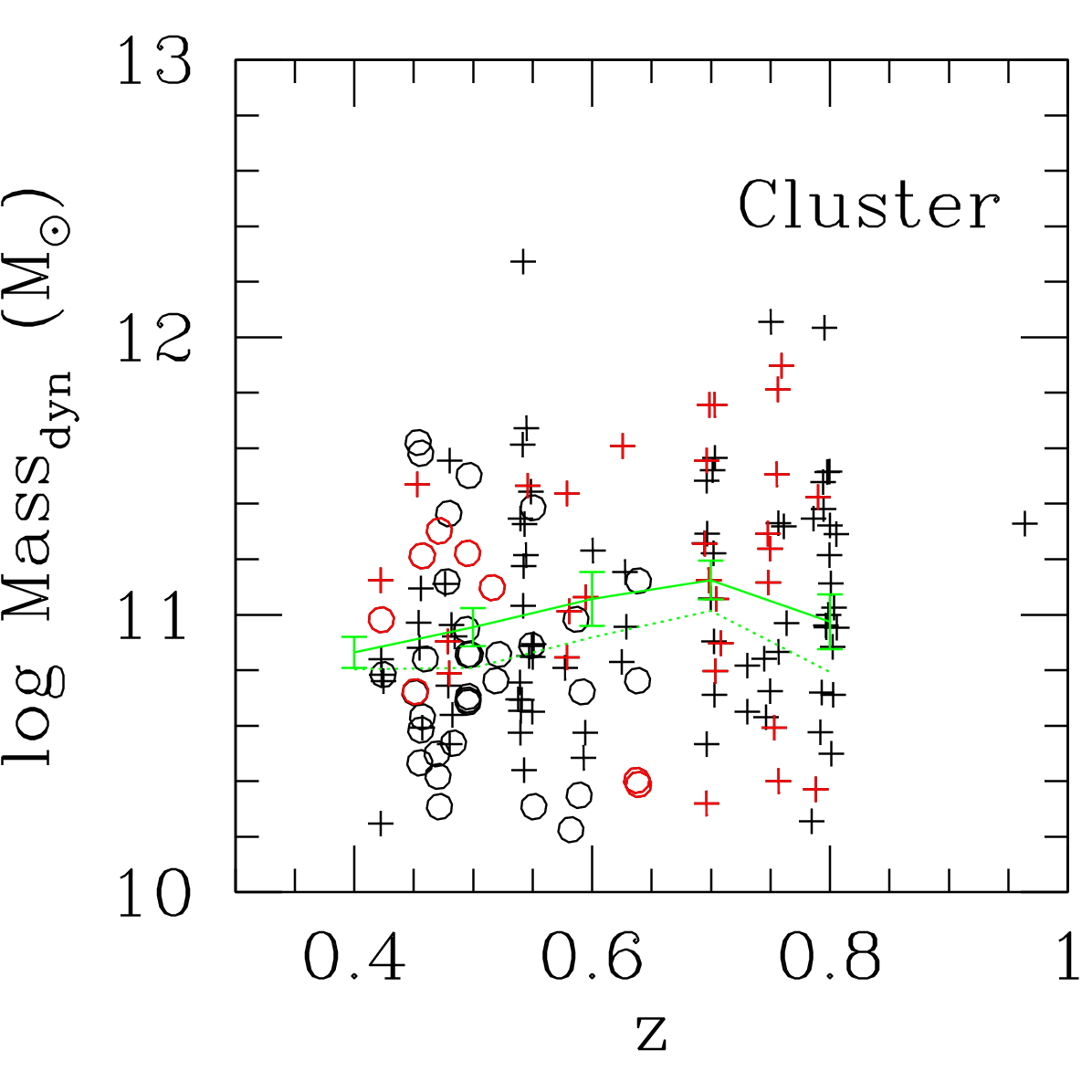,angle=0,width=4cm}}&
\vbox{\psfig{file=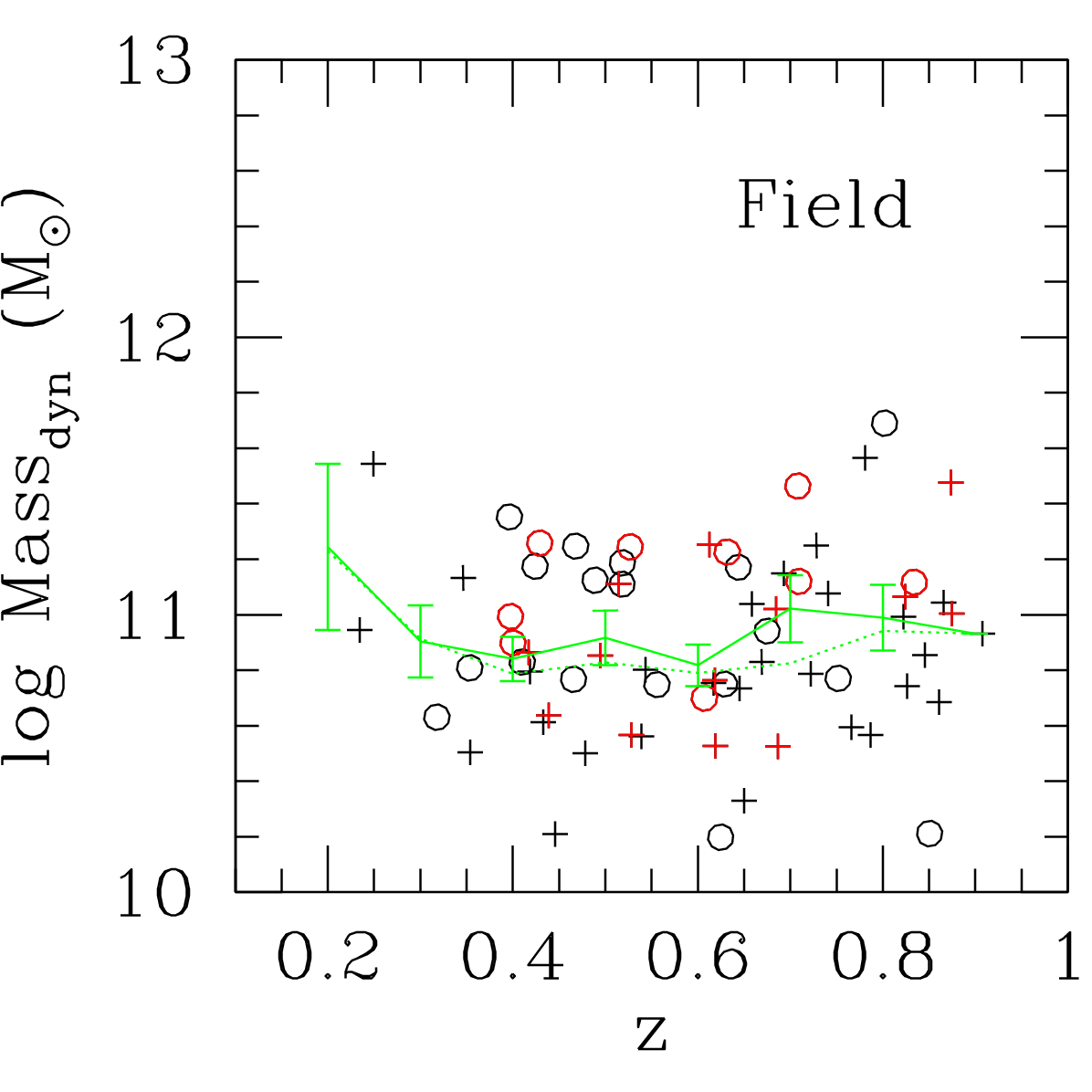,angle=0,width=4cm}}\\
\end{tabular}
\caption{{The distribution with redshift of sizes, surface luminosities, 
and dynamical  masses of the galaxy sample.} We show
the half-luminosity radii $\log R_e$ (top), effective surface 
brightness $\log I_e$ (middle), and dynamical mass $\log M_{dyn}$ (bottom) 
as a function of redshift for cluster
(left) and field (right) galaxies. Black and red points show 
spectroscopic types 1 and 2, respectively. Crosses and circles
 show galaxies with HST and VLT photometry, respectively. The
solid green lines show the mean values in 0.1 
redshift bins with the errors. The dotted lines show the averages 
obtained by weighting each galaxy with the inverse of its selection value. 
The blue lines show the mean 
luminosity evolution derived from Fig. \ref{fig_fieldML}: 
$\log I_e=2.4+1.66\log\frac{1+z}{1.4}/0.83$ for cluster galaxies and 
$\log I_e=2.4+2.27\log\frac{1+z}{1.4}/0.83$ for the 
field. 
\label{fig_ReIez}}
\end{figure}

Figure \ref{fig_ReIez} shows $\log R_e$, $\log I_e$, and dynamical mass
$\log M_{dyn}$  as a function of
redshift.  Following \citet{Vandokkum07}, we compute dynamical masses to be
\begin{equation}
\label{eq_mass}
M_{dyn}=5R_e\sigma^2/G=1.16\times10^6(R_e/kpc),
\times(\sigma/kms^{-1})^2M_\odot
\end{equation}
(see also Sect. \ref{sec_mass}). The mean size of the half-luminosity
radius remains approximately constant at values of $\approx 2.5$ kpc.
In contrast, the surface brightness at $R_e$ increases on average by a
factor 2 from redshift 0.4 (where it is $\approx 250 L_\odot/pc^2$) to
redshift 0.8. This matches the differential luminosity evolution
inferred from the FP zero point evolution with redshift
(see Sect.  \ref{sec_FPclus}). Weighting each galaxy with the inverse
of its selection value to correct for incompleteness (see Sect.
\ref{sec_selection}) pushes the sample averages of $\log R_e$ and
$\log I_e$ to slightly lower and higher values, respectively. As for
the velocity dispersions, the effect is however on the order of the
error in the averages. We note that the situation changes when we
consider the size evolution of mass-selected samples (see Sect.
\ref{sec_sizeev}). We study cluster galaxies with dynamical masses
higher than $1.5\times10^{10} M_\odot$ and field galaxies with
dynamical masses higher than $2.5\times10^{10} M_\odot$.  Both cluster
and field galaxies have on average a dynamical mass of
$10^{11}M_\odot$.

\subsection{Selection function}
\label{sec_selection}

\begin{figure}
\begin{tabular}{c}
\vbox{\psfig{file=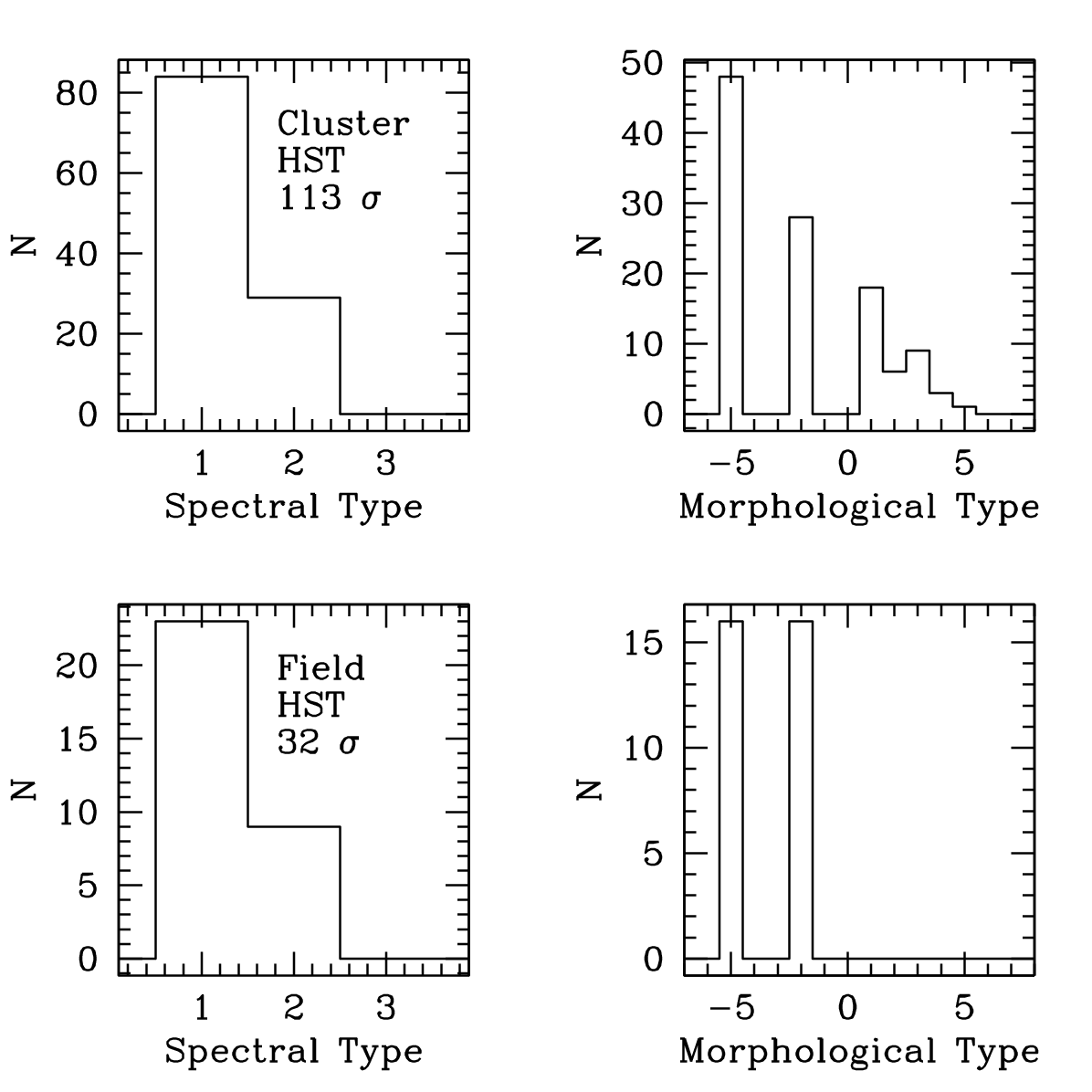,angle=0,width=8cm}}\\
\vbox{\psfig{file=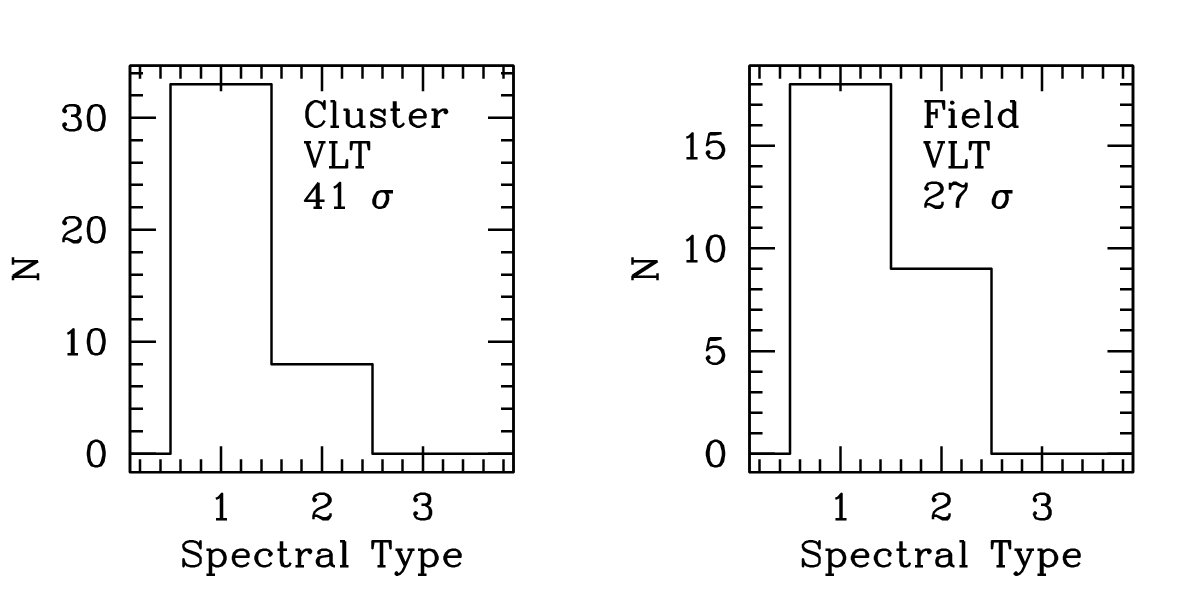,angle=0,width=8cm}}\\
\end{tabular}
\caption{Statistics of the sample of galaxies with measured velocity
  dispersions and photometric parameters.\label{fig_sigmastat}}
\end{figure}

Figure \ref{fig_sigmastat} describes the final sample. We measured
velocity dispersions for 113 cluster and 41 field spectral early-type
galaxies with HST photometry, and 41 cluster and 27 field galaxies
with only VLT good photometry.  A large fraction of galaxies with HST
photometry also have early-type morphology: 67\% of the objects in
clusters and 78\% in the field have been classified as either Es or
S0. Moreover, 77\% of galaxies in clusters and 68\% in the field do
not exhibit [OII] emission, being of spectral type 1.

\begin{figure}
\psfig{file=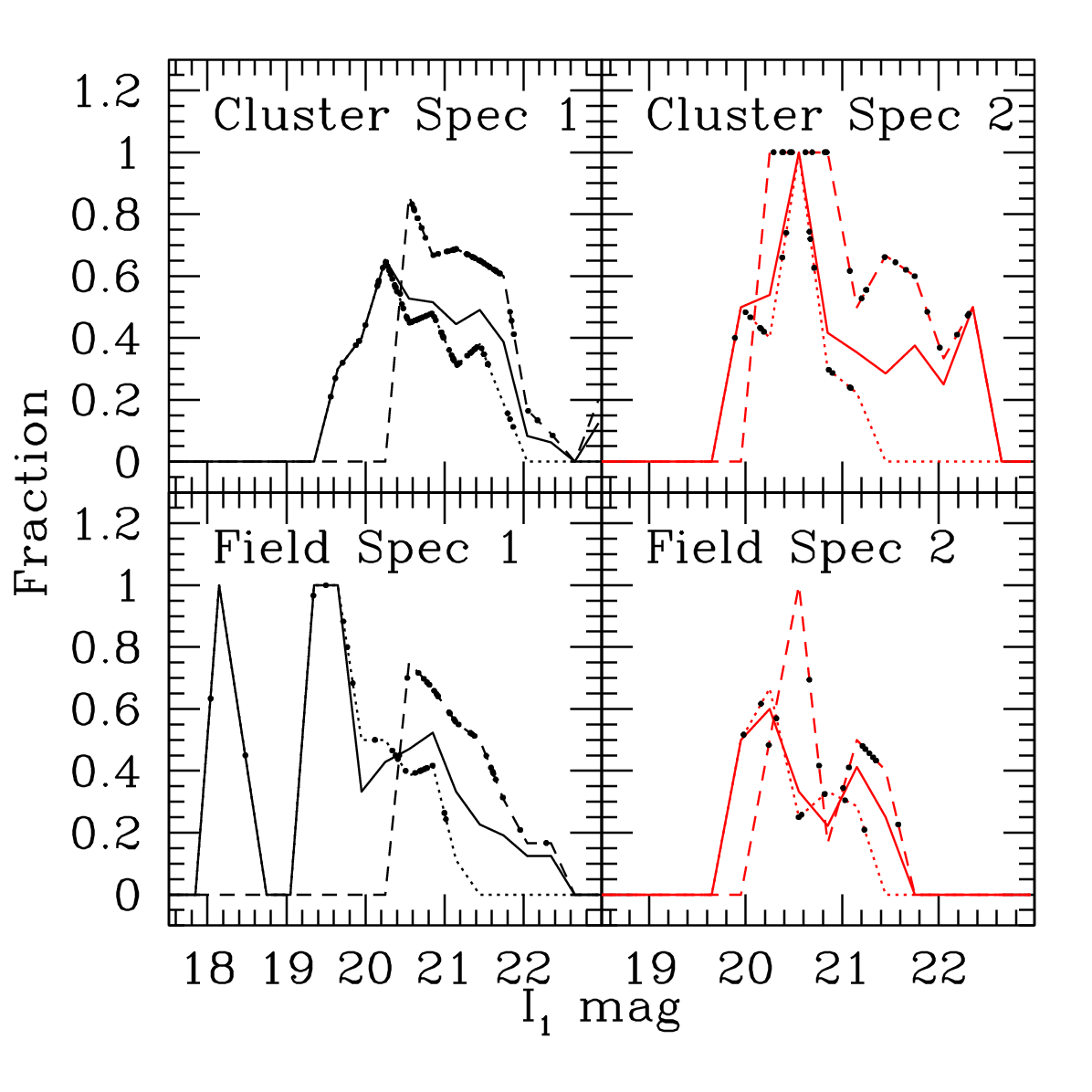,angle=0,width=8cm}
\caption{{The relative completeness functions.}
The fraction of galaxies with an observed spectrum of spectroscopic 
type 1 or 2 for which we could measure velocity dispersions and obtain 
reliable photometric structural parameters. This relative
 completeness is shown for the 
clusters (top row) and the field (bottom row) as a function of galaxy 
magnitude in the I band in a 1 arcsec radius aperture. Colors code the spectral 
type (black: 1; red: 2). 
The full lines show the full redshift range, the dotted lines galaxies 
with $z<0.6$, the dashed lines galaxies with $z\ge0.6$. The dots show 
the magnitudes of the single galaxies and the assigned completeness weight.
\label{fig_completeness}}
\end{figure}

To quantify the selection function of our sample, we assign a
selection probability $P_S$ to each galaxy. This is computed in two
steps.  First, the $\sigma$-completeness probability $P_\sigma$ of the
velocity dispersion measurements is determined. This is shown in Fig.
\ref{fig_completeness}.  For each given spectral type, we compute the
ratio of the number of galaxies with a measured velocity dispersion
and reliable photometric structural photometry (see above) to the
number of galaxies with a spectrum in a given magnitude bin. In a way
similar to \citet{Milvang08}, we use the I band magnitude in a 1
arcsec radius aperture $I_1$. We compute these curves separately for cluster
and field galaxies, and for galaxies with redshifts either lower than or 
equal to
or higher than 0.6.  Finally, we assign the probability
$P_\sigma(I_1,z,ST,F/C)$ to each galaxy by linearly interpolating the
appropriate curve for its redshift $z$, spectral type $ST$, and field
or cluster environment (F/C) as a function of magnitude.  The
$\sigma$-completeness is high at bright magnitudes and declines toward
fainter objects. In this regime, the $\sigma$ completeness is also slightly
higher for higher redshift galaxies, where the exposure times are
longer. The differences between cluster and field galaxies are not as
pronounced.

As a second step, following \citet{Milvang08} we consider the total
number of spectroscopically targeted galaxies $N_T$ { \citep[drawn
  from a photometric magnitude-limited sample far deeper than that considered
  here; see][]{Milvang08}} in a given magnitude bin, separately for
each of the 19 fields we observed.  In the given field, we then
consider the number of galaxies for which we were able to derive a
secure redshift $N_{R}$ {\citep[with a success rate of essentially
  100\%; see][]{Milvang08}}, the number of galaxies spectroscopically
found to be members of any cluster $N_C$, and the number of galaxies
found in the field, $N_F=N_R-N_C$. We construct the ratio functions
$R_C=\frac{N_C}{N_T}$ and $R_F=\frac{N_R-N_C}{N_T}$ and interpolate
them at the magnitude of each galaxy. Finally, we assign to each
galaxy the selection probability $P_S(Cluster)=P_\sigma\times R_C$ or
$P_S(Field)=P_\sigma\times R_F$ if the galaxy belongs to a cluster or
to the field.

\begin{figure}
\psfig{file=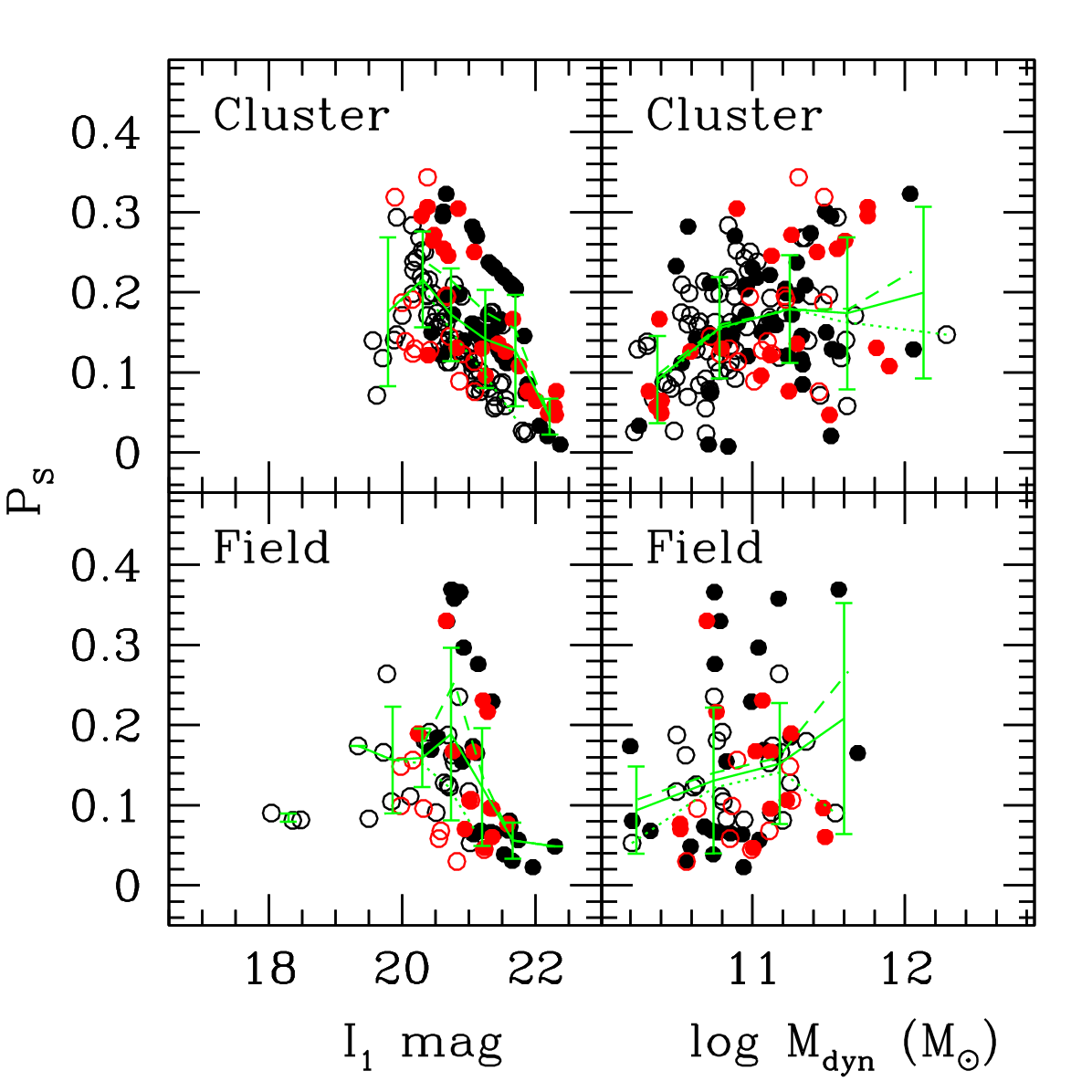,angle=0,width=8cm}
\caption{{The completeness function of the galaxy sample.} 
The completeness weight for the galaxies with a velocity dispersion 
for clusters (top row) and the field (bottom row). Left: as a function of 
galaxy magnitude in the I band in a 1 arcsec radius aperture; 
Right: as a function 
of dynamical mass. Colors code the spectral type (black: 1; red: 2). 
Filled circles 
show galaxies with redshift either equal or 
higher than 0.6, open circles galaxies with 
redshift lower than 0.6. The green full lines with error bars show the bin 
averages and rms over the full redshift range. The dotted lines refer to 
the sample  with $z<0.6$, the dashed lines to the sample with $z\ge0.6$. 
\label{fig_finalcompleteness}}
\end{figure}

Figure \ref{fig_finalcompleteness} shows the resulting probabilities
as a function of $I_1$ and dynamical mass (see Eq. \ref{eq_mass} and
Sect. \ref{sec_mass}). In clusters, we sample 10 to 30\% of the
spectral early-type population. {The selection probability is
  almost flat as a function of mass for $M_{dyn}\ge 4\times 10^{10}
  M_\odot$. This is above the stellar mass completeness limit of our
  parent stellar catalogue. In this mass range, the selection
  probability has no dependence on the galaxy colors. We become
  progressively more incomplete at lower masses, where we sample
  just 10\% of the population.} The effect is less pronounced at
higher redshifts. In the field, the average completeness is lower
($\approx 15$ \%) and similar trends are observed. In general,
$P_\sigma$ traces $P_S$ quite well, with $P_S\approx (0.29\pm0.12)
P_\sigma$. In the abstract and in the following, we quote first
results obtained ignoring selection effects, and then illustrate the effect
of the selection correction.

Tables \ref{tab_cluster} and \ref{tab_field} summarize the velocity
dispersions and the structural parameters of the cluster and field
galaxies, respectively. For each galaxy, we list its name
\citep{White05}, the number of the cluster to which it belongs (if it
is a cluster galaxy, see Table \ref{tab_ZP} for the correspondence
between cluster name and number), spectroscopic redshift and type
\citep{Halliday04, Milvang08}, raw and aperture-corrected velocity
dispersion $\sigma_{mes}$ and $\sigma_{cor}$ with estimated
statistical error, circularized half-luminosity radius $R_e$, surface
brightness $\log I_e$ in the rest-frame B-band, and, when HST images
are available, morphological type.  When VLT-only images are
available, the morphological flag is set to be $*$ when the SExtractor
flag is equal to 3, i.e., when the photometric parameters are expected
to be contaminated by companions.  Moreover, we list the selection
probabilities $P_S$ and the stellar masses (see Sect.
\ref{sec_mass}).  In addition, Table \ref{tab_SerVLT} gives the
circularized $R_e$ and $\log I_e$ derived from Sersic fits (to HST
images) and bulge+disk fits to VLT images for the galaxies for which
both HST and VLT images are available.

\begin{table*}[h!]
\caption{The FP parameters of cluster galaxies. \label{tab_cluster}}
\begin{tabular}{ccccccccccccc}
\hline
Name & $N_{clus}$ & z & S & $\sigma_{mes}$ & $\sigma_{cor}$ & d$\sigma$ &
$R_e$ & $\log I_e$ & T & $P_S$ & $\log M_*$ &d$\log M_*$ \\
    &    &  &  Type    & (km/s)         & (km/s)           & (km/s)                &
(kpc) & $\log L_\odot/pc^2$ & Type & & $\log M_\odot$ & $\log M_\odot$   \\
\hline
 EDCSNJ1040403-1156042           &           9           &       0.702           &           1           &       161.3           &       167.1           &        13.6           &       5.147           &       2.092           &          -5           &       0.199           &       11.54           &        0.06          \\        
 EDCSNJ1040407-1156015           &           9           &       0.703           &           1           &       155.9           &       161.5           &        15.4           &       1.698           &       2.807           &          -2           &       0.149           &       11.28           &        0.05          \\        
 EDCSNJ1040346-1157566           &           9           &      0.7024           &           1           &       148.9           &       154.3           &        17.8           &         2.9           &       2.404           &          -5           &       0.161           &       11.25           &        0.08          \\        
 EDCSNJ1040396-1155183           &           9           &      0.7046           &           2           &       226.2           &       234.4           &        13.4           &       1.789           &        2.68           &          -5           &       0.096           &       11.09           &        0.09          \\        
 EDCSNJ1040356-1156026           &           9           &      0.7081           &           2           &       172.7           &       178.9           &         8.4           &       2.123           &       2.703           &          -5           &       0.304           &       11.35           &        0.12          \\
\hline
\end{tabular}
\end{table*}

\addtocounter{table}{-1}
\begin{table*}[h!]
\caption{Continued}
\begin{tabular}{ccccccccccccc}
\hline
Name & $N_{clus}$ & z & S & $\sigma_{mes}$ & $\sigma_{cor}$ & d$\sigma$ &
$R_e$ & $\log I_e$ & T & $P_S$ & $\log M_*$ &d$\log M_*$ \\
    &    &   & Type    & (km/s)         & (km/s)           & (km/s)                &
(kpc) & $\log L_\odot/pc^2$ & Type & & $\log M_\odot$ & $\log M_\odot$   \\
\hline
 EDCSNJ1232313-1250327           &           4           &      0.5496           &           1           &       162.6           &       167.7           &        18.5           &       1.373           &       2.549           &          -5           &       0.084           &       10.82           &        0.12          \\        
 EDCSNJ1232317-1249275           &           4           &       0.542           &           1           &       138.2           &       142.5           &        13.1           &       4.564           &       2.143           &           1           &       0.238           &       11.21           &        0.10          \\        
 EDCSNJ1232309-1249408           &           4           &      0.5485           &           1           &       251.6           &       259.5           &        22.9           &       3.559           &       2.485           &          -2           &       0.071           &       11.57           &        0.05          \\        
 EDCSNJ1232303-1251092           &           4           &      0.5428           &           1           &       140.8           &       145.2           &        25.1           &       1.128           &       2.613           &          -2           &       0.085           &       10.59           &        0.11          \\        
 EDCSNJ1232303-1251441           &           4           &        0.55           &           1           &       169.9           &       175.2           &        19.8           &       1.976           &       2.331           &           3           &       0.103           &        10.8           &        0.11          \\
\hline
\end{tabular}
\end{table*}

\addtocounter{table}{-1}
\begin{table*}[h!]
\caption{Continued}
\begin{tabular}{ccccccccccccc}
\hline
Name & $N_{clus}$ & z & S & $\sigma_{mes}$ & $\sigma_{cor}$ & d$\sigma$ &
$R_e$ & $\log I_e$ & T & $P_S$ & $\log M_*$ &d$\log M_*$ \\
    &    &    & Type & (km/s)         & (km/s)           & (km/s)                &
(kpc) & $\log L_\odot/pc^2$ & Type & & $\log M_\odot$ & $\log M_\odot$   \\
\hline
 EDCSNJ1059107-1253020           &          13           &      0.4552           &           1           &       169.2           &       173.8           &          21           &      0.8354           &       2.711           &           -           &       0.080           &       10.53           &        0.11          \\        
 EDCSNJ1059096-1253197           &          13           &       0.455           &           2           &         214           &       219.8           &        17.6           &       8.433           &       1.665           &           *           &       0.000           &       11.46           &        0.10          \\        
 EDCSNJ1059053-1255535           &          13           &      0.4572           &           2           &       204.4           &         210           &        19.2           &       3.187           &       2.057           &           -           &       0.196           &       10.84           &        0.10          \\        
 EDCSNJ1059046-1251583           &          13           &      0.4561           &           1           &       115.9           &       119.1           &        14.3           &       2.335           &       2.243           &           -           &       0.199           &       10.87           &        0.11          \\        
 EDCSNJ1059093-1253065           &          13           &      0.4537           &           1           &       270.4           &       277.8           &        30.9           &       4.667           &       1.405           &           -           &       0.058           &       10.49           &        0.14          \\
\hline
\end{tabular}
\end{table*}

\addtocounter{table}{-1}
\begin{table*}[h!]
\caption{Continued}
\begin{tabular}{ccccccccccccc}
\hline
Name & $N_{clus}$ & z & S & $\sigma_{mes}$ & $\sigma_{cor}$ & d$\sigma$ &
$R_e$ & $\log I_e$ & T & $P_S$ & $\log M_*$ &d$\log M_*$ \\
    &    &    & Type   & (km/s)         & (km/s)           & (km/s)                &
(kpc) & $\log L_\odot/pc^2$ & Type & & $\log M_\odot$ & $\log M_\odot$   \\
\hline
 EDCSNJ1420098-1233566           &          15           &      0.4958           &           1           &       143.6           &       147.8           &        11.5           &       1.909           &         2.6           &           -           &       0.213           &       10.99           &        0.10          \\        
 EDCSNJ1420164-1235291           &          15           &      0.4958           &           1           &       119.9           &       123.4           &        17.7           &        2.87           &       2.099           &           -           &       0.127           &       10.78           &        0.13          \\        
 EDCSNJ1420201-1236297           &          15           &      0.4969           &           1           &       268.4           &       276.3           &           7           &       11.91           &       1.639           &           *           &       0.000           &       11.67           &        0.08          \\        
 EDCSNJ1420219-1237051           &          15           &      0.4956           &           1           &       212.6           &       218.8           &        28.5           &      0.8876           &       2.781           &           -           &       0.055           &       10.61           &        0.13          \\        
 EDCSNJ1420235-1237178           &          15           &      0.4957           &           1           &       148.8           &       153.2           &        14.8           &       2.639           &       2.209           &           -           &       0.133           &       10.81           &        0.12          \\
\hline
\end{tabular}
\end{table*}

\begin{table*}[h!]
\caption{The FP parameters of field galaxies \label{tab_field}}
\begin{tabular}{cccccccccccc}
\hline
Name & z & S & $\sigma_{mes}$ & $\sigma_{cor}$ & d$\sigma$ &
$R_e$ & $\log I_e$ & T & $P_S$ & $\log M_*$ &d$\log M_*$ \\
    &       & Type   & (km/s)         & (km/s)           & (km/s)                &
(kpc) & $\log L_\odot/pc^2$ & Type & & $\log M_\odot$ & $\log M_\odot$   \\
\hline
 EDCSNJ1040391-1155167           &       0.766           &           1           &       177.8           &       184.4           &        29.9           &      0.9953           &       2.842           &          -5           &       0.049           &       10.78           &        0.07          \\                                
 EDCSNJ1040343-1155414           &      0.7807           &           1           &       290.4           &       301.3           &        17.1           &       3.494           &       2.562           &          -5           &       0.369           &       11.58           &        0.05          \\                                
 EDCSNJ1040476-1158184           &      0.6171           &           1           &       161.9           &       167.4           &        10.4           &       1.745           &       2.561           &          -2           &       0.276           &       10.99           &        0.08          \\                                
 EDCSNJ1054253-1148349           &      0.8657           &           1           &       231.8           &       240.9           &        43.6           &       1.646           &       2.916           &          -5           &       0.057           &       11.11           &        0.07          \\                                
 EDCSNJ1054289-1146428           &      0.2491           &           1           &       258.6           &       261.5           &         5.5           &       4.418           &       2.225           &          -2           &       0.090           &       11.51           &        0.55          \\
\hline
\end{tabular}
\end{table*}
\addtocounter{table}{-1}
\begin{table*}[h!]
\caption{Continued}
\begin{tabular}{cccccccccccc}
\hline
Name & z & S & $\sigma_{mes}$ & $\sigma_{cor}$ & d$\sigma$ &
$R_e$ & $\log I_e$ & T & $P_S$ & $\log M_*$ &d$\log M_*$ \\
    &       & Type   & (km/s)         & (km/s)           & (km/s)                &
(kpc) & $\log L_\odot/pc^2$ & Type & & $\log M_\odot$ & $\log M_\odot$   \\
\hline
 EDCSNJ1354130-1230263           &      0.8223           &           1           &       251.3           &       260.9           &        17.6           &       1.244           &       3.174           &           3           &       0.229           &       11.09           &        0.05          \\                                
 EDCSNJ1054143-1144503           &      0.3976           &           2           &       243.9           &       249.8           &        34.4           &       1.361           &       2.142           &           -           &       0.045           &       10.64           &        0.11          \\                                
 EDCSNJ1054499-1247587           &       0.802           &           1           &       248.3           &       257.7           &        13.1           &       6.362           &       2.041           &           -           &       0.165           &       11.39           &        0.04          \\                                
 EDCSNJ1216435-1203502           &      0.6693           &           2           &       231.1           &       239.2           &        21.3           &        2.57           &       2.296           &           *           &       0.000           &       11.08           &        0.07          \\                                
 EDCSNJ1232293-1254348           &      0.7518           &           1           &         175           &       181.5           &        25.7           &       1.543           &       2.885           &           -           &       0.066           &       11.69           &        0.26          \\
\hline
\end{tabular}
\end{table*}

\begin{table*}[h!]
\caption{The structural parameters of galaxies with measured 
velocity dispersions and HST photometry derived from Sersic fits to 
HST images and bulge+disk fits to VLT images. \label{tab_SerVLT}}
\begin{tabular}{ccccc}
\hline
Name & $R_e(Sersic)$ & $\log I_e(Sersic)$ & $R_e(VLT)$ & $\log I_e(VLT)$\\
     & (kpc)         & $\log L_\odot/pc^2$& (kpc)      & $\log L_\odot/pc^2$\\
\hline
 EDCSNJ1040403-1156042           &       6.153           &       1.986           &       9.822           &       1.682          \\                                                                                                                                                                                                        
 EDCSNJ1040407-1156015           &       1.698           &       2.808           &       2.923           &       2.448          \\                                                                                                                                                                                                        
 EDCSNJ1040346-1157566           &       3.348           &       2.319           &       5.857           &       1.885          \\                                                                                                                                                                                                        
 EDCSNJ1040396-1155183           &       2.244           &       2.549           &       2.624           &       2.411          \\                                                                                                                                                                                                        
 EDCSNJ1040356-1156026           &       2.345           &       2.659           &       2.543           &       2.606          \\                                                                                                                                                                                                        
 EDCSNJ1054244-1146194           &       5.945           &       2.155           &        12.3           &       1.648          \\
\hline
\end{tabular}
\end{table*}
\addtocounter{table}{-1}
\begin{table*}[h!]
\caption{Continued}
\begin{tabular}{ccccccccc}
\hline
Name & $R_e(Sersic)$ & $\log I_e(Sersic)$ & $R_e(VLT)$ & $\log I_e(VLT)$\\
     & (kpc)         & $\log L_\odot/pc^2$& (kpc)      & $\log L_\odot/pc^2$\\
\hline
 EDCSNJ1232313-1250327           &        1.44           &       2.523           &       19.95           &      0.6675          \\                                                                                                                                                                                                        
 EDCSNJ1232317-1249275           &       5.449           &       2.049           &       5.643           &       1.964          \\                                                                                                                                                                                                        
 EDCSNJ1232309-1249408           &        3.62           &       2.476           &       4.613           &       2.269          \\                                                                                                                                                                                                        
 EDCSNJ1232303-1251092           &       1.087           &       2.637           &       1.262           &       2.472          \\                                                                                                                                                                                                        
 EDCSNJ1232303-1251441           &       2.704           &        2.14           &       2.002           &       2.294          \\                                                                                                                                                                                                        
 EDCSNJ1232370-1248239           &       1.831           &       2.605           &       2.289           &       2.428          \\
\hline
\end{tabular}
\end{table*}
\addtocounter{table}{-1}
\begin{table*}[h!]
\caption{Continued}
\begin{tabular}{ccccccccc}
\hline
Name & $R_e(Sersic)$ & $\log I_e(Sersic)$ & $R_e(VLT)$ & $\log I_e(VLT)$\\
     & (kpc)         & $\log L_\odot/pc^2$& (kpc)      & $\log L_\odot/pc^2$\\
\hline
 EDCSNJ1054240-1147364           &       4.464           &       2.263           &        10.7           &       1.664          \\                                                                                                                                                                                                        
 EDCSNJ1054525-1244189           &       4.663           &       2.296           &       6.144           &       2.093          \\                                                                                                                                                                                                        
 EDCSNJ1054353-1246528           &       3.469           &       2.282           &       28.39           &       1.085          \\                                                                                                                                                                                                        
 EDCSNJ1054487-1245052           &       1.629           &       2.695           &           0           &       -1949          \\                                                                                                                                                                                                        
 EDCSNJ1216402-1201593           &       2.423           &       2.402           &       3.819           &       2.033          \\                                                                                                                                                                                                        
 EDCSNJ1216508-1157576           &       1.096           &       3.011           &      0.9015           &       3.123          \\
\hline
\end{tabular}
\end{table*}

\section{The fundamental plane of the EDisCS galaxies}
\label{sec_FP}

\subsection{The FP of EDisCS clusters}
\label{sec_FPclus}

\begin{figure*}
\begin{tabular}{cccc}
\vbox{\psfig{file=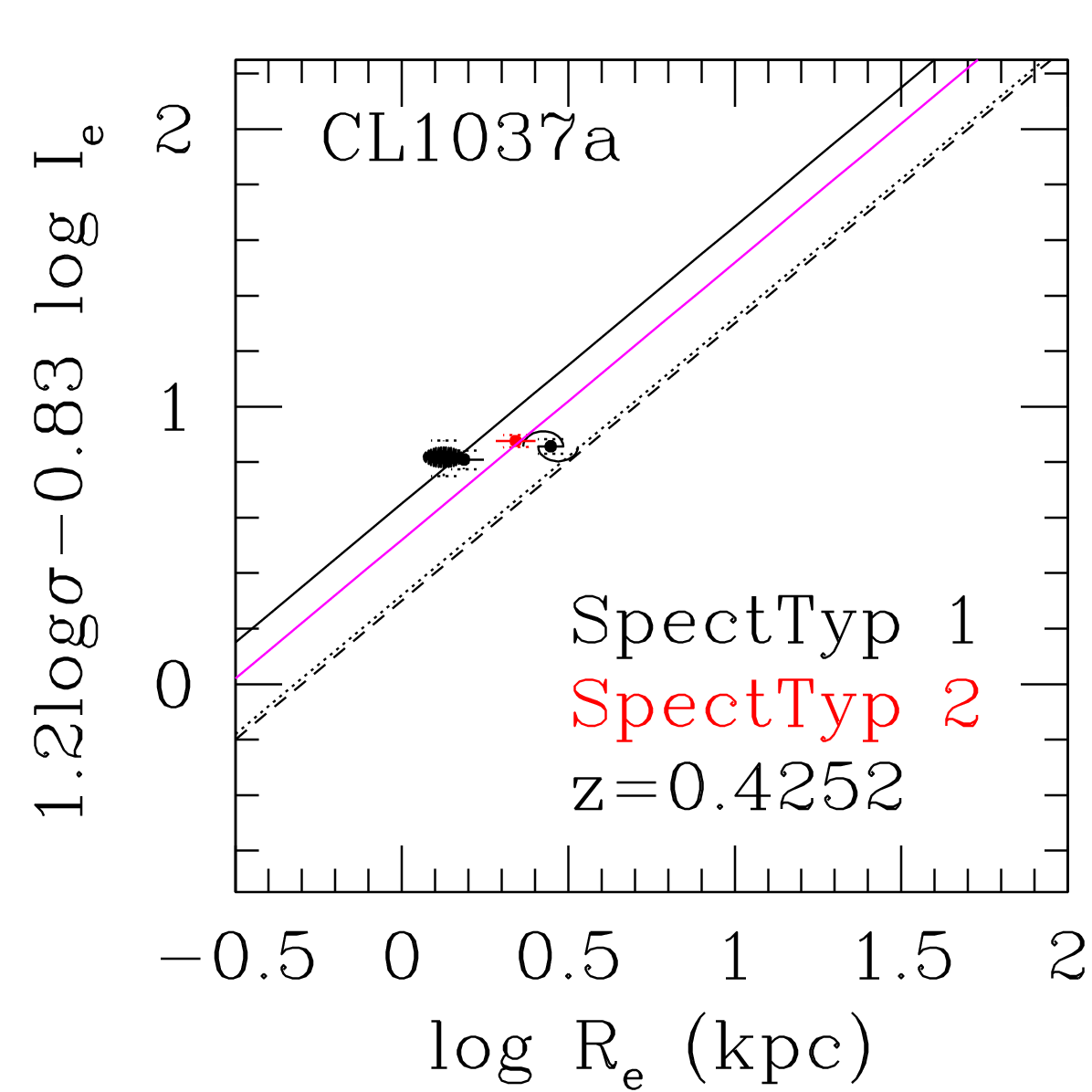,angle=0,width=4.cm}}&    
\vbox{\psfig{file=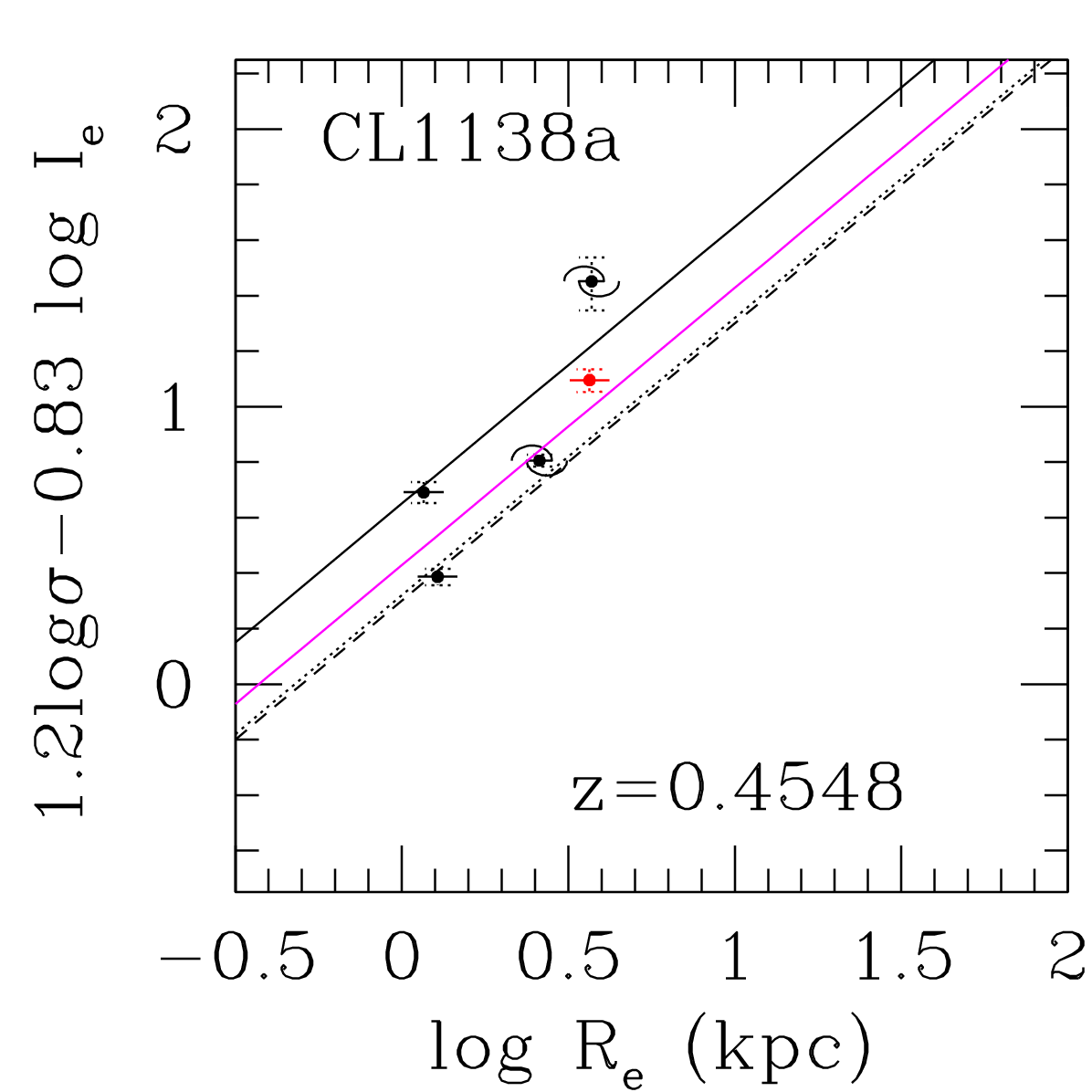,angle=0,width=4.cm}}&   
\vbox{\psfig{file=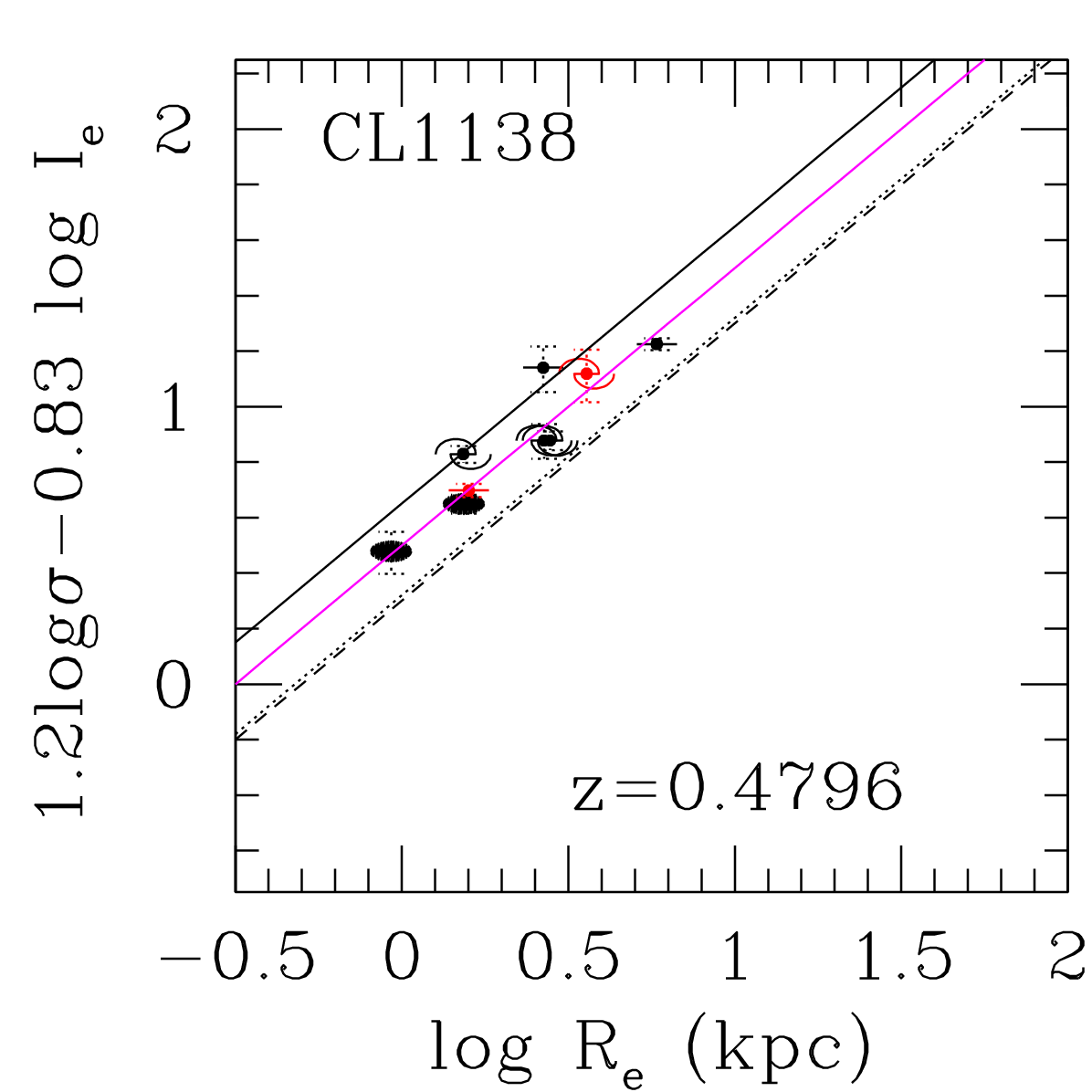,angle=0,width=4.cm}}&   
\vbox{\psfig{file=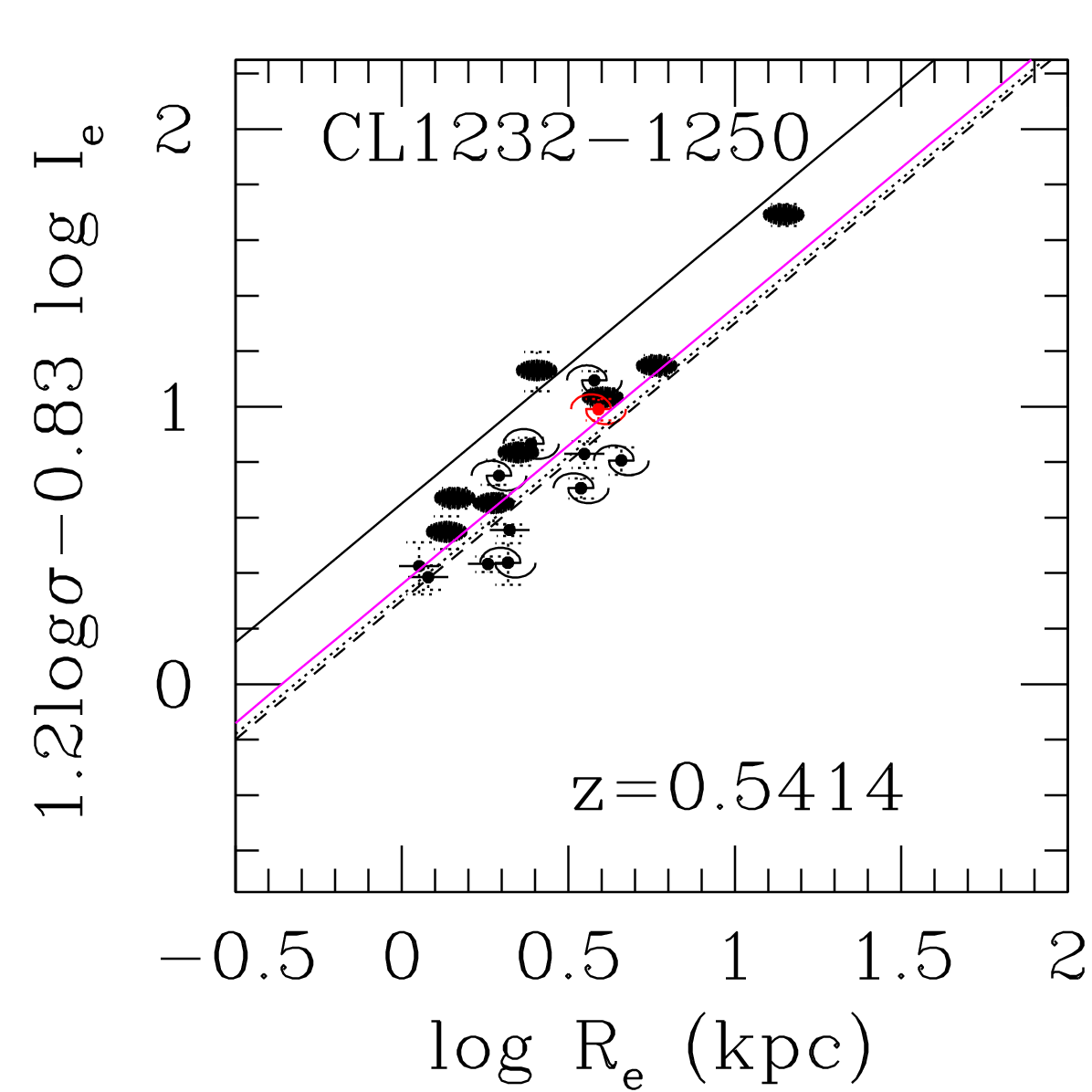,angle=0,width=4.cm}} \\   
\vbox{\psfig{file=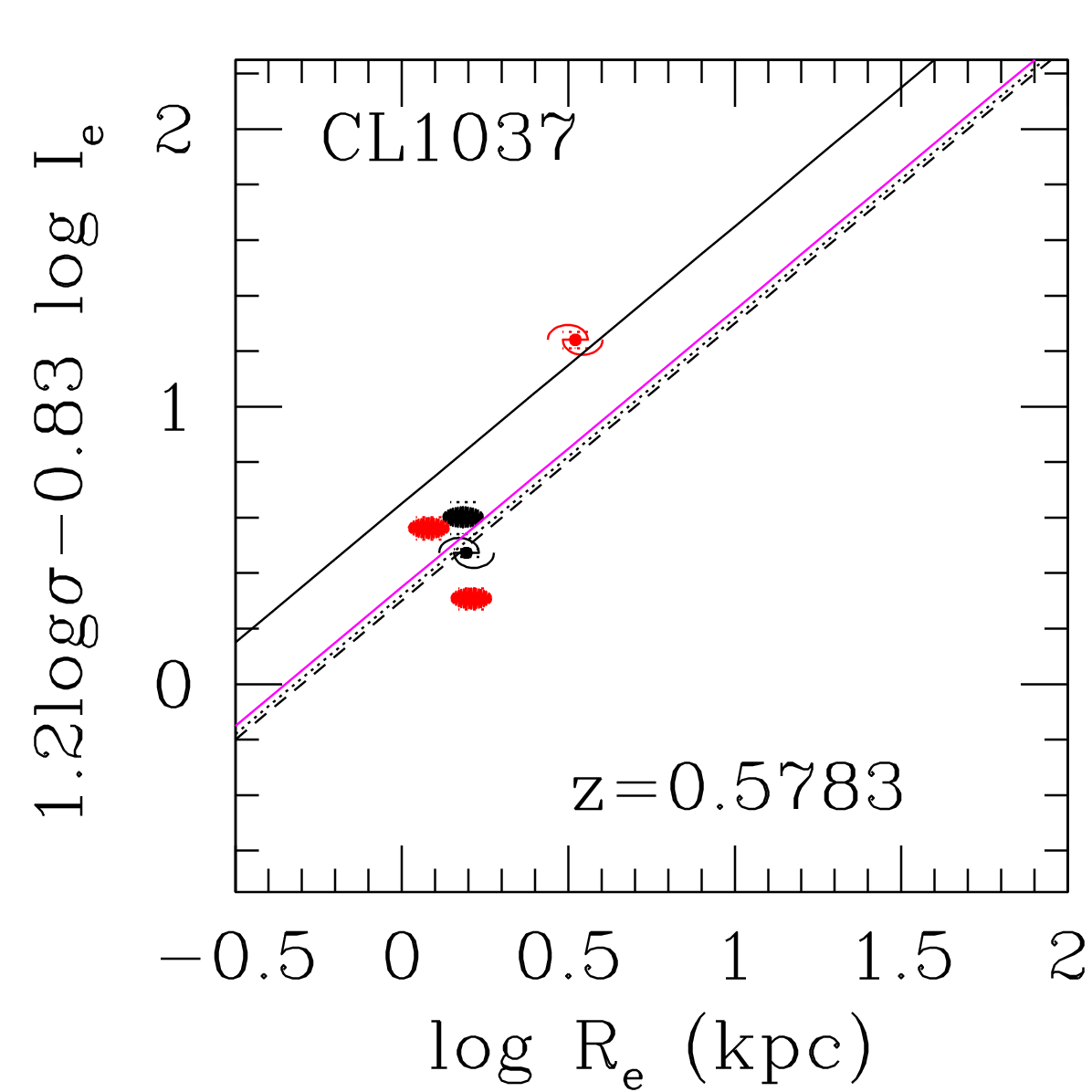,angle=0,width=4.cm}}&   
\vbox{\psfig{file=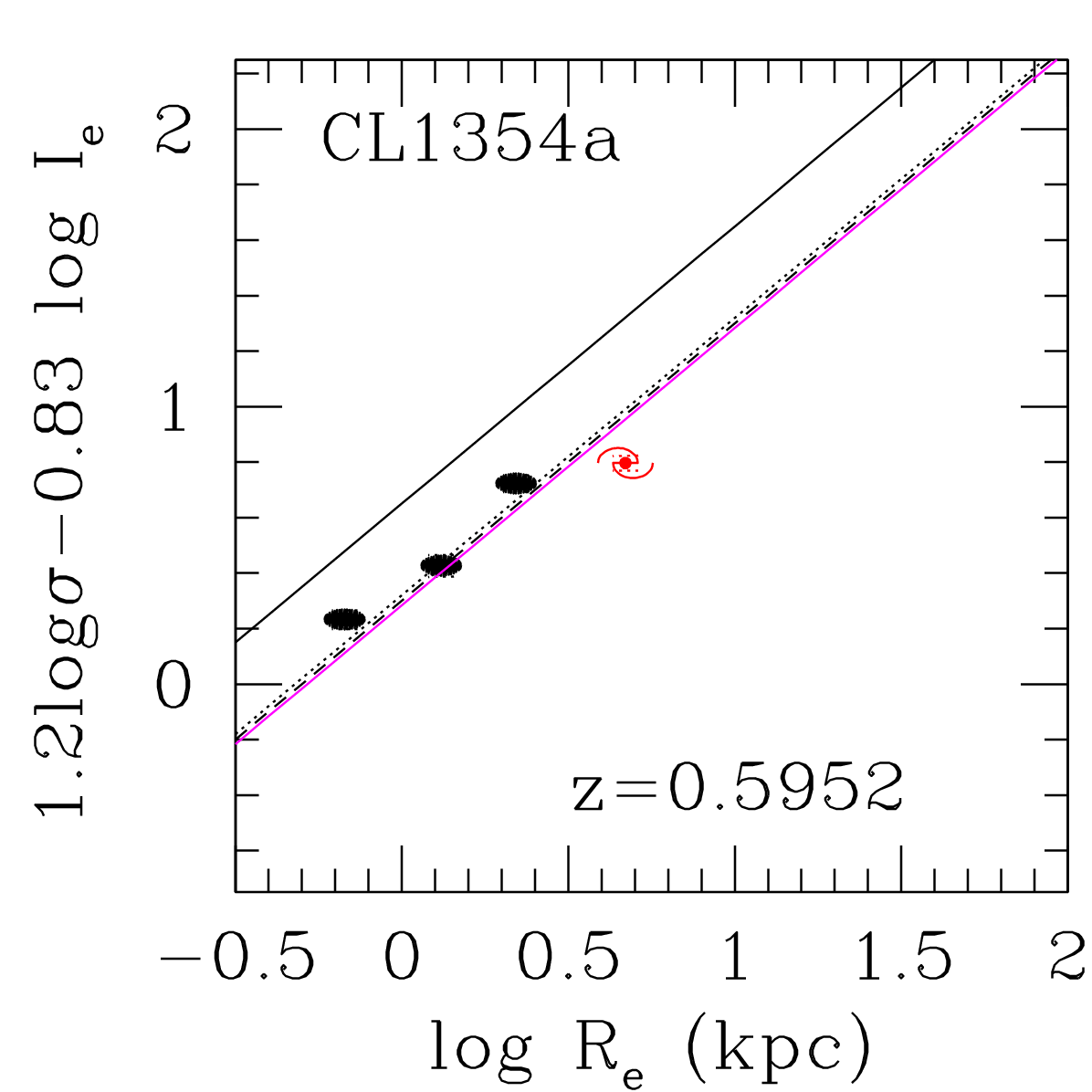,angle=0,width=4.cm}}&   
\vbox{\psfig{file=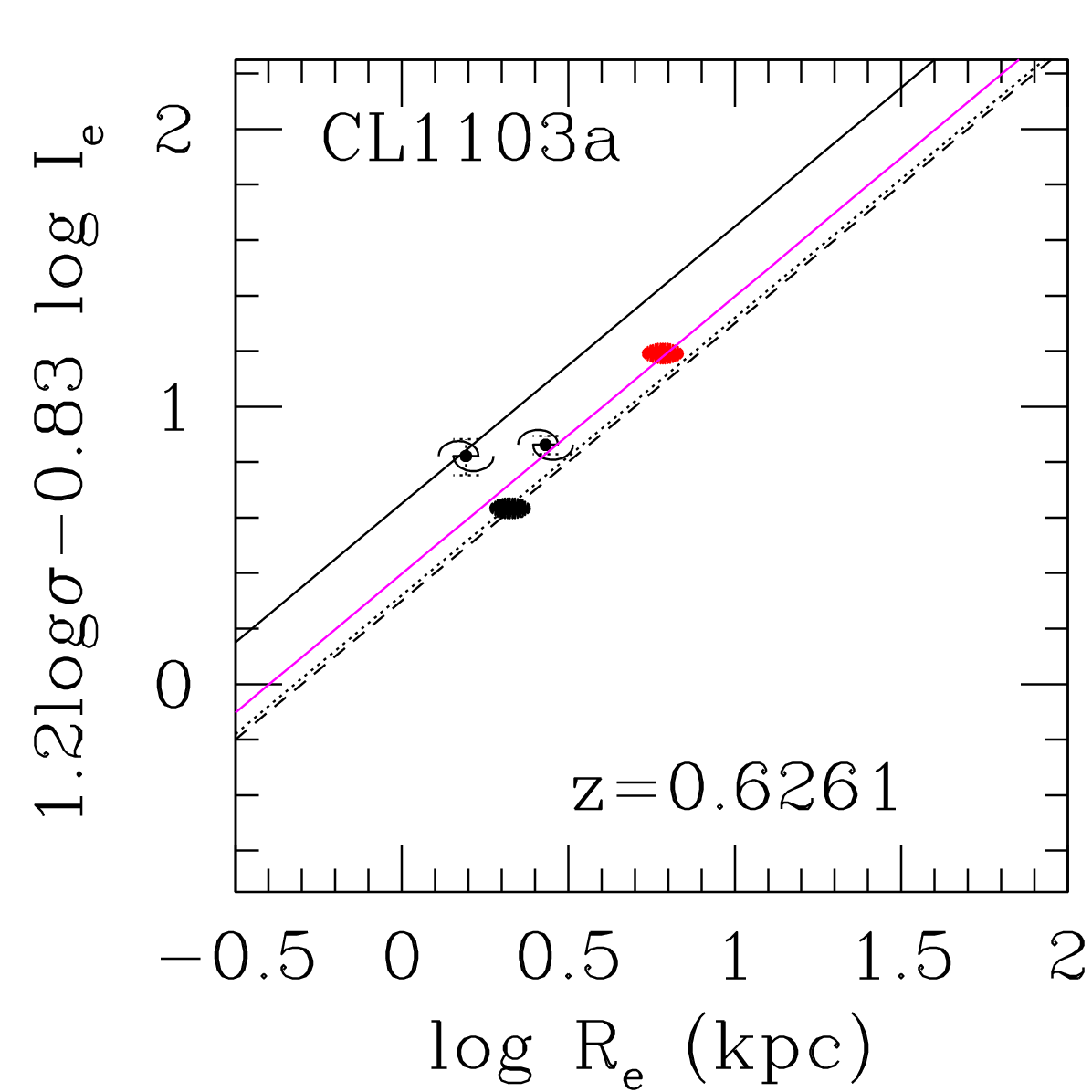,angle=0,width=4.cm}}&  
\vbox{\psfig{file=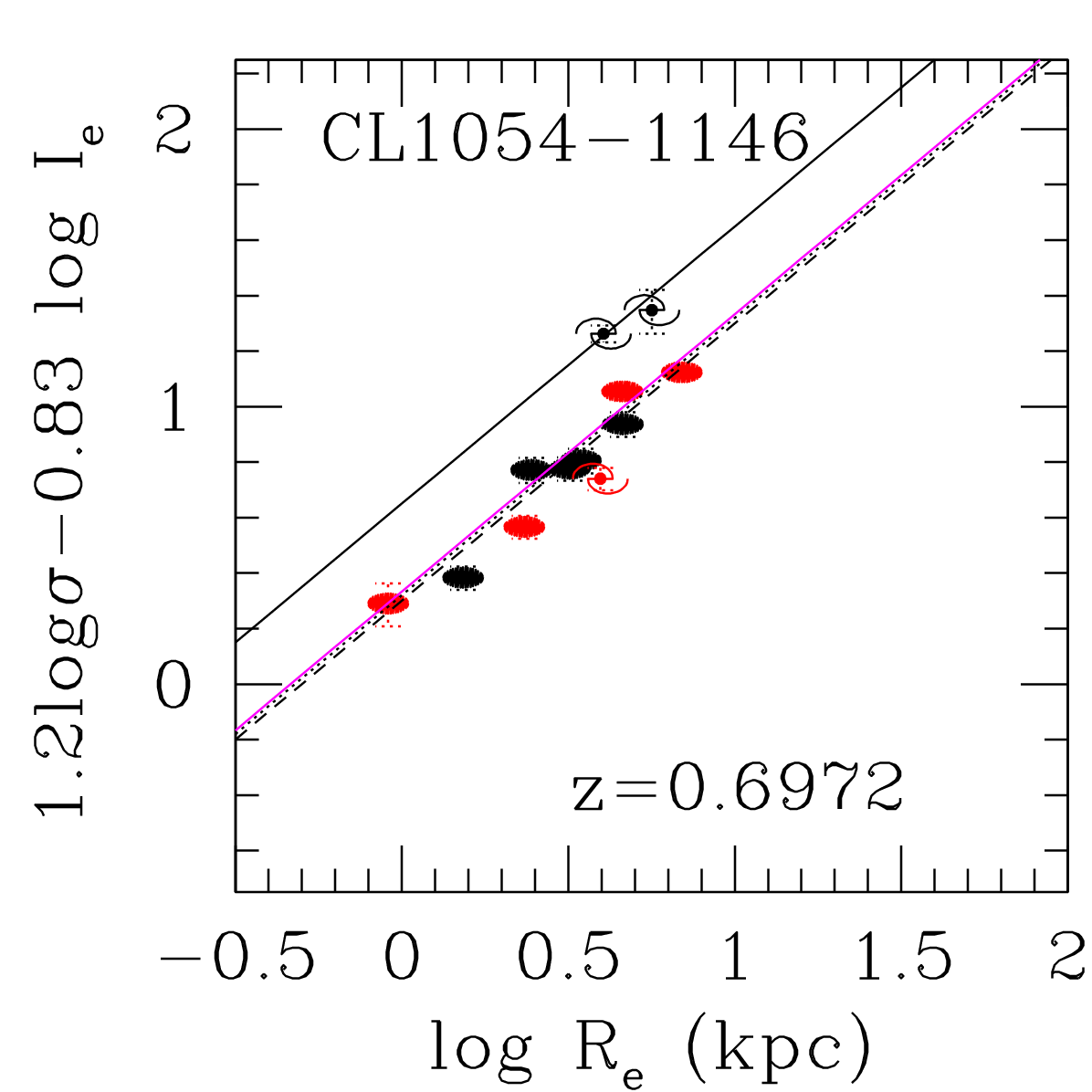,angle=0,width=4.cm}}\\  
\vbox{\psfig{file=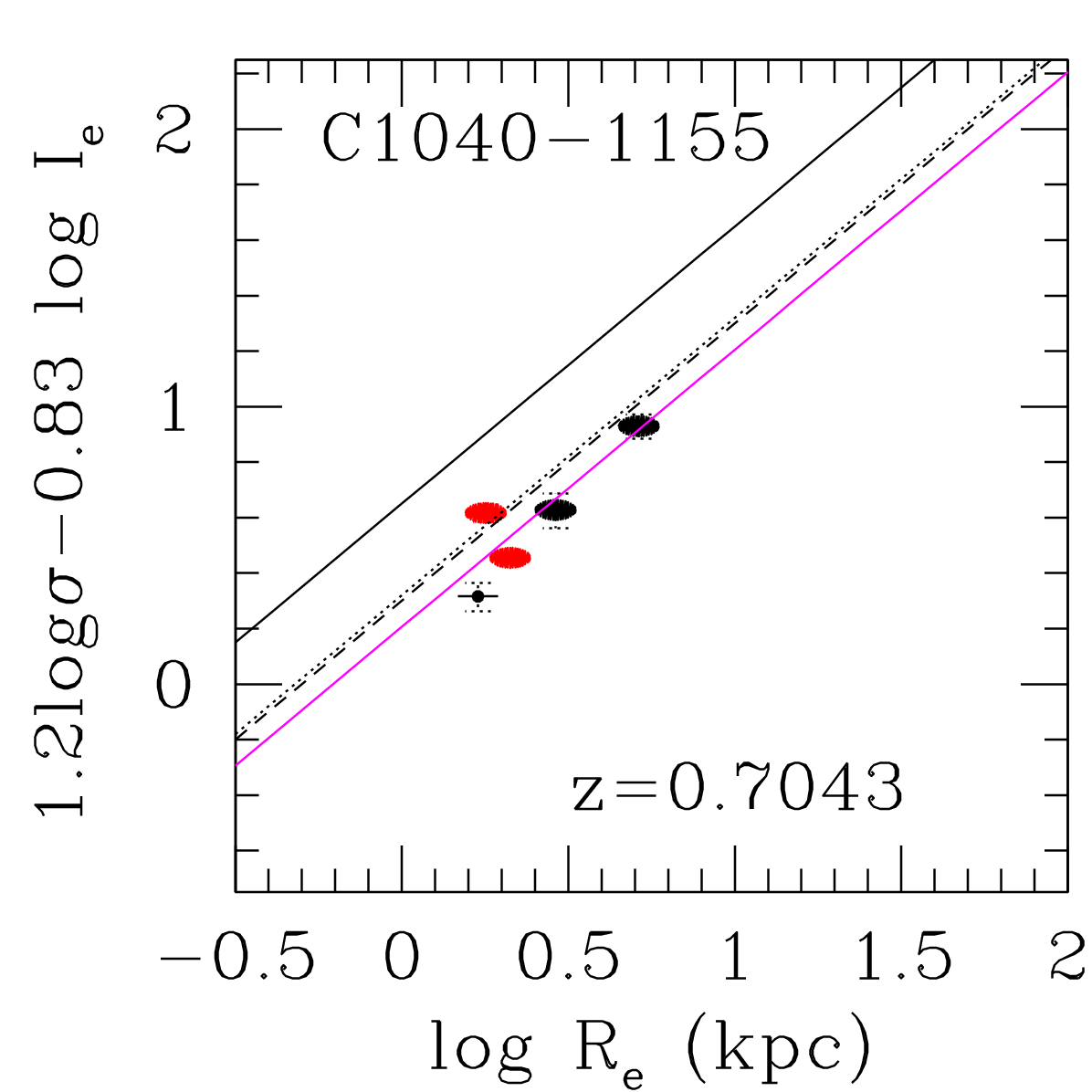,angle=0,width=4.cm}}&    
\vbox{\psfig{file=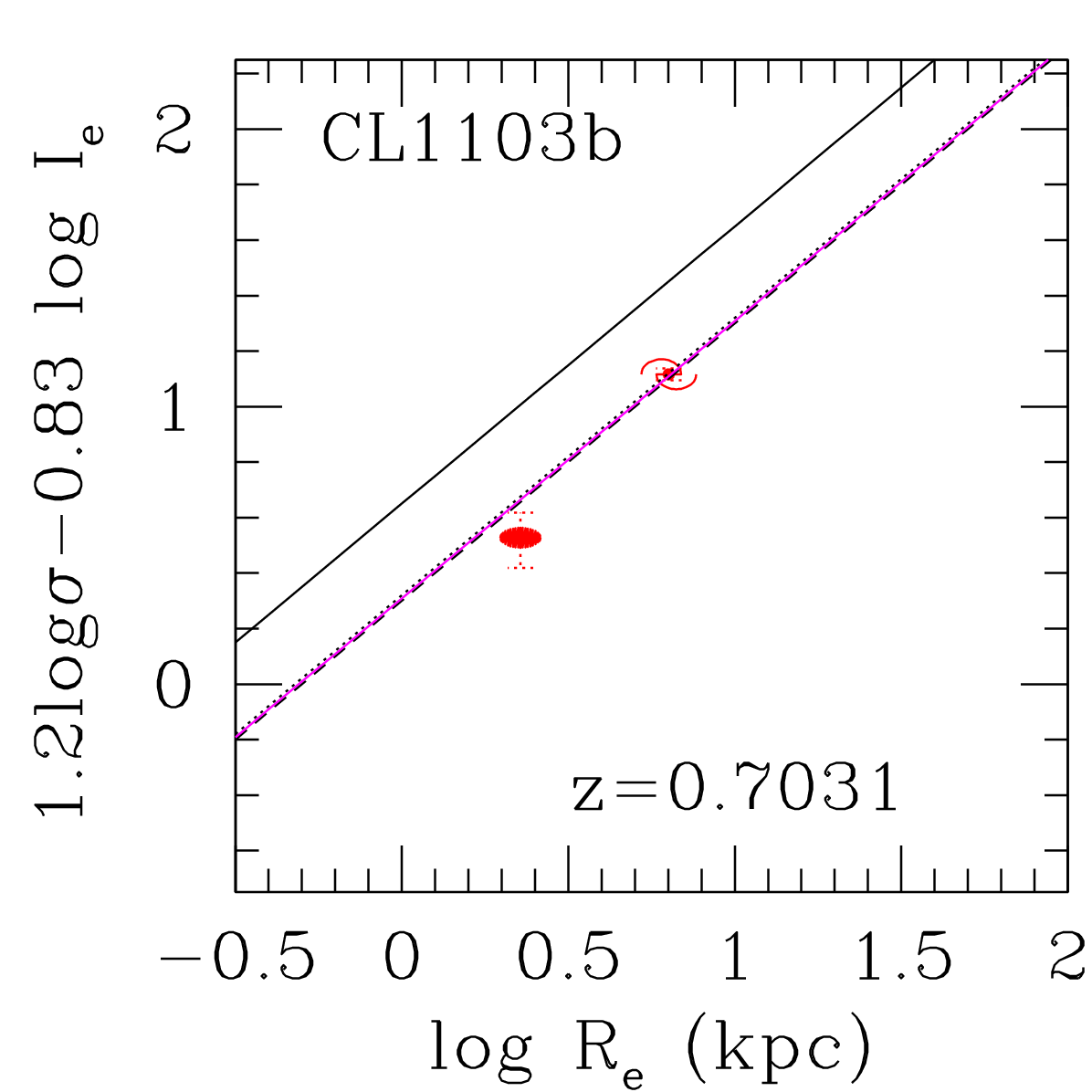,angle=0,width=4.cm}}&  
\vbox{\psfig{file=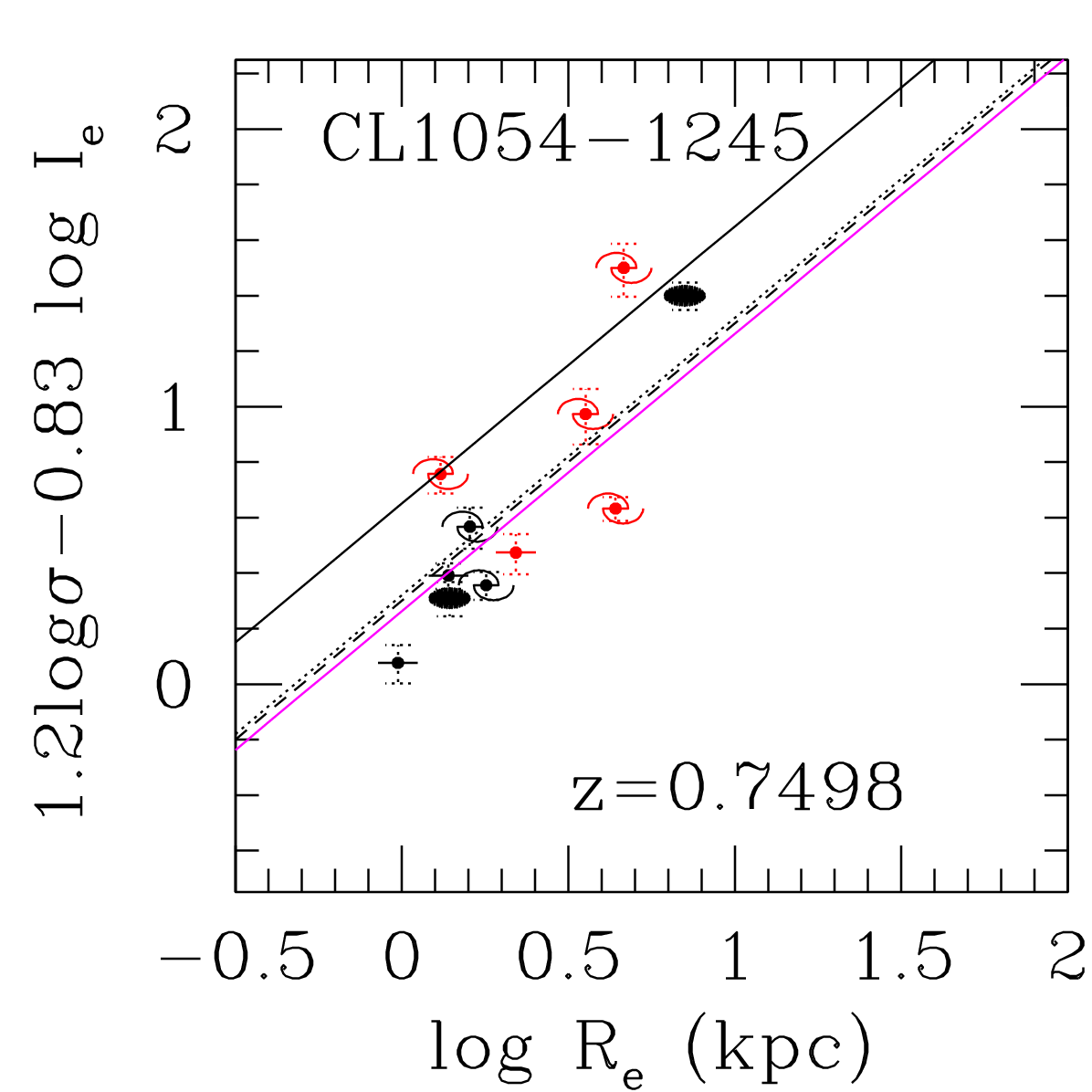,angle=0,width=4.cm}}&  
\vbox{\psfig{file=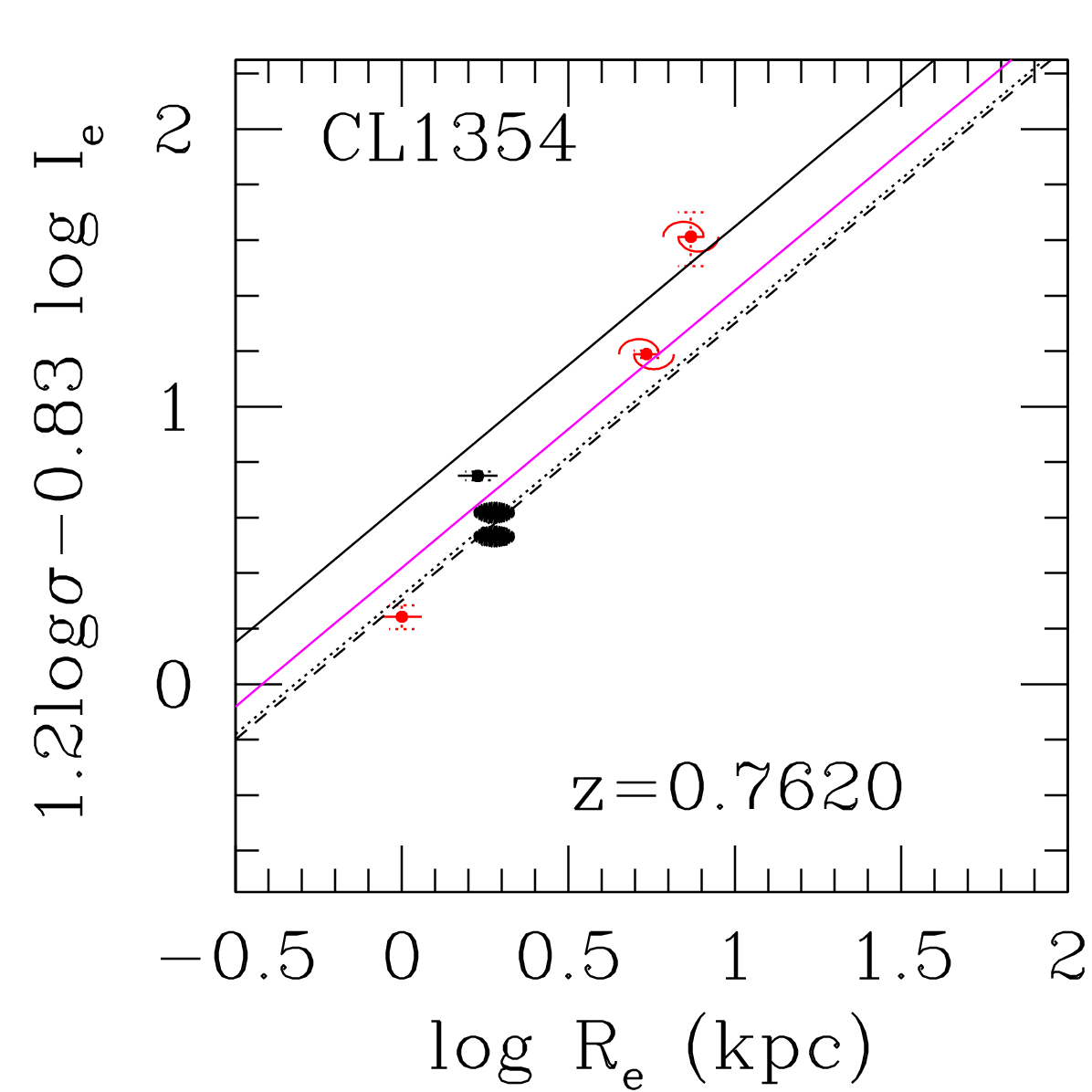,angle=0,width=4.cm}}\\    
\vbox{\psfig{file=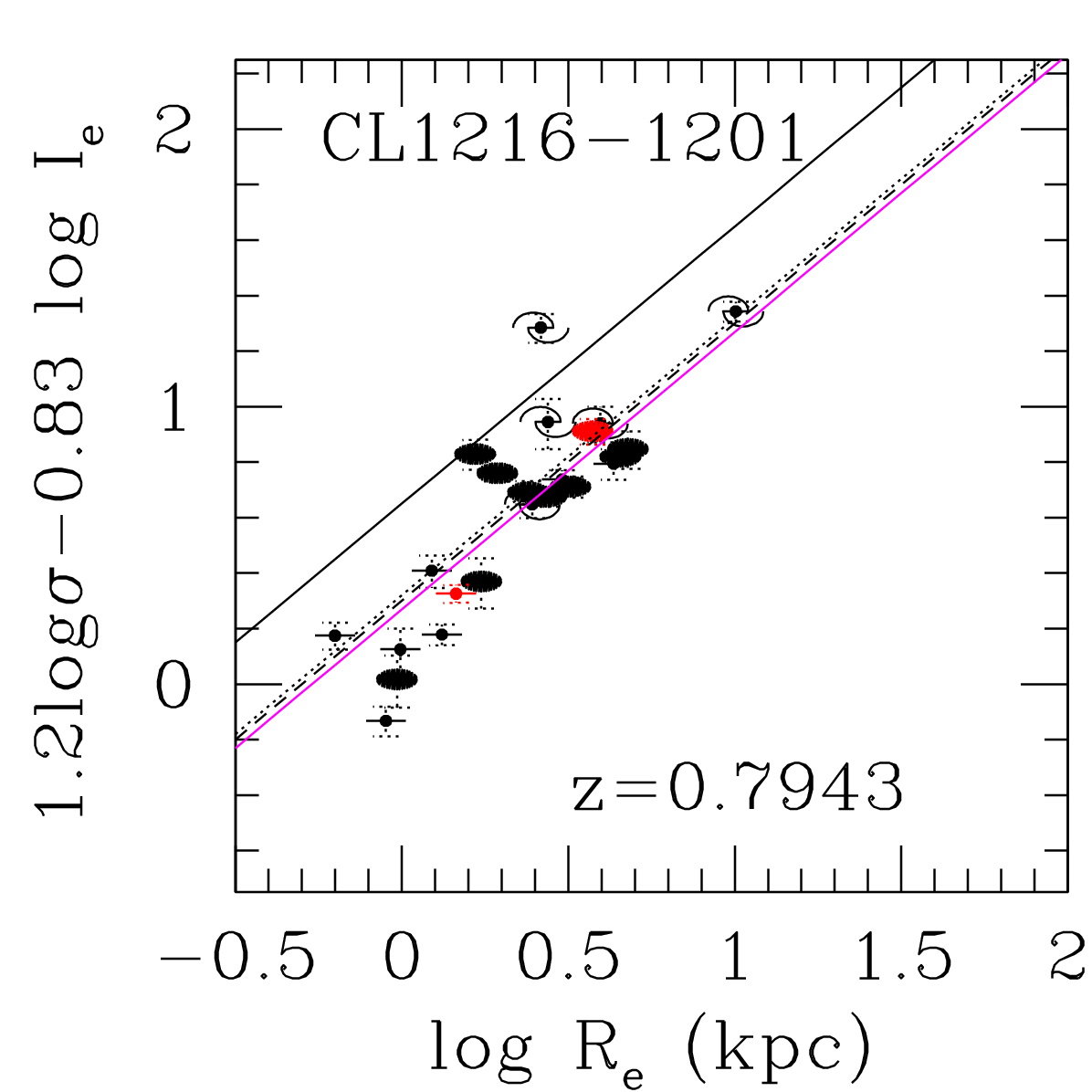,angle=0,width=4.cm}}&    
\vbox{\psfig{file=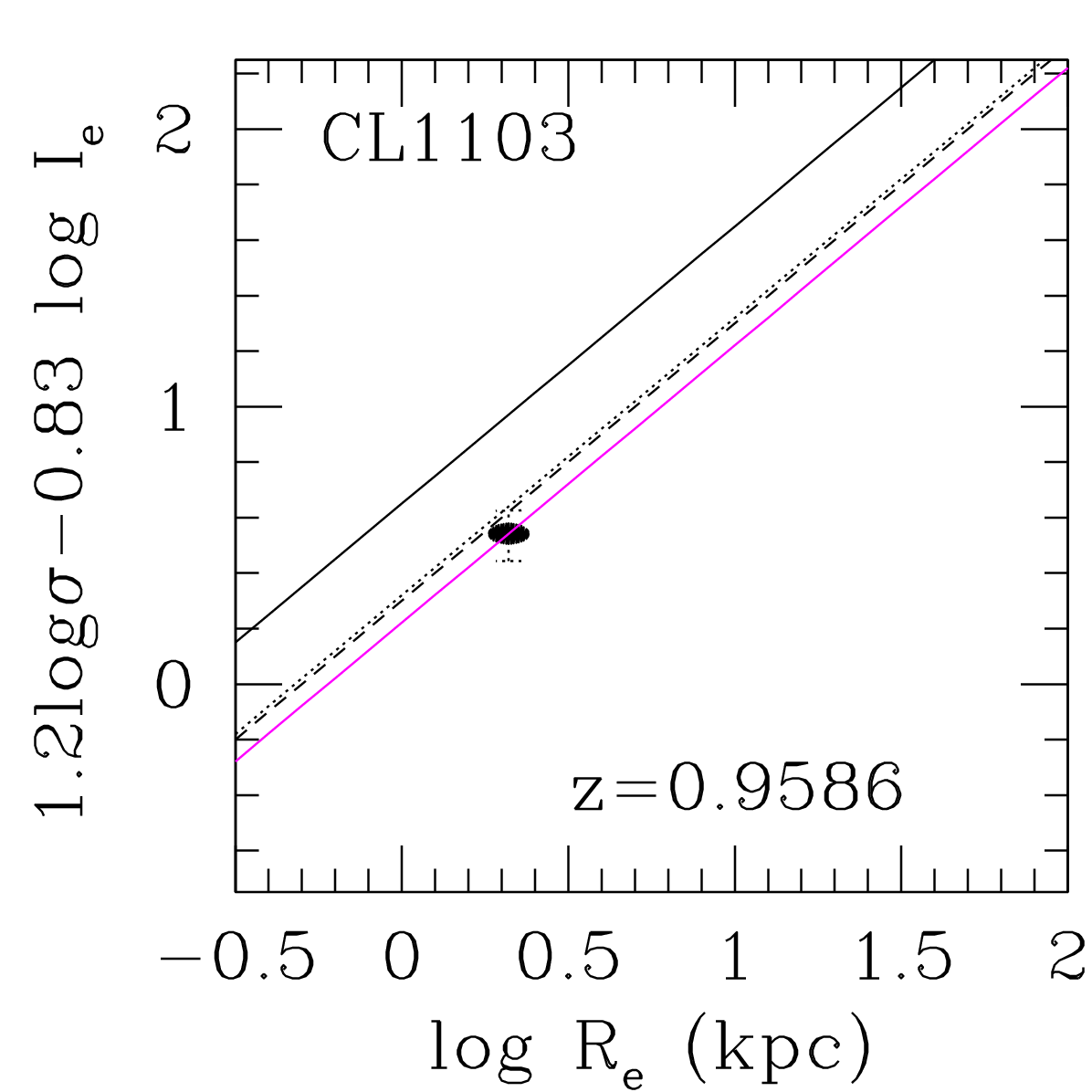,angle=0,width=4.cm}}& &\\  
\end{tabular}
\caption{The FP of the EDisCS clusters with HST photometry. Each
  cluster is identified by its short name for clarity, see Table
  \ref{tab_ZP} for the full name. Colors code the spectroscopic type
  (black = 1, red = 2).  Symbols code the morphology: filled ellipses
  show $T\le-4$, filled circles crossed by a line $-3\le T\le 0$,
  spirals $T>0$.  The magenta line shows the best-fit FP line with no
  selection weighting.  The full line shows the Coma cluster at zero
  redshift.  The black dotted and dashed lines show data for the
  clusters MS2053-04 at $z=0.58$ and MS1054-03 at $z=0.83$,
  respectively, from \citet{Wuyts04}.
\label{fig_FPHSTClus} }
\end{figure*}

\begin{figure*}
\begin{tabular}{cccc}
\vbox{\psfig{file=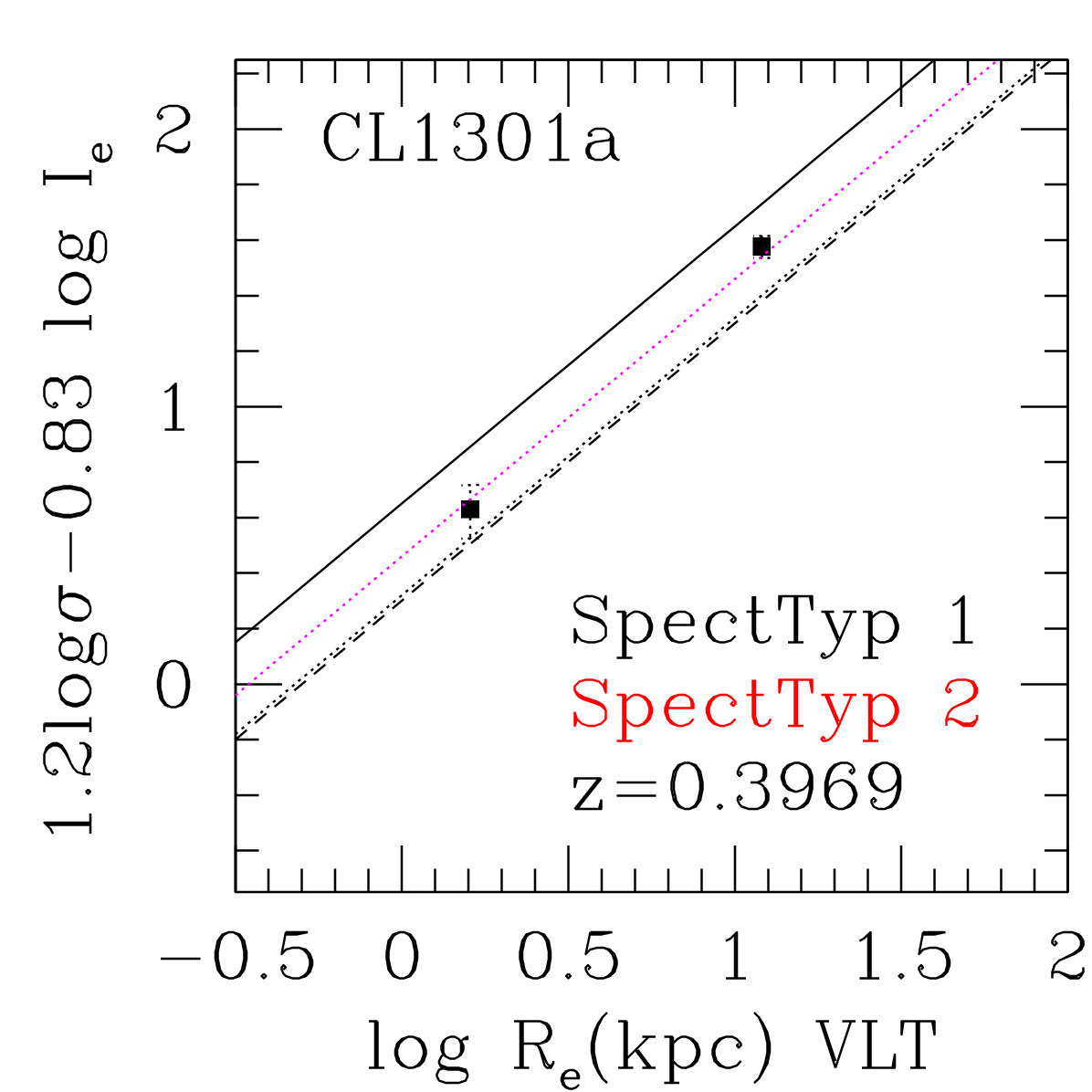,angle=0,width=4cm}}& 
\vbox{\psfig{file=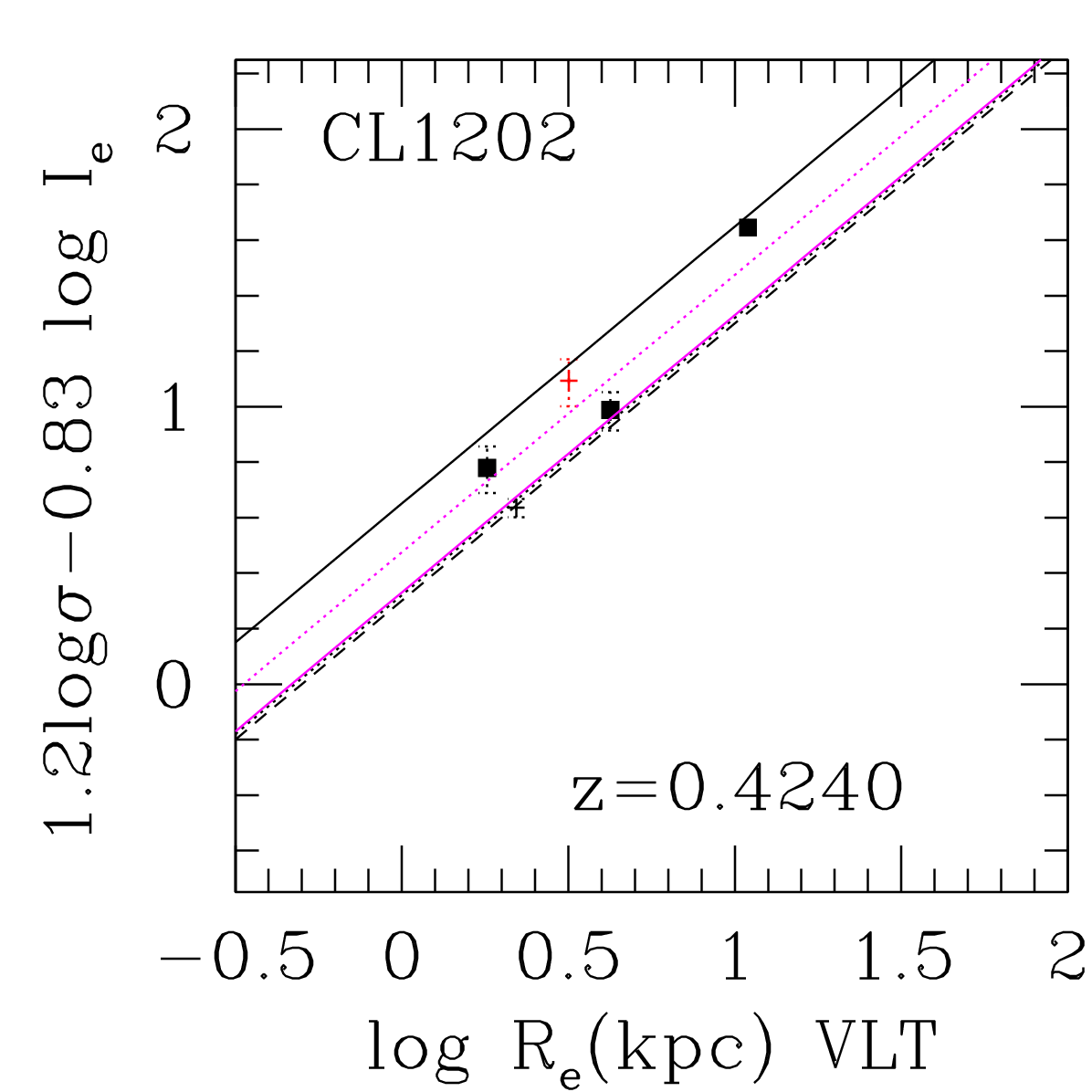,angle=0,width=4cm}}&  
\vbox{\psfig{file=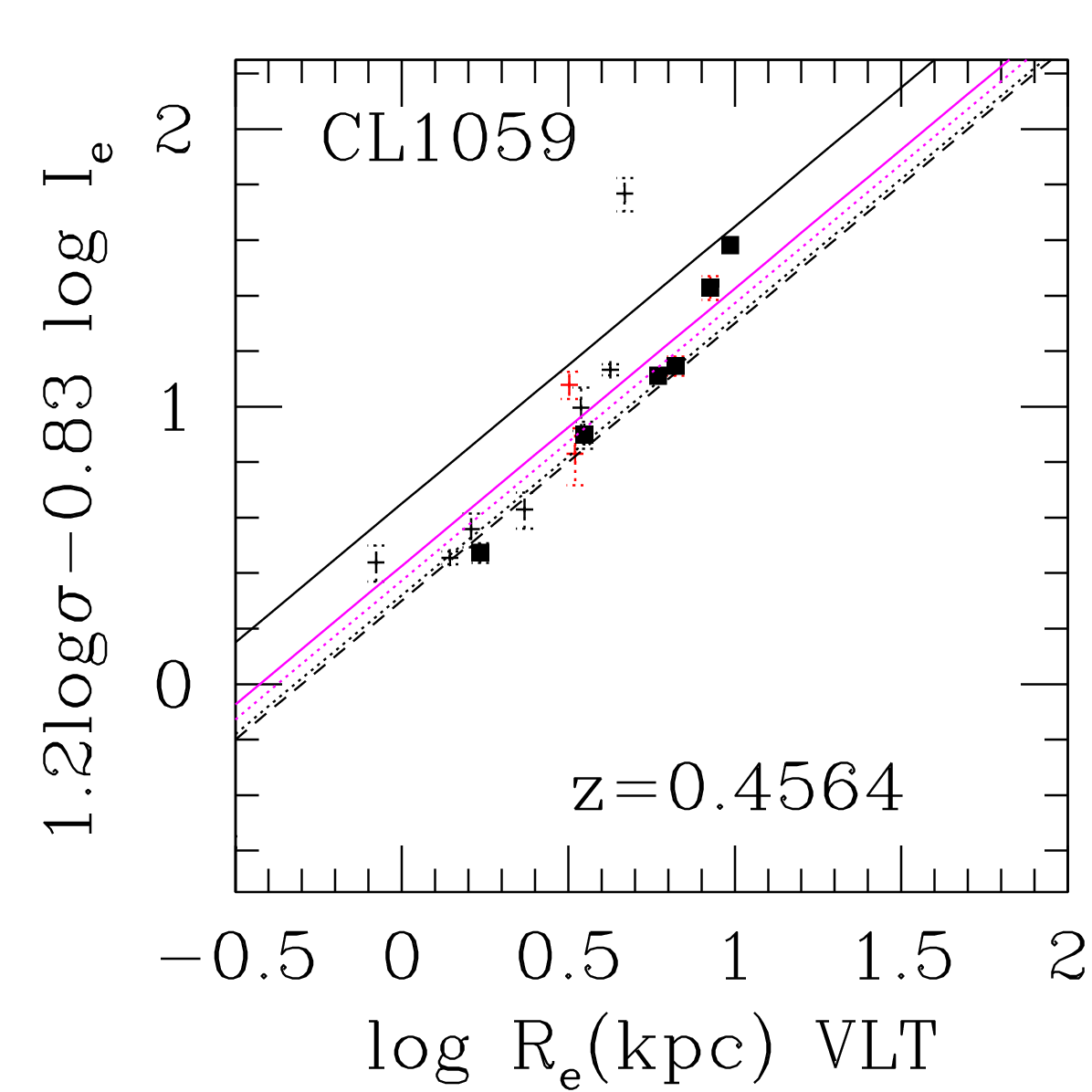,angle=0,width=4cm}}&  
\vbox{\psfig{file=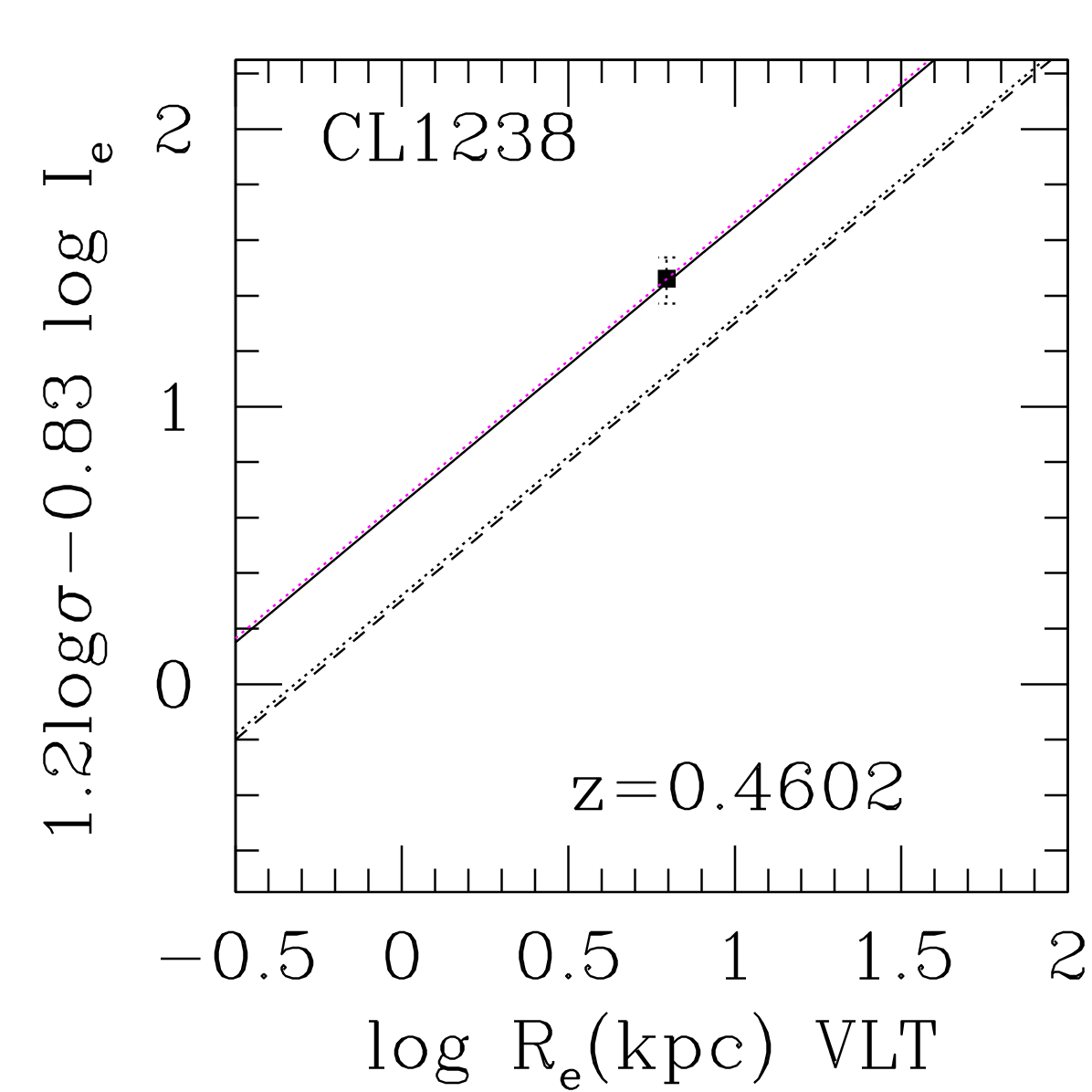,angle=0,width=4cm}}\\ 
\vbox{\psfig{file=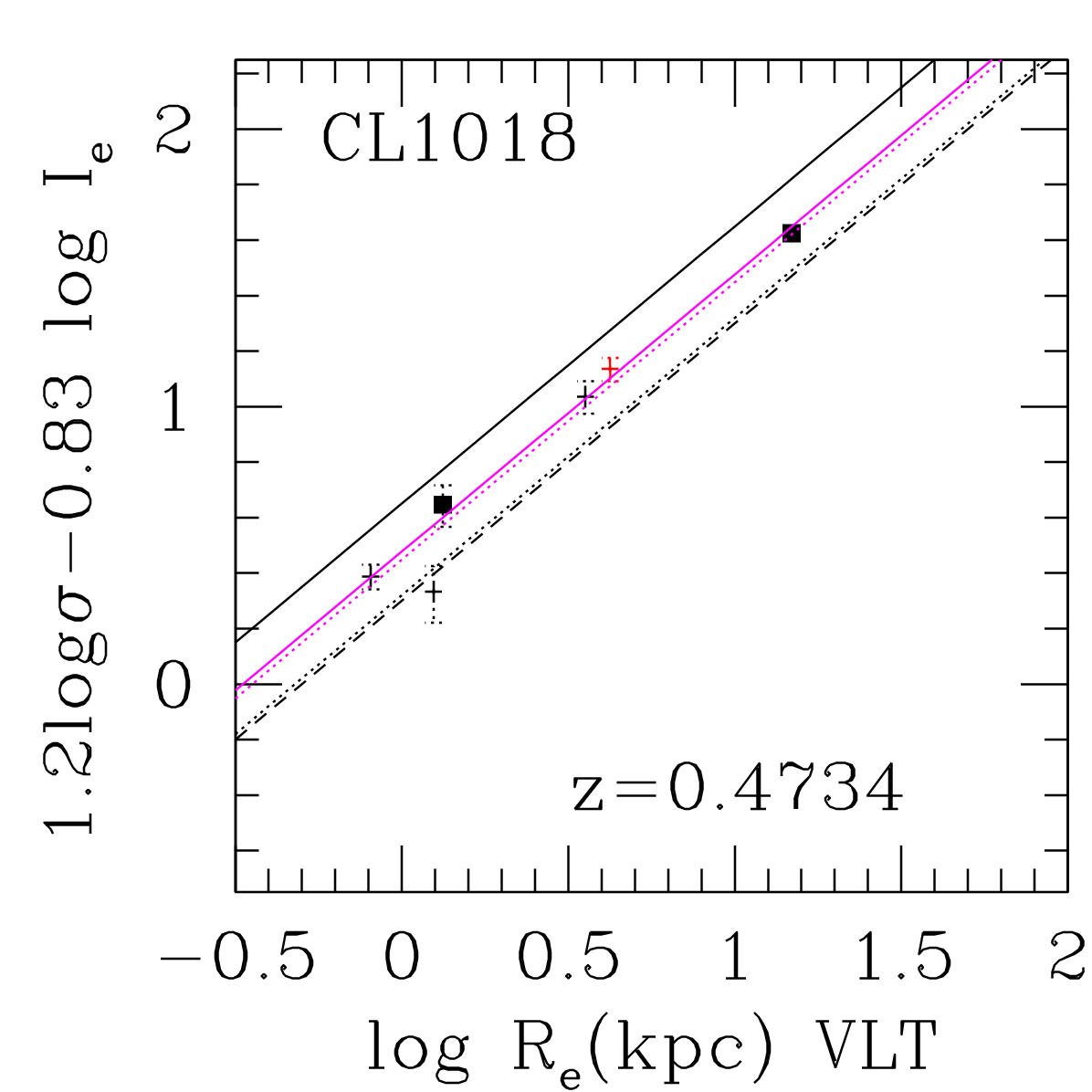,angle=0,width=4cm}}&  
\vbox{\psfig{file=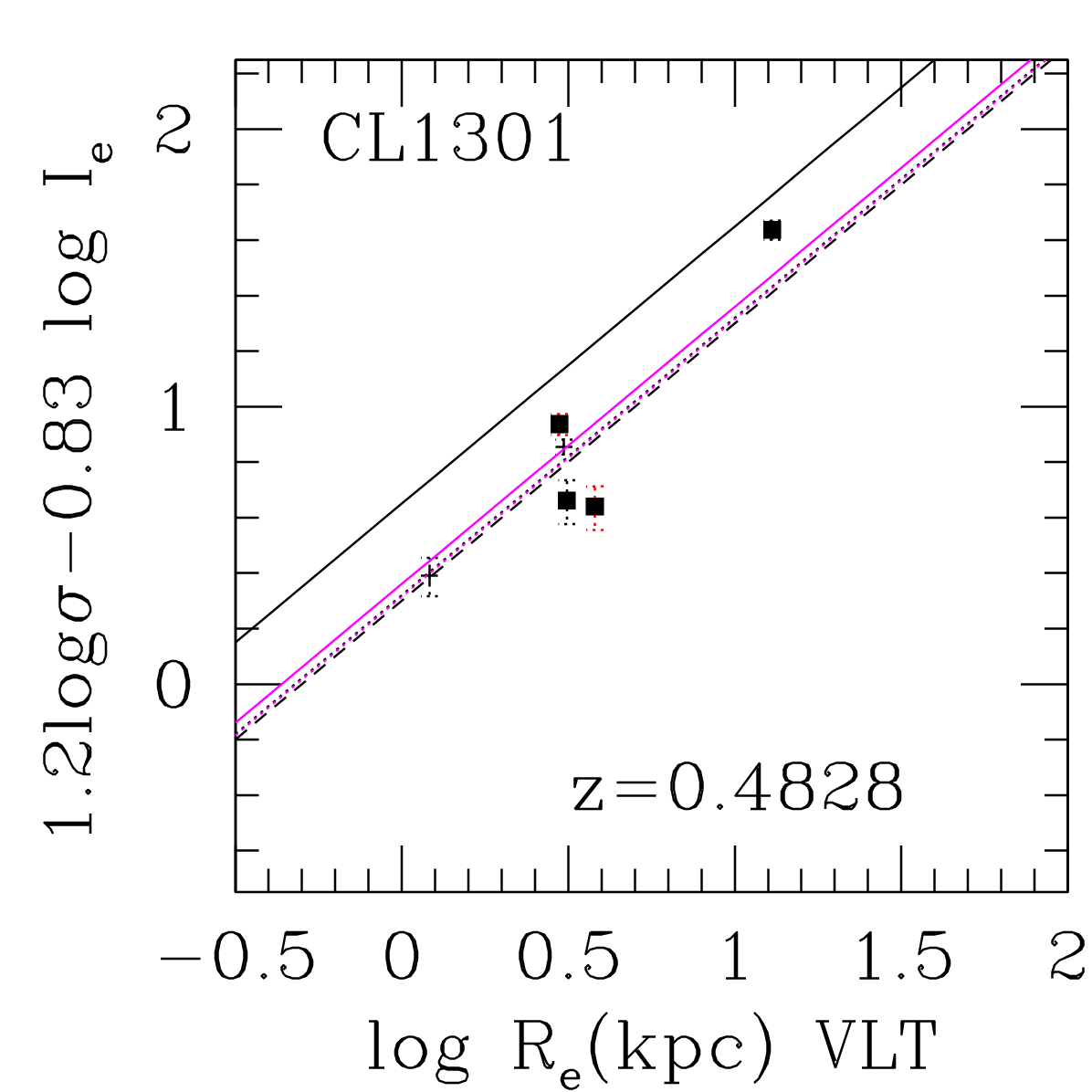,angle=0,width=4cm}}&  
\vbox{\psfig{file=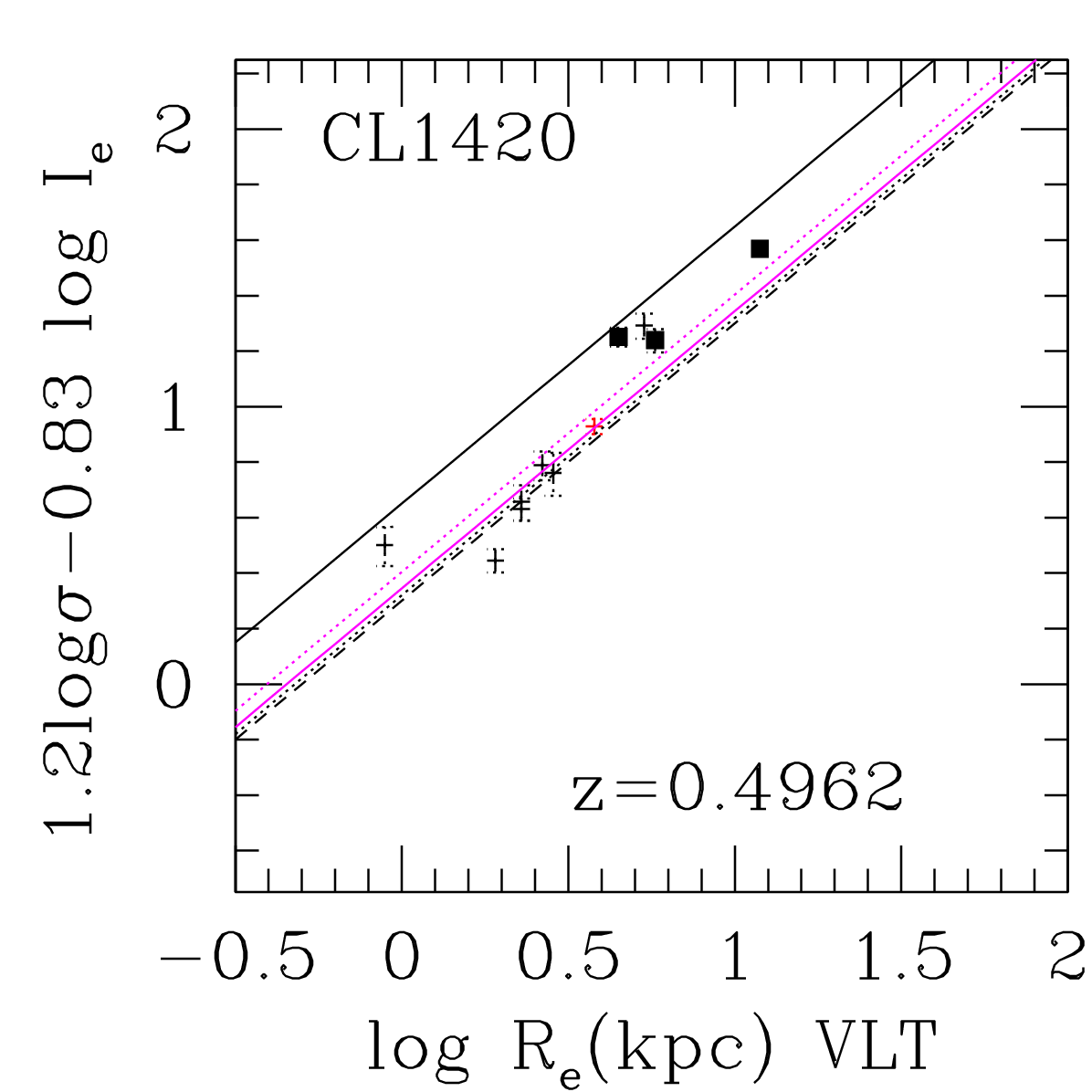,angle=0,width=4cm}}&  
\vbox{\psfig{file=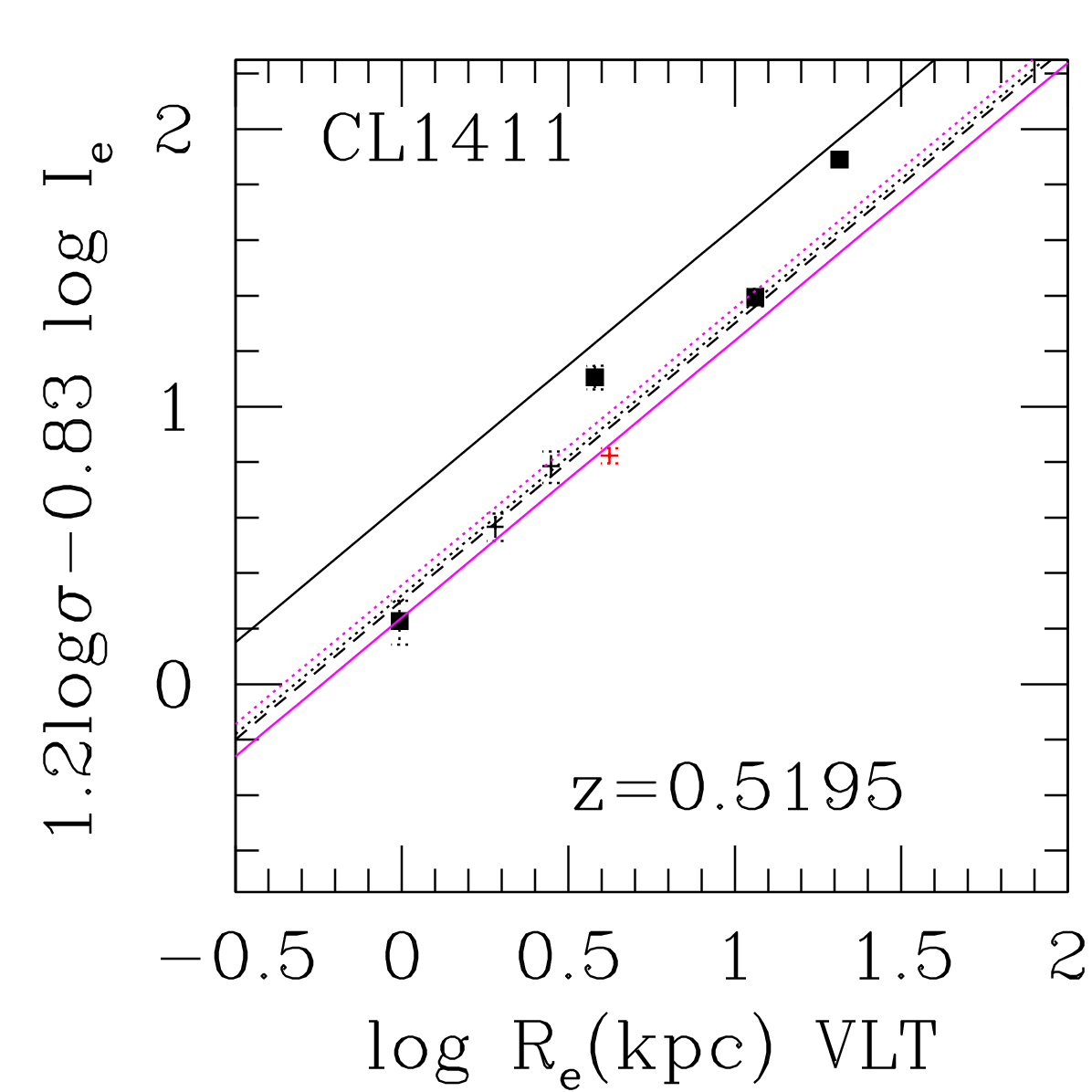,angle=0,width=4cm}}\\ 
\vbox{\psfig{file=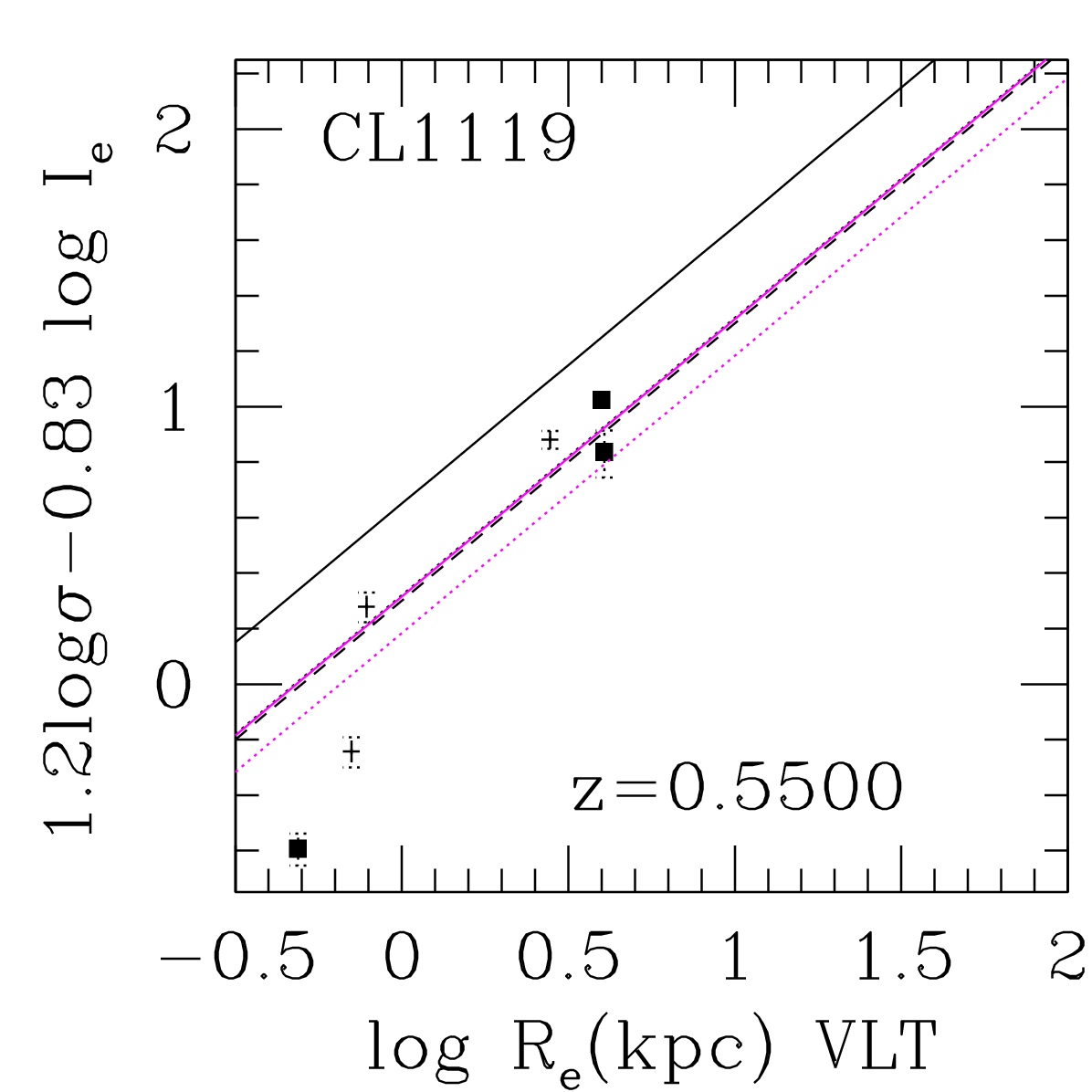,angle=0,width=4cm}}&  
\vbox{\psfig{file=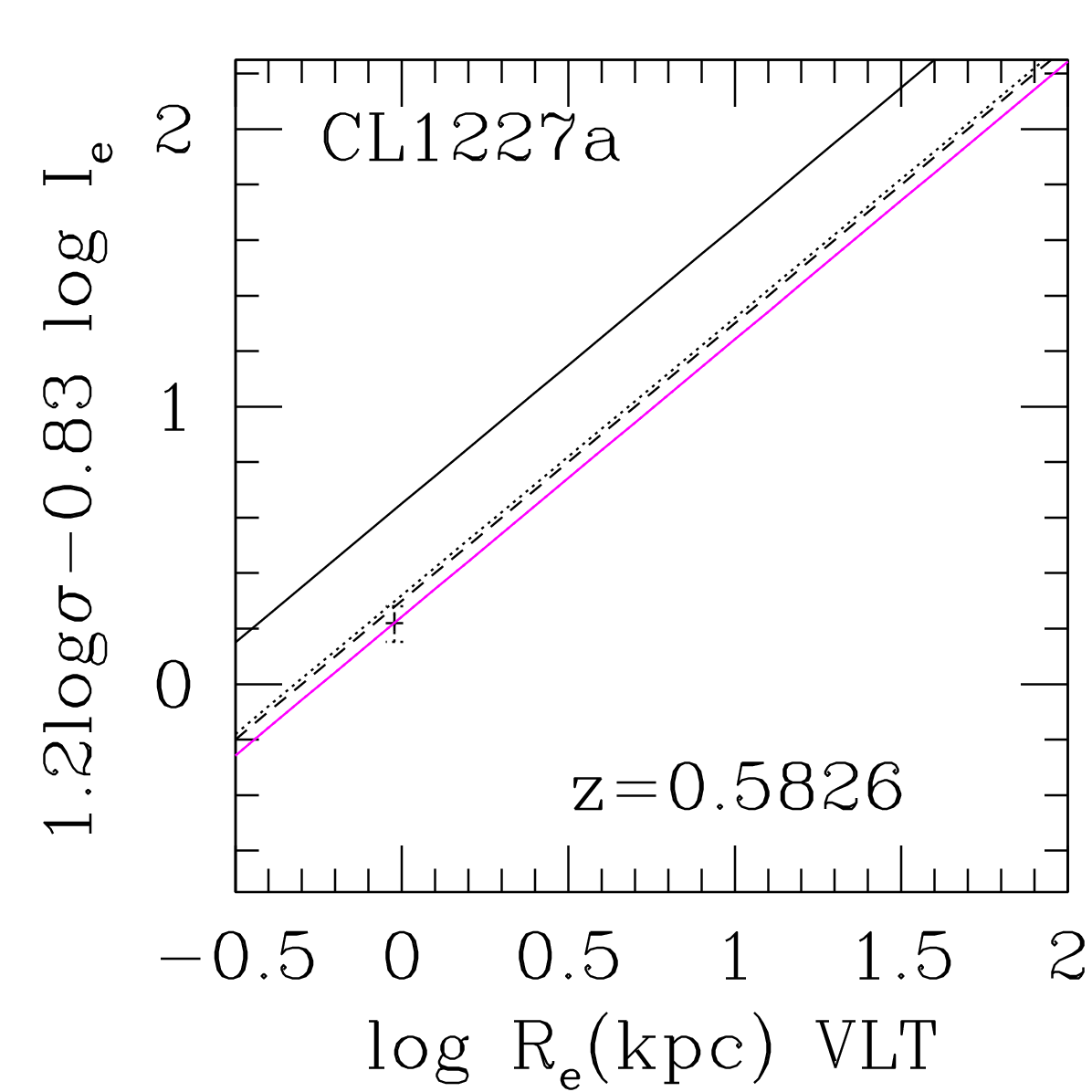,angle=0,width=4cm}}& 
\vbox{\psfig{file=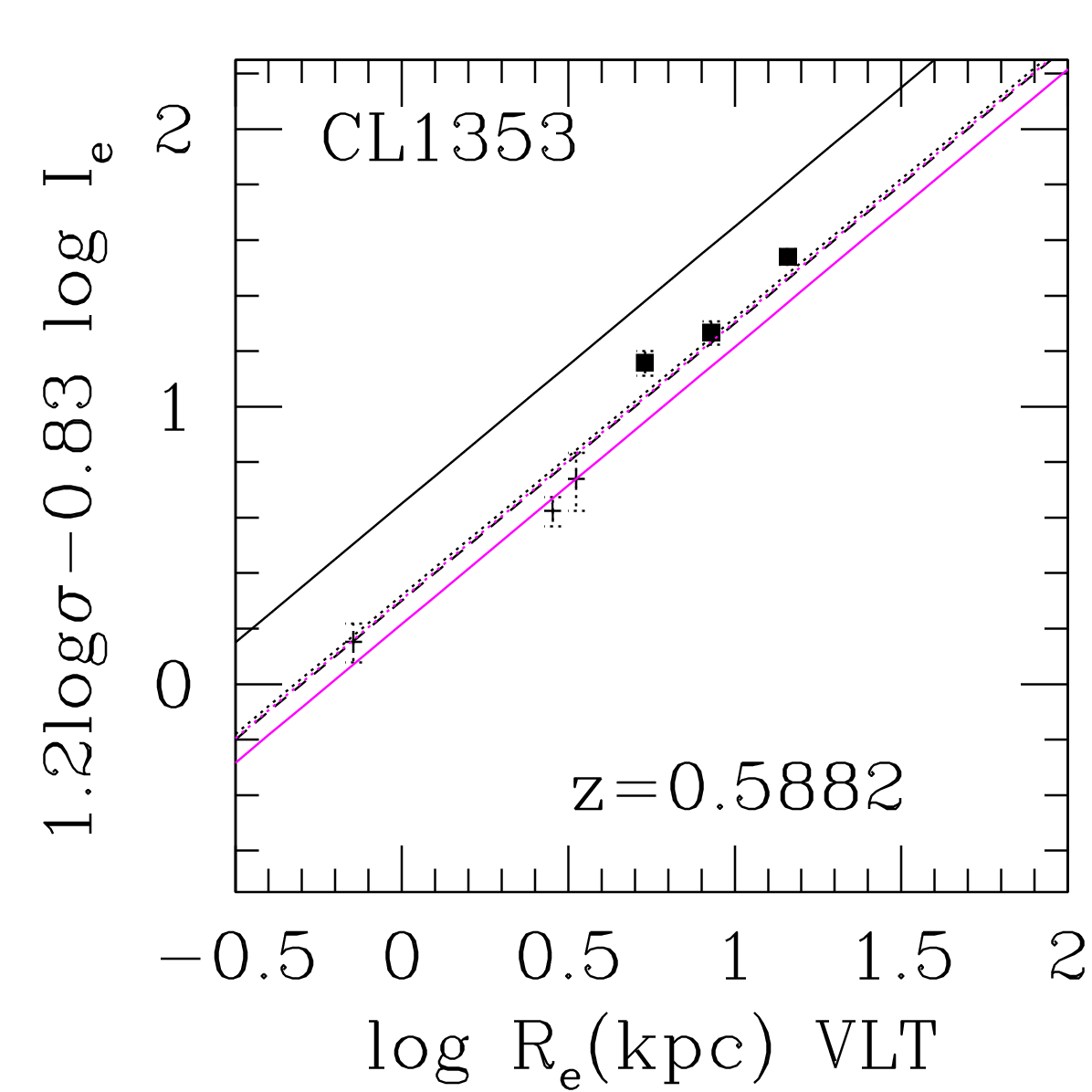,angle=0,width=4cm}}&  
\vbox{\psfig{file=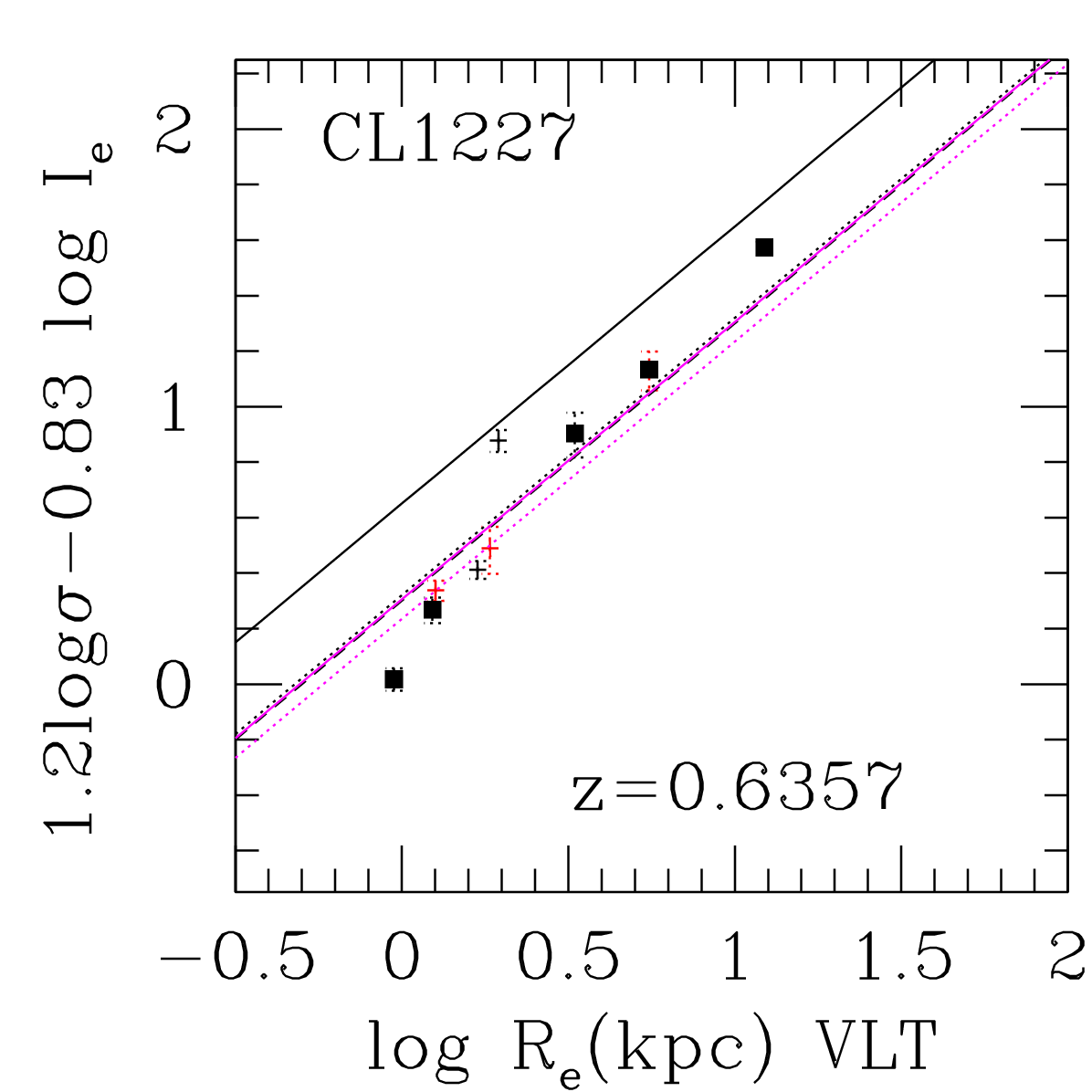,angle=0,width=4cm}}\\  
\end{tabular}
\caption{The FP of the EDisCS clusters with VLT only photometry.  Each
  cluster is identified by its short name for clarity, see Table
  \ref{tab_ZP} for the full name.  Colors code the spectroscopic
  type. The black squares show galaxies with SExtractor flags
  different from 0 or 2 and therefore unreliable photometric parameters.
The dotted magenta line shows the
  best-fitting FP line to all galaxies.
 The solid magenta line shows the best-fitting FP
  line considering only galaxies with spectroscopy type $\le2$ and
  SExtractor flag 0 or 2.    The full line
  shows the Coma cluster at zero redshift. The black dotted and dashed
  lines show the clusters MS2053-04 at $z=0.58$ and MS1054-03 at
  $z=0.83$ from \citet{Wuyts04}, respectively.  \label{fig_FPVLTClus}
}
\end{figure*}

Figure \ref{fig_FPHSTClus} shows the FP of the 14 EDisCS clusters with
HST photometry, while Fig. \ref{fig_FPVLTClus} provides the FP of the
additional 12 clusters with VLT-only photometry. In each cluster, good FP
parameters are available for only a small number of galaxies ($<9$), the
exceptions being cl1232.5-1144, cl1054.4-1146, cl1054.7-1245, and cl1216.8-1201.
Therefore, at this stage we do not attempt to fit the parameters of
the FP except for the zero point, keeping the velocity dispersion and
surface brightness slopes fixed to the local values 
\citep[$\alpha_0=1.2$, $\beta_0=-0.83/(-2.5)=0.33$,][]{Wuyts04}.
In Sect. \ref{sec_rotation}, we argue that this is a good approximation
up to redshift 0.7.  Following \citet{Vandokkum07}, we compute the
zero point as
%
\begin{eqnarray}
\label{eq_FPZP}
ZP& = & \Sigma w (1.2\log \sigma(km/s)-0.83\log I_e(L_\odot/pc^2)\nonumber \\
  &   & -\log R_e(kpc))/\Sigma w,
\end{eqnarray}
%
where the sum comprises all $N$ galaxies in a cluster with measured
velocity dispersion, early spectroscopic type (1 or 2), and (for
clusters with only VLT photometry) SExtractor flag 0 or 2,
irrespective of morphology. At this stage, we weight each point
with $w=(1/1.2d\sigma)^2$, where $d\sigma$ is the error on $\sigma$,
and do not apply selection weighting to be consistent with the
procedures adopted in the literature and minimize scatter. {We note
  that this could generate systematic differences, given that the
  considered surveys have different selection functions. We explore
  the influence of our selection function on the results below}. The
error in the zero point is $\delta ZP=rms(ZP)/\sqrt{N}$. 

\begin{figure*}
\begin{tabular}{cc}
\vbox{\psfig{file=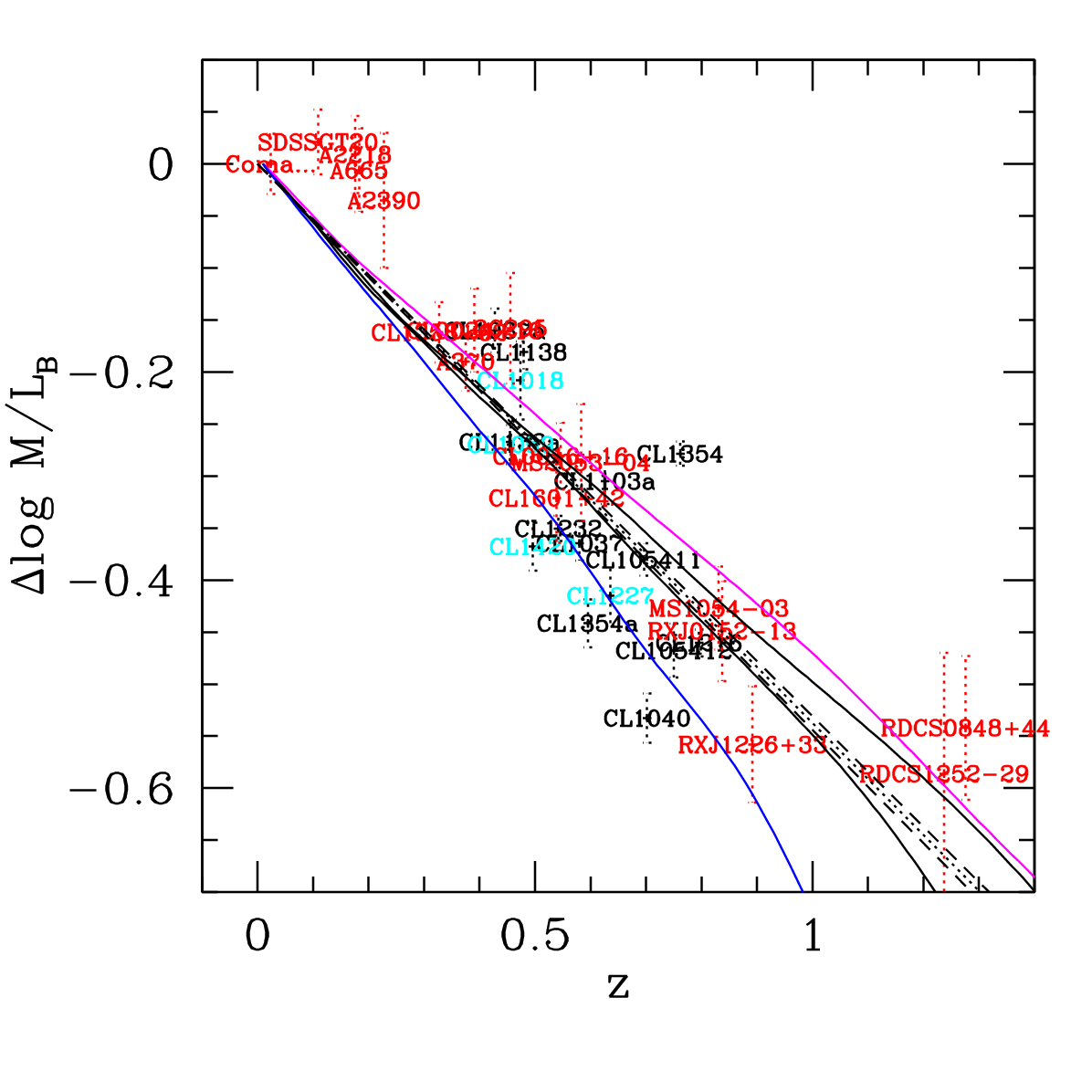,angle=0,width=8cm}}&
\vbox{\psfig{file=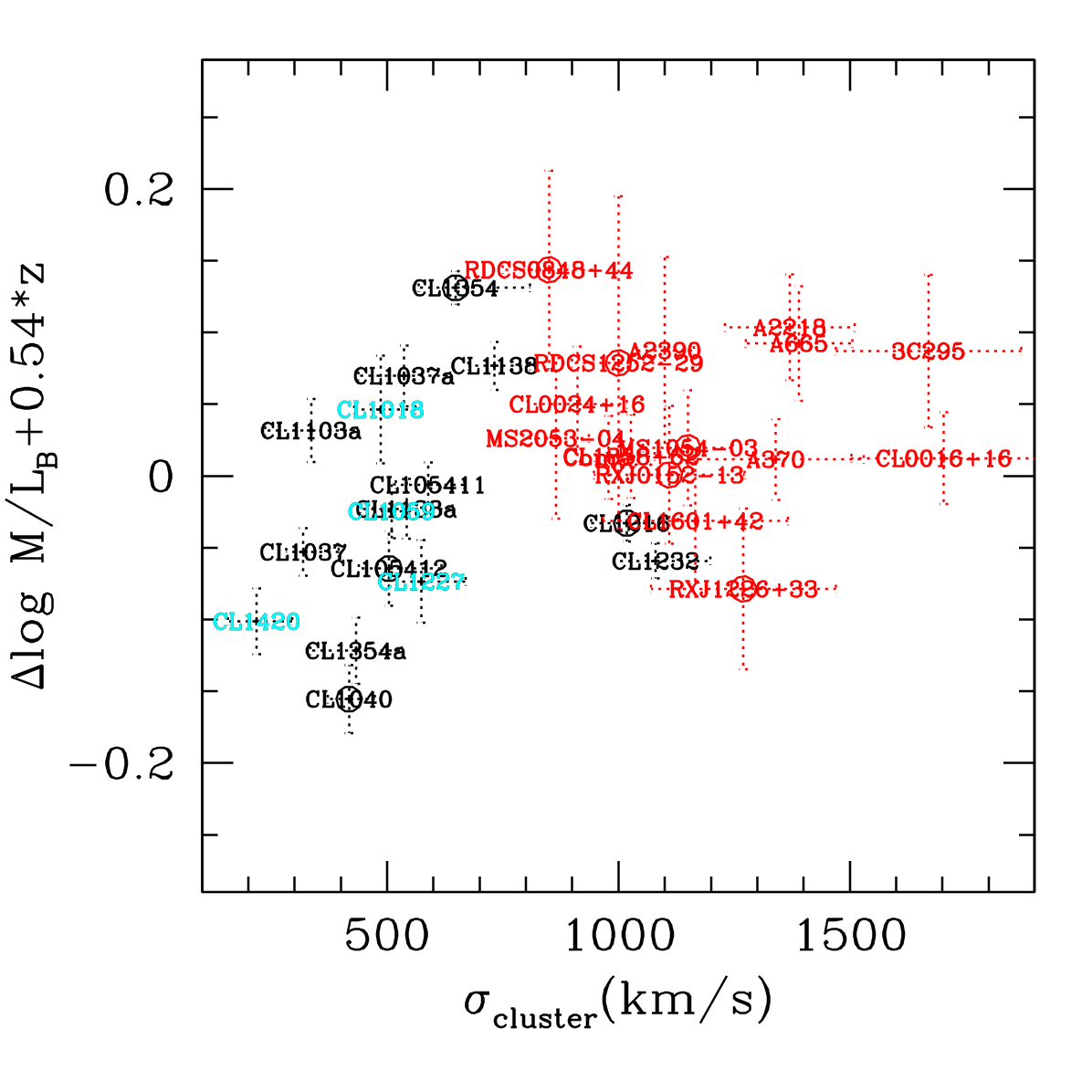,angle=0,width=8cm}}\\
\end{tabular}
\caption{Left: the redshift evolution of the B band mass-to-light
  ratio. The full black lines show the simple stellar population (SSP)
  predictions for a Salpeter IMF and formation redshift of either $z_f=2$
  (lower) or 2.5 (upper curve) and solar metallicity from
  \citet{Maraston05}. The blue line shows the SSP for $z_f=1.5$ and
  twice-solar metallicity, the magenta line the SSP for $z_f=2.5$ and
  half-solar metallicity. The dotted line shows the best-fit
  linear relation and the 1$\sigma$ errors dashed. Right: the (absence
  of) correlation of the M/L residuals $\Delta \log M/L_B+0.54z$ with
  cluster velocity dispersion.  Black points are EDisCS clusters with
  HST photometry, cyan points with VLT photometry. Each
  EDisCS cluster is identified by its short name for clarity, see Table
  \ref{tab_ZP} for the full name. Red points are from
  the literature, \citet{Bender98} and \citet{Vandokkum07}.  Cluster
  velocity dispersions come from \citet{Halliday04} and
  \citet{Milvang08} for EDisCS clusters and from \citet{Edwards02}
  (Coma), \citet{Leborgne92} (A2218), \citet{Gomez00} (A665),
  \citet{Carlberg96} (A2390), \citet{Fisher98} (CL1358+62),
  \citet{Mellier88} (A370), \citet{Poggianti06} (MS1054-03 and
  CL0024+16), \citet{Vandokkum07} (3C295, CL1601+42, CL0016+16),
  \citet{Tran05} (MS2053-04), \citet{Jorgensen05} (RXJ0152-13), and
  \citet{Jorgensen06} (RXJ1226+33) for the literature clusters. We
  estimate $\sigma_{clus}$ for RDCS1252-29 and RDCS084+44 from their
  bolometric X-ray luminosity and the relation of \citet{Johnson06}. 
Circles mark cluster at redshift $>0.7$.
\label{fig_HSTVLTdML}}
\end{figure*}

Following \citet{Wuyts04}, we use the Coma cluster as a reference
point for the whole sample with $ZP=0.65$. All past studies measuring
the peculiar motions of the local universe of early-type galaxies
\citep[][and references therein]{LyndenBell88, Colless01, Hudson04}
agree with the conclusion that Coma, the richest and, in the FP context,
the most well-studied local cluster, is at rest with respect to the cosmic
microwave background and therefore the best suited as a reference.  We
convert the variation in the FP zero point into a variation in the
mean mass-to-light ratio of galaxies in the B band with respect to
Coma using the relation $\Delta\log M/L_B= (ZP-0.65)/0.83$ (where
0.83=$\beta_0\times 2.5$, see Eqs.  \ref{eq_DeltaL} and
\ref{eq_DeltaLdyn}). We note that at this stage we still implicitly
assume, as in the past, that no evolution in size or velocity
dispersion is taking place. Figure \ref{fig_HSTVLTdML}, left, shows
$\Delta\log M/L_B$ as a function of redshift. Only clusters with 4 or
more ($N\ge 4$) galaxies are considered. Table \ref{tab_ZP} gives the
relevant quantities: cluster number (Col. 1), cluster name
\citep[Col. 2, from][]{Milvang08}, cluster short name (Col. 3), type
of photometry used (HST or VLT, Col. 4), cluster velocity dispersion
(Col. 5), {$\Delta \log M/L_B$} (Col. 6), scatter (Col. 7), and
number of galaxies considered (Col. 8).  Table \ref{tab_ZP} also lists
the first six columns for the remaining clusters without FP
ZPs. {If we compute $\Delta \log M/L_B$ using the VLT photometry
  for the 12 clusters with both HST and VLT photometry, we derive a mean
  value $\Delta \log M/L_B(VLT-HST)=-0.04$ (-0.02 if
two outliers, CL1354 and
  CL1138, are not considered) with an rms of 0.06 or an
  error in the mean of 0.02 (see also Sect. \ref{sec_mass}).}

\begin{table*}
\caption{The parameters of the EDisCS clusters with measured FP 
zero points $\Delta \log M/L_B$, without selection weighting.\label{tab_ZP}} 
\begin{tabular}{cllcccccc}
\hline
$N_{clus}$ & Cluster & Short Name &  Phot$^a$ & $z_{clus}$ & $\sigma_{clus}$ & $\Delta \log M/L_B$ & Scatter & N\\
          &         &            &           &          & (km/s)          & (dex) & (dex) & \\
\hline
           1           &  cl1037.9-1243a &   CL1037a           &           1           &       0.4252           &           $         537           ^           {           +          46           }           _           {           -          48           }           $           &           $     -0.16         \pm     0.02           $           &      0.13           &           4          \\
           2           &  cl1138.2-1133a &   CL1138a           &           1           &      0.4548           &           $         542           ^           {           +          63           }           _           {           -          71           }           $           &           $     -0.27         \pm     0.02           $           &      0.27           &           5          \\
           3           &  cl1138.2-1133 &    CL1138           &           1           &      0.4796           &           $         732           ^           {           +          72           }           _           {           -          76           }           $           &           $     -0.18         \pm     0.02           $           &      0.08           &           9          \\
           4           &  cl1232.5-1144 &    CL1232           &           1           &      0.5414           &           $        1080           ^           {           +         119           }           _           {           -          89           }           $           &           $     -0.35         \pm     0.01           $           &      0.17           &          20          \\
           5           &  cl1037.9-1243 &    CL1037           &           1           &        0.5783           &           $         319           ^           {           +          53           }           _           {           -          52           }           $           &           $     -0.36         \pm     0.02           $           &      0.27           &           5          \\
           6           &  cl1354.2-1230a &   CL1354a           &           1           &      0.5952           &           $         433           ^           {           +          95           }           _           {           -         104           }           $           &           $     -0.44         \pm     0.02           $           &      0.14           &           4          \\
           7           & cl1103.7-1245a &    CL1103a           &           1           &      0.6261           &           $         336           ^           {           +          36           }           _           {           -          40           }           $           &           $     -0.30         \pm     0.02           $           &      0.15           &           4          \\
           8           & cl1054.4-1146 &  CL105411           &           1           &      0.6972           &           $         589           ^           {           +          78           }           _           {           -          70           }           $           &           $     -0.38         \pm     0.02           $           &      0.17           &          12          \\
           9           &  cl1040.7-1155 &    CL1040           &           1           &       0.7043           &           $         418           ^           {           +          55           }           _           {           -          46           }           $           &           $     -0.53         \pm     0.02           $           &      0.11           &           5          \\
          10           &  cl1054.7-1245 &  CL105412           &           1           &      0.7498           &           $         504           ^           {           +         113           }           _           {           -          65           }           $           &           $     -0.47         \pm     0.03           $           &       0.30           &          11          \\
          11           &  cl1354.2-1230 &    CL1354           &           1           &       0.762           &           $         648           ^           {           +         105           }           _           {           -          110           }           $           &           $     -0.28         \pm     0.01           $           &      0.22          &           6          \\
          12           &  cl1216.8-1201 &    CL1216           &           1           &      0.7943           &           $        1018           ^           {           +          73           }           _           {           -          77           }           $           &           $     -0.46         \pm     0.01           $           &       0.23           &          23          \\
          13           &  cl1059.2-1253 &    CL1059           &           0           &      0.4564           &           $         510           ^           {           +          52           }           _           {           -          56           }           $           &           $     -0.27         \pm     0.02           $           &      0.10           &           8          \\
          14           &  cl1018.8-1211 &    CL1018           &           0           &      0.4734           &           $         486           ^           {           +          59           }           _           {           -          63           }           $           &           $     -0.21         \pm      0.04           $           &      0.12           &           4          \\
          15           & cl1420.3-1236 &     CL1420           &           0           &      0.4962           &           $         218           ^           {           +          43           }           _           {           -          50           }           $           &           $     -0.37         \pm     0.02           $           &       0.14           &           8          \\
          16           &  cl1227.9-1138 &    CL1227           &           0           &      0.6357           &           $         574           ^           {           +          72           }           _           {           -          75           }           $           &           $     -0.42         \pm     0.03           $           &      0.21           &           4          \\

17 & cl1103.7-1245  & CL1103  & 1 & 0.9586 & $534^{+101}_{-120}$ & - & - & -\\
18 & cl1103.7-1245b & CL1103b & 1 & 0.7031 & $252^{+65}_{-85}  $ & - & - & -\\
19 & cl1119.3-1129  & CL1119  & 0 & 0.5500 & $166^{+27}_{-29 } $ & - & - & -\\
20 & cl1202.7-1224   & CL1202  & 0 & 0.424 & $518^{+92}_{-104} $ & - & - & -\\
21 & cl1227.9-1138a & CL1227a & 0 & 0.5826 & $341^{+42}_{-46}  $ & - & - & -\\
22 & cl1238.5-1144  & CL1238  & 0 & 0.4602 & $447^{+135}_{-181}$ & - & - & -\\
23 & cl1301.7-1139  & CL1301  & 0 & 0.4828 & $687^{+81}_{-86}  $ & - & - & -\\
24 & cl1301.7-1139a & CL1301a & 0 & 0.3969 & $391^{+63}_{-69}  $ & - & - & -\\
25 & cl1353.0-1137  & CL1353  & 0 & 0.5882 & $666^{+136}_{-139}$ & - & - & -\\
26 & cl1411.1-1148  & CL1411  & 0 & 0.5195 & $710^{+125}_{-133}$ & - & - & -\\
\hline
\multicolumn{7}{l}{(a) 1: with HST photometry, 0: with VLT photometry.}\\
\end{tabular}
\end{table*}

We add to the EDisCS sample 15 clusters from the literature
\citep{Vandokkum07}, plus A370 from \citet{Bender98}. They span the
redshift range $z=0.109-1.28$ and sample the high cluster velocity
dispersion ($\sigma_{clus}>800 km/s$) regime only. Moreover, as a
common zero-redshift comparison we add the Coma cluster. A linear
weighted fit to the whole sample gives $\Delta \log
M/L_B=(-0.54\pm0.01)z$.  Applying selection weighting reduces the
slope to $-0.47$. A fit restricted to the literature sample alone
gives $-0.49\pm0.02$. \citet{Wuyts04} derive $-0.47$,
whereas \citet{Vandokkum07} find $-0.555\pm 0.042$.  In view of the size
evolution discussion of Sect.  \ref{sec_sizeev}, where dependencies of
$\log(1+z)$ are considered, we also fit the slope $\eta$ of the form  $\Delta
\log M/L_B=\eta\log(1+z)$. The results are summarized in Table
\ref{tab_FPslope}. 

The residuals of the EDisCS cluster sample have an rms of 0.08 dex.
The literature sample, which does not probe clusters with small
velocity dispersions (see below), has an rms of 0.06 dex, the
clusters at low redshift ($z\le 0.2$) having systematically positive
residuals. The combined sample has an rms scatter of 0.07. Taking into
account the measurement errors, this implies an intrinsic scatter of
0.06 dex or 15\% in M/L.  The best-fit line closely matches the
prediction of simple stellar population models \citep{Maraston05} with
high formation redshift ($2\le z_f\le 2.5$) and solar metallicities.
Here and below we make use of \citet{Maraston05} models to translate
mass-to-light or luminosity variations into formation ages or
redshifts.  Similar conclusions would be obtained using other models
\citep[e.g.,][]{BC03}, see for example \citet{Jaffe10}. However, we
bear in mind that systematic errors still affect the SPP approach
\citep[see][for the difficulties in reproducing the colors of real
galaxies]{Maraston09, Conroy10}.

Trimming the sample to high-precision data only (for example,
considering only velocity dispersions determined to a precision higher
than 10\%) does not change the overall picture.  We discuss the
effects of cutting the sample according to mass, spectroscopic type,
or morphology in Sect. \ref{sec_mass}, where we consider the sample on
a galaxy by galaxy basis, since any selection drastically reduces the
number of clusters with at least 4 galaxies.

Figure \ref{fig_HSTVLTdML} (right panel) shows the residuals $\Delta
\log M/L_B+0.54z$ as a function of the cluster velocity dispersion. No
convincing correlation is seen (the Pearson coefficient is 0.21, the
Spearman coefficient 0.39 with a probability of 2.5\% that a
correlation exists), confirming that cluster massive early-type
galaxies follow passive evolution up to high redshifts not only in
massive clusters, as has been established (see discussion in the
Introduction), but also in lower mass structures down to the group
size. There is a hint that the scatter could increase in the low
velocity dispersion clusters: while the combined EDisCS+Literature
sample of high velocity dispersion clusters ($\sigma_{clus}>800$ km/s)
exhibit an rms of the residuals $\Delta \log M/L_B+0.54z$ of 0.06 dex,
the lower $\sigma_{clus}$ EDisCS clusters exhibit an rms of 0.08
dex. We note that the scatter in M/L measured in each cluster is
larger (up to 0.3 dex) and intrinsic (i.e., not caused by measurement
errors).

\subsection{Environment and mass dependence}
\label{sec_mass}

\begin{figure*}
\begin{tabular}{cc}
\vbox{\psfig{file=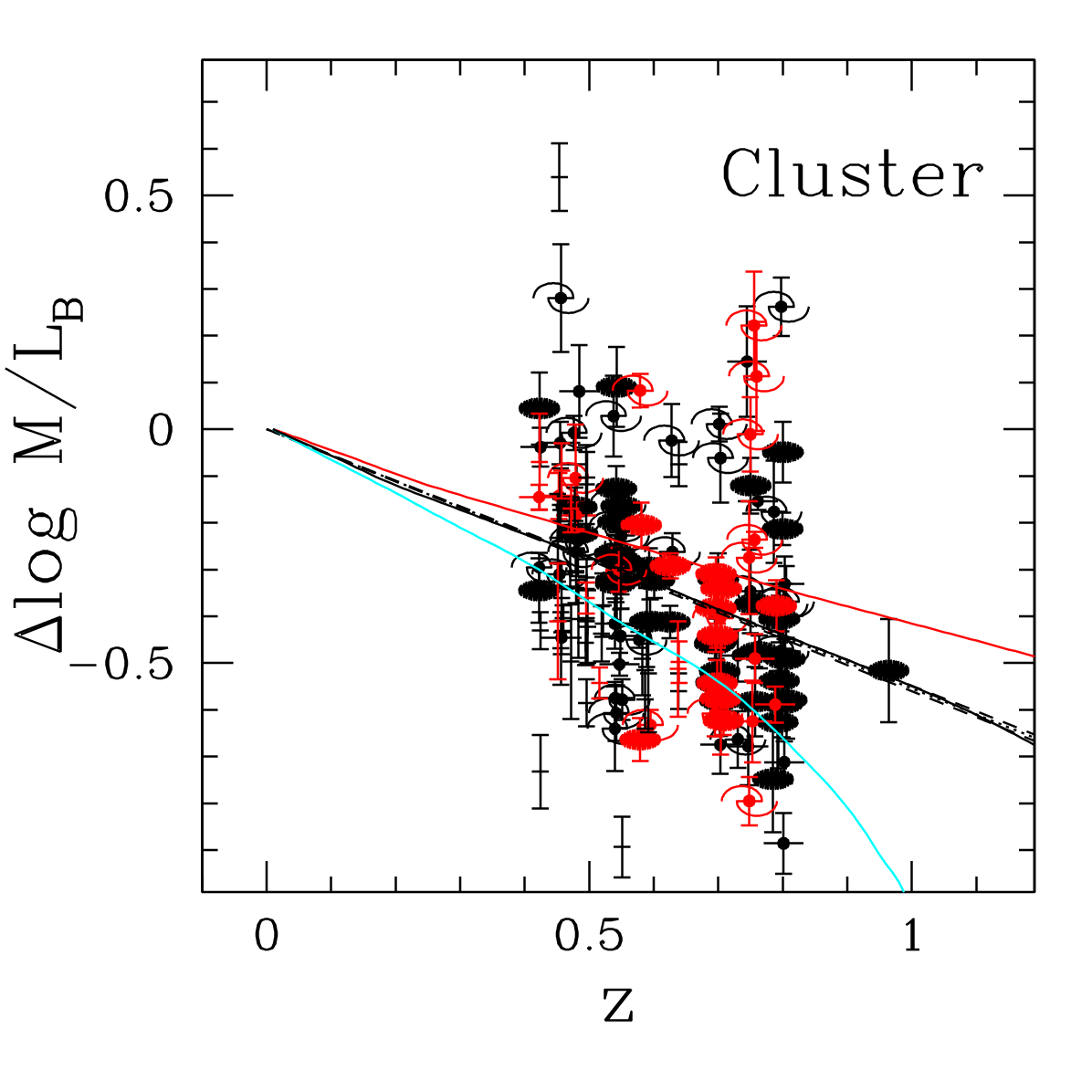,angle=0,width=8cm}}&
\vbox{\psfig{file=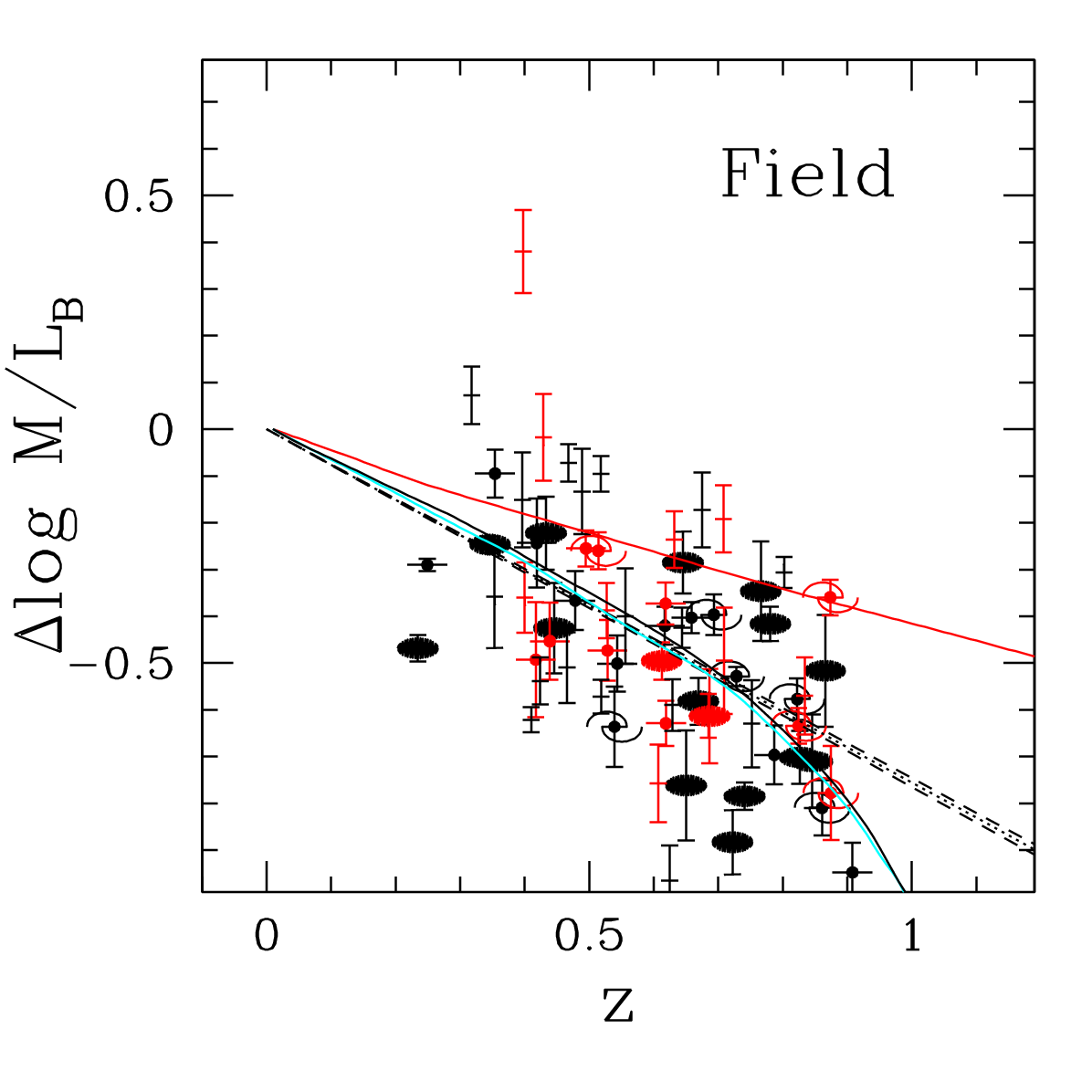,angle=0,width=8cm}}\\
\vbox{\psfig{file=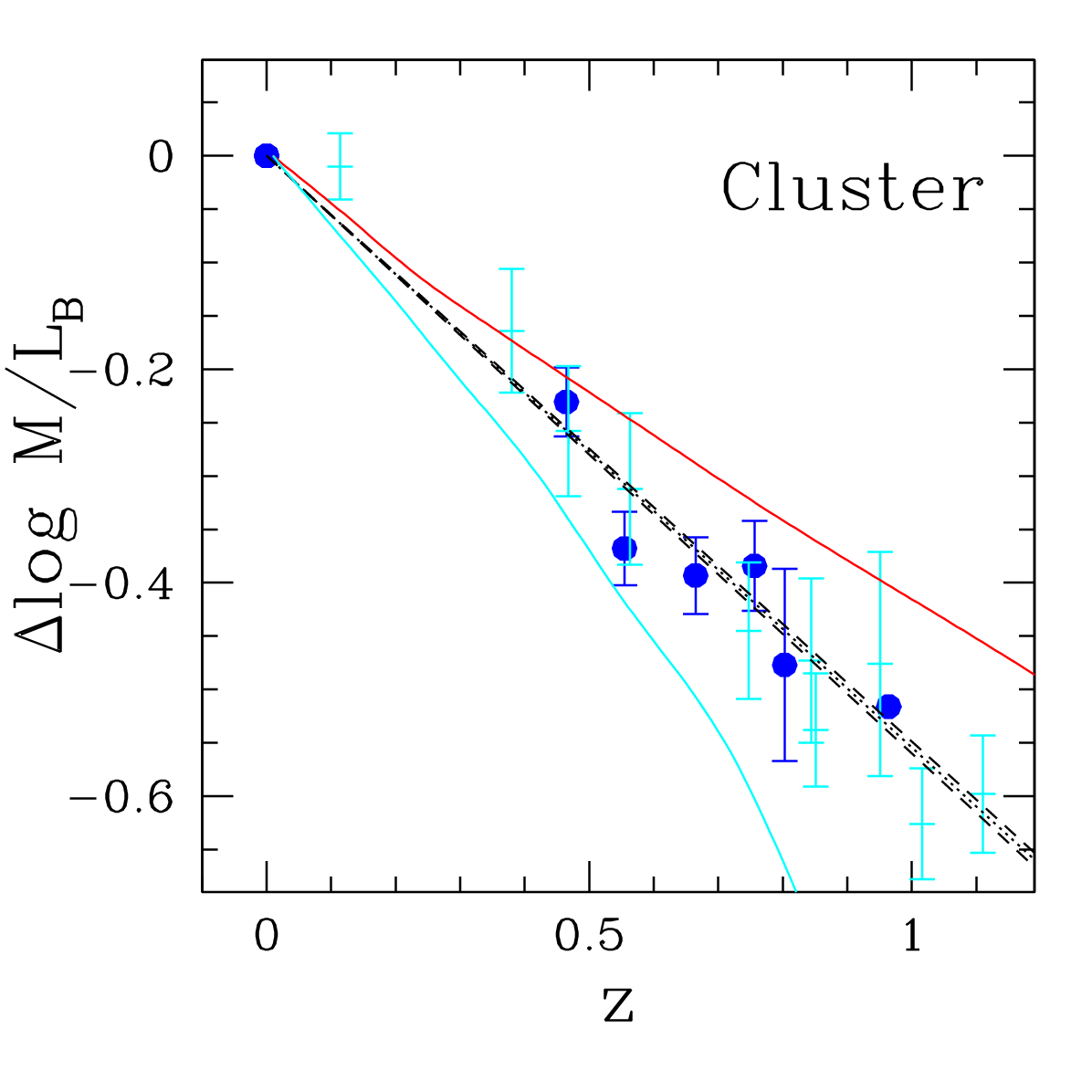,angle=0,width=8cm}}&
\vbox{\psfig{file=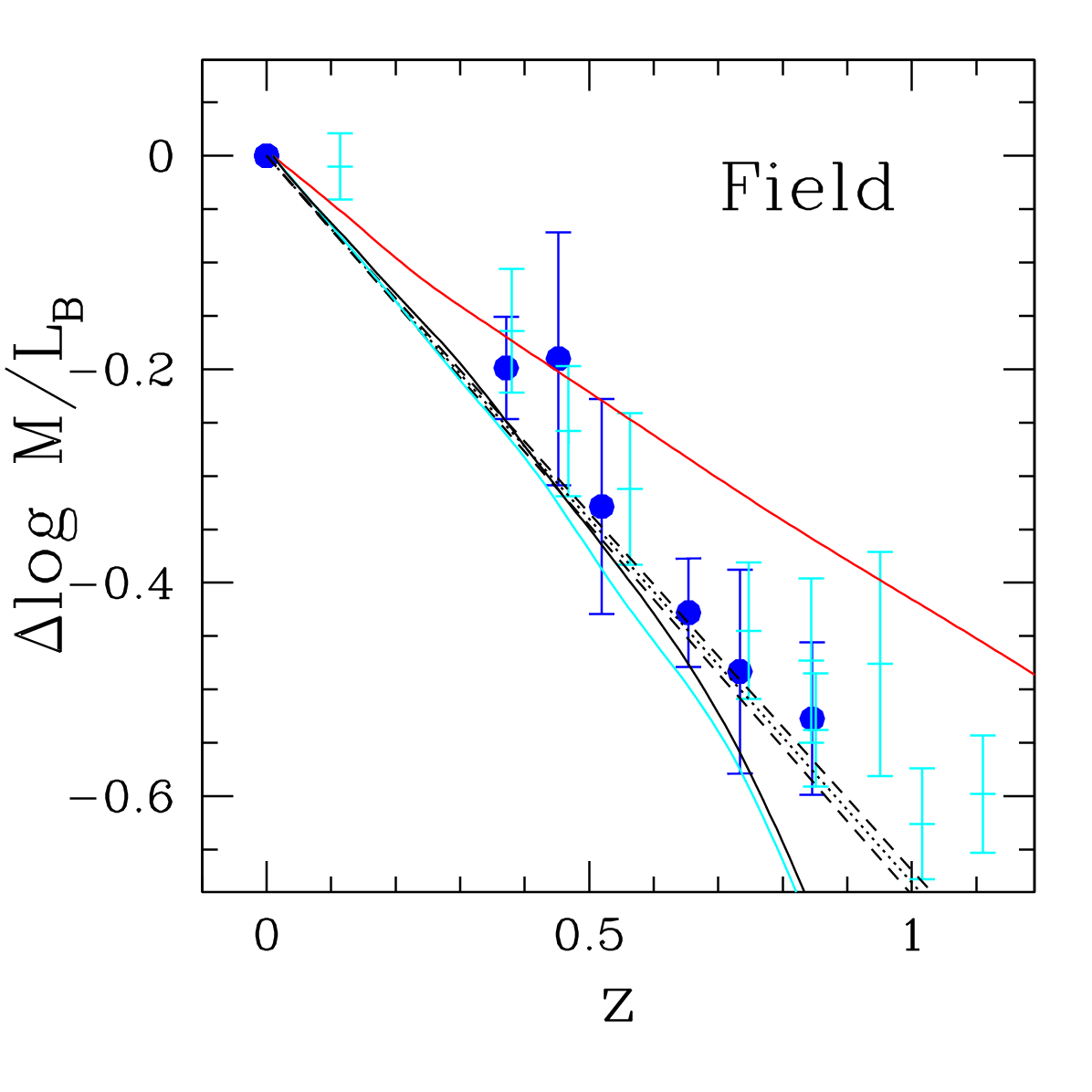,angle=0,width=8cm}}\\
\end{tabular}
\caption {The redshift evolution of the mass-to-light ratio for cluster (left)
  and field (right) galaxies. Top: Black and red indicate galaxies
  with spectroscopic types 1 and 2, respectively. Morphologies for
  galaxies with HST photometry are coded as in
  Fig. \ref{fig_FPHSTClus}. Galaxies with VLT photometry only are
  shown as crosses. Only galaxies with good VLT photometry
  (i.e., SExtractor flag 0 or 2) are plotted.  
The solid black lines show the solar metallicity 
SSP for $z_f=2$ (cluster) and  $z_f=1.2$ (field). The solid red line shows
  the SSP for $z_f=3.5$ and half-solar metallicity, the cyan line
  shows the SSP for $z_f=1.2$ and twice-solar metallicity. The dotted
  line shows the best-fit linear relation (-0.55z for cluster and
  -0.76z for field galaxies) and the 1$\sigma$ errors dashed.  Bottom:
  The blue points show averages over redshift bins 0.1 wide. The cyan
  points are average field galaxies from \citet{Vandokkum07}. Only
  field galaxies (plot to the right) with dynamical masses higher than
  $10^{11}M_\odot$ are considered.
\label{fig_fieldML}
}
\end{figure*}

We now consider the sample on a galaxy by galaxy basis. As in Eq.
\ref{eq_FPZP}, in Fig.  \ref{fig_fieldML} we show the evolution with
redshift of $\Delta \log M/L_B=(1.2\log \sigma -0.83 \log I_e-\log
R_e-0.65)/0.83$ for the EDisCS cluster (left) and field (right)
galaxies.  {For the 74 galaxies with both HST and VLT photometry,
  we derive a mean difference $\Delta \log M/L_B(VLT-HST)=-0.02$ with
  an rms of 0.06 or an error in the mean of 0.02, similar to that
  quoted for clusters in Sect. \ref{sec_FPclus}.}  In general, there
is scatter in the galaxy data that falls even below the SPP model line
for a formation redshift $z_f=1.2$ with twice-solar metallicity, or to
positive values that are impossible to explain with simple stellar
population models. Many of these deviant points are galaxies with
late-type morphology. Their measured velocity dispersion might not be
capturing their dynamical state dominated by rotation.

First, we turn our attention to galaxies belonging to clusters.
Averaging the points in redshift bins 0.1 wide shows that cluster
galaxies closely follow the mean linear fit derived for clusters as a
whole.  This corresponds to a solar metallicity SPP model with
formation redshift $z_f=2$ or formation lookback time of 10 Gyr (see
Sect. \ref{sec_ages} for a detailed discussion). The average values do
not change within the errors if a cut either in mass
($M_{dyn}>10^{11}M_\odot$) or morphology ($T\le 0$) is
applied. Table \ref{tab_FPslope} lists the slope $\eta$ and $\eta'$ of
$\Delta \log M/L=\eta\log(1+z)=\eta' z$ derived by cutting the sample
in a progressively more selective way. In general, $P_S$ selection
weighting produces shallower slopes. Shallower slopes are also
obtained when only massive galaxies or spectral types ST=1 are
considered. The steepest slope ($\eta' =-0.56$) is obtained by
considering only galaxies with HST early-type morphologies, no
restrictions on spectral type or mass, and no selection weighting. The
shallowest slope ($\eta' =-0.32$) is obtained considering only galaxies
more massive than $10^{11} M_\odot$, with spectral type ST=1, no
constraints on morphology and $P_S$ weighting. Finally,
  considering galaxies with HST photometry and no constraints on
  morphology or mass, but with ellipticity less than $1-b_e/a_e\le0.6$ changes
  the slopes only minimally, from $\eta'=-0.53$ (for 113 objects) to
  $\eta'=-0.56$ (for 88 objects).

In contrast, galaxies in the field have values of $\Delta \log M/L_B$
more negative than the corresponding cluster bins starting from
$z\approx 0.45$. For our sample, a solar metallicity SSP model with
formation redshift $z_f=1.2$ is an accurate representation of the
data. This corresponds to a formation age of 8.4 Gyr or a mean age
difference of 1.6 Gyr between cluster and field galaxies (see
Sect. \ref{sec_ages} for a detailed discussion).  The slopes $\eta$
listed in Table \ref{tab_FPslope} for field galaxies are always
steeper than the ones derived for cluster galaxies. The shallowest
($\eta'=-0.67$) is obtained when considering only galaxies more
massive than $10^{11} M_\odot$ with ST=2. Here we approach the result
of \citet{Vandokkum07}, who detect only a very small age difference
between cluster and field galaxies of these masses and morphologies.
Still, our shallowest slope for field galaxies is steeper than
  the steepest slope for cluster galaxies.

\begin{table*}[h!]
\caption{The slopes of the zero point evolution of the FP
$\Delta \log M/L = 0.4(ZP(z)-ZP(0))/\beta_0=\eta' z=\eta \log(1+z)$.
\label{tab_FPslope}} 
\begin{tabular}{lrlllcclcc}
\hline
Type             & $N_{gal}$ & $P_S$ & HST & VLT & ST & Morph & $M_{dyn}$            & $\eta'$          & $\eta$   \\
\hline
Clusters         & 132      & No    & Yes & Yes & 2     & 10    & All              & $-0.54\pm 0.01$ & $-1.61\pm0.01$\\
Clusters         & 132      & Yes   & Yes & Yes & 2     & 10    & All              & $-0.47\pm0.003$ & $-1.43\pm0.01$\\
Cluster galaxies & 154      & No    & Yes & Yes & 2     & 10    & All              & $-0.55\pm0.006$ & $-1.66\pm0.02$\\
Cluster galaxies & 154      & Yes   & Yes & Yes & 2     & 10    & All              & $-0.48\pm0.01$ & $-1.45\pm0.03$\\
Cluster galaxies & 67       & No    & Yes & Yes & 2     & 10    & $10^{11}M_\odot$ & $-0.44\pm0.01$   & $-1.34\pm0.02$\\
Cluster galaxies & 67       & Yes   & Yes & Yes & 2     & 10    & $10^{11}M_\odot$ & $-0.36\pm0.01$   & $-1.10\pm0.02$\\
Cluster galaxies & 43       & No    & Yes & Yes & 1     & 10    & $10^{11}M_\odot$ & $-0.41\pm0.01$   & $-1.24\pm0.03$\\
Cluster galaxies & 43       & Yes   & Yes & Yes & 1     & 10    & $10^{11}M_\odot$ & $-0.32\pm0.01$   & $-0.97\pm0.03$\\
Cluster galaxies & 76       & No    & Yes & No  & 2     & 0     & All              & $-0.56\pm0.01$ & $-1.70\pm0.02$\\
Cluster galaxies & 76       & Yes   & Yes & No  & 2     & 0     & All              & $-0.51\pm0.01$ & $-1.54\pm0.04$\\
Cluster galaxies & 33       & No    & Yes & No  & 2     & 0     & $10^{11}M_\odot$ & $-0.47\pm0.01$    & $-1.44\pm0.03$\\
Cluster galaxies & 33       & Yes   & Yes & No  & 2     & 0     & $10^{11}M_\odot$ & $-0.44\pm0.01$   & $-1.34\pm0.03$\\
Cluster galaxies & 24       & No    & Yes & No  & 1     & 0     & $10^{11}M_\odot$ & $-0.46\pm0.01$    & $-1.41\pm0.03$\\
Cluster galaxies & 24       & Yes   & Yes & No  & 1     & 0     & $10^{11}M_\odot$ & $-0.43\pm0.01$   & $-1.32\pm0.03$\\
Field galaxies   & 68       & No    & Yes & Yes & 2     & 10    & All              & $-0.76\pm0.01$  & $-2.27\pm0.03$\\
Field galaxies   & 68       & Yes   & Yes & Yes & 2     & 10    & All              & $-0.76\pm0.01$ & $-2.28\pm0.03$\\
Field galaxies   & 28       & No    & Yes & Yes & 2     & 10    & $10^{11}M_\odot$ & $-0.68\pm0.01$    &$-2.05\pm0.03$ \\
Field galaxies   & 28       & Yes   & Yes & Yes & 2     & 10    & $10^{11}M_\odot$ & $-0.67\pm0.01$   &$-1.99\pm0.04$ \\
Field galaxies   & 16       & No    & Yes & Yes & 1     & 10    & $10^{11}M_\odot$ & $-0.70\pm0.01$   &$-2.10\pm0.04$ \\
Field galaxies   & 16       & Yes   & Yes & Yes & 1     & 10    & $10^{11}M_\odot$ & $-0.73\pm0.02$   &$-2.16\pm0.04$ \\
Field galaxies   & 32       & No    & Yes & No  & 2     & 0     & All              & $-0.83\pm0.01$ &$-2.46\pm0.04$\\
Field galaxies   & 32       & Yes   & Yes & No  & 2     & 0     & All              & $-0.87\pm0.02$ &$-2.58\pm0.05$\\
Field galaxies   & 8        & No    & Yes & No  & 2     & 0     & $10^{11}M_\odot$ & $-0.83\pm0.02$    &$-2.43\pm0.06$ \\
Field galaxies   & 8        & Yes   & Yes & No  & 2     & 0     & $10^{11}M_\odot$ & $-0.90\pm0.02$    &$-2.59\pm0.07$ \\
Field galaxies   & 6        & No    & Yes & No  & 1     & 0     & $10^{11}M_\odot$ & $-0.82\pm0.02$    &$-2.40\pm0.06$\\
Field galaxies   & 6        & Yes   & Yes & No  & 1     & 0     & $10^{11}M_\odot$ & $-0.91\pm0.02$    &$-2.59\pm0.07$\\
\hline
\end{tabular}
\end{table*}

We compute dynamical masses as in Eq. \ref{eq_mass}. As discussed in
the Introduction, the validity of this equation can be questioned in
many respects. The value of the appropriate structural constant need
not to be the same for every galaxy. If ordered motions dominate the
dynamics of a galaxy, as must be the case for disk galaxies, the
use of velocity dispersion is inappropriate. Moreover, we also
assume that the structure proportionality constant does not vary with
redshift, which might not be true.  Nevertheless, on average Eq.
\ref{eq_mass} delivers values that compare reasonably with stellar
masses. We compute the (total) stellar masses from ground-based,
rest-frame absolute photometry derived from SED fitting
\citep{Rudnick09}, adopting the calibrations of \citet{BdJ01}, with a
'diet' Salpeter IMF (with constant fractions of stars of mass less
than 0.6$M_\odot$) and B-V colors, and renormalized using the
corrections for an elliptical galaxy given in \citet{deJongBell07}.
The method to calculate the rest-frame luminosities and colors
  is described in \citet{Rudnick03}, and the rest-frame filters have
  been taken from \citet{Bessel90}. Although the photometric redshifts
  and rest-frame SEDs have been computed from the matched aperture
  photometry of \citet{White05}, the rest-frame luminosities have been
  adjusted to total values, as described in \citet{Rudnick09}.

In general, the dynamical masses are somewhat lower than the stellar
ones ($M_{dyn}/M_*=0.91$ for cluster galaxies, 0.75 for field
galaxies), with an intrinsic scatter of a factor of two, on the order
of the typical combined precision achieved for dynamical and
  stellar masses.  If we consider only galaxies with HST morphology
$T<0$, the ratio $M_{dyn}/M_*$ drops to 0.74 for cluster and 0.56 for
field galaxies.  Moreover, a possible decreasing trend with redshift
of the ratio $M_{dyn}/M_*$ is seen at the $2-\sigma$ level, which is
not unexpected given the size and velocity dispersion evolution
discussed in Sect. \ref{sec_ReSigmaEv}.  To conclude, the tendency to
have $M_{dyn}/M_*<1$ may indicate that the structural constant used in
Eq. \ref{eq_mass} is too low.  However, we note that the
structural constant is the one that dynamical studies at low redshifts
prefer \citep{Cappellari06, Thomas10}.  Alternatively, our adopted IMF
contains too high a fraction of low mass stars \citep{Baldry08}.
  Finally, we refer to \citet{Thomas10} for a discussion of the role
  of dark matter in the estimation of $M_{dyn}$.

In the following, we consider relations as a function of
both dynamical and stellar masses to assess the robustness of each
result.

\begin{figure*}
\begin{tabular}{ccc}
\vbox{\psfig{file=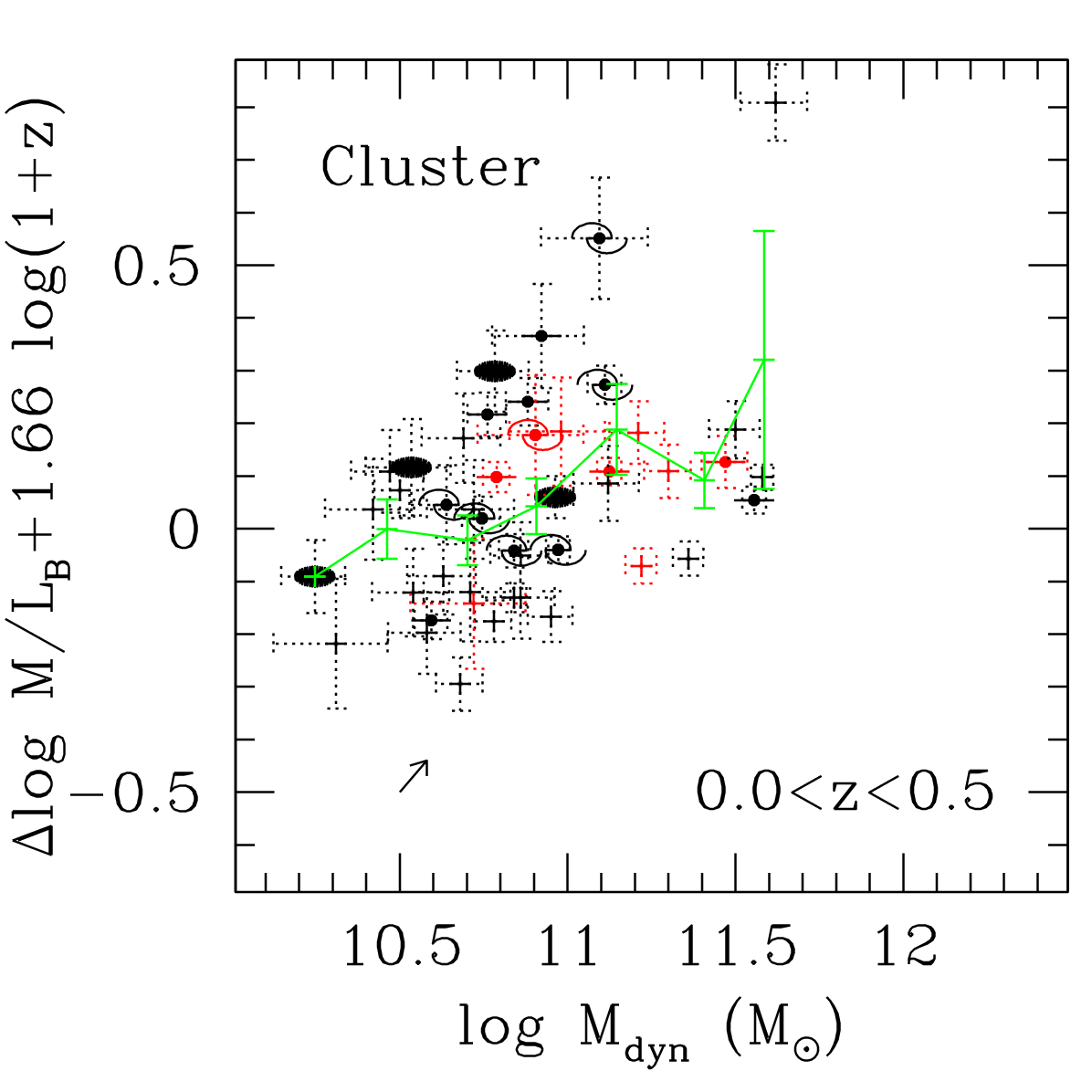,angle=0,width=5.3cm}}& 
\vbox{\psfig{file=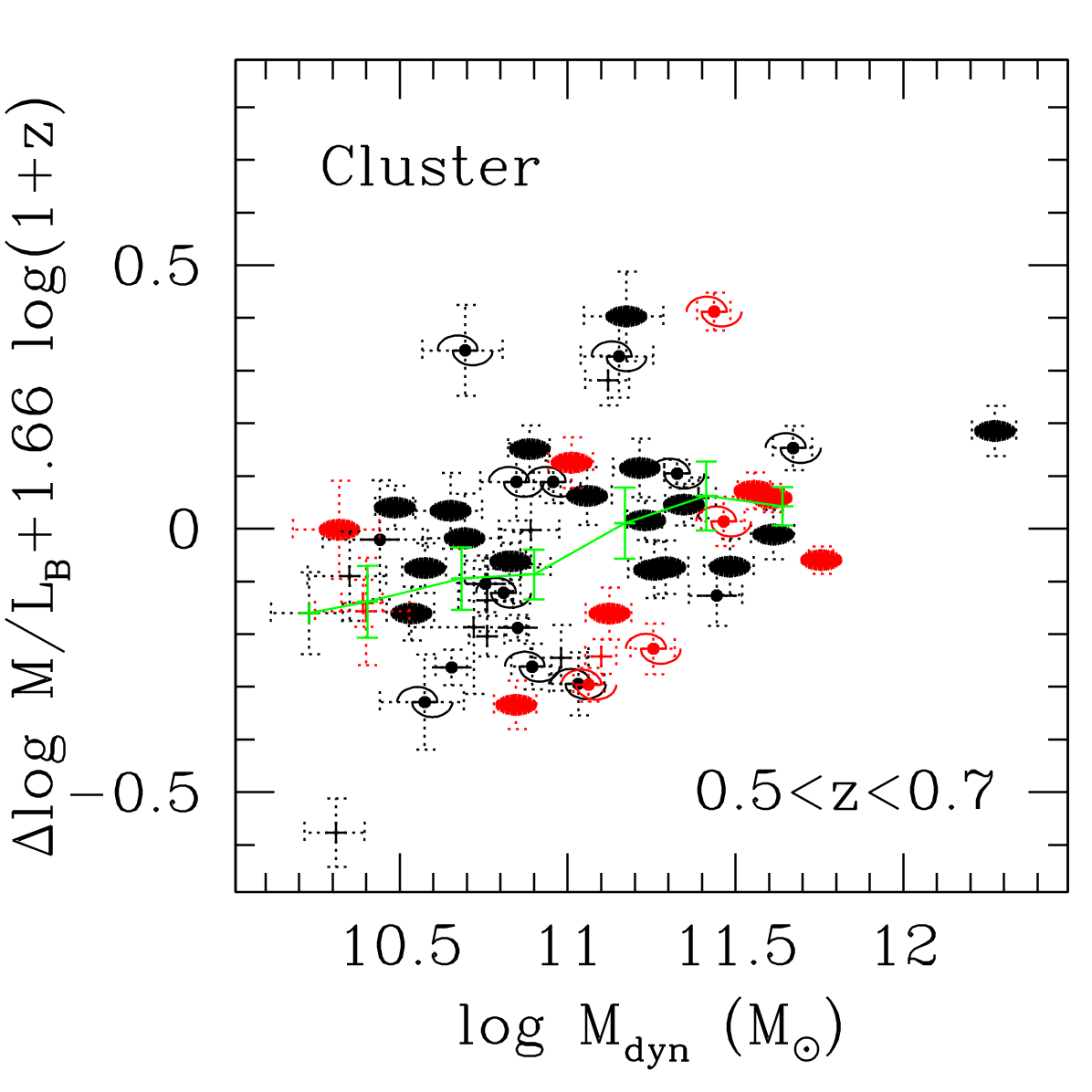,angle=0,width=5.3cm}}& 
\vbox{\psfig{file=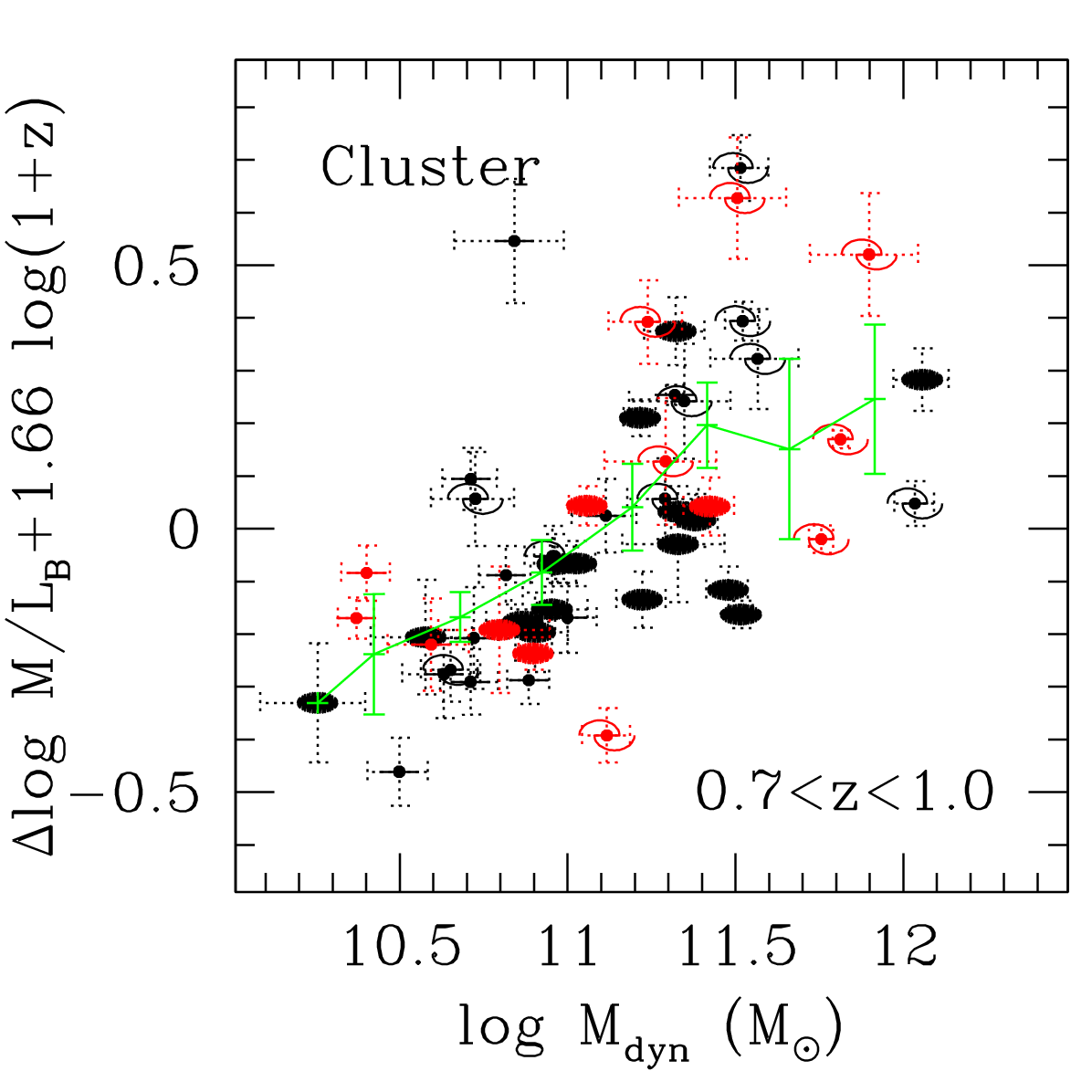,angle=0,width=5.3cm}}\\
\vbox{\psfig{file=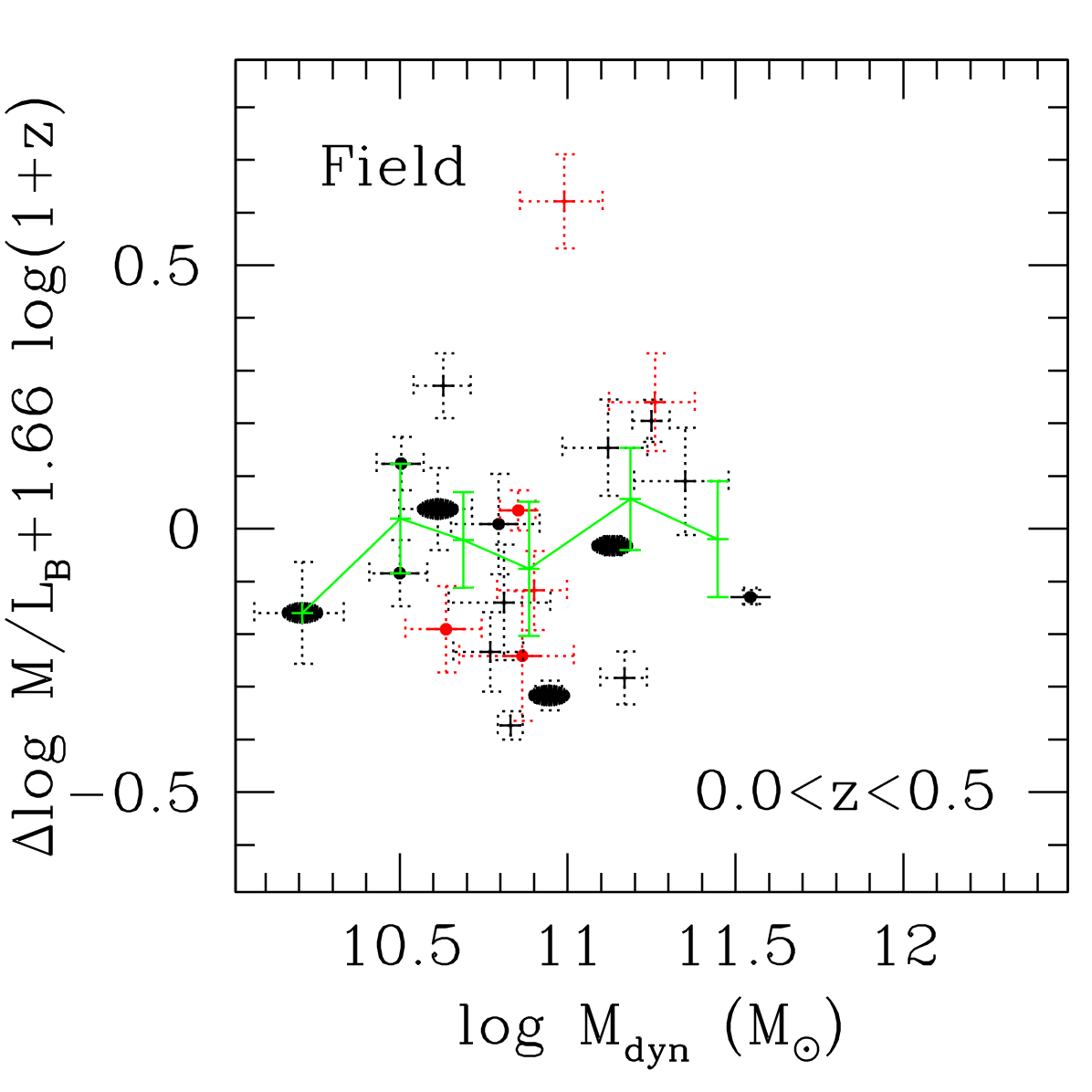,angle=0,width=5.3cm}}&
\vbox{\psfig{file=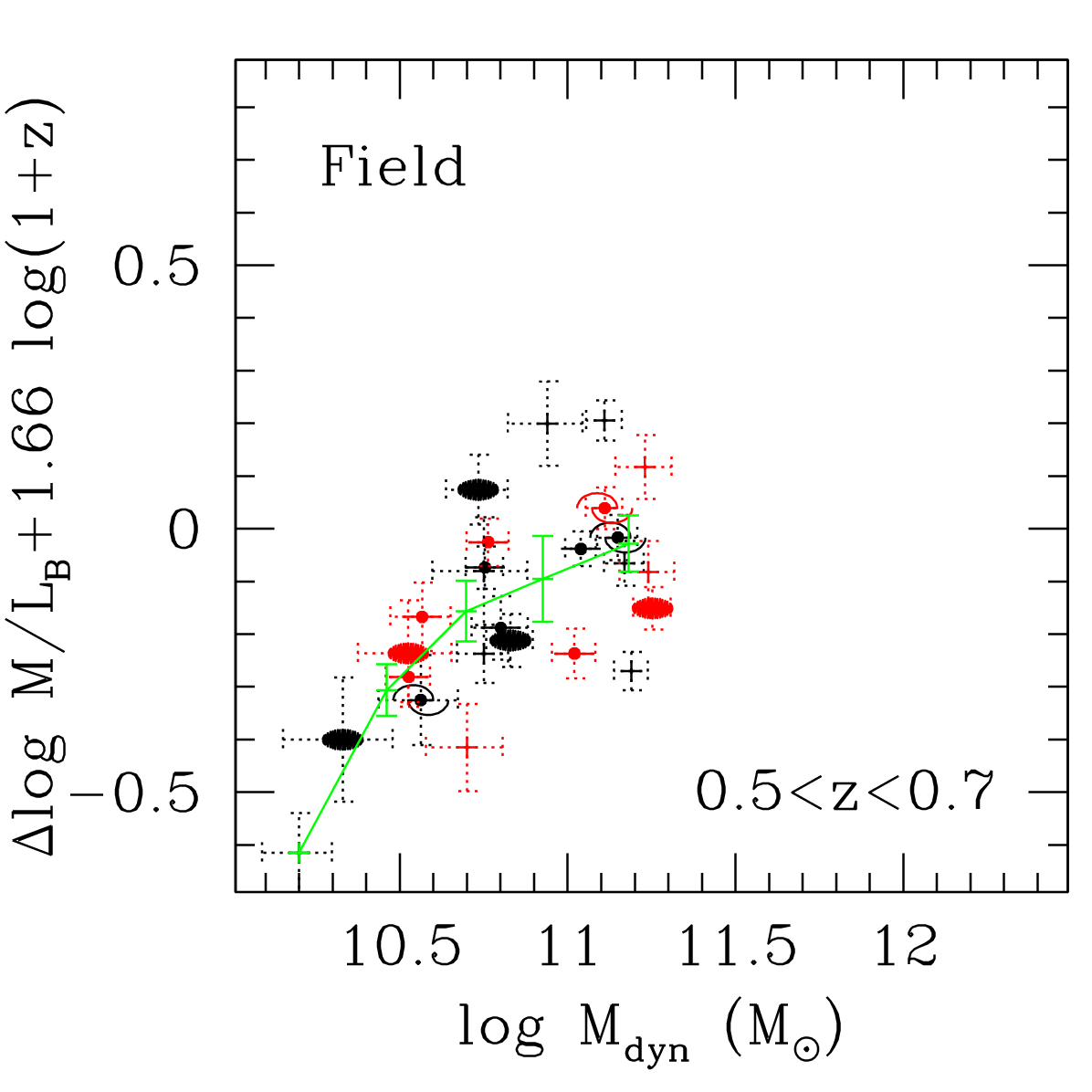,angle=0,width=5.3cm}}&
\vbox{\psfig{file=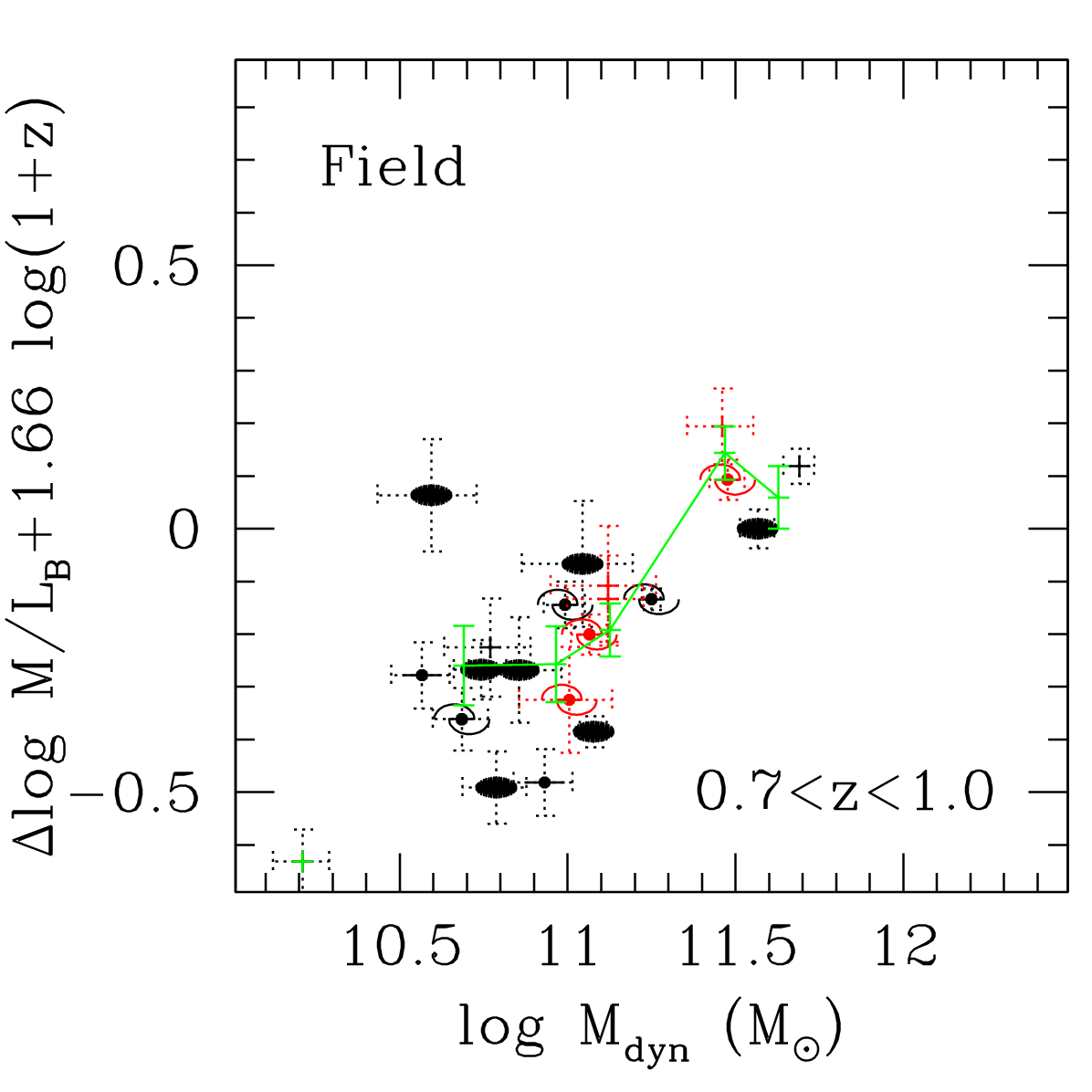,angle=0,width=5.3cm}}\\
\end{tabular}
\caption{{The mass dependence of FP mass-to-light ratios.} 
Top: the residuals $\Delta \log M/L_B+1.66\log(1+z)$ as a
  function of galaxy mass for cluster galaxies at low (left, $z<0.5$), 
intermediate (middle, $0.5\le z<0.7$), 
  and high (right, $z> 0.7$) redshift. {The arrow in the top left panel shows the how 
points change due to the typical 10\% error in velocity dispersion.} Bottom: the residuals 
$\Delta \log M/L_B+1.66\log(1+z)$ as a function of galaxy dynamical mass for field
  galaxies at low (left, $z<0.5$),  intermediate (middle, $0.5\le z<0.7$), and 
high (right, $z\ge 0.7$)
  redshift. Colors and symbols as in Fig.  \ref{fig_fieldML}. The
  green points show averages over $\log M_{dyn}$ bins 0.25 dex wide.
  \label{fig_clusmassdML}}
\end{figure*}

Figure \ref{fig_clusmassdML} shows the residuals $\Delta \log
M/L_B+1.66\log(1+z)$ as a function of galaxy dynamical mass, for
cluster (top) and field galaxies (bottom), at low (left) and high
(right) redshifts.  We divided the sample into three redshift bins of
$z<0.5$, $0.5\le z<0.7$, and $z\ge0.7$.  Averaging the points in mass
bins 0.25 dex wide, one derives the following (see also
Fig. \ref{fig_Ages}).  At low redshifts ($z<0.5$), there is no
convincing systematic trend between mass and residuals from the
passively evolved FP, for both cluster (where the Pearson coefficient
is 0.55 with a 2.5 $\sigma$ deviation from the no correlation hypothesis)
and field galaxies (where the Pearson coefficient is 0.15 for a
t-value of 0.58 in agreement with the absence of a
correlation). Within the errors, the solar metallicity SSP
model with $z_f=2$ provides a reasonable description of the evolution
of luminosity of all cluster and field early-type galaxies more
massive than $10^{10}M_\odot$. At intermediate redshifts ($0.5\le
z<0.7$), field (and to a lower extent cluster) galaxies with dynamical
masses lower than $10^{11}M_\odot$ show systematically negative mean
residuals.  At higher redshifts ($z\ge0.7$), both cluster and field
galaxies with masses lower than $10^{11}M_\odot$ show systematically
negative mean residuals, i.e., are brighter than predicted by the
passively evolved FP at zero redshift, with Spearman correlation
coefficients between mass and residuals larger than 0.66 and a t-value
of 6.3 for cluster galaxies.  The trends are stronger if we restrict
the sample to galaxies with HST early-type ($T<0$) morphology.  {We
  note that down to masses $\approx 4\times 10^{10} M_\odot$ we 
  sample a constant fraction ($\approx 20$\%) of the existing galaxy
  population (see Fig. \ref{fig_finalcompleteness}). At lower masses,
  however, this drops to just 10\% and we might expect residual
  selection effects to play a role, as discussed in
  \citet{vanderWel05}.}  We do not detect any additional dependence on
cluster velocity dispersion.

\subsection{The rotation of the fundamental plane} 
\label{sec_rotation} 

As discussed by \citet{Alighieri05}, a mass dependence of the $\Delta
\log M/L$ residuals implies a rotation of the FP as a
function of redshift. Here we investigate the effect by assuming that
the zero point variation $\Delta \log M/L_B=-1.66\times \log (1+z)$ for cluster
and $\Delta \log M/L_B=-2.27\times \log(1+z)$ for field galaxies is caused
entirely by pure luminosity evolution. Accordingly, we correct the
surface brightnesses of cluster galaxies by applying the offset
$\Delta\log I_e=-1.66\times \log(1+z)/0.83$ and of field galaxies by applying
$\Delta\log I_e=-2.27\times \log (1+z)/0.83$.  This agrees with the
observed evolution of the average effective surface brightness (see
dotted line in Fig.  \ref{fig_ReIez}), except for the highest redshift
bins. We then fit the parameters $\alpha$ and $\beta$ of Eq.
\ref{eq_FP} using the maximum likelihood algorithm of
\citet{Saglia01}, which uses multi-gaussian functions to describe the
distribution of data points, taking into account the full error
covariance matrix and selection effects {\citep[for a Bayesian approach to 
the modeling of systematic effects see][]{Treu01}}. To ensure 
uniformity with the
procedures adopted in the literature, the results are derived with and
without taking into account selection effects, but the differences
between the two approaches are always smaller than the large
statistical errors. We fit three redshift ranges for cluster galaxies
and two for field galaxies.  The results are shown in Table
\ref{tab_FPcoef}. The errors are computed as the 68\% percentiles of
the results of Monte Carlo simulations of each fitted sample as in
\citet{Saglia01}.  The low redshift bins (up to z=0.7) infer $\alpha$
coefficients that are compatible with local values ($\alpha\approx
1.2$) and $\beta$ coefficients ($\beta\approx 0.23-0.3$) slightly
smaller than the local value ($\beta\approx 0.33$). In contrast, the
highest redshift bins produce shallower $\log \sigma$ slopes. Given
the relatively low number of galaxies per bin, {especially in 
the low velocity dispersion regime,} the statistical
significance is just $\approx 1\sigma$, but the trend confirms the
claims of the literature (see Sect. \ref{sec_Intro}). In particular,
both values of $\alpha$ and $\beta$ decrease at high redshift, as
observed by \citet{Alighieri05}, a consequence of the flattening with
redshift of the power-law relation between luminosity and mass (see
Sect.  \ref{sec_sizeev}).

\begin{table*}[th!]  
\caption{The coefficients
    $\alpha$ and $\beta$ of the EDisCS FP as a function redshift and
    the derived quantities $\epsilon=\frac{2-\alpha}{2\alpha}$ (with
    $M/L\propto L^\epsilon$), $\lambda=\frac{1}{1+\epsilon}$ (with
    $L\propto M^\lambda$), $A=\frac{10\beta-2-\alpha}{5\beta}$.
    \label{tab_FPcoef}} 
\begin{tabular}{cclcccccl}
\hline
    z range  & $N_{gal}$ & $P_S$ & $\alpha$    & $\beta$         & $\epsilon$     & $\lambda$     & $A$            & Environment\\
    \hline
    0        &          &  -  & 1.2           & 0.33          & 0.33           & 0.75          & 0.06           & Local Sample\\
    0.4-0.5  & 46       & No  & $1.09\pm0.33$ & $0.23\pm0.02$ & $0.42\pm0.30$  & $0.71\pm0.14$ & $-0.69\pm0.37$ & Cluster\\
    0.5-0.7  & 57       & No  & $1.04\pm0.24$ & $0.30\pm0.02$ & $0.46\pm0.24$  & $0.68\pm0.11$ & $-0.03\pm0.21$ & Cluster\\
    0.7-1.0  & 50       & No  & $0.69\pm0.42$ & $0.23\pm0.02$ & $0.95\pm1.07$  & $0.51\pm0.24$ & $-0.35\pm0.42$ & Cluster\\
    0.4-0.7  & 39       & No  & $1.05\pm0.6$  & $0.23\pm0.02$ & $0.45\pm0.63$  & $0.69\pm0.27$ & $-0.67\pm0.58$ & Field\\
    0.7-1.0  & 21       & No  & $0.37\pm0.27$ & $0.23\pm0.02$ & $2.20\pm2.09$  & $0.31\pm0.20$ & $-0.06\pm0.30$ & Field\\
    0.4-0.5  & 46       & Yes & $1.06\pm0.25$ & $0.27\pm0.02$ & $0.44\pm0.24$  & $0.69\pm0.11$ & $-0.29\pm0.25$ & Cluster\\
    0.5-0.7  & 57       & Yes & $1.09\pm0.30$ & $0.32\pm0.03$ & $0.42\pm0.27$  & $0.70\pm0.13$ & $+0.07\pm0.26$ & Cluster\\
    0.7-1.0  & 50       & Yes & $0.44\pm0.20$ & $0.25\pm0.03$ & $1.77\pm1.21$  & $0.36\pm0.13$ & $+0.02\pm0.28$ & Cluster\\
    0.4-0.7  & 39       & Yes & $1.00\pm0.7$  & $0.23\pm0.02$ & $0.50\pm0.75$  & $0.67\pm0.32$ & $-1.03\pm0.76$ & Field\\
    0.7-1.0  & 21       & Yes & $0.22\pm0.7$  & $0.20\pm0.02$ & $4.04\pm3.42$  & $0.20\pm0.13$ & $+0.15\pm0.20$ & Field\\
    \hline 
\end{tabular} 
\end{table*} 

\section{Size and velocity dispersion evolution}
\label{sec_sizeev} 

\subsection{Setting the stage}
\label{sec_stage}

Up to this point, we have analyzed and interpreted the ZP
variations of the FP based on the assumption that it is caused mainly
by a variation in the luminosity. As discussed in the Introduction,
there is growing evidence that early-type galaxies evolve not only in
terms of luminosity, but also in size and velocity dispersion. Here we
examine the consequences of these findings.

In general, if sizes were shrinking with increasing redshift, we would
expect the surface brightness to increase. Therefore, if the velocity
dispersions do not increase a lot, the net effect will be to reduce
the net amount of brightening with redshift caused by stellar
population evolution. In detail, setting $\Delta ZP=ZP(z)-ZP(0)$ and
using the fact that $\langle SB_e\rangle =-2.5\log (L/2\pi R_e^2)$, we derive
\begin{equation}
 \Delta ZP=\alpha_0\Delta \log \sigma
-2.5\beta_0\Delta \log L+(5\beta_0-1)\Delta \log R_e,
\label{eq_deltazp}
\end{equation}
where $\Delta \log R_e=\overline{\log R_{e}(z)}-\overline{\log
  R_{e}(0)}$ and $\Delta \log \sigma =\overline{\log
  \sigma(z)}-\overline{\log \sigma(0)}$ are the variations with
redshifts in the mean half-luminosity radius and average surface
brightness. Therefore, the redshift variation in the luminosity,
taking into account the size and velocity dispersion evolution of
galaxies is:
%
%
\begin{equation}
\Delta \log L =
  \frac{10\beta_0-2}{5\beta_0}\Delta \log R_e+\frac{2\alpha_0}{5\beta_0}
\Delta \log \sigma-\frac{2\Delta ZP}{5\beta_0}.  
\label{eq_DeltaL} 
\end{equation}
 We note that the ZP variations have been determined by
assuming constant $\alpha_0$ and $\beta_0$ coefficients, which is
probably not true at the high redshift end of our sample (see Sect.
\ref{sec_rotation}).

If the variations are computed at constant dynamical mass, then
$\Delta \log \sigma=-\Delta \log R_e/2$, as in the ``puffing''scenario
of \citet{Fan08}, see below, Eq. \ref{eq_DeltaL} becomes
\begin{equation} 
\Delta \log L_{pu} =
  \frac{10\beta_0-2-\alpha_0}{5\beta_0}\Delta \log R_e-\frac{2\Delta
    ZP}{5\beta_0}.  
\label{eq_DeltaLdyn} 
\end{equation} 
In this case, the contribution of the size evolution to the luminosity
evolution at constant mass derived from the FP is zero
if
\begin{equation}
  A_0=\frac{10\beta_0-2-\alpha_0}{5\beta_0}, \label{eq_A} 
\end{equation} 
is zero, i.e., $\alpha_0=10\beta_0-2$. This is the expected relation between
$\alpha$ and $\beta$ if the mass-to-light ratio $M/L$ varies as a
power law of the luminosity $M/L\propto L^\epsilon$, in which case one
has $L\propto M^{\frac{1}{1+\epsilon}}=M^\lambda$,
$\alpha=\frac{2}{1+2\epsilon}$, and 
$\beta=\frac{2}{5}\frac{1+\epsilon}{1+2\epsilon}$. 
Table \ref{tab_FPcoef} lists the values of $\epsilon$,
$\lambda$, and $A$ implied by the fits of the FP coefficients
performed in Sect. \ref{sec_rotation}. 

If we parametrize all variations as a function of $\log(1+z)$ as
$\Delta \log R_e =\nu\log(1+z)$, $\Delta \log \sigma= \mu\log(1+z)$, and 
$\Delta ZP=\kappa\log(1+z)$, we find that
%
%
\begin{equation}
\Delta \log L = (\frac{10\beta_0-2}{5\beta_0}\nu+\frac{2\alpha_0}{5\beta_0}\mu
-\frac{2}{5\beta_0}\kappa)\log(1+z)+\phi z,
\label{eq_DeltaLz}
\end{equation}
where $\phi z$ is the correction for progenitor bias estimated
by \citet{vanDokkum01b} to be $\phi=+0.09$. Their result can be applied to
our work directly, since our redshift dependence of the FP ZP matches
closely that considered there.

As discussed in the Introduction, the size and $\sigma$ evolution of
galaxies is usually interpreted as a result of the merging history of
galaxies. The merger models of \citet{Hopkins09} predict $\nu_{me}
\approx-0.5$ and $\mu_{me}=0.1$ for galaxies with constant stellar
mass $M_*\approx 10^{11}$ with $(M_{halo}/R_{halo})/(M_*/R_e)\approx
2$. This means that $\Delta \log R_e=-0.2\Delta \log \sigma$.  As an
alternative explanation, \citet{Fan08} proposed the 'puffing'
scenario, where galaxies grow in size conserving their mass as a
result of quasar activity. In this case, one has $\sigma_{pu}\propto
R_e^{-1/2}$.  We note, however, that this mechanism should already
have come to an end at redshift 0.8. Moreover, the strong velocity
dispersion evolution predicted by the puffing scenario at redshifts
higher than 1 was ruled out by \citet{Cenarro09}. 

%
Using $\nu=-0.5$, $\mu=+0.1$, the change in the slope $\Delta
\tau=\frac{10\beta_0-\bf{2}}{5\beta_0}\nu+\frac{2\alpha_0}{5\beta_0}\mu$ of
the luminosity evolution $\Delta \log L=\tau\log(1+z)$ (see
Eq. \ref{eq_DeltaL}) is $\approx {\bf -0.25}$ units.

\subsection{The redshift evolution of $R_e$ and $\sigma$}
\label{sec_ReSigmaEv}

Following \citet{vanderWel08}, we investigated the size evolution of
EDisCS galaxies by considering the $Mass-R_e$ relation for objects
with masses higher that $3\times 10^{10} M_\odot$. In Fig.
\ref{fig_ReMass}, we divided our sample into 8 redshift bins (centered
on redshifts from 0.25 to 0.95 of bin size $\Delta z=0.1$) and
fit the relation $R_e=R_c(M/M_c)^b$. We considered both dynamical
($M_{dyn}$, left) and stellar ($M_*$, right) masses, and we weighted
each galaxy with $1/P_S$. Within the errors, $b$ does not vary much
and is compatible with the values $b=0.56$ found locally.  
In Fig. \ref{fig_ReMass} we therefore keep its value fixed and determine $R_c$
at the mass $M_c=2\times 10^{11} M_\odot$.  We fitted the function
$R_c(z)=R_c(0)\times(1+z)^\nu$ and summarize the values of the
parameters resulting from the fits in Table \ref{tab_massfits}.  
  As becomes clear below, this does not necessarily describe 
  the evolution in size of a galaxy of fixed mass, but rather at any
  given redshift the mean value of the size of the evolving population
  of galaxies with this given mass.

{Given the larger uncertainties in the $R_e$ values derived from
  VLT photometry, we first fitted the HST dataset alone (entries 1 and 2
  of Table \ref{tab_massfits}).  Within the errors, both $R_c(0)$ and
  the slope are very similar to the values reported by
  \citet[$R_c(0.06)=4.8$ kpc, $\nu=0.98\pm0.11$]{vanderWel08} for
  both dynamical and stellar mass fits. Our results do not change
  within the errors if we separately fit galaxies belonging to
  clusters or to the field.  If we add the galaxies with VLT
  photometry only (entries 3 and 4 of Table \ref{tab_massfits}), we
  derive larger $R_c$ and steeper slopes.}

Figure \ref{fig_meanRc} shows $R_c$ as a function of redshift when we
apply a correction for progenitor bias as in
\citet{Valentinuzzi10a}. The EDisCS galaxies considered here are a
sample of spectroscopically selected passive objects. In contrast, a
morphologically selected local sample of early-type galaxies contains
objects with relatively young ages that, when evolved to EDisCS
redshifts, would not be recognized as being spectroscopically passive.
\citet{Valentinuzzi10a} analyze the WINGS sample of local galaxies and
determine their ages by means of a spectral analysis. They select
objects that were already passive (i.e., have an age $\ge 1.5$ Gyr) at
the cosmic time of the redshifts $z=0, 0.25, 0.5, 0.75$ and 1, and
compute the median half-luminosity radii of massive galaxies.  The
resulting $R_e$ vary as $R_e=(4.1\pm 0.1) -(0.8\pm0.2)z$ (kpc), when
selecting galaxies with dynamical masses $10^{11}<M_{dyn}/M_\odot<
3\times 10^{11}$, and as $(4.3\pm0.1)-(0.9\pm0.2)z$ when selecting
galaxies with stellar masses $10^{11}<M_*/M_\odot< 3\times 10^{11}$.
Therefore, we multiply the $R_c(z)$ derived at
$M_{dyn}=10^{11}M_\odot$ by $\frac{4.1}{4.1-0.8z}$, and those at
$M_*=10^{11}M_\odot$ by $\frac{4.3}{4.3-0.9z}$. With these
corrections, the residual evolution is small, and even compatible with
no evolution up to redshift $z\approx 0.7$ with dynamical masses, and
0.5 with stellar masses.  {Similar results are derived if we also
  consider the galaxies with VLT photometry, with the caveats
  discussed above.   Our correction for progenitor bias is of course
somewhat model-dependent, since objects might cross the boundaries
between populations.  For example, there might be $z\sim 0.6$ passive
galaxies that produce $z=0$ descendants with some younger stars, after
accreting gas or gas-rich objects.

\begin{figure*}
  \begin{tabular}{cc}
    \vbox{\psfig{file=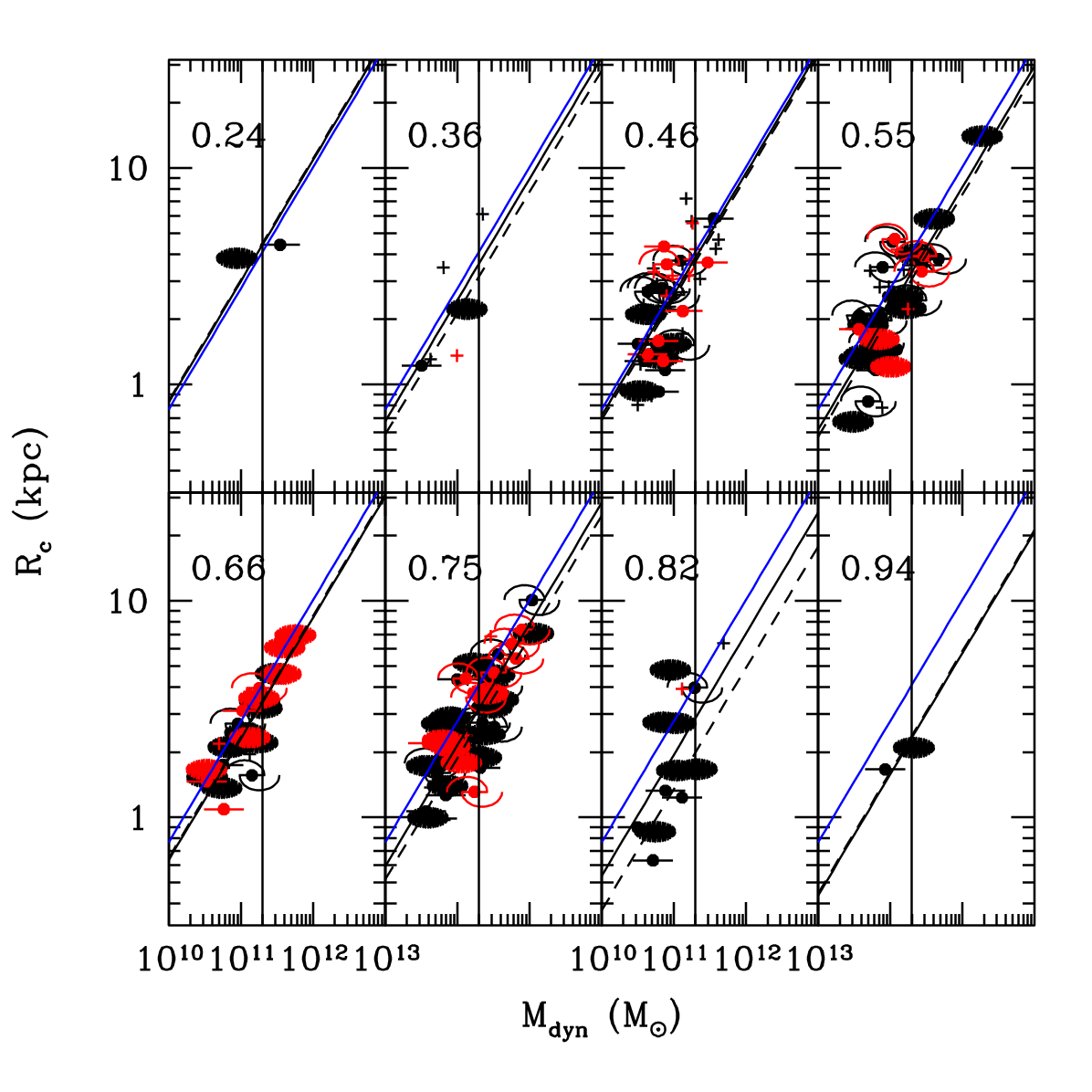,angle=0,width=8.5cm}}&
    \vbox{\psfig{file=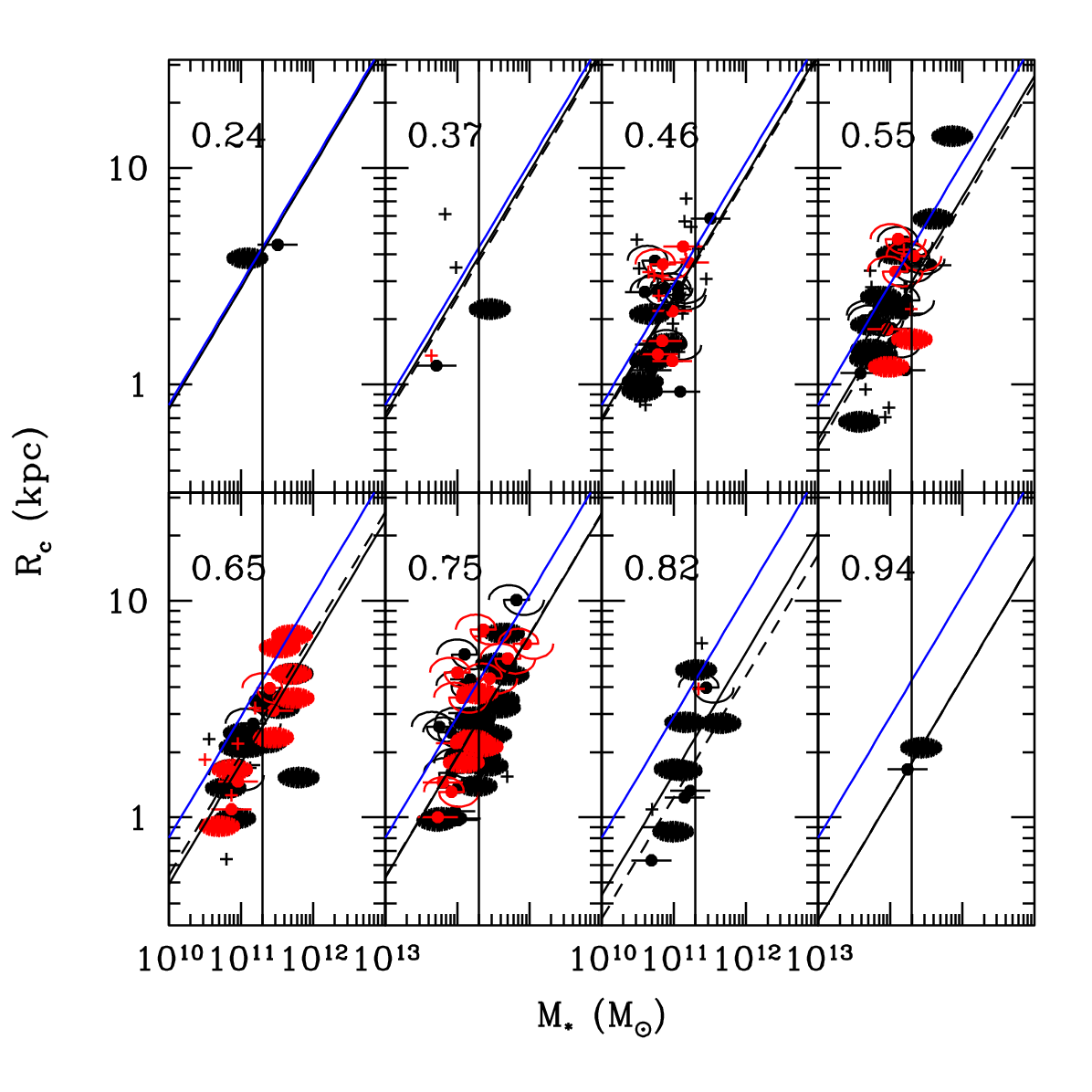,angle=0,width=8.5cm}}\\
  \end{tabular} \caption{The evolution in the $R_e-mass$ relation with
    redshift. Left: dynamical masses. Right: stellar masses.  Colors
    and symbols are as in Fig.  \ref{fig_fieldML}. The numbers give the
    average redshift in each bin. The full lines show the best-fit
    relation $R_e=R_c(M/2\times10^{11} M_\odot)^{0.56}$ with uniform
    galaxy weighting, the dashed lines with selection weighting. 
The blue lines show
the reference line at zero redshifts. The
    vertical lines show the $2\times10^{11} M_\odot$ mass.
    \label{fig_ReMass}}
\end{figure*} 

\begin{figure*}
  \begin{tabular}{cc}
    \vbox{\psfig{file=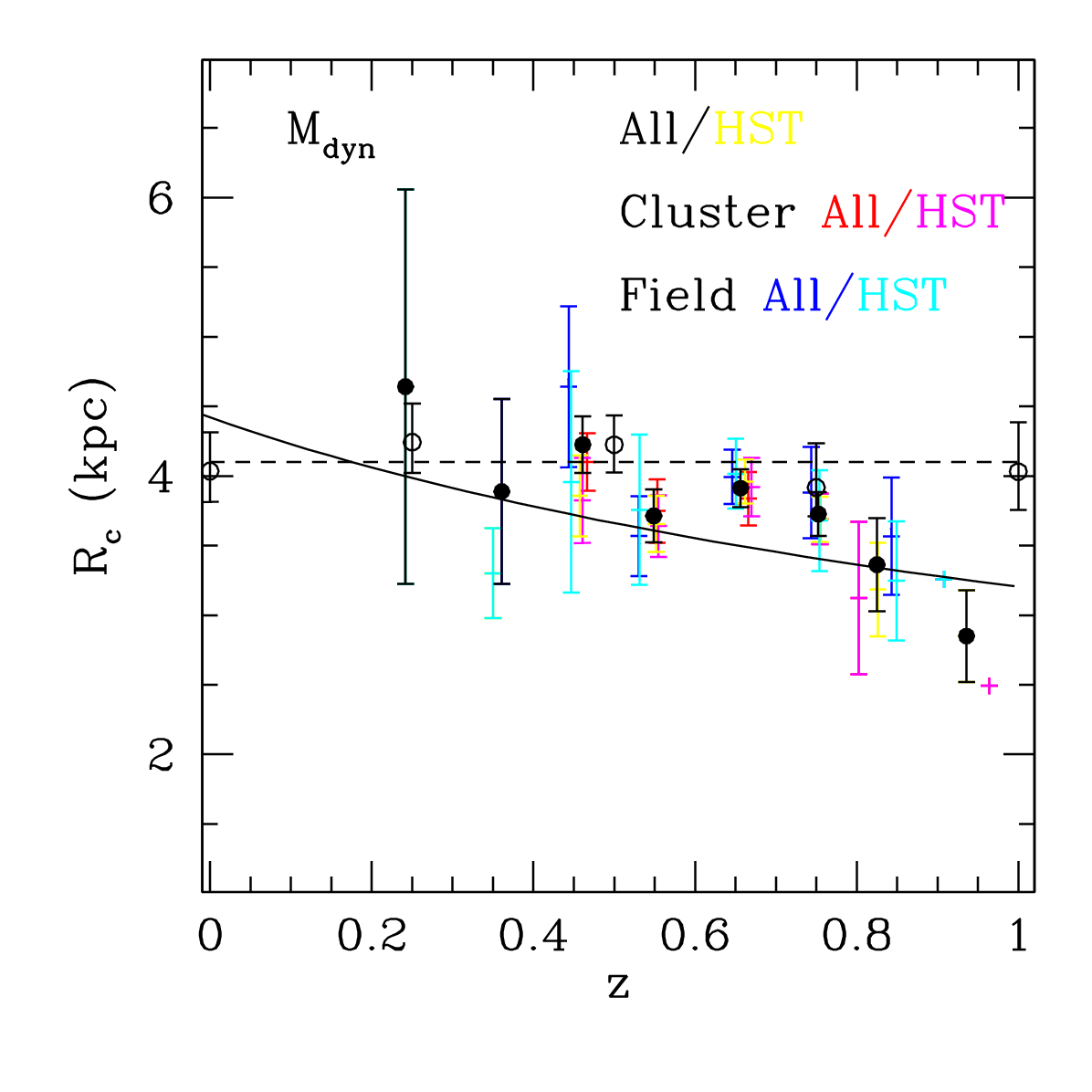,angle=0,width=8cm}}&
    \vbox{\psfig{file=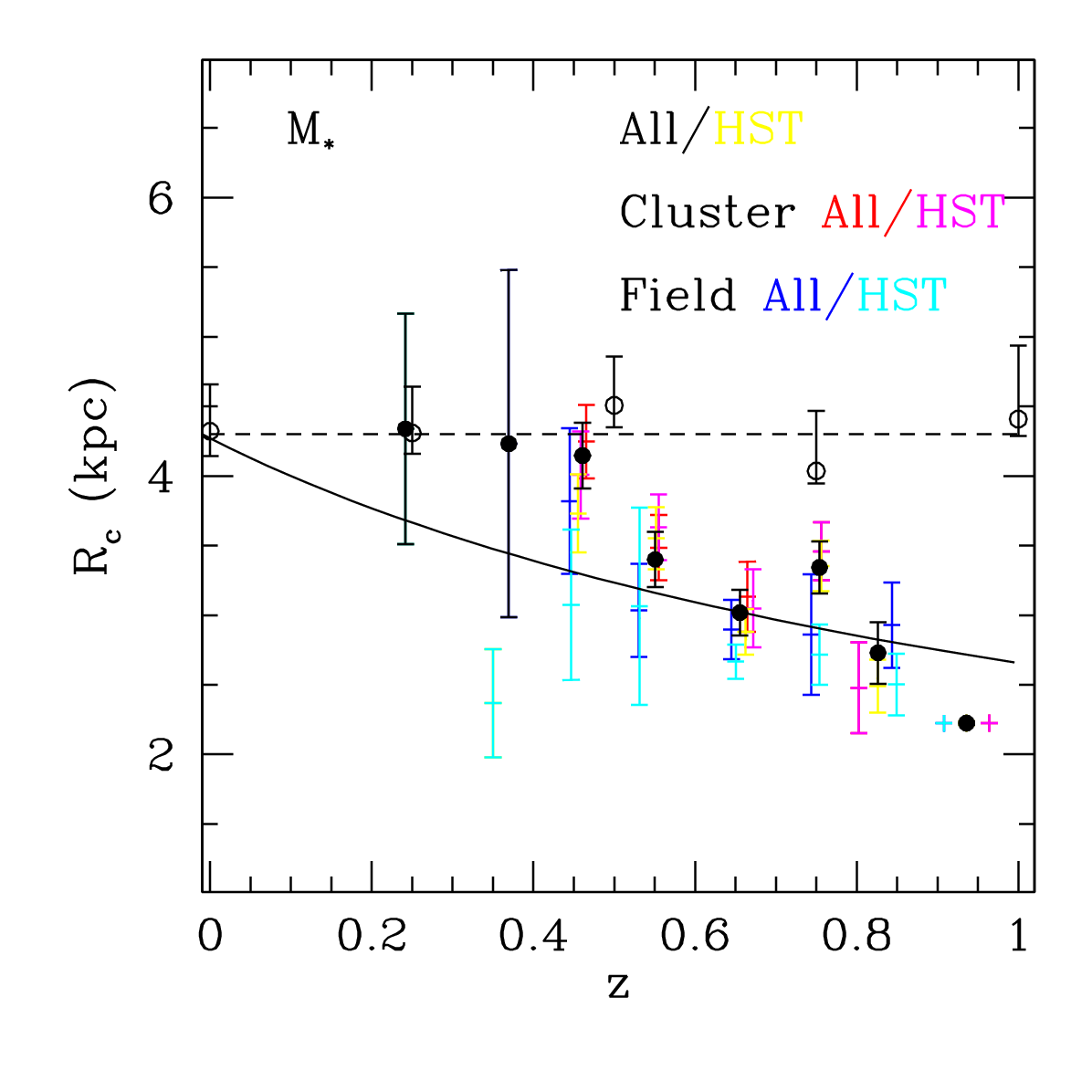,angle=0,width=8cm}}\\
  \end{tabular} \caption{The size evolution  with redshift
 of EDisCS galaxies corrected for
    progenitor bias (see text). Left:
    $R_c$ as a function of redshift {at $2\times 10^{11}M_\odot$ (see Fig. \ref{fig_ReMass}, left)} derived using $M_{dyn}$. Right:
    $R_c$ {at $2\times 10^{11}M_\odot$  (see Fig. \ref{fig_ReMass}, right)} as a function of redshift derived using $M_{*}$. 
The full lines show the best-fit
    function to the galaxies with HST photometry (yellow points) $R_c=R_c^0\times (1+z)^{-\nu}$ without selection
  weighting (see Table \ref{tab_massfits}, case 1).
    The open dots show the local sample of \citet{Valentinuzzi10a} evolved at 
the redshifts 0, 0.25, 0.5, 0.75, and 1, after having applied the progenitor 
bias 
correction. The dashed line shows the local value.  
\label{fig_meanRc}}
\end{figure*} 

\begin{table*}[h!]
\caption{The redshift evolution of the mass correlation fits. Selection 
weighting and progenitor bias correction are applied when $P_S$=Yes and PB=Yes.
  \label{tab_massfits}} 
\begin{tabular}{clccccccc}
\hline
Case & Parameter       &$P_S$& PB  & $\log M$ slope  & \multicolumn{2}{c}{$M_{dyn}$}           &  \multicolumn{2}{c}{$M_*$}\\
\hline
                &     &     &      & & z=0          & (1+z)\^~Slope   & z=0          & (1+z) \^~Slope \\
\hline
1  & $R_e (kpc)$ HST & No  & No  & 0.56 & $5.1\pm 0.7$ & $-1.0\pm 0.3$   & $4.3\pm 1.1$ & $-1.0\pm0.6$\\
2  & $R_e (kpc)$ HST & Yes & No  & 0.56 & $5.5\pm 0.9$ & $-1.3\pm 0.4$   & $4.6\pm 1.3$ & $-1.2\pm0.7$\\
3  & $R_e (kpc)$     & No  & No  & 0.56 & $5.7\pm 0.5$ & $-1.2\pm 0.2$   & $6.1\pm 0.6$ & $-1.6\pm0.2$\\
4  & $R_e (kpc)$     & Yes & No  & 0.56 & $5.7\pm 0.8$ & $-1.3\pm 0.3$   & $6.4\pm 0.9$ & $-1.7\pm0.4$\\
5  & $R_e (kpc)$ HST & No  & Yes & 0.56 & $4.4\pm 0.4$ & $-0.46\pm0.2$   & $4.3\pm 0.7$ & $-0.68\pm0.4$\\
6  & $R_e (kpc)$ HST & Yes & Yes & 0.56 & $4.6\pm 0.5$ & $-0.67\pm0.3$   & $4.3\pm 0.7$ & $-0.84\pm0.4$\\
7  & $R_e (kpc)$     & No  & Yes & 0.56 & $4.6\pm 0.4$ & $-0.5\pm0.2$    & $4.8\pm 0.4$ & $-0.75\pm0.2$\\
8  & $R_e (kpc)$     & Yes & Yes & 0.56 & $4.6\pm 0.5$ & $-0.65\pm0.2$   & $4.9\pm 0.5$ & $-0.86\pm0.3$\\
9  & $\sigma (km/s)$ & No  & No  & 0.23 & $175\pm 8 $  & $+0.59\pm 0.10$ & $185\pm 13$  & $+0.34\pm 0.14$\\
10  & $\sigma (km/s)$ & Yes & No  & 0.23 & $175\pm 14 $ & $+0.68\pm 0.17$ & $189\pm 23$  & $+0.39\pm 0.24$\\
11  & $\sigma (km/s)$ & No  & Yes & 0.23 & $188\pm 7 $  & $+0.41\pm 0.08$ & $199\pm 9$   & $+0.19\pm 0.1$\\
12 & $\sigma (km/s)$ & Yes & Yes & 0.23 & $188\pm 10 $ & $+0.49\pm 0.11$ & $201\pm 15$  & $+0.27\pm 0.16$\\
13 & $L (10^{10} L_\odot)$ Cluster & No   & No & 0.75 & $2.1\pm 0.4$ & $+2.1\pm0.4$    & $2.8\pm 0.7$ & $+1.2\pm0.4$\\
14 & $L (10^{10} L_\odot)$ Cluster & Yes  & No & 0.75 & $1.9\pm 1.4$ & $+1.9\pm1.2$    & $2.5\pm 0.6$ & $+1.4\pm0.4$\\
15 & $L (10^{10} L_\odot)$ Field   & No   & No & 0.75 & $2.4\pm 0.2$ & $+2.4\pm0.2$    & $2.3\pm 0.5$ & $+1.9\pm 0.4$\\
16 & $L (10^{10} L_\odot)$ Field   & Yes  & No & 0.75 & $2.0\pm 0.7$ & $+2.7\pm0.6$    & $2.0\pm 0.8$ & $+2.1\pm 0.8$\\
\hline
\end{tabular}
\end{table*}

In Figs. \ref{fig_SigMass} and \ref{fig_meanSigc}, we show the
analogous plots and fits for the velocity dispersion. Table
\ref{tab_massfits} lists the relative results. We find that $\sigma$
scales as $\approx M^{0.23}_{dyn}$.  The trend weakens at low masses
and high redshifts, especially when stellar masses are
considered. {The fit at constant dynamical mass is just a
  consistency check, which should infer a slope of the redshift
  dependence of opposite sign to and half the value of the one
  measured for $R_c$ ($\mu=+0.68$) and this is the case.}  In
contrast, a weaker redshift evolution ($\mu=0.39$) is derived if
stellar masses are considered, in agreement with
\citet{Cenarro09}. This is expected, since, as discussed
above, $\sigma_c$ at fixed $M_*$ should certainly be smaller than
$\sigma_c$ at fixed $M_{dyn}$ given that $M_{dyn}/M_*<1$.

Following the procedure of \citet{Valentinuzzi10a} described above, we
construct the local sample of WINGS galaxies with velocity
dispersions, trimmed to have only massive spectroscopically passive
galaxies at the redshifts $z=0$, 0.25, 0.5, 0.75 and 1, and compute the median
$\sigma$ of massive galaxies.  The resulting $\sigma$ vary as
$\sigma=(197\pm 2) +(6\pm4)z$ km/s, when selecting galaxies with
dynamical masses $10^{11}<M_{dyn}/M_\odot< 3\times 10^{11}$, and is constant at
$(210\pm1.5)$ km/s when selecting galaxies with stellar masses
$10^{11}<M_*/M_\odot< 3\times 10^{11}$.  Therefore, we correct the
measured $\sigma_c$ for the progenitor bias by multiplying the values
derived at $M_{dyn}=10^{11}M_\odot$ by $\frac{197}{197+6z}$, and no
correction is applied at constant stellar mass. The residual
redshift evolution after the correction and fitting the point at zero
redshift is small.

\begin{figure*}
\begin{tabular}{cc}
\vbox{\psfig{file=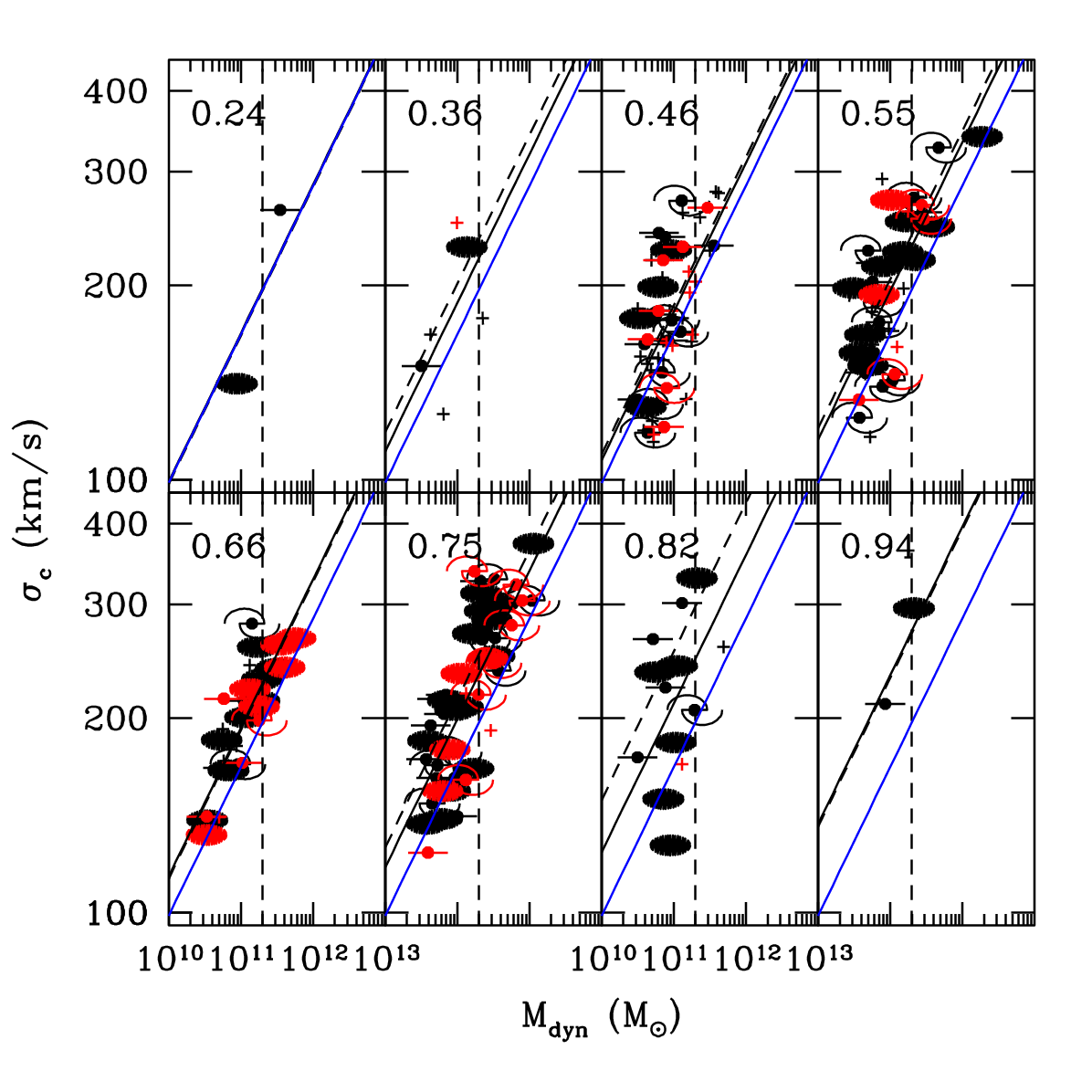,angle=0,width=8.5cm}}& 
\vbox{\psfig{file=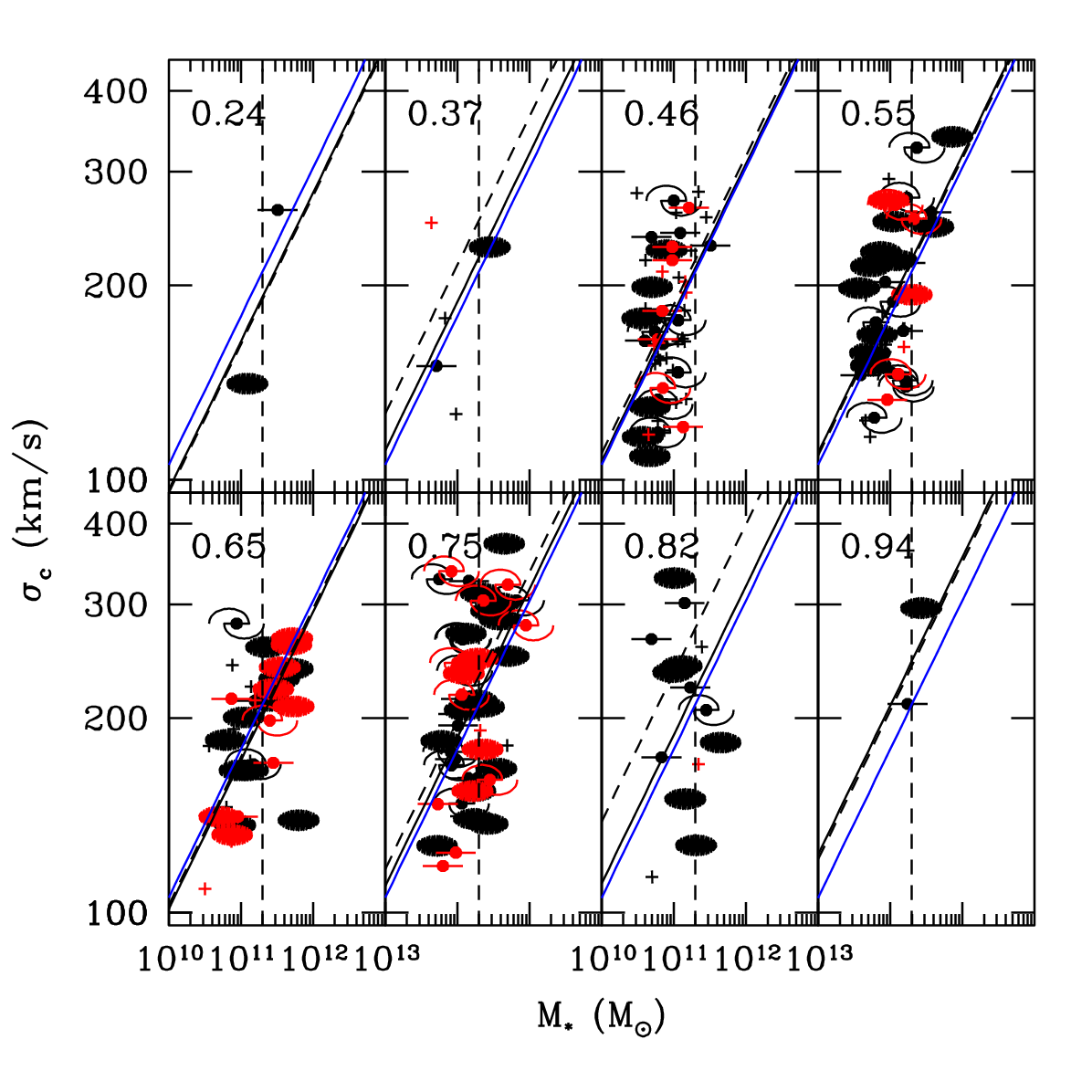,angle=0,width=8.5cm}}\\
\end{tabular}
\caption{The evolution of the $\sigma$-mass relation with redshift. 
Left: dynamical masses. Right: stellar masses. Colors and symbols as
  in Fig.  \ref{fig_fieldML}. The numbers give the average redshift in
  each bin. The full lines show the best-fit relation
  $\sigma=\sigma_C(M/M_C)^{0.23}$ with uniform galaxy weighting, 
the dashed line with selection weighting.  The blue lines show
the reference line at zero redshifts. The vertical line marks the
  $2\times 10^{11} M_\odot$ mass.
 \label{fig_SigMass}}
\end{figure*} 

\begin{figure*}
\begin{tabular}{cc}
\vbox{\psfig{file=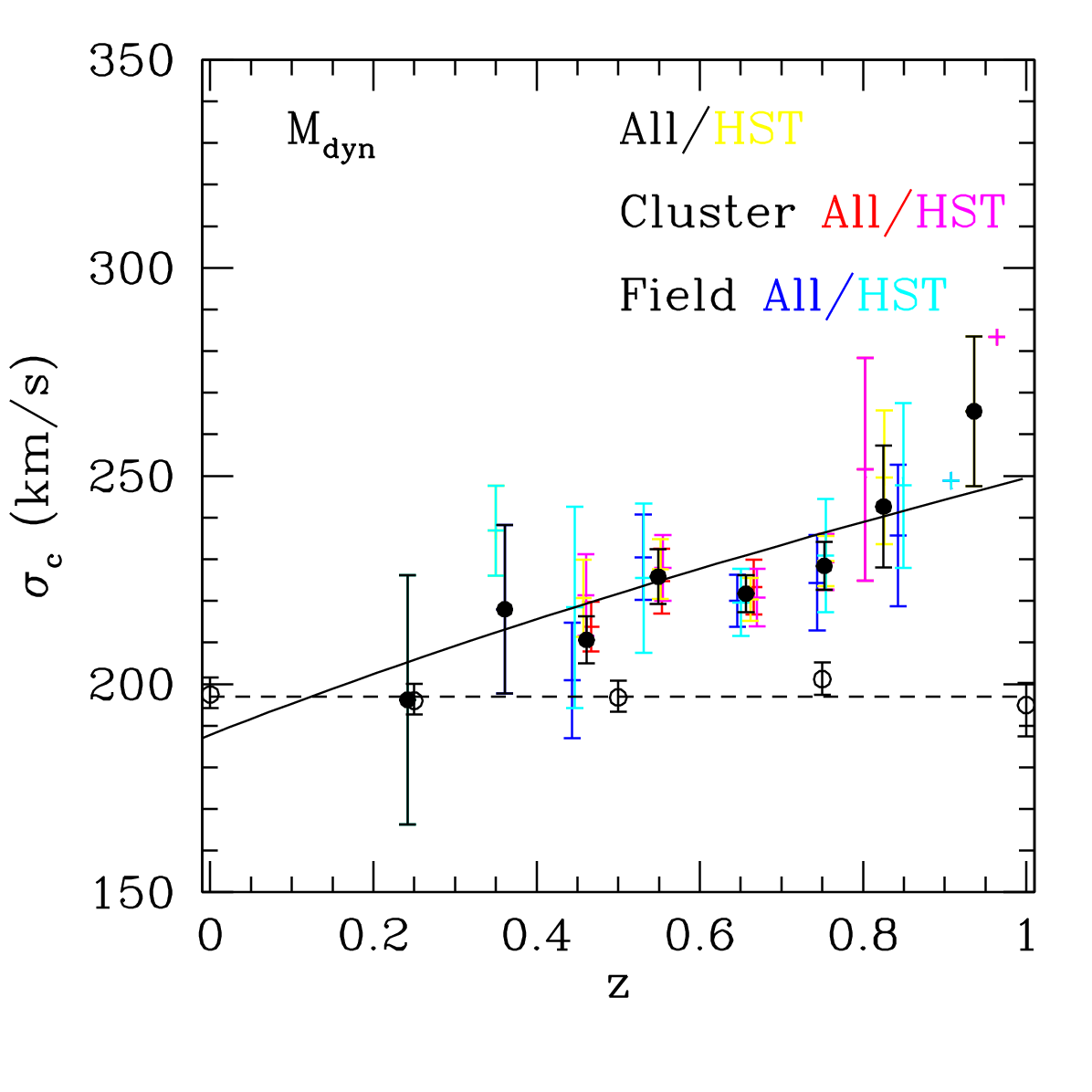,angle=0,width=8cm}}& 
\vbox{\psfig{file=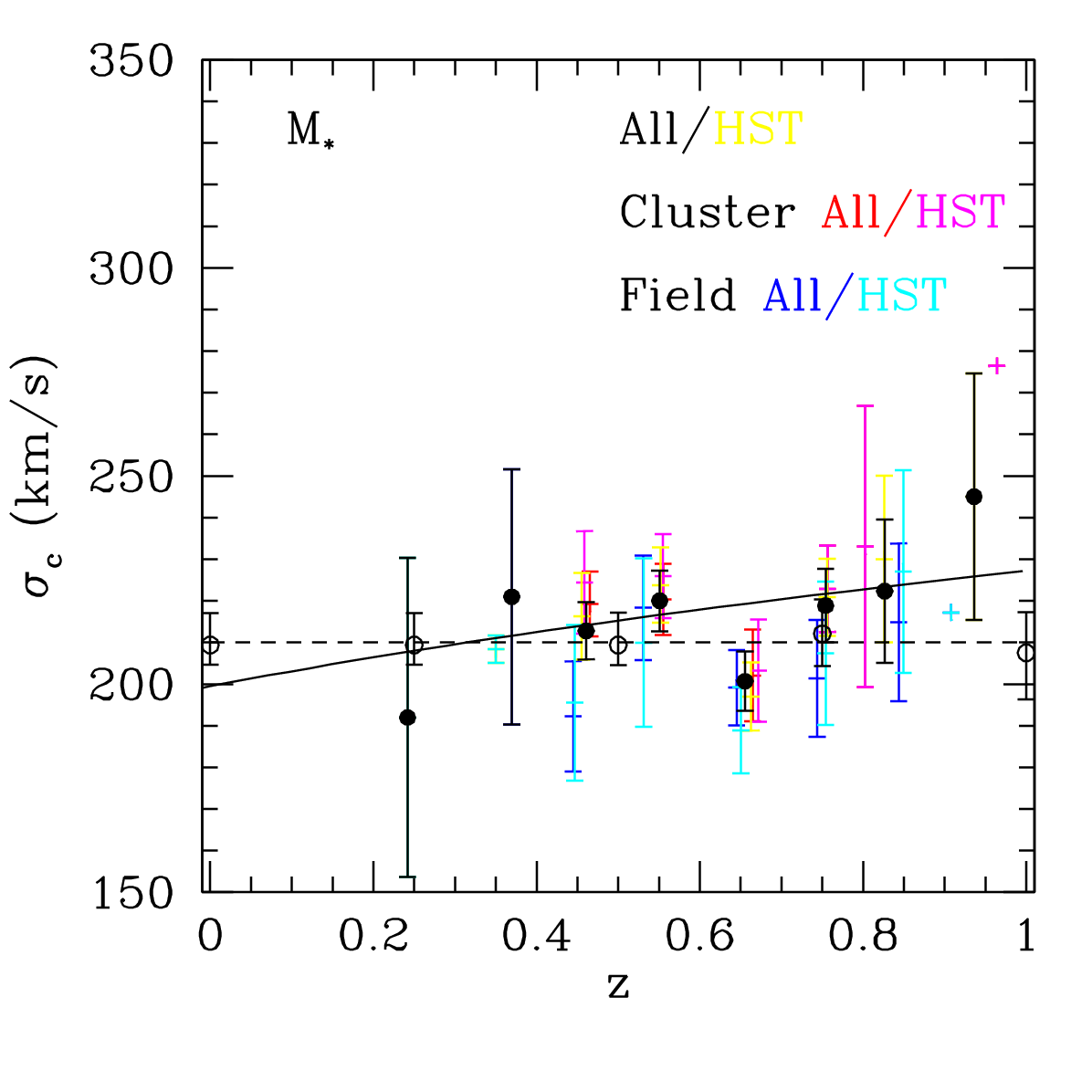,angle=0,width=8cm}}\\
\end{tabular}
\caption{The $\sigma$ evolution {with redshift} of EDisCS galaxies
  corrected for progenitor bias (see text). Left: $\sigma_c$ as a
  function of redshift derived using $M_{dyn}$. Right: $\sigma_c$ as a
  function of redshift derived using $M_{*}$. The full lines show the
  best-fit function $\sigma_c=\sigma_c^0\times (1+z)^\mu$ without
  selection weighting (see Table \ref{tab_massfits}). The open dots
  show the local sample of \citet{Valentinuzzi10a} evolved to the
  redshifts 0, 0.25, 0.5, 0.75, and 1, and after having applied the progenitor 
bias
  correction. The dashed line shows the local
  value.  \label{fig_meanSigc}}
\end{figure*} 

Table \ref{tab_lumev} lists the changes in the slope $\Delta
\tau=\frac{10\beta_0-2}{5\beta_0}\nu+\frac{2\alpha_0}{5\beta_0}\mu$ 
of the luminosity evolution $\Delta \log L=\tau\log(1+z)$ (see
Eq. \ref{eq_DeltaL}) derived from the measured variation in the FP ZP
caused by the size and velocity dispersion evolution, coding the
different cases listed in Table \ref{tab_massfits}. For example,
``case 1+9 $M_{dyn}$'' uses the value of $\nu$ derived for the
redshift evolution of $R_c$ constructed using the $R_e-M_{dyn}$
relation with HST data, without both selection weighting and
progenitor bias correction (case 1 of Table \ref{tab_massfits}), and
the value of $\mu$ derived from the redshift evolution of $\sigma$
inferred in turn from the $\sigma-M_{dyn}$ relation, without both
selection weighting and progenitor bias correction (case 9 of Table
\ref{tab_massfits}), 
obtaining $\Delta \tau ={\bf +0.07}$. This implies that
the luminosity evolution inferred from the ZP evolution of the EDisCS
clusters without selection weighting ($L\sim (1+z)^{1.61}$, see Table
\ref{tab_FPslope}) becomes $L\sim (1+z)^{1.68}$.

To summarize, {none of the values of $\Delta \tau$ listed in Table
\ref{tab_lumev} differ statistically from zero.  However, without
taking into account the progenitor bias (rows two to five of Table
\ref{tab_lumev}), the values of $\nu$ and $\mu$ are much larger than
inferred by the merger scenario of \citet{Hopkins09} and when used
in Eq. \ref{eq_DeltaLz} 
change the predicted luminosity evolution samewhat. The same applies 
 by taking into account the progenitor bias (rows six to nine of Table 
\ref{tab_lumev}).

\subsection{Luminosity evolution: the direct fit}
\label{sec_trueLum}

To close the loop, in Fig. \ref{fig_LMass} we directly considered the
relation between total luminosity $L_B$ and dynamical mass as a
function of redshift. In general, the power law $L=L_C(M/M_C)^{0.75}$
provides a reasonable fit to the data.  Without selection weighting, 
we derived $L_C$ as a function of redshift as shown in Fig.
\ref{fig_meanLc} for dynamical and stellar masses.  Fitting the
power-law relation $L_C=L_C(0)(1+z)^\tau$ to $z>0.4$ data points, separately
for cluster and field galaxies, we derived the results listed in Table
\ref{tab_massfits}.  The zero-redshift extrapolations compare well to
the local values derived by considering the sample of \citet{Faber89}
($L_C=2.2\times 10^{10} L_\odot$ and $M/L_B=8.9 M_\odot/L_\odot$). As
for the $\sigma_c-z$ relation, we inferred a shallower luminosity evolution
when measuring $L_c$ at constant $M_*$. The luminosity evolution with
redshift is steeper for field galaxies.

Given the large errors in the luminosity fits, the derived luminosity
evolution agrees with the ones derived from the FP analysis.  At face
value, the FP ZPs without size and velocity dispersion evolution
corrections slightly overestimate the luminosity evolution at constant
stellar mass and underestimate that at constant dynamical mass. This
corroborates the conclusion that the corrections $\Delta \tau$ for
size and velocity dispersion evolution must be small, as one finds 
when the progenitor bias is taken into account.

\subsection{Ages}
\label{sec_ages}

As a final step, we translated the observed evolution in the FP zero
points, into an age estimate. We examined three cases that define the
realistic range of possible luminosity evolutions: 
(1) minimal luminosity
evolution, using $\Delta \tau={\bf -0.38}$ ($M_*$, {\bf case 2+10, 
of Table 7 and 9}) and $\phi=0$;
(2) $\Delta \tau=0$ and $\phi=0$,
where the small size and velocity dispersion correction compensates
the progenitor bias of \citet{vanDokkum01b}; (c) $\Delta \log L = \phi
z$, where the size and velocity dispersion correction is zero and we
take into account the progenitor bias of \citet{vanDokkum01b}. We
convert the mean $\Delta \log M/L_B$ for cluster and field galaxies
measured in the mass bins of Fig.  \ref{fig_clusmassdML} into an age,
by considering the various options for size evolution discussed above.
We use the solar-metallicity (motivated by the analysis of the
averaged line indices discussed below), Salpeter IMF SSP models of
\citet{Maraston05} at the appropriate mean redshift of the bin.  Ages
older than the age of the universe at that redshift are set to the age
of the universe. Figure \ref{fig_Ages} shows the results, Table
\ref{tab_ages} gives the average values for $M_{dyn}<10^{11}M_\odot$
and $M_{dyn}>10^{11}M_\odot$. 

Cluster galaxies more massive than
$10^{11} M_\odot$ are 6 to 8 Gyr old, with formation redshifts higher
than 1.5, while galaxies of lower masses are some 3-4 Gyr
younger. This parallels the findings of \citet{Sanchez09}, where the
analysis of spectral indices of EDisCs cluster galaxies with velocity
dispersion larger than 175 km/s assigns them formation redshifts
$>1.4$. Galaxies with lower velocity dispersions have instead younger
ages, compatible with continuous low levels of star
formation. Alternatively, the low-mass, spectroscopic early-type
cluster sample is building up progressively with the acquisition of
new and young objects, as discussed in \citet{Delucia04, Delucia07}
and \citet{Rudnick09}.
The result also agrees with the analysis of the scatter in the
color-magnitude relation of EDisC clusters of \citet{Jaffe10}.

Field galaxies are slightly younger than cluster galaxies at the same
redshift and mass.  
Taking into account the size and velocity dispersion evolution considered above
in the case $M_*$, 2+10 of Table 7 and 9
 pushes all formation ages upwards by up to 2 Gyr. Taking into account the
progenitor bias of \citet{vanDokkum01b} reduces the ages by 1 to 2
Gyr. Table \ref{tab_ages} lists mean ages for the HST sample of
galaxies with morphologies $T<0$. The differences between low and high
mass galaxies are smaller.

We next correlated the FP ages of Fig.
\ref{fig_Ages} with those derived from the analysis of the spectral
indices. As performed in \citet{Sanchez09}, we averaged the spectra of
the galaxies appearing in each mass bin shown in Fig.
\ref{fig_clusmassdML}, measured the Fe4383, HdA, and CN2 indices, and
recovered the ages and metallicities of SSP models that reproduce
their values best. The derived metallicity averaged over the sample is
solar, which justifies the choice above. We also estimated
luminosity-weighted and mass-weighted ages directly by fitting the
spectra with a library of model spectra. Within the large errors,
there is overall agreement with the ages derived from the indices. The
optimal match is achieved when considering the minimal evolution ages.
As a final check, we derived the rest-frame U-B and B-V colors
corresponding to a SSP of solar metallicity and age derived as above
and compared them to the measured averaged
colors. The agreement was fair, but either the
colors predicted using the FP ages are too red or the spectral ages
appear to be too high. The discrepancy is exacerbated when size evolution is
taken into account. This could simply reflect the known difficulties
for stellar population synthesis models in reproducing the colors of
real galaxies \citep{Maraston09}.

\begin{table*}[h!]
\caption{The ages derived from the evolution of the FP ZP, averaged 
for $M_{dyn}<10^{11}M_\odot$ and $M_{dyn}>10^{11}M_\odot$. The variations 
for the case of maximal evolution and the progenitor bias 
of \citet{vanDokkum01b} are also given.
\label{tab_ages}} 
\begin{tabular}{lllllccc}
\hline
        &     &     &       &      & \multicolumn{3}{c}{Age (Gyr)}\\
Type    & HST & VLT & Morph & Mass & $z<0.5$ & $0.5<z<0.7$ & $z>0.7$\\
\hline
Cluster & Yes & Yes & 10    & $<$  & $4.7^{\bf +2.3}_{-1.2}$  & $2.4^{ +1.1}_{0.4}$  & $1.8^{+0.8}_{-0.3}$\\
Cluster & Yes & Yes & 10    & $>$  & $8.6^{+0}_{-0.3}$       & $5.7^{+2.0}_{-1.7}$    & $6.3^{+0.5}_{-0.3}$\\ 
Cluster & Yes & No  & 0     & $<$  & $7.2^{\bf +0.3}_{-0.1}$  & $3.2^{ +1.6}_{-0.7}$  & $1.8^{+0.7}_{-0.3}$\\  
Cluster & Yes & No  & 0     & $>$  & $8.5^{+0.2}_{-1.0}$     &  $4.4^{ +2.3}_{-1.1}$ & $4.6^{+2.1}_{-1.3}$\\
Field   & Yes & Yes & 10    & $<$  & $3.9^{\bf +2.0}_{-0.8}$  & $1.7^{ +0.64}_{-0.3}$  & $1.4^{+0.5}_{-0.2}$ \\
Field   & Yes & Yes & 10    & $>$  & $6.7^{\bf +1.7}_{2.0}$   & $3.8^{ +2.3}_{-0.9}$  & $3.6^{+1.4}_{-1.0}$\\
Field   & Yes & No  & 0     & $<$  & $3.4^{+2.8}_{-0.7}$     & $2.2^{+2.3}_{-0.4}$     & $1.4^{+1.1}_{-0.3}$\\  
Field   & Yes & No  & 0     & $>$  & $3.3^{\bf +1.3}_{-0.4}$  &$2.1^{ +0.9}_{-0.3}$   & $2.4^{+1.6}_{-0.6}$\\  
\hline
\end{tabular}
\end{table*}

 \section{Conclusions}
\label{sec_conclusions} 

We have examined the FP of EDisCS spectroscopic
early-type galaxies, in both the cluster and the field. Combining
structural parameters from HST and VLT images and velocity dispersions
from VLT spectra, we have compiled a catalogue of 154 cluster and 68
field objects in the redshift range 0.2-0.9. For the first time, we
have explored the FP of galaxy clusters of medium-to-low velocity
dispersion in the redshift range 0.4-0.9. At face-value, on average,
the evolution of the zero point follows the predictions of simple
stellar population models with high ($\approx 2$) formation redshift
for all clusters, independent of their velocity dispersion, with a
slight increase (from 15\% to 18\%) in the scatter in mass-to-light
ratios for clusters with low ($\sigma_{clus}<600$ km/s) velocity
dispersions.  The FP zero point of field galaxies follows similar
tracks up to redshift $\approx 0.5$, but implies brighter
luminosities, or lower formation redshifts at higher redshifts.

We have determined dynamical and stellar masses for our galaxies. The
ratio $M_{dyn}/M_*$ is $\approx 0.9$ with a scatter of a factor 2 and
a tendency to decrease with redshift. We investigated the FP residuals
as a function of galaxy mass. At high redshifts ($z>0.7$ for cluster
galaxies, slightly below for field galaxies), galaxies with mass lower
than $\approx 10^{11}M_\odot$ have lower mass-to-light ratios than a
passive evolution of the ZP predicts. This implies that there is a
rotation in the FP: we confirm that for cluster
galaxies the velocity dispersion coefficient $\alpha$ is compatible
with the local value up to a redshift $z=0.7$ and decreases to
$\alpha\approx 0.7\pm0.4$ at higher redshifts, 
{but this detection is of low statistical significance}. 

We have investigated the size and velocity dispersion evolution of our
sample.  At a given mass, galaxy sizes decrease and velocity
dispersions increase at increasing redshift. {We fitted the relations
 $R_e\approx
(1+z)^{-1.0\pm 0.3}$,} and $\sigma\approx (1+z)^{0.59\pm0.1}$ and $\sigma\approx
(1+z)^{0.34\pm0.14}$ at a constant dynamical or stellar mass of $2\times
10^{11}M_\odot$, respectively, for both cluster and field galaxies.
However, after taking into account the progenitor bias affecting our
sample (large galaxies that joined the local early-type class only
recently will progressively disappear in higher redshift samples), the
effective size and velocity dispersion evolution reduced
substantially (to $R_e\propto (1+z)^{-0.5\pm0.2}$ and $\sigma\propto
(1+z)^{0.41\pm0.08}$ for dynamical masses and $R_e\propto
(1+z)^{-0.68\pm0.4}$ and $\sigma\propto (1+z)^{0.19\pm0.10}$ for
stellar masses).

We computed the luminosity evolution predicted by the ZP variation with
redshift of the FP when the size and velocity dispersion evolution are
taken into account. 
The corrections computed at constant dynamical masses without a
progenitor bias correction almost cancel out; at constant stellar 
masses they reduce the slope of the $(1+z)$ dependence of
luminosity by $-0.38$ units (case 2+10 of Table \ref{tab_massfits}).
 Fitting directly the luminosity-mass
relation, we derived a luminosity evolution that agrees with the one
derived from the FP analysis and does not allow for large size and
velocity dispersion corrections, as indeed it is always the case.

Using simple stellar population models, we translated the variations in
the FP ZP into formation ages as a function of redshift and galaxy
mass. Massive ($M>10^{11}M_\odot$) cluster galaxies are old, with
formation redshifts $z_f>1.5$. In contrast, lower mass galaxies are
just 2 to 3 Gyr old.  This agrees with the EDisCS results presented in
\citet{Delucia04, Delucia07}, \citet{Poggianti06}, \citet{Sanchez09},
and \citet{Rudnick09}, who argue from different points of view that
the lower luminosity, lower mass population of early-type galaxies
comes in place only at later stages in clusters. Field galaxies at all
masses are somewhat younger (by $\approx 1$ Gyr) than the cluster ones
with similar masses and redshifts. In general, the FP ages agree reasonably 
well with those derived from spectral indices. 

To conclude, our analysis of the FP, size, and velocity dispersion
evolution of EDisCS galaxies points towards a picture where a large
fraction of the population became passive only fairly recently. The
high redshift passive galaxies are a biased subset of all the present
passive galaxies. At any probed redshift, from 0 to 1, passive
galaxies are an inhomogeneous population in terms of their  formation
paths, and as redshift increases, a subset of the population
leaves the sample, with less massive galaxies dropping out of the
sample more rapidly with redshift than the more massive ones, and with a 
somewhat
accelerated pace in the field. Only when these effects are taken into
account may coherent estimates of the luminosity evolution of early-type
galaxies from the colors, indices, and the FP zero point be
derived.

\begin{acknowledgements} This work was supported by the
  Sonderforschungsbereich 375 of the German Research Foundation.  The
  Dark Cosmology Centre is funded by the Danish National Research
  Foundation. GDL acknowledges financial support from the European
  Research Council under the European Community's Seventh Framework
  Programme (FP7/2007-2013)/ERC grant agreement n. 202781. {The anonymous
 referee report helped us improve the presentation of the results.}
\end{acknowledgements}

\begin{table}[h!]
\caption{
The change in slope $\Delta \tau=\frac{10\beta_0-2}{5\beta_0}\nu+\frac{2\alpha_0}{5\beta_0}\mu$ 
of the luminosity evolution $\Delta \log L=\tau\log(1+z)$ 
(see Eq. \ref{eq_DeltaL}) derived from the measured variation in the FP ZP
caused by the size and velocity dispersion evolution for the different
cases listed in Table \ref{tab_massfits}.
\label{tab_lumev}} 
\begin{tabular}{cccc}
\hline
Case                  & $\nu$ & $\mu$ & $\Delta \tau$\\
\hline
1+9 $M_{dyn}$          & -1.0  & +0.59 &{ 0.07}$\pm$  { 0.28}\\
1+9 $M_*$             & -1.0  & +0.34 &{ -0.29}$\pm$ { 0.52}\\
2+10 $M_{dyn}$         & -1.3  & +0.68 &{  -0.03}$\pm${ 0.40} \\
2+10 $M_*$            & -1.2  & +0.39 &{ -0.38}$\pm$ { 0.66} \\
5+11 $M_{dyn}$         & -0.46 & +0.41 &{  +0.23}$\pm${ 0.20}\\ 
5+11 $M_*$            & -0.68 & +0.19 &{ -0.26}$\pm$ { 0.36}\\
6+12 $M_{dyn}$         & -0.67 & +0.49 &{  +0.18}$\pm${ 0.28}\\
6+12 $M_*$            & -0.84 & +0.27 &{ -0.27}$\pm$ { 0.39}\\
\hline
\end{tabular}
\end{table}

\begin{figure*}
\begin{tabular}{cc}
\vbox{\psfig{file=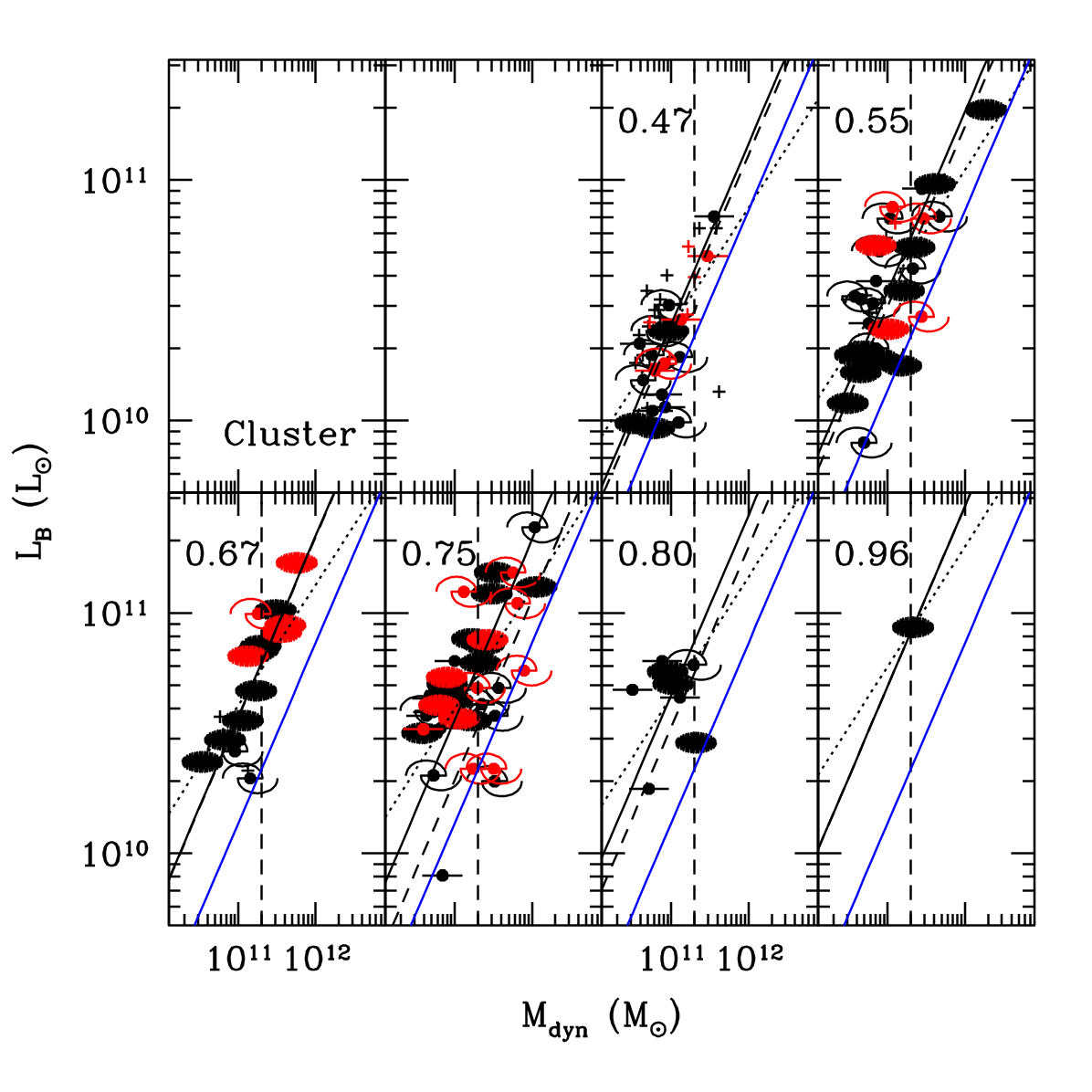,angle=0,width=8.5cm}}& 
\vbox{\psfig{file=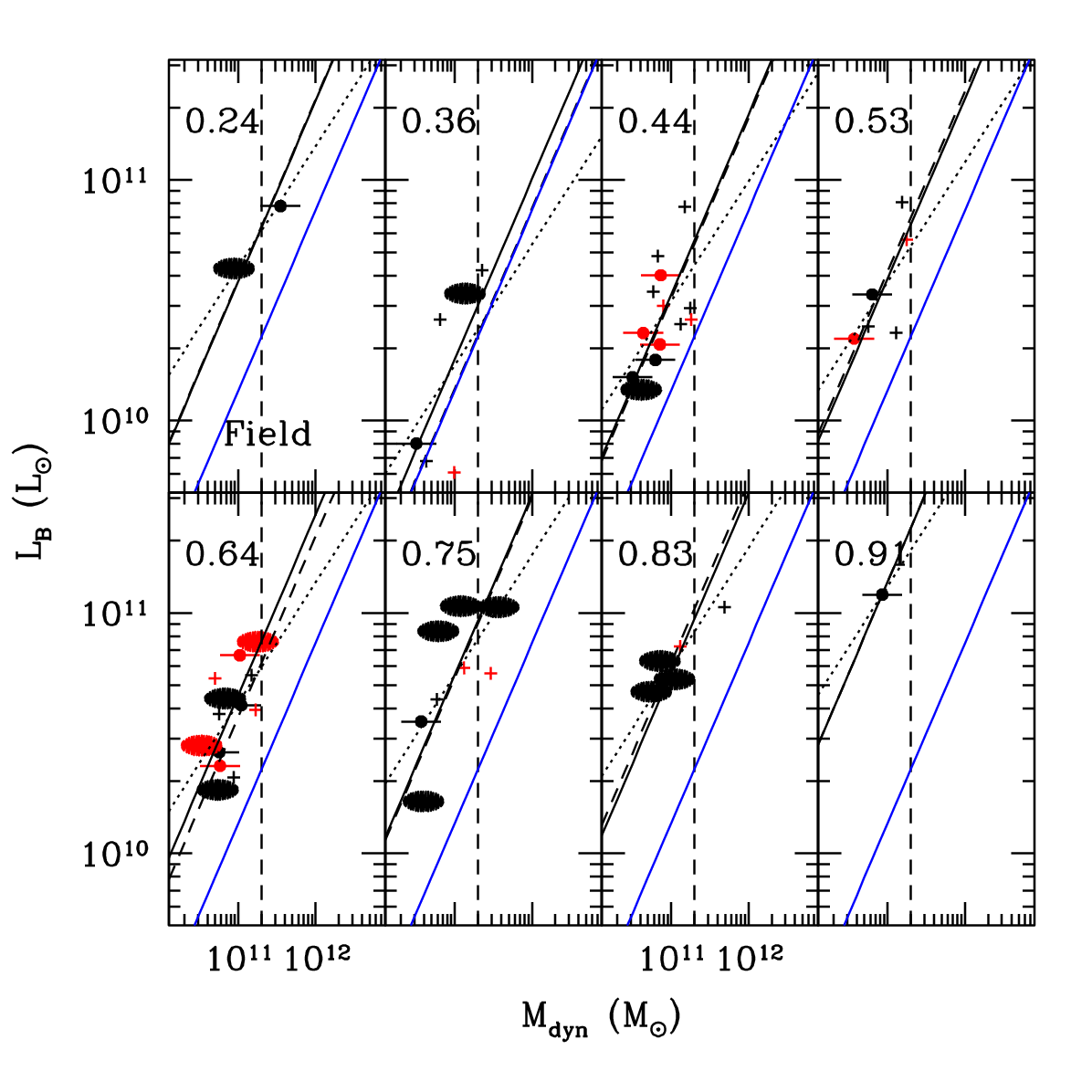,angle=0,width=8.5cm}}\\
\end{tabular}
\caption{The evolution of the luminosity-mass relation with redshift
  for cluster (left) and field (right) galaxies. Colors and symbols as
  in Fig.  \ref{fig_fieldML}. The numbers give the average redshift in
  each bin. The full lines show the best-fit relation
  $L_B=L_c(M/M_C)^{0.75}$, abd the dotted lines $L=L_c(M/M_C)^{0.5}$, both
  relations with uniform galaxy weighting. The dashed lines show the 0.75
  power law with selection weighting.  The blue lines show
the reference line at zero redshifts. The vertical line indicates the
  $2\times 10^{11} M_\odot$ mass.
 \label{fig_LMass}}
\end{figure*} 

\begin{figure*}
\begin{tabular}{cc}
\vbox{\psfig{file=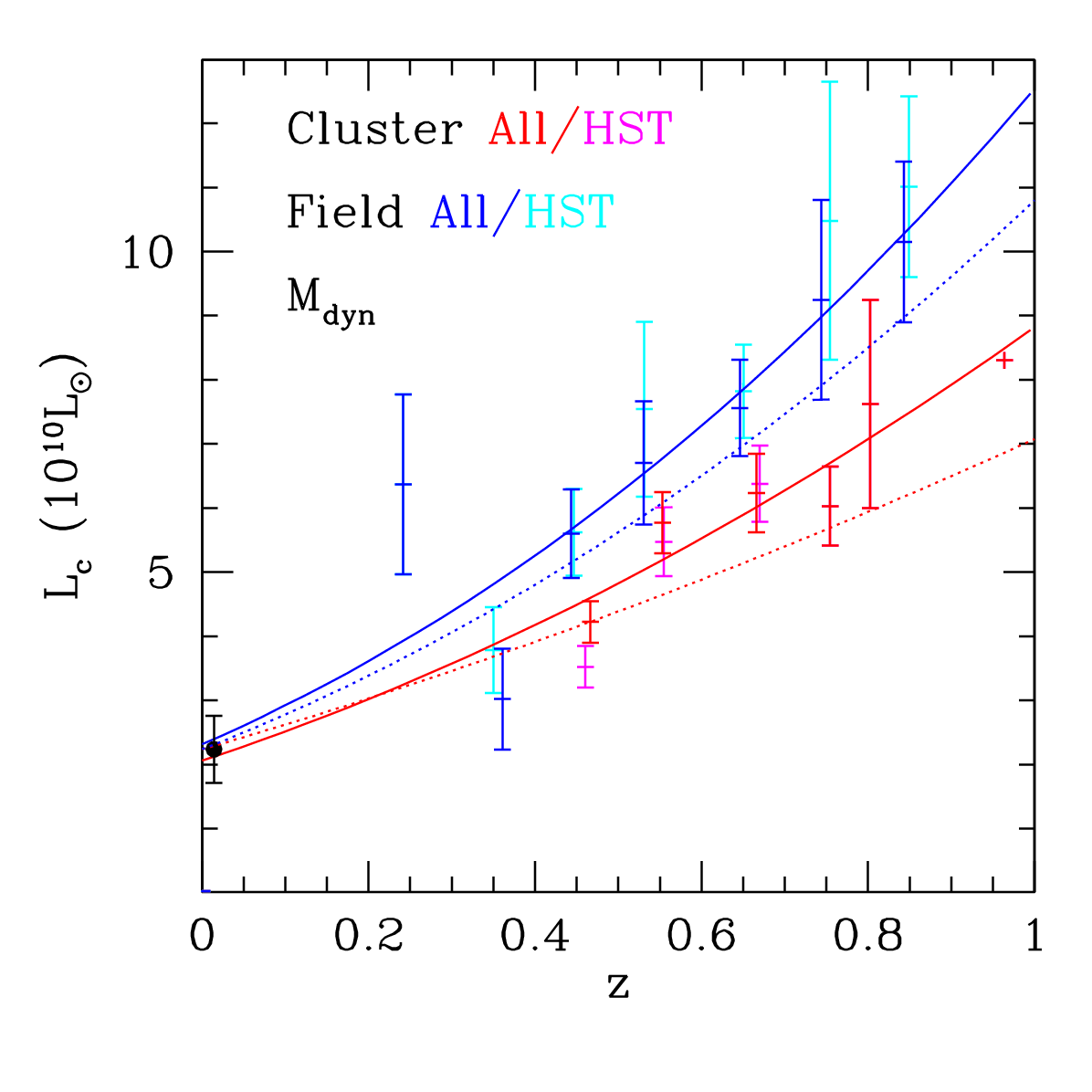,angle=0,width=8cm}}& 
\vbox{\psfig{file=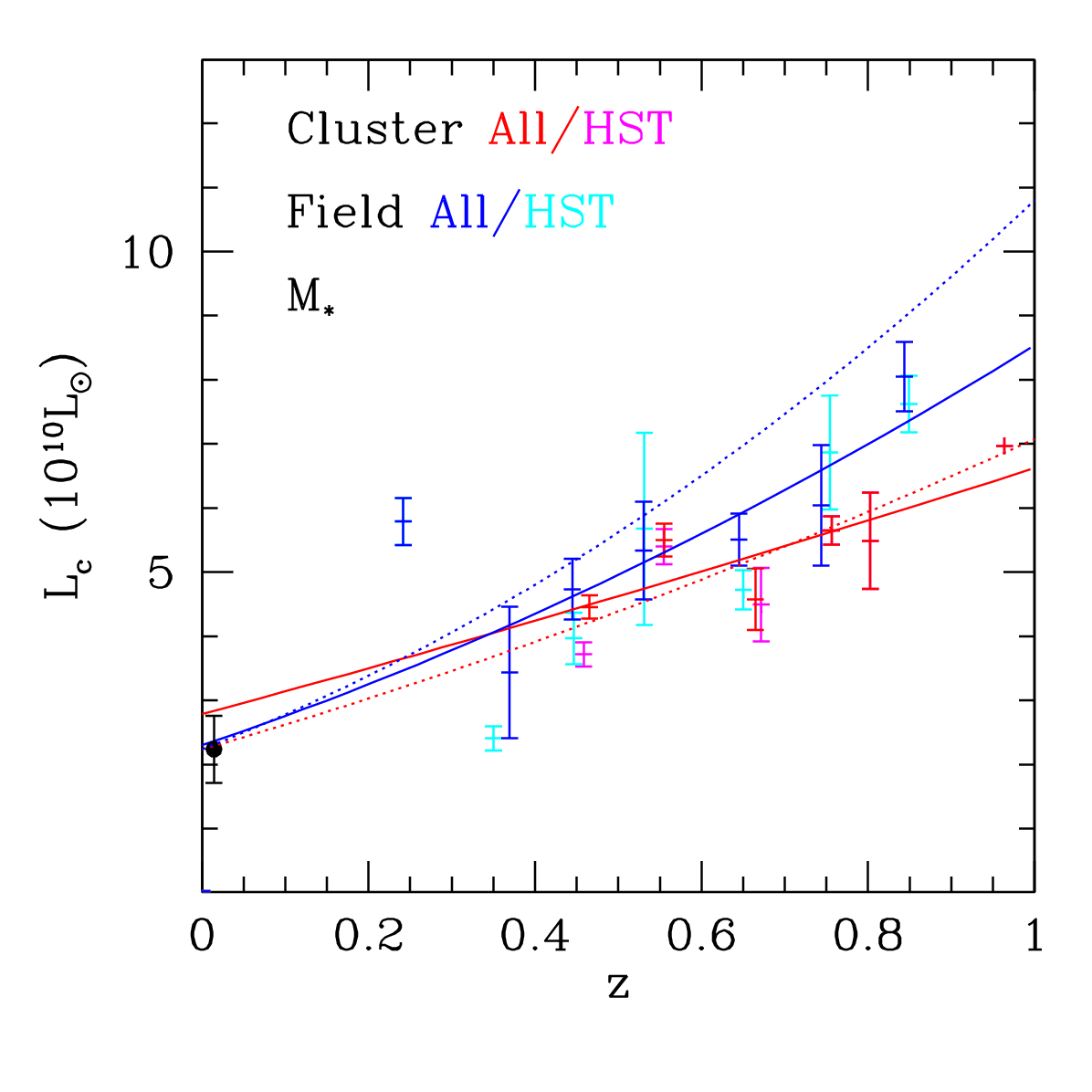,angle=0,width=8cm}}\\
\end{tabular}
\caption{The luminosity evolution {with redshift} of EDisCS galaxies. Left: $L_c$ as a
  function of redshift derived using $M_{dyn}$. Right: $L_c$ as a
  function of redshift derived using $M_{*}$. The full lines show the
  best-fit function $L_c=L_c^0\times (1+z)^\tau$, red for cluster and
  blue for field galaxies, derived for the full sample without selection
  weighting.  The dotted lines show the corresponding redshift
  dependences derived from the FP analysis with $\Delta \tau=0$ (Table
  \ref{tab_FPslope}).  The black dot shows the local value derived
  from \citet{Faber89}.
\label{fig_meanLc}}
\end{figure*} 

\begin{figure*}
\psfig{file=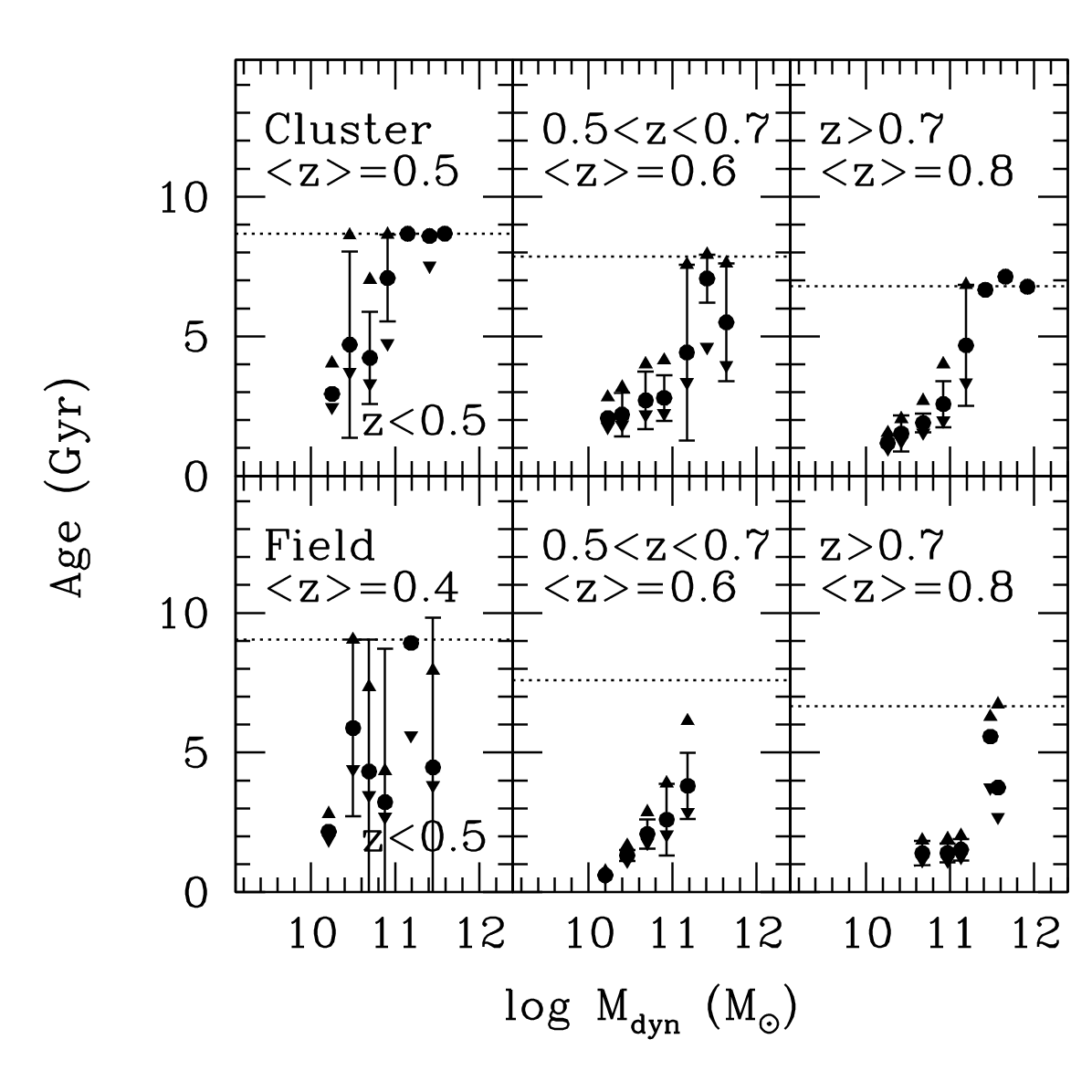,angle=0,width=15cm}
\caption{The ages of cluster (top) and field (bottom) galaxies at low
  (left, $z<0.5$ left), medium ($0.5<z<0.7$, middle), and high ($z>0.7$, right) 
redshifts as a function of dynamical mass. 
The circles show the ages as derived from the bare 
FP zero point evolution. The triangles pointing upwards take into account 
size evolution at constant $M_*$, case 2+10 (see Table \ref{tab_lumev}). 
The triangles pointing downwards take into account the progenitor bias 
of \citet{vanDokkum01b}. 
  The median redshifts are given and the corresponding
  ages of the universe are shown by the dotted lines.
  \label{fig_Ages}}
\end{figure*}


\bibliographystyle{aa}


\appendix
\section{Circularized half-luminosity radii}
\label{app_ReCir}

GIM2D delivers bulge $a_e$ and disk $a_h$ scale lengths along the major
axis, bulge apparent flattening $(b/a)_B$ and disk inclination angles
$i$ (corresponding to an apparent flattening $(b/a)_D=1-\cos i$), and
bulge-to-total ratios $B/T$. When fitting Sersic profiles, GIM2D
delivers the $n$ Sersic index, the major axis $a_e^{Ser}$ and the
flattening $(b/a)^{Ser}$.  We compute the circularized half-luminosity
radius $R_e$ of the resulting galaxy model as follows.  We determine
the flux inside a circular aperture of radius R (the so-called curve
of growth) of a model of apparent flattening $b/a$ and surface density
distribution constant on ellipses $f(x,y)=f(\sqrt{x^2/a^2+y^2/b^2})$
as
\begin{equation}
\label{eq_coggen}
F(R) = \int_0^{2\pi}\int_0^R f(\sqrt{(R'\cos \phi)^2/a^2+(R'\sin \phi)^2/b^2}R'dR'd\phi.
\end{equation} 
Using  $F_c(R)=2\pi\int_0^R rf(r)dr$ we get
\begin{equation}
F(R) =  4b^2\int_0^{\pi/2}\frac{F_c(r/b\sqrt{1-(1-b^2/a^2)\cos^2\phi})}
{1-(1-b^2/a^2)\cos^2 \phi}d\phi. \\
\label{eq_angint}
\end{equation} 
We perform the angular integration numerically, using
$F_c^{deVauc}(z)=1-(1+\sum_{i=1}^7z^i/i!)e^{-z}$, with
$z=7.67(r/R_{eB})^{1/4}$, and $F_c^{exp}(x)=1-(1+x)e^{-x}$, with
$x=R/h$ for the normalized de Vaucouleurs and exponential density laws
respectively. For a Sersic profile of given n, we use $F_c^n=P(2n,X)$,
where $P$ is the incomplete $\Gamma$ function and
$X=k(r/R_{eSer})^{1/n}$ and $k=1.9992n-0.3271$ \citep{Simard02}. 
We determine $R_e$ by solving the equation
\begin{equation}
\label{eq_half}
B/T\times F^{deVauc}(R_e)+(1-B/T)F^{exp}(R_e)=0.5
\end{equation}
for the bulge plus disk models, and
\begin{equation}
\label{eq_halfser}
F^{n}(R_e)=0.5
\end{equation}
for the Sersic fits numerically. In general, the resulting $R_e$ agree
within 1\% with the half-luminosity radii derived by measuring the
curves of growth directly from (ACS HST like) images generated by
GIM2D with the fit parameters and no PSF convolution, but the
image-based method overestimates $R_e$ by up to 10 \% when it is
smaller than 4 pixels (0.2 arcsec). 

{Figure \ref{fig_Sersic} illustrates that a more accurate approximation 
of the
  circularized radius $R_e(Ser)$ of Sersic profiles, more accurate 
  than 2\%, is obtained by taking the simple mean
  $R_{ave}=(a_e+b_e)/2$ of the major and minor axis scale lengths
  $a_e$ and $b_e$ instead of the harmonic mean
  $R_{har}=\sqrt{a_e\times b_e}$. This is surprising only at a first sight,
  since $R_{har}$ goes to zero as the flattening increases,
  while $R_{ave}$ does not. Therefore $R_{ave}$ is bound to more closely
  approximate the half-luminosity radius derived from circular
  curves of growth at high ellipticities.  On the other hand, the
  effective surface brightness within the ellipse of semi-major and
  minor axis $a_e$ and $b_e$ is constant whatever the flattening,
  while this is not true for the surface brightness within the circle
  of radius $R_e(Ser)$. Since in this exercise the total luminosity
  $L$ is kept constant, we have $\log R_e(Ser)/R_{har}=0.2(\langle
  SB_e\rangle-\langle SB_e^{har}\rangle$, with $\langle
  SB_e\rangle=-2.5\log \frac{L}{2\pi R_e(Ser)^2}$ and $\langle
  SB_e^{har}\rangle=-2.5\log \frac{L}{2\pi R_{har}^2}$. This is almost
  orthogonal to the FP (see Eq. \ref{eq_FP}), making the choice of
  method unimportant, as far as not too many disks seen edge-one
  (i.e. of very high flattening) are present in the sample (see
  Fig. \ref{fig_checkRe} and discussion in
  Sect. \ref{sec_photometry}).  }

\begin{figure}
\psfig{file=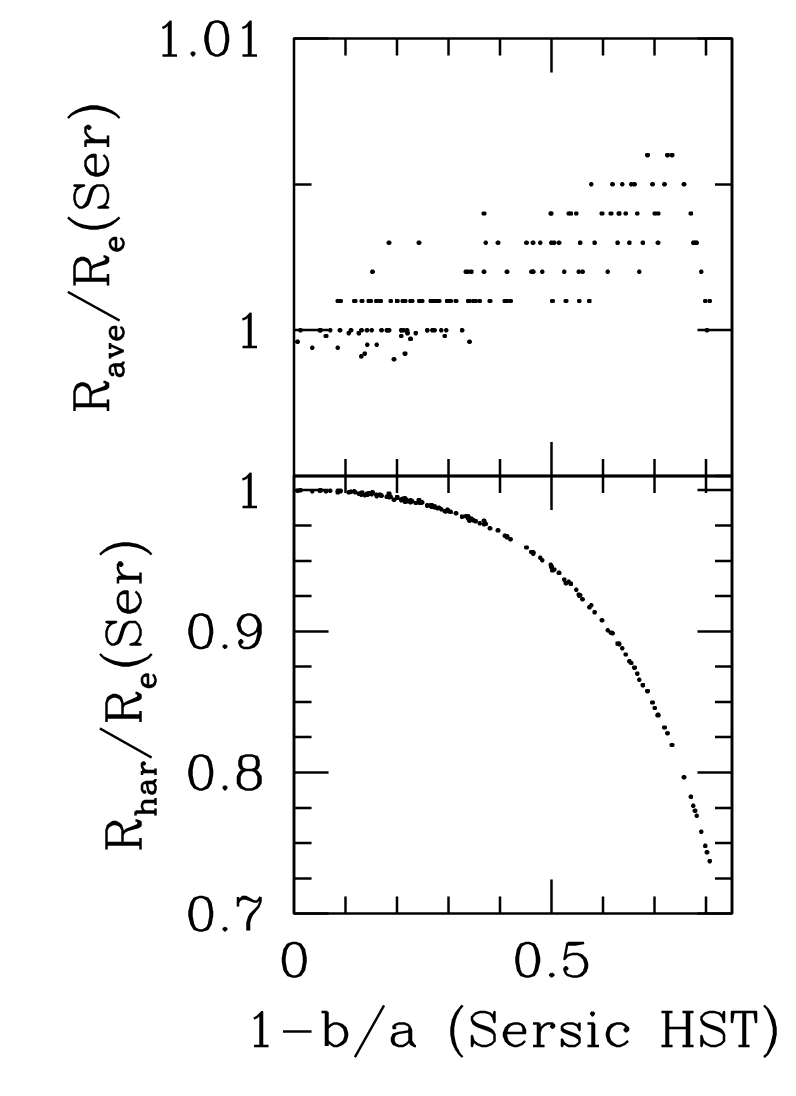,angle=0,width=7cm}
\caption{The circularized half-luminosity radius $R_e(Ser)$  of
    the sample of EDisCS galaxies with HST photometry and velocity
    dispersions computed according to Eqs. \ref{eq_angint} and
  \ref{eq_halfser} compared to the simple mean $R_{ave}=0.5(a_e+b_e)$
  (top) and harmonic mean $R_{har}=\sqrt{a_e\times b_e}$ (bottom) as a
  function of the ellipticity $1-b_e/a_e$. The simple mean
  approximates $R_e(Ser)$ better.  \label{fig_Sersic}}
\end{figure} 
\vfill*
\end{document}